\documentclass[a4paper,12pt]{report}

\usepackage[T1]{fontenc}
\usepackage[english]{babel}

\usepackage{amsmath,amssymb,amsfonts}
\usepackage{mathtools}
\usepackage{mathrsfs}
\usepackage{bbold}
\usepackage{slashed}
\usepackage[f]{esvect}

\usepackage{graphicx}
\usepackage[dvipsnames]{xcolor}
\usepackage{caption}
\usepackage{subcaption}
\usepackage{pdfpages}
\usepackage{pifont}
\usepackage{xcolor}
\usepackage{tikz}
\usetikzlibrary{arrows.meta,positioning}
\usepackage{quantikz}
\usepackage{subcaption}
\usepackage{xcolor}

\usepackage{multicol}
\usepackage{csquotes}
\usepackage{url}
\usepackage{lipsum}
\usepackage{cite}

\usepackage[colorlinks=true,citecolor=blue, linkcolor=black, urlcolor=blue]{hyperref}
\usepackage{braket}

\usepackage{tikz}
\usetikzlibrary{arrows.meta,positioning,calc,fit,shapes.geometric}
\usepackage{graphicx} 

\usepackage{subcaption}
\usepackage{blochsphere}
\tikzset{
  box/.style={draw, rounded corners, align=center, minimum height=7mm, minimum width=14mm},
  op/.style={box, fill=gray!10},
  stage/.style={box, fill=gray!6},
  arrow/.style={-Latex, line width=0.6pt},
  boxc/.style={draw,rounded corners, align=center,
  inner sep=6pt, fill=blue!10},
  boxcc/.style={draw,rounded corners, align=center,
  inner sep=6pt, fill=red!20},
  dashedarrow/.style={-Latex, line width=0.6pt, dashed},
  plus/.style={circle, draw, inner sep=0pt, minimum size=4mm, font=\bfseries},
}

\usepackage[a4paper,top=3cm,bottom=3cm,left=3cm,right=3cm]{geometry}
\usepackage{bibentry}
\nobibliography*

\usepackage{fancyhdr}

\fancypagestyle{plain}{%
  \fancyhf{}
  \fancyhead[LE,RO]{\thepage}
  \fancyhead[LO,RE]{\leftmark}
  
}

\begin{document}

\pagestyle{empty}

\begin{figure}[!t]
  \centering
  \begin{subfigure}{.35\textwidth}
    \centering
    \raisebox{0.02cm}{\includegraphics[height=1.9cm]{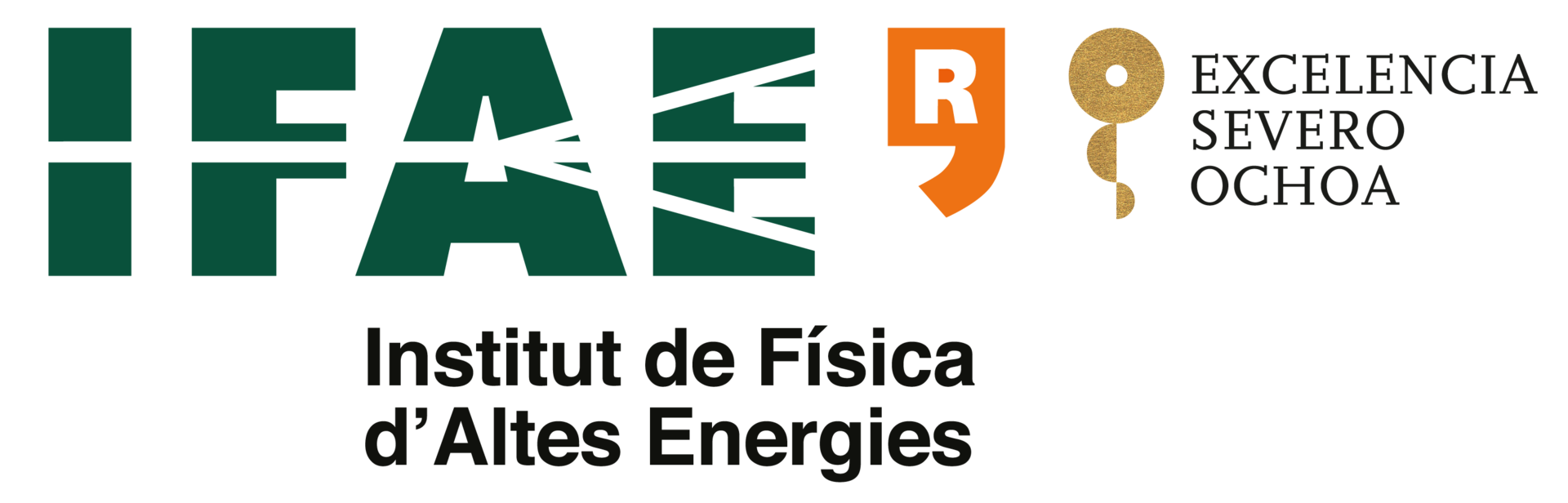}}
  \end{subfigure}%
  \hfill
  \begin{subfigure}{.3\textwidth}
    \centering
    \includegraphics[height=1.8cm]{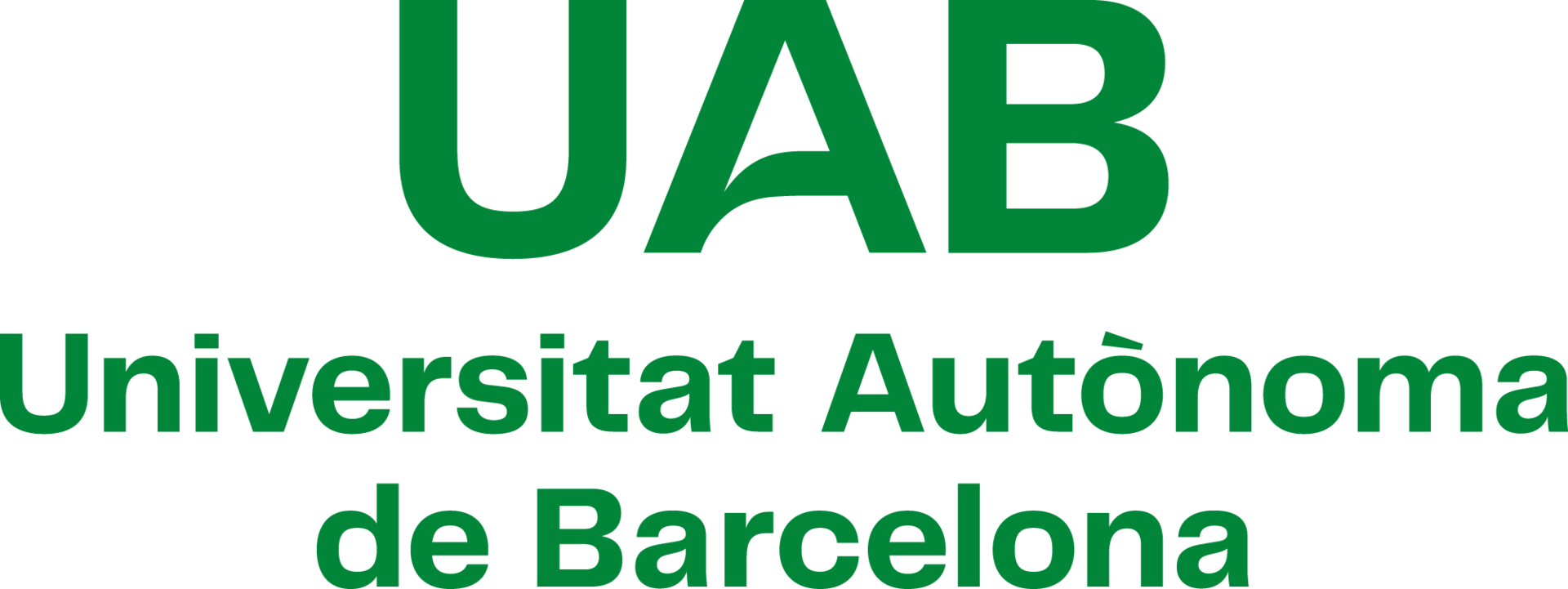}
  \end{subfigure}
  \hfill
  \begin{subfigure}{.3\textwidth}
    \centering
    \raisebox{0.85cm}{\includegraphics[height=0.9cm]{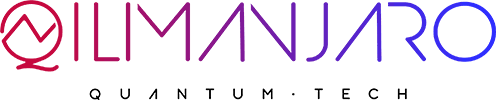}}
  \end{subfigure}%
  \end{figure}
\begin{center}

\vspace*{1.cm}
\begin{center}
  \rule{\textwidth}{0.4pt}
\end{center}

{\Large MACHINE LEARNING DEVELOPMENT FOR QUANTUM COMPUTING AND NEUTRINO PHYSICS\\}

\vspace{-0.5cm}
\begin{center}
  \rule{\textwidth}{0.4pt}
\end{center}

\vspace*{1cm}

{\large
A thesis presented for the degree of\\
Doctor of Physics by:
}

\vspace*{1cm}

{\Large
\textbf{Annalisa De Lorenzis}\\
}
\vspace*{3.5cm}
\begin{flushright}
\begin{minipage}{0.65\textwidth}
{\large
\begin{tabular}{@{}ll}
\textbf{Supervisors:} & M. Pilar Casado Lechuga\\
                      & Thorsten Lux\\
                      & Arnau Riera Graells\\[0.6cm]
\textbf{\hspace{1.45cm}Tutor:}       & M. Pilar Casado Lechuga
\end{tabular}
}
\end{minipage}
\end{flushright}
\vspace*{1.5cm}
\end{center}
{\large May 29, 2026}
\clearpage
\pagestyle{fancy}
\pagenumbering{roman}
\setcounter{page}{1}
\thispagestyle{empty}
\vspace*{2.2cm}
\begin{flushright}
\textit{
To my family,  
for their love and support.\\
To the memory of my grandparents and uncles.\\
To Luigi, for bringing light to the darkest days.\\
To my past self,  
for not giving up.
}
\end{flushright}
\vspace*{\fill}
\clearpage
\chapter*{Abstract}
\addcontentsline{toc}{chapter}{Abstract}
This thesis investigates the application of machine-learning methods in the context of quantum computing and neutrino physics, with
particular emphasis on the construction of effective representations for
complex, high-dimensional data.

The first part of the work is devoted to Quantum Extreme Learning
Machines (QELMs), a hybrid quantum--classical framework in which
classical data are encoded into quantum states and processed through
fixed quantum dynamics, while learning is performed by a classical
readout layer. Within this framework, we analyze the role of encoding
strategies, feature-reduction methods, Hamiltonian structure, and
measurement, with particular focus on the relationship between quantum
dynamics, expressivity, entanglement, and classical simulability.

The second part of the thesis concerns the application of deep learning to the analysis of images produced by water Cherenkov detectors in neutrino physics. Convolutional architectures, including residual networks, are developed for the classification of complex events in realistic simulated datasets, showing that such models can effectively extract relevant information from detector data.

Taken together, these results highlight the potential of machine learning, in both its classical and quantum forms, as a powerful framework for the analysis of complex data in fundamental physics, while also outlining relevant challenges and directions for future research.
\newpage
\chapter*{List of publications}
\addcontentsline{toc}{chapter}{List of publications}
The main results of this thesis are presented in these two papers:

\begin{enumerate}
    \item A. De Lorenzis, M. P. Casado, M. P. Estarellas, N. Lo Gullo, T. Lux, F. Plastina, A. Riera, J. Settino, ``Harnessing Quantum Extreme Learning Machines for Image Classification,'' \textit{Physical Review Applied} \textbf{23}, 044024 $-$ April, 2025. The article was later selected for inclusion in \textit{Quantum Frontiers}, a curated collection by \textit{Physical Review Applied}.
    \item A. De Lorenzis, M. P. Casado, N. Lo Gullo, T. Lux, F. Plastina, A. Riera, ``Entanglement and Classical Simulability in Quantum Extreme Learning Machines,'' \textit{Physical Review Applied}, \textbf{25}, 054052, $-$ May, 2026.\\
\end{enumerate}

In addition, I am co-author of the following publications:

\begin{enumerate}
    \item Hyper-Kamiokande Collaboration, K. Abe \textit{et al.}, ``Sensitivity of the Hyper-Kamiokande experiment to neutrino oscillation parameters using acceleration neutrinos'', Eur.Phys.J.C 86 (2026) 2, 170.
    \item Hyper-Kamiokande Collaboration, K. Abe \textit{et al.}, ``The Hyper-Kamiokande experiment: input to the update of the European Strategy for Particle Physics'', arXiv: 2506.16641 [hep-ex], $-$ June, 2025.
    \item D. Attié \textit{et al.}, ``Analysis of test beam data taken with a prototype of TPC with resistive Micromegas for the T2K Near Detector upgrade'', Nucl.Instrum.Meth.A 1052, 168248 $-$ July, 2023.
    \item L. Ambrosi \textit{et al.}, ``Characterization of charge spreading and gain of encapsulated resistive Micromegas detectors for the upgrade of the T2K Near Detector Time Projection Chambers'', Nucl.Instrum.Meth.A 1056, 168534 $-$ November, 2023.
\end{enumerate}
\newpage

\tableofcontents
\pagestyle{plain}
\cleardoublepage 
\pagenumbering{arabic}
\setcounter{page}{1}
\chapter{Introduction}
In the last few decades, many areas of science, technology, and society
have undergone a profound and remarkably rapid digital transformation.
Whereas in the past datasets were often relatively small and could be
analyzed using simple, problem-specific tools, modern applications
routinely generate massive, heterogeneous, and high-dimensional streams
of data. Examples range from sensor networks, medical imaging, and
financial markets to social media, recommendation systems, and
large-scale scientific experiments. In this context, it has become
essential to develop efficient techniques capable of handling large
volumes of data and extracting from them reliable, interpretable, and
actionable information.

A particularly powerful class of such techniques is provided by
\emph{machine learning} \cite{bishop2006, Goodfellow-et-al-2016, geron2022}, a broad family of methods designed to learn
patterns and predictive rules directly from data. Rather than relying
exclusively on handcrafted features or explicit physical models,
machine-learning algorithms adapt their internal parameters to examples,
thereby discovering structures that may be difficult or impossible to
encode by hand. These methods now play a central role in a wide range of
applications, including image and speech recognition, natural language
processing, anomaly detection, and data-driven scientific discovery.

In parallel to these developments, the increasing complexity of physical systems and data representations raises fundamental computational challenges. The intrinsic difficulty of simulating quantum systems on classical computers was first emphasized by Richard Feynman \cite{feynman1982simulating, Nielsen_Chuang_2010, Montanaro_2016, Preskill:2018jim}, who suggested that quantum devices may provide a natural framework for modeling physical processes that are otherwise computationally intractable \cite{Arute48651, Huang:2021pei}.

Within this broader context, quantum machine learning (QML) \cite{Biamonte:2016ugo, Schuld:2018gao, Schuld:2018uel, book, Congarticle, Liu:2022bpb, Lloyd:2013lby, Schuld2014, Havlicek:2018nqz} has emerged as an active research area at the interface between quantum computing and machine learning, with the goal of exploiting quantum systems and algorithms for data analysis, representation learning, and classification tasks. Among the different QML approaches, \emph{Quantum Extreme Learning Machines} (QELMs) \cite{DeLorenzis_1, DeLorenzis_2, Sakurai:2022ala, Hayashi:2022xab, Hayashi:2025gwk, Sakurai_simpleH, QELMSuprano, Vetrano:2024vbh, Innocenti2023} provide a particularly attractive hybrid framework, since they combine quantum feature generation with a simple classical readout layer.

The present thesis is situated within this broader landscape and lies at the intersection of machine learning, quantum computing, and neutrino physics. On the one hand, it investigates QELMs for supervised classification tasks, with particular emphasis on the role of data encoding, feature reduction, quantum dynamics, and measurement in the construction of effective representations of classical data. On the other hand, it develops and applies deep-learning techniques to the analysis of neutrino detector images. Despite the apparent difference between these two research directions, they are united by a common objective: the construction of effective representations of complex, high-dimensional data. In both settings, the central challenge is to design models that transform raw inputs into feature spaces in which relevant patterns can be extracted in an efficient, robust, and physically meaningful way.

More specifically, the work presented in this thesis is organized around two complementary research directions.

The first research direction concerns QELMs, a class of hybrid quantum--classical models inspired by reservoir computing \cite{Sakurai:2022ala, AngelatosPhysRevX.11.041062, Mujal_2021, Domingo:2022fre, Martinez2021, Fujii2017, chenPhysRevApplied.14.024065} and by the extreme learning machine (ELM) paradigm \cite{HUANG2006489, articleHuang2011, wang2022review, markowska-kaczmar_extreme_2021, 10.1007/s10462-013-9405-z, Huang_extreme2}. In the framework studied here, the processing pipeline is decomposed into four stages: a classical preprocessing step that compresses the input data, a quantum encoding map that prepares an $N$-qubit state, a fixed unitary evolution generated by a many-body Hamiltonian, and a measurement stage whose outcomes define a classical feature vector processed by a shallow readout network. Unlike variational quantum models \cite{Peruzzo:2013bzg, Farhi:2014ych, Cerezo:2020jpv}, the quantum part of the architecture is not optimized through an iterative training loop: the encoding, the Hamiltonian, the evolution time, and the measurement prescription are fixed in advance, while all trainable parameters are confined to the final classical classifier. This makes QELMs a useful setting in which to study the representational role of quantum dynamics itself, disentangled from the optimization difficulties of deep parameterized circuits.

A central technical issue in this framework is the embedding of high-dimensional classical data into a small quantum register without discarding the information most relevant for classification. For this reason, a substantial part of the thesis is devoted to the interplay between classical compression and quantum encoding. We consider both linear and nonlinear feature-reduction strategies, in particular Principal Component Analysis and convolutional autoencoders, and we compare several encoding schemes, including simple angle, dense-angle, uniform Bloch-sphere, general, and amplitude encodings. We also analyse different interacting Hamiltonians for the quantum layer and investigate how the corresponding measurement statistics can be exploited by a shallow classical readout.

These ingredients are first studied in a systematic image-classification analysis on standard benchmarks such as MNIST \cite{mnist} and Fashion-MNIST \cite{fashionmnistnovelimagedataset}, and later revisited in a more controlled setting including CIFAR-10 \cite{cifar10}. The first study shows that the preprocessing stage is already decisive in the strongly compressed regime relevant to near-term quantum models: nonlinear latent representations produced by autoencoders consistently outperform PCA, and this advantage persists on the more challenging Fashion-MNIST dataset. The comparison between encoding strategies further shows that dense-angle and uniform Bloch-sphere encodings provide the most favourable balance between accuracy and implementability, since they exploit two Bloch-sphere parameters per qubit while remaining compatible with relatively shallow state-preparation circuits. Amplitude encoding is more compact in terms of qubit resources, but becomes competitive only when sufficiently many latent features are injected into the amplitudes, which in turn makes state preparation more demanding.

The same study also clarifies the role of the quantum evolution. Once the latent representation and the encoding are fixed, the classification accuracy increases rapidly with the number of qubits and reaches a plateau around $N \simeq 10$. In this regime, the performance becomes largely insensitive to the microscopic details of the interacting Hamiltonian, indicating that the relevant ingredient is not a finely tuned model of dynamics, but rather the ability of the evolution to redistribute the encoded information into a measurement space more suitable for the final classifier. With this architecture, QELMs reach accuracies comparable to, and in some cases better than, several previously proposed hybrid quantum--classical models for image classification \cite{DeLorenzis_1}.

The second QELM study focuses on the relation between classification performance, entanglement growth, and classical simulability. In this part of the thesis, the architecture is fixed more rigidly: the latent classical features are encoded through dense-angle encoding, the quantum layer is implemented by a nearest-neighbour XX Hamiltonian with periodic boundary conditions, and the only continuous hyperparameter varied is the evolution time $t$. This setup allows one to analyse, in a controlled way, how the dynamics of the quantum layer correlates with the quality of the generated features across MNIST, Fashion-MNIST, and CIFAR-10.

A central result is that the QELM accuracy displays a sharp transition as a function of $t$: after an initial low-performance region, both training and test accuracies rise rapidly around a characteristic time $t^\ast$ and then saturate to a plateau. Remarkably, the location of this transition is essentially independent of system size, suggesting that the relevant timescale is set by local dynamical processes rather than by the total number of qubits. Moreover, the plateau accuracy coincides, within numerical uncertainties, with that obtained by replacing the XX evolution with Haar-random unitaries acting on the same encoded states.

This observation motivates a more refined interpretation of the role of quantum dynamics in QELMs. The onset of useful feature generation does not require fully scrambled states or highly nonlocal entanglement. Instead, the relevant regime is associated with the build-up of local entanglement and with the spreading of information over a finite range, consistent with a local dynamical timescale. The thesis further shows that this regime remains compatible with efficient tensor-network simulation and can be reproduced, to a large extent, by shallow random circuits with local gates. The results therefore do not provide evidence for a generic quantum advantage, but they do identify a physically meaningful intermediate regime in which quantum dynamics enriches the representation of classical data while remaining structurally close to models that are still classically tractable \cite{DeLorenzis_2}.

The second research direction of the thesis concerns the application of machine learning to neutrino physics, with particular emphasis on the classification of complex event topologies in water Cherenkov detectors. The experimental context is the Hyper-Kamiokande \cite{Hyper-Kamiokande:2018ofw, Hyper_2015, Hyper-Kamiokande:2025fci, Hyper-Kamiokande:2025asb} programme and, more specifically, its Intermediate Water Cherenkov Detector (IWCD), a cylindrical detector instrumented with multi-PMT optical modules. In this setting, neutrino interactions are not observed directly; rather, they are inferred from the charge and timing patterns produced by Cherenkov light on the detector surface. A crucial technical step is therefore the construction of a suitable image representation of the detector response. In this thesis, the cylindrical surface is unwrapped into a rectangular grid in $(\phi,z)$ coordinates, and the charge and timing information associated with the photosensors is organised into image channels, yielding multi-channel detector images that can be processed by convolutional architectures.

The supervised task considered in detail is the discrimination between single-vertex and pile-up events. Single-vertex events contain one neutrino interaction in the relevant time window, but may still exhibit non-trivial multi-ring topologies because of the multiplicity of charged particles in the final state. Pile-up events instead contain two or more independent interactions within the same detector readout window, producing overlapping Cherenkov patterns that are more difficult to disentangle. This is therefore a physically relevant classification problem for high-rate near detectors, where one seeks robust discrimination based not only on total light yield but also on the spatial and temporal structure of the event.

To address this problem, the thesis develops a ResNet-18--based convolutional classifier \cite{he2016deep} tailored to IWCD images. The adopted network preserves the residual structure of the standard ResNet-18 architecture, with four residual stages, but modifies the input stem to better preserve fine spatial information in detector images.

The results show that deep convolutional models are highly effective for this task. When both charge and timing information are used in the non-compact representation, the classifier reaches Area Under the Curve (AUC) values close to unity on both simplified topological samples and more realistic generator-level event samples. Timing information emerges as especially informative for separating single-vertex from pile-up events, while compact representations remain competitive with the full non-compact input. 

Additional studies as a function of event-level variables show that the classification becomes more difficult when the interaction vertices are close in space and time, or when the event develops near detector boundaries, in agreement with the physical picture of strongly overlapping or partially truncated Cherenkov patterns.

Finally, the thesis includes an exploratory transfer of the QELM framework to realistic neutrino-image data. Because present QELM implementations require strong compression of the classical input, this study is intentionally preliminary. Detector images are reduced to low-dimensional feature vectors, encoded into 10-qubit states through Bloch-sphere encoding, evolved with an XX-type Hamiltonian, and classified through a single sigmoid readout. Even in this constrained setting, the QELM extracts non-trivial discriminating information from IWCD events, with the mixed charge-plus-timing representation giving the best performance among the tested configurations. Although the results remain significantly below those obtained with ResNet-18, they show that the QELM pipeline can be meaningfully transferred from standard image benchmarks to realistic detector data.
In addition to the analyses presented in this part of the thesis, I also contributed, together with other collaborators, to the development of the open-source WatChMaL framework \cite{WatChMaL}, in particular to the introduction of timing information in the analysis pipeline.

\paragraph{Main contributions of the thesis.}
The main contributions of this thesis can be summarized as follows. First, it develops and formalizes a QELM pipeline for supervised image classification in which the effects of feature reduction, encoding, Hamiltonian choice, and measurement can be analysed separately. Second, it shows that, in the compressed regime relevant to near-term quantum models, nonlinear latent representations obtained through autoencoders are more effective than PCA, and that dense-angle-type encodings provide a particularly favourable trade-off between performance and implementability. Third, it demonstrates that high QELM performance can be reached without highly specific Hamiltonian engineering, and that in the XX-based setting the relevant operating regime is associated with local entanglement growth and remains compatible with efficient classical simulation. Fourth, it develops a ResNet-18--based framework for the classification of IWCD single-vertex and pile-up events from multi-channel charge-and-timing images, achieving robust performance across multiple simulated datasets. Fifth, it presents a first proof-of-concept application of the QELM framework to realistic neutrino-image data, thereby establishing a concrete connection between the two main parts of the thesis.

\paragraph{Structure of the thesis.}
The remainder of this thesis is organized as follows. Chapter~2 introduces the machine-learning concepts used throughout the thesis, including supervised learning, neural networks, optimization, evaluation metrics, autoencoders, and convolutional architectures. Chapter~3 reviews the aspects of quantum computing most relevant for the present work, with emphasis on Hilbert-space representations, unitary dynamics, and measurement as ingredients of quantum feature generation. Chapter~4 presents the general QELM framework and describes in detail the classical preprocessing stage, the encoding strategies, the Hamiltonians used for the quantum layer, the measurement stage, and the classical readout. Chapter~5 applies this framework to image classification and analyses the impact of feature reduction, encoding, and Hamiltonian choice. Chapter~6 studies in greater depth the relationship between performance, entanglement growth, information spreading, and classical simulability in XX-based QELMs. Chapter~7 introduces the physical context of neutrino detection, Cherenkov radiation, and water Cherenkov detectors, with particular attention to Hyper-Kamiokande and the IWCD. Chapter~8 develops and evaluates deep-learning models for neutrino detector images and includes an exploratory QELM application to the same classification problem. Finally, Chapter~9 summarizes the main results and discusses possible future directions.
 
\newpage
\chapter{Machine Learning Background}
\label{ch:ml_background}
\section{Historical overview}
The term \emph{machine learning} was introduced in the late 1950s by
Arthur Samuel \cite{Samuel}, a pioneer in computer gaming and artificial intelligence,
who developed one of the first programs capable of improving its own
performance at the game of checkers through experience. The conceptual
origins of the field, however, are broader and lie in the study of human
cognitive processes. In 1949, Donald Hebb \cite{hebb1949} proposed a theoretical model
of an artificial neuron and a learning mechanism based on the
strengthening of synaptic connections, which would profoundly influence
the development of neural networks. A few years earlier, Walter Pitts and Warren McCulloch \cite{mcculloch1943} introduced an early mathematical formalization of networks of
logical neurons, showing how simple binary units can, in principle,
implement complex computations.

In the following decades, the first devices and algorithms able to learn
from data were developed. In the 1960s, so-called \emph{learning
machines} appeared, designed to recognise patterns in signals such as
sonar traces, electrocardiograms, or speech waveforms, often using
elementary reinforcement mechanisms. Interest in pattern recognition and
automatic classification continued throughout the 1970s and 1980s, with
the development of statistical models and techniques such as principal
component analysis for dimensionality reduction.

An important turning point occurred between the 1980s and 1990s, with
the rediscovery and widespread adoption of the \emph{backpropagation}
algorithm \cite{rumelhart1986} for training multi-layer neural networks and the introduction
of kernel-based methods such as \emph{support vector machines} \cite{cortes1995}. During
this period, an operational definition of machine learning \cite{bishop2006, geron2022}, often
attributed to Tom Mitchell, became established: a program is said to
“learn” if its performance on a given task improves with experience,
with respect to a quantitative performance measure. At the same time,
the distinction between supervised, unsupervised, and reinforcement
learning took shape and remains at the core of modern taxonomies.

Starting from the late 2000s, so-called \emph{deep learning} \cite{lecun2015, he2016deep} brought
about a further transformation. The combination of more powerful
hardware (in particular graphics processing units GPUs), large annotated datasets, and new neural
architectures made it possible to train very deep networks with millions
of parameters, capable of achieving state-of-the-art performance in
tasks such as image classification, speech recognition, and natural
language processing. Among the most emblematic results of this period
are the introduction of \emph{generative adversarial networks} (GANs) \cite{Goodfellow-et-al-2016},
able to synthesise realistic data, and the successes of deep
reinforcement-learning systems such as AlphaGo \cite{Silver2016MasteringTG}, which have surpassed the
best human players in highly complex board games.

In this thesis we do not attempt to provide an exhaustive historical
survey, but rather focus on the fundamental concepts that are most
relevant for the applications to quantum computing and neutrino physics
discussed in the following chapters.

\section{Supervised and unsupervised learning}
\label{sec:basics_ml}
At a high level, a machine-learning problem can be described in terms of
\emph{data}, \emph{models}, and a \emph{learning objective}. We are given a dataset with input $\mathbf{x}^{(i)}$ with $i=1, \ldots, m$ (e.g. an image, a time series, or
a feature vector) and, depending on the task, a target $\mathbf{y}^{(i)}$ (e.g.\ a class
label or a real-valued output). We then choose a family of models
$\mathcal{F} = \{f_\theta\}$, such as a neural network, parametrized by $\theta$ and an objective
function $\mathcal{L}(\theta)$ that quantifies how well
$f_\theta$ matches the data. Learning consists in adjusting $\theta$ so
as to minimize $\mathcal{L}$ on the training data, while still
performing well on unseen examples.

In \emph{supervised learning}, each input $\mathbf{x}^{(i)}$ is
associated with an explicit target $\mathbf{y}^{(i)}$. The goal is to
learn a mapping $f_\theta:\mathbf{x}\mapsto \hat{\mathbf{y}}$ that
predicts the correct output for new, unseen inputs. Typical supervised
tasks include:
\begin{itemize}
  \item \emph{classification}, where $\mathbf{y}$ is a discrete class
        label (e.g.\ neutrino vs.\ background event);
  \item \emph{regression}, where $\mathbf{y}$ is a real-valued quantity
        (e.g.\ an energy or a continuous parameter).
\end{itemize}

In \emph{unsupervised learning}, only the inputs
$\{\mathbf{x}^{(i)}\}$ are given, and there are no explicit targets.
The objective is to identify structure in the data, such as clusters or
low-dimensional manifolds. Examples include clustering algorithms
($k$-means, Gaussian mixture models) and dimensionality reduction
methods (Principal Component Analysis, autoencoders). In this thesis,
unsupervised methods are mainly used for feature extraction and
dimensionality reduction, for instance to compress high-dimensional
images into a latent representation.

\section{Neural Networks}

Over the past decade, neural networks have become one of the central tools in
modern machine learning. They are at the core of many applications in computer vision, natural language processing, game
playing and scientific data analysis. Their success stems from their ability
to approximate complex nonlinear functions directly from data, with minimal feature engineering.

At a high level, a neural network \cite{Goodfellow-et-al-2016} is a function $f_\theta$ with
parameters $\theta$ (usually called weights and biases) adjusted (trained) so that $f_\theta(\mathbf{x})$
produces a desired output $y$ when given an input $\mathbf{x}$. The training
process is guided by data and an objective (loss) function that quantifies how
well the network performs on a particular task. By minimizing this loss over a
training dataset, the network learns patterns present in the data
and can then be used to make predictions on unseen examples.

Although the name ``neural network'' originates from analogies with biological neurons, modern neural networks are best viewed as compositions of simple mathematical operations. In what follows, we first introduce the basic computational unit, the neuron, and the way neurons are combined to form a network. We then describe the loss function, the optimization procedure used for training, and the metrics used to evaluate performance.

\subsection{The neuron}

The fundamental building block of a neural network is the \emph{neuron}. Conceptually, a neuron receives one or more
real-valued inputs, linearly combines them, and passes the result through a
nonlinear activation function to produce an output.

\begin{figure}
	\centering
\includegraphics[width=1\linewidth]{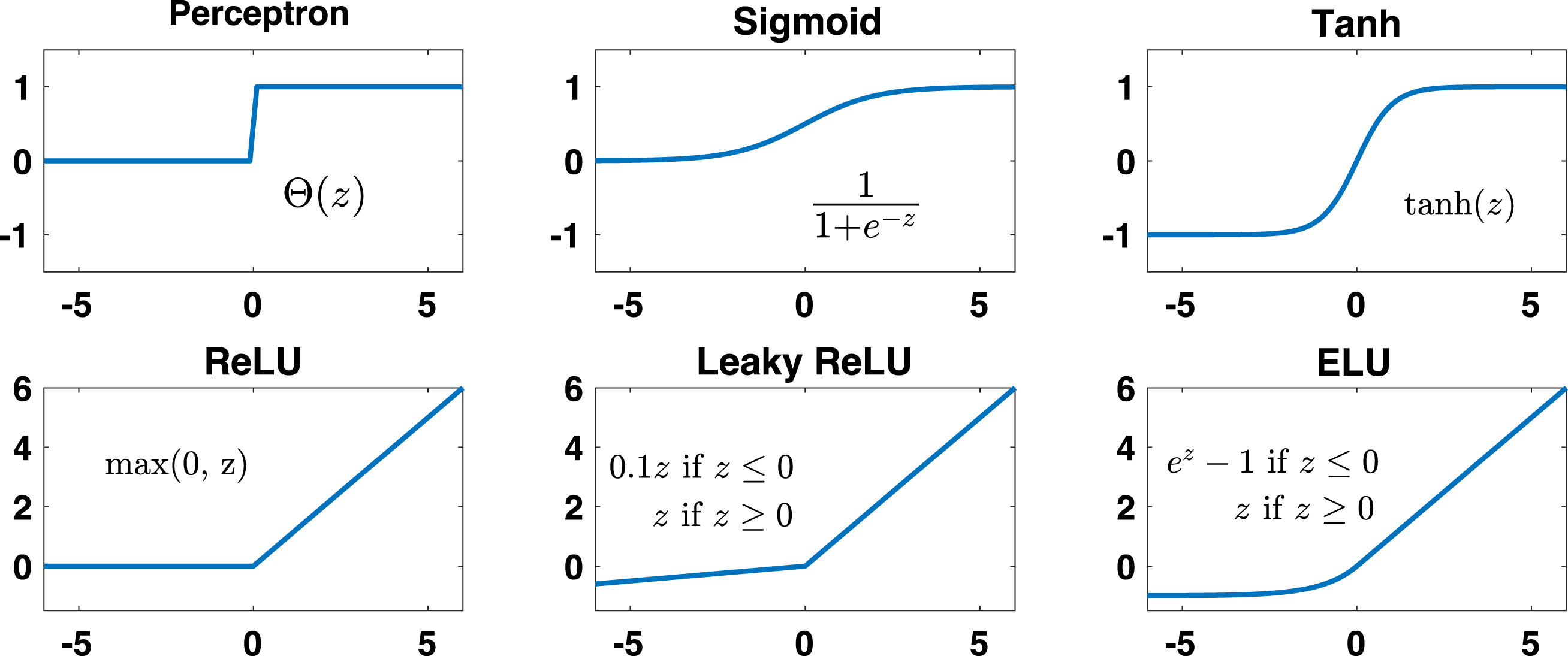}
	\caption{Some non-linear activation functions. Both saturating (top row) and non-saturating (bottom row) functions are shown \cite{Mehta:2018dln}.
\label{fig:activations}}
\end{figure}

Let $\mathbf{x} = (x_1, x_2, \dots, x_d)^\top \in \mathbb{R}^d$ denote the input
vector to a neuron. The neuron is associated with a weight vector
$\mathbf{w} = (w_1, w_2, \dots, w_d)^\top$ and a bias term $b \in \mathbb{R}$.
Its output $a$ is given by
\begin{equation}
    a = \sigma(z),
    \qquad \textrm{with} \qquad
    z = \mathbf{w}^\top \mathbf{x} + b,
\end{equation}
where $\sigma$ is a nonlinear \emph{activation function}. Common choices
include:
\begin{itemize}
    \item The rectified linear unit (ReLU), $\sigma(z) = \max(0, z)$. It is piecewise linear, computationally cheap, and does not saturate in the positive region, which helps maintain stronger gradients in deep models. For these reasons, ReLU (and its variants, such as Leaky ReLU and ELU) has become the default choice for hidden layers in many deep-learning architectures.
    \item The hyperbolic tangent,
    $\sigma(z) = \tanh(z)$. Historically, $\tanh$ was widely used in hidden layers, but it has been largely superseded by ReLU, since the hyperbolic tangent saturates for large $|z|$.
    \item The logistic (sigmoid) function, $\sigma(z) = \frac{1}{1 + e^{-z}}$. It has historically been used in the output layer for binary classification, as it can be interpreted as a probability. However, in hidden layers it tends to suffer from saturation (very small gradients for large $|z|$), which can slow down training in deep networks.
\end{itemize}
The activation functions described above, together with some other examples such as the perceptron, are shown in Fig.~\ref{fig:activations}.

The perceptron (or binary step) activation function,
\[
\sigma(z)=
\begin{cases}
1, & z \geq 0, \\
0, & z < 0,
\end{cases}
\]
played an important historical role in early neural-network models. However, it is rarely used in modern deep networks because it is non-differentiable and has zero gradient almost everywhere, making gradient-based optimization ineffective.

The nonlinearity of $\sigma$ is crucial. Without it, a network of neurons would
collapse to a single linear transformation, unable to model complex
relationships. The combination of linear operations and nonlinear activations
allows neural networks to approximate highly nonlinear functions.

\subsection{Layers and feed-forward networks}
\label{subsec:layers_feedforward}

\begin{figure}
	\centering
\includegraphics[width=1\linewidth]{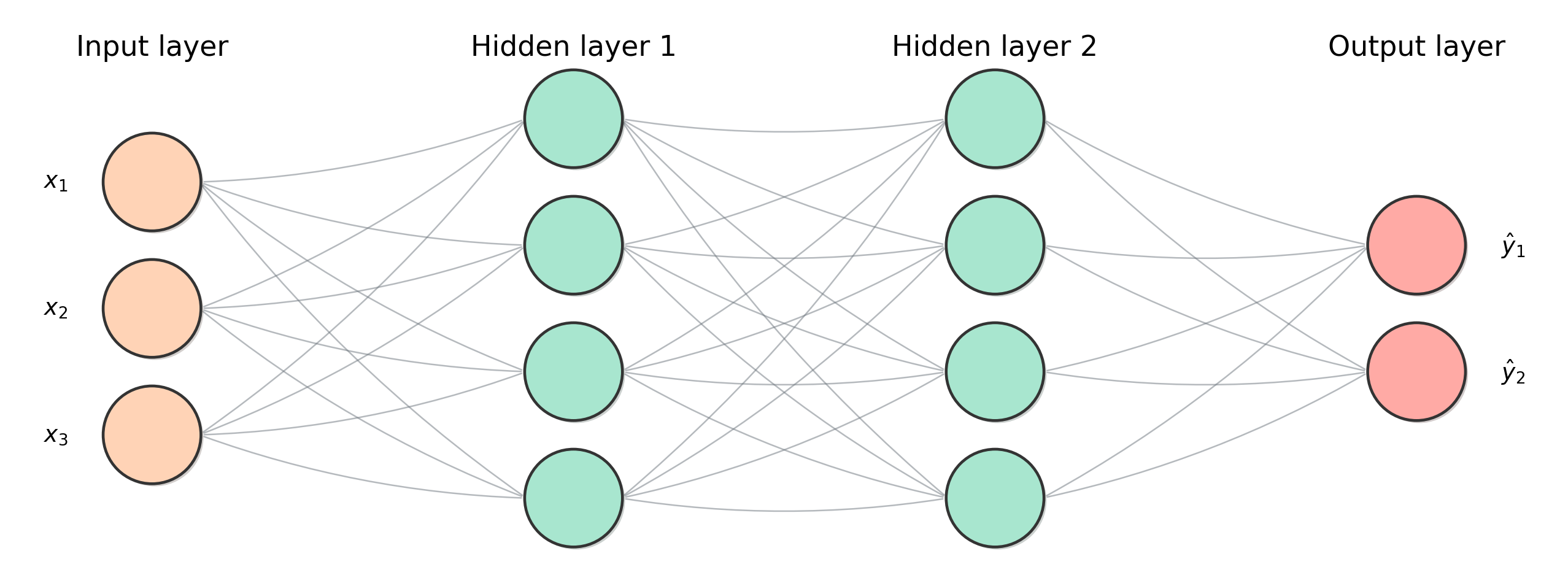}
	\caption{A representation of a feed-forward neural network with two hidden layers. The network takes input data characterized by three features and outputs a two-dimensional vector.
\label{fig:feedforwardNN}}
\end{figure}

In practice, neurons are organized into \emph{layers}. A layer takes as input a
vector of activations from the previous layer and outputs a new vector of
activations. The simplest and most common architecture is the
\emph{feed-forward} (or fully connected, or dense) network. An example of a feed-forward architecture is shown in Fig.~\ref{fig:feedforwardNN}. 

Consider a network with $L$ layers (excluding the input). Let
$\mathbf{a}^{(0)} = \mathbf{x}$ be the input, and for $\ell = 1, \dots, L$,
\begin{equation}
    \mathbf{a}^{(\ell)} = \sigma^{(\ell)}\big(\mathbf{z}^{(\ell)}\big),
    \qquad \textrm{with} \qquad
    \mathbf{z}^{(\ell)} = W^{(\ell)} \mathbf{a}^{(\ell-1)} + \mathbf{b}^{(\ell)},
\end{equation}
where:
\begin{itemize}
    \item $W^{(\ell)}$ is the weight matrix of layer $\ell$,
    \item $\mathbf{b}^{(\ell)}$ is the bias vector,
    \item $\sigma^{(\ell)}$ is applied element-wise to the pre-activation vector
    $\mathbf{z}^{(\ell)}$.
\end{itemize}
The vector $\mathbf{a}^{(L)}$ is the output of the network. Depending on the
problem, the output layer may use a specific activation:
\begin{itemize}
    \item For regression, a linear output ($\sigma^{(L)}(z) = z$) is often used.
    \item For binary classification, a sigmoid activation is common.
    \item For multi-class classification, the \emph{softmax} function is typically used:
          \begin{equation}
              \mathrm{softmax}(\mathbf{z})_k
              = \frac{e^{z_k}}{\sum_{j} e^{z_j}}.
          \end{equation}
\end{itemize}

Networks with many hidden layers (layers between input and output) are
called \emph{deep neural networks}. A central theoretical result, the universal
approximation theorem \cite{cybenko1989, hornik1989}, states that, under mild conditions, such networks can
approximate any continuous function on a compact domain, provided they have
enough units. In practice, depth and width must be chosen carefully to balance
expressivity, computational cost and the risk of overfitting.

While fully connected networks are the conceptual baseline, many specialized
architectures exist, such as, for example, \emph{Convolutional Neural Networks} (CNNs) which exploit spatial structure and are well suited for image data; \emph{Recurrent Neural Networks} (RNNs), designed to process sequential data; \emph{autoencoders} that learn low-dimensional representations by reconstructing their input, etc. We will provide more details on some of these architectures later in this thesis.

\section{Loss Functions}

To train a neural network, one must quantify how well the network performs on a
given task. This is the role of the \emph{loss function} (also called the cost
or objective function) \cite{bishop2006}. Given an input--output pair $(\mathbf{x}, y)$ and
network prediction $\hat{y} = f_\theta(\mathbf{x})$, the loss function
$\mathcal{L}(y, \hat{y})$ measures the discrepancy between $y$ and $\hat{y}$.
Training aims to find parameters $\theta$ that minimize the average loss over a
training dataset.

The loss function on the entire training set is given by
\begin{equation}
    \mathcal{L}(\theta)
    = \frac{1}{m} \sum_{i=1}^{m} \mathcal{L}\big(y_i, f_\theta(\mathbf{x}_i)\big), 
\end{equation}
and the explicit choice of $\mathcal{L}(y,\hat y)$ depends on the particular task.
\begin{itemize}
\item For regression problems, where $y \in \mathbb{R}$ (or $\mathbb{R}^k$), a common
choice is the \emph{mean squared error} (MSE), defined as
\begin{equation}
    \mathcal{L}_{\mathrm{MSE}}(y, \hat{y})
    = \frac{1}{2} \left\| y - \hat{y} \right\|^2.
    \label{eq:loss_mse}
\end{equation}
MSE penalizes large
errors more strongly than small ones and is differentiable everywhere.

\item For classification, losses are typically based on likelihood. In the Binary classification case, 
for $y \in \{0,1\}$, with output $\hat{y} \in (0,1)$ interpreted as the probability of class 1, a common choice is the \emph{binary cross-entropy} loss:
\begin{equation}
    \mathcal{L}_{\mathrm{BCE}}(y, \hat{y})
    = -\big[ y \log \hat{y} + (1-y)\log(1 - \hat{y}) \big].
\end{equation}

\item In the multi-class classification case, instead,
with $C$ classes and one-hot target vector $\mathbf{y}$ (with $y_c = 1$ for the
true class $c$, zero otherwise), one often uses the \emph{categorical
cross-entropy}:
\begin{equation}
    \mathcal{L}_{\mathrm{CE}}(\mathbf{y}, \hat{\mathbf{y}})
    = - \sum_{c=1}^{C} y_c \log \hat{y}_c,
\end{equation}
where $\hat{\mathbf{y}}$ is the softmax output of the network. This loss is
equivalent to the negative log-likelihood of the correct class under the predicted distribution of the model.
\end{itemize}

\section{Optimization}

Training a neural network corresponds to minimizing the loss
$\mathcal{L}(\theta)$ with respect to the parameters $\theta$. Because
$\mathcal{L}(\theta)$ is typically non-convex and high-dimensional, we rely on iterative, gradient-based optimization methods.

\subsection{Gradient descent and backpropagation}

The basic idea of \emph{gradient descent} is to update the parameters in the direction of steepest descent of the loss:
\begin{equation}
    \theta \leftarrow \theta - \eta \, \nabla_\theta \mathcal{L}(\theta),
\end{equation}
where $\eta > 0$ is the learning rate. The correct choice of this hyperparameter can be crucial for an efficient optimization. Indeed, a small learning rate may
lead to slow convergence, while a learning rate that is too large can
cause divergence or oscillations.

For neural networks, computing the gradient $\nabla_\theta \mathcal{L}$ efficiently is made possible by the \emph{backpropagation} algorithm, which applies the chain rule of calculus layer by layer, starting from the output.

However, computing the gradient using all $m$ training examples at each step
(batch gradient descent) can be computationally expensive. Instead, we
could use
\begin{itemize}
    \item \emph{Stochastic Gradient Descent (SGD)}, in which case gradients are estimated on
          a single example at a time;
    \item \emph{Mini-batch gradient descent} in which case gradients are computed on small batches of
          size $m_B$, balancing stability and speed:
          \begin{equation}
              \theta \leftarrow \theta - \eta \,
              \nabla_\theta \mathcal{L}_{\mathcal{B}}(\theta),
          \end{equation}
          where $\mathcal{B}$ is a mini-batch and
          \begin{equation}
              \mathcal{L}_{\mathcal{B}}(\theta)
              = \frac{1}{m_B} \sum_{i \in \mathcal{B}}
                \mathcal{L}(y_i, f_\theta(\mathbf{x}_i)).
          \end{equation}
\end{itemize}
One full pass over the entire training
set is called an \emph{epoch}. Training thus proceeds over multiple
epochs, during which the network parameters are gradually updated.
\subsection{Advanced optimization algorithms}
\label{subsec:advanced_optimizers}

Several extensions of basic gradient descent methods have been proposed to improve convergence, such as:
\begin{itemize}
    \item \emph{Momentum}, accelerating learning by accumulating a velocity
          vector in parameter space, smoothing out noisy gradient fluctuations.
    \item \emph{Adam} (Adaptive Moment estimation), adapting the learning rate for
          each parameter based on estimates of first and second moments of the
          gradients. It is widely used due to its robustness and ease of use.
    \item \emph{RMSProp},  adapting learning rates based
          on gradient history.
\end{itemize}

The choice of optimizer and learning rate schedule can significantly affect
training speed and final performance. Often, a learning rate that decays over time (step decay or exponential decay) improves convergence. In this thesis we mainly employ the Adam optimizer, as implemented in
standard deep-learning libraries, both for classical networks and for
the classical readout of the QELM.

\subsection{Training, validation and test sets}

A central concept in machine learning is that of \emph{generalization}:
a good model should not only fit the training data, but also perform
well on new inputs drawn from the same underlying distribution. To
evaluate generalization, the available data are usually split into
non-overlapping subsets:
\begin{itemize}
  \item a \emph{training set}, used to optimize the model parameters;
  \item a \emph{validation set}, used for model selection and
        hyperparameter tuning;
  \item a \emph{test set}, used only at the end of the analysis to
        estimate the final performance.
\end{itemize}
The optimization uses only the training set; the validation set guides
decisions such as model architecture, regularization strength and early
stopping. \\

Two important failure modes in the training of supervised learning models are overfitting and underfitting. A model \emph{underfits} when it is too simple or trained
for too few epochs, failing to capture the relevant structure in the
data (high training and validation error). A model \emph{overfits} when
it fits the training data extremely well but fails to generalize to new
examples (low training error but high validation error). Regularization
techniques such as weight decay, dropout and early stopping, together
with appropriate model capacity and careful monitoring of validation
metrics, are essential to strike a balance between these two extremes.

In the remainder of the thesis, we will highlight overfitting/underfitting
effects whenever relevant, in particular when comparing different
architectures or training regimes.

\section{Evaluation Metrics}
\label{sec:metrics}

While the loss function guides optimization, it does not always correspond
directly to the quantity we ultimately care about. For example, in
classification we may minimize cross-entropy but evaluate performance using
accuracy, precision or other task-specific metrics. It is therefore important
to define appropriate \emph{metrics} \cite{bishop2006} to assess model quality on validation and
test data. \\ 
For regression tasks, a commonly used metric is the \emph{Mean Squared Error (MSE)}, defined as in Eq.~\ref{eq:loss_mse}.
For classification tasks, the simplest metric is \emph{accuracy}, defined as
\begin{equation}
    \mathrm{Accuracy}
    = \frac{\text{number of correctly classified examples}}{m}.
\end{equation}
In particular, for binary classification problems, it is useful to introduce
the \emph{confusion matrix}, which counts the number of:
\begin{itemize}
  \item true positives (TP): positive samples correctly classified;
  \item true negatives (TN): negative samples correctly classified;
  \item false positives (FP): negative samples misclassified as positive;
  \item false negatives (FN): positive samples misclassified as negative.
\end{itemize}
The overall accuracy is then defined as
\begin{equation}
  \mathrm{Accuracy}
  = \frac{\mathrm{TP} + \mathrm{TN}}
         {\mathrm{TP} + \mathrm{TN} + \mathrm{FP} + \mathrm{FN}}.
\end{equation}

However, when classes are imbalanced, accuracy may be misleading. In such
cases, additional metrics are informative:
\begin{itemize}
    \item \emph{Precision} (for a given class, say ``positive''):
          \begin{equation}
              \mathrm{Precision}
              = \frac{\text{true positives}}
                     {\text{true positives} + \text{false positives}},
          \end{equation}
          answering the question ``of all predicted positives, how many were correct?''

    \item \emph{Recall} (or sensitivity):
          \begin{equation}
              \mathrm{Recall}
              = \frac{\text{true positives}}
                     {\text{true positives} + \text{false negatives}},
          \end{equation}
          answering the question ``of all actual positives, how many did we find?''

    \item \emph{F1 score}, the harmonic mean of precision and recall:
          \begin{equation}
              \mathrm{F1}
              = 2 \cdot
              \frac{\mathrm{Precision} \cdot \mathrm{Recall}}
                   {\mathrm{Precision} + \mathrm{Recall}}.
          \end{equation}
\end{itemize}

Other tools include confusion matrices, Receiver Operating Characteristic
(ROC) curves, and the Area Under the ROC Curve (AUC), which assess performance
across different decision thresholds. \\
Many classifiers output a continuous score or probability
$\hat{p} \in [0,1]$ for the positive class. By varying a decision
threshold $\tau$ and assigning samples with $\hat{p} \ge \tau$ to the
positive class, one obtains different trade-offs between:
\begin{itemize}
  \item the true positive rate (TPR or sensitivity),
        \(
          \mathrm{TPR} = \frac{\mathrm{TP}}{\mathrm{TP} + \mathrm{FN}},
        \)
  \item the false positive rate (FPR),
        \(
          \mathrm{FPR} = \frac{\mathrm{FP}}{\mathrm{FP} + \mathrm{TN}}.
        \)
\end{itemize}
The ROC curve plots TPR versus FPR
as the threshold $\tau$ is varied. The AUC is a threshold-independent summary of classifier performance: an
AUC of 0.5 corresponds to random guessing, whereas an AUC of 1.0
indicates perfect classification. \\

For multi-class classification, the confusion matrix is a $C \times C$
table, where $C$ is the number of classes. The entry in row $i$ and
column $j$ records how many samples of true class $i$ were predicted as
class $j$. Visualizing the confusion matrix provides insight into which
classes are systematically confused with each other.

Throughout this thesis we will report accuracy and AUC values, as well
as confusion matrices, to assess and compare the performance of the
classical and quantum-inspired models introduced in the following
chapters.

\section{Convolutional Neural Networks}
\label{sec:cnn}

The fully connected networks introduced in the previous sections treat their
inputs as generic vectors in $\mathbb{R}^d$, ignoring any particular structure
the data might possess. However, many types of data, in particular images (such those in the MNIST \cite{mnist} and CIFAR-10 \cite{cifar10}
datasets or the detector images in neutrino physics),
exhibit strong local correlations and are
naturally organized on regular grids. \emph{Convolutional Neural Networks}
(CNNs) \cite{lecun98} are specifically designed to exploit such structure, and they form the
backbone of most modern architectures in computer vision and related domains.

The key idea behind CNNs is to replace generic linear transformations with
\emph{convolutional} layers, in which small learnable filters (or kernels) are
applied locally and repeatedly across the input. This induces parameter
sharing (the same filter is used at all spatial locations) and sparse
connectivity (each output unit depends only on a small neighborhood), which
significantly reduces the number of parameters and improves statistical
efficiency.

\subsection{Convolutions and feature maps}
\label{subsec:cnn_convolutions}

For concreteness, consider a 2D input (e.g.\ a grayscale image) represented by
a matrix $X \in \mathbb{R}^{H \times W}$, where $H$ and $W$ are the height and
width. A convolutional layer applies a set of $K$ learnable filters
$\{F^{(k)}\}_{k=1}^{K}$, each of size $h \times w$, to the input. The output of
filter $k$ is a \emph{feature map} $Y^{(k)}$, whose entries are given by the
discrete convolution
\begin{equation}
    Y^{(k)}_{i,j}
    = \sigma \!\left(
        \sum_{u=1}^{h} \sum_{v=1}^{w}
        F^{(k)}_{u,v} \,
        X_{i+u-1,\, j+v-1}
        + b^{(k)}
      \right),
\end{equation}
where $\sigma$ is a nonlinear activation function (e.g.\ ReLU) and $b^{(k)}$ is
a bias term associated with filter $k$. In practice, convolutions may be
implemented with a \emph{stride} greater than 1 (skipping some positions) and
with \emph{padding} (zero-padding the input at the boundaries) to control the
spatial dimensions of the output.

\begin{figure}
	\centering
\includegraphics[width=1\linewidth]{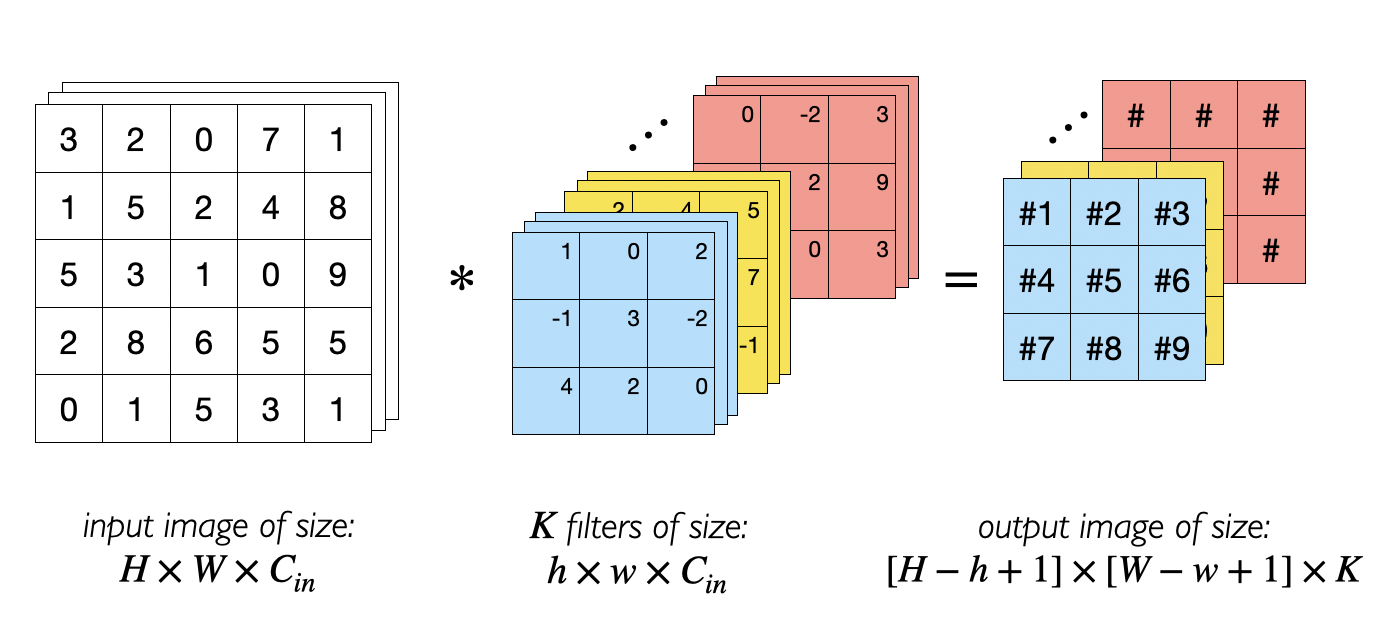}
	\caption{Convolution of a multi-channel image with $K$ filters. 
\label{fig:convolution}}
\end{figure}

For multi-channel inputs (e.g.\ RGB images or feature maps from previous
layers), each filter has the form
$F^{(k)} \in \mathbb{R}^{C_{\mathrm{in}} \times h \times w}$, where
$C_{\mathrm{in}}$ is the number of input channels. A schematic illustration of this operation is shown in Fig.~\ref{fig:convolution}. The convolution then sums
the contributions over channels:
\begin{equation}
    Y^{(k)}_{i,j}
    = \sigma \!\left(
        \sum_{c=1}^{C_{\mathrm{in}}}
        \sum_{u=1}^{h} \sum_{v=1}^{w}
        F^{(k)}_{c,u,v} \,
        X_{c,\, i+u-1,\, j+v-1}
        + b^{(k)}
      \right).
\end{equation}
Stacking all $K$ feature maps along a new channel dimension yields the output
tensor of the convolutional layer, with shape
$C_{\mathrm{out}} \times H' \times W'$ (where $C_{\mathrm{out}} = K$ and
$H', W'$ depend on stride and padding).

Intuitively, each filter learns to respond to a particular local pattern
(e.g.\ an edge or a texture) and deeper layers combine these
local patterns into more complex and abstract features.

\subsection{Pooling and upsampling}
\label{subsec:cnn_pooling}

\begin{figure}
	\centering
\includegraphics[width=0.9\linewidth]{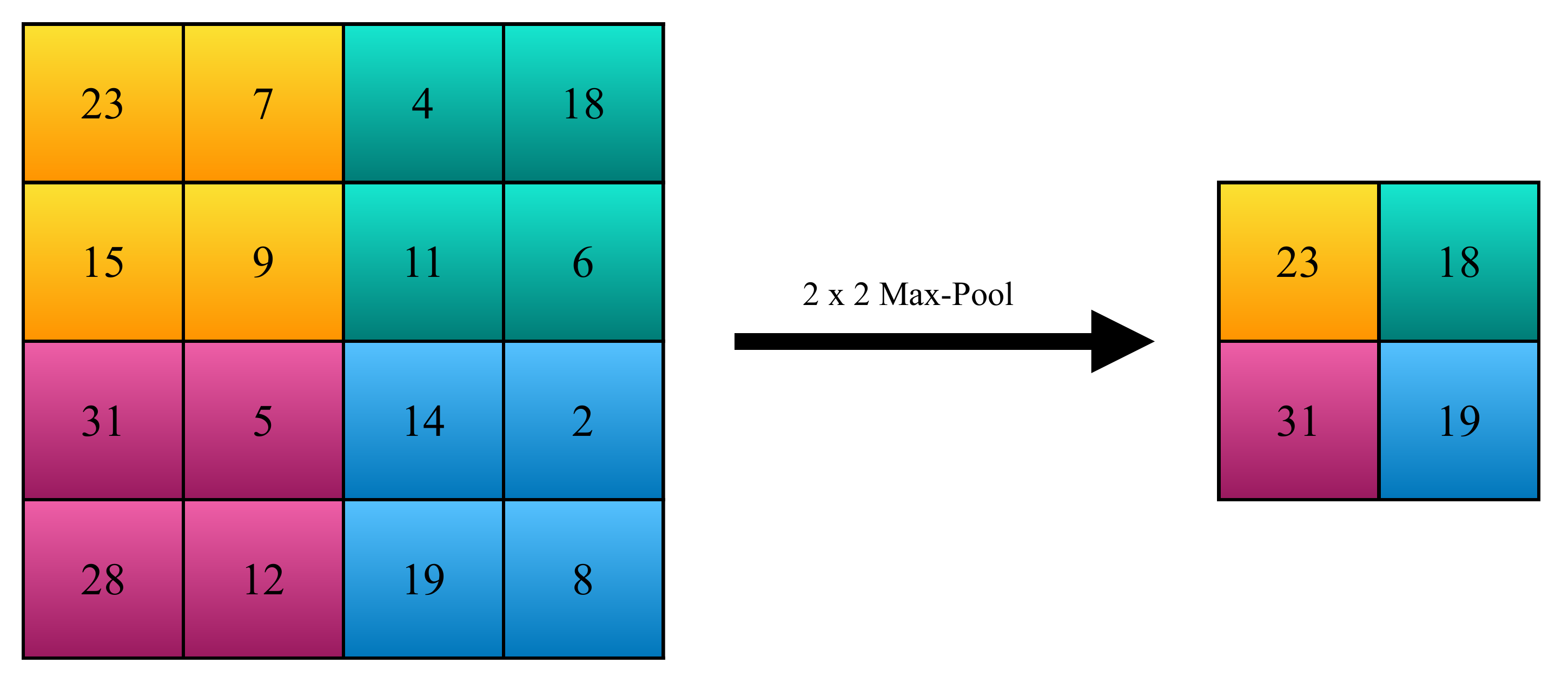}
	\caption{Example of a $2\times 2$ max-pooling operation. Each $2\times 2$ region of the input feature map is replaced by its maximum value, producing a downsampled output.
\label{fig:max_pool}}
\end{figure}

Convolutional layers are often interleaved with \emph{pooling} (or subsampling)
layers, whose role is to reduce the spatial resolution while preserving the
most salient information. A common choice is \emph{max pooling}, which
partitions each feature map into non-overlapping (or partially overlapping)
regions and outputs the maximum value in each region, as illustrated in Fig.~\ref{fig:max_pool}. In particular, a
$2 \times 2$ max-pooling operation reduces the height and width of the feature
map by a factor of 2:
\begin{equation}
    Z_{i,j}
    = \max_{(u,v) \in \{1,2\}^2}
      Y_{2i+u-2,\, 2j+v-2}.
\end{equation}
Other pooling operations include average pooling (taking the mean instead of
the maximum) or more general aggregation functions.

Pooling has two main effects:
\begin{itemize}
    \item it reduces the dimensionality of the feature maps, lowering
          computational cost and the risk of overfitting;
    \item it provides a degree of translational invariance, as small shifts of
          the input have limited effect on the pooled outputs.
\end{itemize}

In modern architectures, pooling is sometimes replaced or complemented by
strided convolutions, which jointly perform feature extraction and downsampling.\\

In decoder-like parts of convolutional architectures, such as the decoder of
a convolutional autoencoder, one often needs to \emph{increase} the spatial
resolution of the feature maps, effectively inverting the action of pooling or
strided convolutions. This is achieved through \emph{upsampling} layers. A common choice consists in a simple interpolation-based upsampling (e.g.\ nearest
neighbour or bilinear interpolation). In a typical encoder-decoder architecture,
downsampling layers in the encoder compress the input into a compact latent
representation, while corresponding upsampling layers in the decoder
progressively restore the original resolution, allowing the network to produce
detailed reconstructions in the output space.

\subsection{Typical CNN architectures}
\label{subsec:cnn_architectures}

A typical CNN for image processing consists of a sequence of convolutional
layers (possibly grouped into blocks), each followed by nonlinear activations
and optionally pooling or normalization layers, progressively transforming the
input into higher-level feature maps. In many architectures, the spatial
resolution is gradually reduced while the number of channels increases,
trading spatial detail for high-level feature extraction.

\begin{figure}
	\centering
\includegraphics[width=1\linewidth]{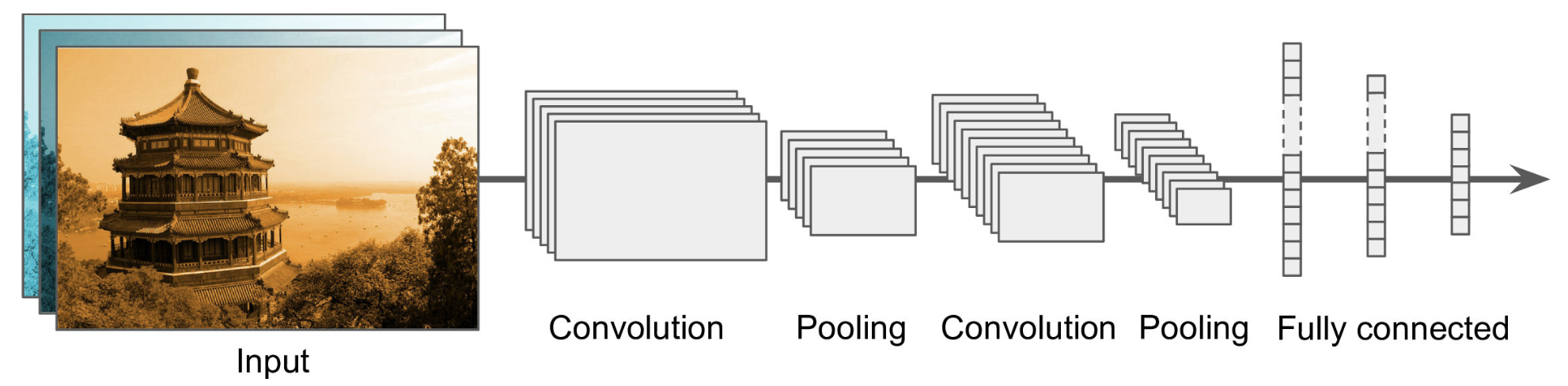}
	\caption{ Architecture of a typical CNN. Neurons in the convolutional layers calculate the convolution of the image with a filter. Pooling layers perform a coarse graining to give a smaller height and width while preserving the number of channels. The convolutional and pooling layers are then followed by fully connected layers \cite{geron2022}. 
\label{fig:cnn}}
\end{figure}

Eventually, the high-level features are fed into one or more fully connected
layers (sometimes after global pooling operations) to produce the final
output, such as class scores for classification or real-valued predictions for
regression, see for instance Fig.~\ref{fig:cnn}. The entire network is trained end-to-end using the same
gradient-based optimization techniques described in
the previous section, with a suitable loss function depending on the
task.\\
Among the many convolutional architectures developed in deep learning, residual networks (ResNets) are of particular importance, as they enable the effective training of deeper models. We briefly review their main ideas in the next subsection.

\subsection{Residual neural networks}
\label{sec:resnet}
\subsubsection{Residual Networks}
\label{subsec:residual}

Empirically, very deep CNNs, with many convolutional layers, tend to be more
expressive and capable of learning increasingly complex features. However, in
practice, simply stacking more layers often leads to optimization difficulties
and degraded performance. 

Residual Networks (ResNets) \cite{he2016deep} were proposed to address this problem by introducing
\emph{shortcut connections} that allow information and gradients to flow more
easily through very deep architectures. ResNets make it possible to train
networks with even hundreds of layers using standard optimization
techniques, and they have become a standard building block in modern deep
learning.

In practice, training very deep plain CNNs (without
shortcut connections) often leads to a phenomenon known as the
\emph{degradation problem}. 

As the depth increases beyond some point, the training error of a plain network
starts to increase, even though the network has strictly more parameters. This
degradation is not primarily due to overfitting since it appears on the training
error, but rather to optimization difficulties: gradients can  become very small (vanishing gradient problem) or very large (exploding gradient problem) as they propagate through many nonlinear layers, making it hard
for standard gradient-based methods to find good solutions.

Residual networks address this issue by explicitly parameterizing the layers to learn deviations from the identity mapping, rather than the full mapping
itself and this simple idea greatly
facilitates the training of very deep models.\\

\noindent
\textbf{Residual blocks} \\
Consider a stack of layers that takes an input feature vector
$\mathbf{x}$ and produces an output $\mathcal{H}(\mathbf{x})$. In a traditional
CNN, the network directly learns the mapping
\begin{equation}
    \mathbf{y} = \mathcal{H}(\mathbf{x}).
\end{equation}
Residual learning instead reparameterizes the problem in terms of a
\emph{residual function} $\mathcal{F}$:
\begin{equation}
    \mathbf{y} = \mathcal{F}(\mathbf{x}, \mathbf{W}) + \mathbf{x},
    \label{eq:residual_block}
\end{equation}
where $\mathcal{F}$ is typically a small stack of convolutional layers with
parameters collectively denoted by $\mathbf{W}$. The term $\mathbf{x}$ is
added to the output through an explicit \emph{shortcut connection}.
Equation~\eqref{eq:residual_block} can be interpreted as learning the
difference between the desired mapping and the identity:
\begin{equation}
    \mathcal{F}(\mathbf{x}, \mathbf{W}) = \mathcal{H}(\mathbf{x}) - \mathbf{x}.
\end{equation}
If the identity mapping is optimal for some layers, then it
should be easier to drive the residual $\mathcal{F}$ towards zero than to
force a stack of nonlinear layers to approximate the identity itself.
\begin{figure}
    \centering
    \includegraphics[width=0.5\linewidth]{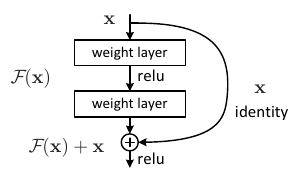}
    \caption{Residual learning: a building block. Figure adapted from \cite{he2016deep}. }
    \label{fig:residual_block}
\end{figure}
Figure~\ref{fig:residual_block} typically depicts a residual
block as two or three convolutional layers whose output is summed with the
original input through a shortcut path. A nonlinearity (such as ReLU) is often
applied after the addition. \\

\noindent
\textbf{Identity and projection shortcuts}\\
In the simplest case, the input and output of the residual block have the same
dimension, and the shortcut connection is a pure identity:
\begin{equation}
    \mathbf{y} = \mathcal{F}(\mathbf{x}, \mathbf{W}) + \mathbf{x}.
\end{equation}
This is referred to as an \emph{identity shortcut}. Identity shortcuts have no
additional parameters and negligible computational cost.

In deeper stages of the network, it is often necessary to change the spatial
resolution or the number of channels (for example, to downsample the feature
maps and increase the number of filters). In those cases, the shortcut performs
a \emph{projection} so that the dimensions match:
\begin{equation}
    \mathbf{y} = \mathcal{F}(\mathbf{x}, \mathbf{W}) + W_s \mathbf{x},
\end{equation}
where $W_s$ is typically a convolution with stride greater than
one (for spatial downsampling) and an appropriate number of output channels.
This is known as a \emph{projection shortcut}.

Both identity and projection shortcuts ensure that the gradient can flow
directly from deeper layers to shallower ones, reducing vanishing/exploding-gradient
effects and making optimization more stable.

\subsubsection{ResNet Architectures}
\label{subsec:resnet_architectures}

Residual blocks can be used as building blocks to construct deep architectures
of varying depth. 
The basic residual block consists of two $3 \times 3$ convolutional layers
with the same number of output channels, each followed by batch normalization
and a nonlinearity (commonly ReLU). The residual function has the form
\begin{equation}
    \mathcal{F}(\mathbf{x})
    =  \mathrm{BN}_2 ( W_2 * \sigma( \mathrm{BN}_1 ( W_1 * \mathbf{x} ) ) ),
\end{equation}
where $*$ denotes convolution, $\mathrm{BN}$ batch normalization, and $\sigma$
the activation function. The output of $\mathcal{F}$ is then added to the input
$\mathbf{x}$ through a shortcut connection, as in
Eq.~\eqref{eq:residual_block}, and then the nonlinearity function is applied.

Complete ResNet architectures are constructed by stacking many residual blocks
in stages of constant spatial resolution. A typical architecture begins with a
stem consisting of a convolution and pooling layer, followed by several stages
where the feature maps are progressively downsampled and the number of channels
increases. \\

\noindent
\textbf{Batch Normalization} \\
Several of the layers appearing in residual blocks make use of \emph{batch
normalization} (BN), a technique introduced to stabilize and accelerate the
training of deep neural networks. The main idea is to normalize the
pre-activation values within each mini-batch so that they have approximately
zero mean and unit variance, and then to allow the network to learn an
appropriate rescaling and shifting.

Consider a mini-batch of activations
$\{u_1, u_2, \dots, u_m\}$ for a given neuron, where
$m$ is the batch size. Batch normalization first computes the batch mean and
variance
\begin{equation}
    \mu_{\mathcal{B}} = \frac{1}{m} \sum_{i=1}^m u_i,
    \qquad
    \sigma_{\mathcal{B}}^2 = \frac{1}{m} \sum_{i=1}^m (u_i - \mu_{\mathcal{B}})^2,
\end{equation}
and then normalizes each activation:
\begin{equation}
    \hat{u}_i = \frac{u_i - \mu_{\mathcal{B}}}
                    {\sqrt{\sigma_{\mathcal{B}}^2 + \varepsilon}},
\end{equation}
where $\varepsilon$ is a small constant introduced for numerical stability.
To preserve the representational power of the network, BN introduces two
learnable parameters, $\gamma$ and $\beta$, and outputs the transformed
activations
\begin{equation}
    y_i = \gamma \hat{u}_i + \beta.
\end{equation}
The parameters $\gamma$ and $\beta$ are learned jointly with the rest of the
network and allow the layer to recover any necessary scale and shift if that is
beneficial for the task.

During training, the batch statistics $(\mu_{\mathcal{B}}, \sigma_{\mathcal{B}}^2)$
are computed on each mini-batch. For inference, running estimates of the mean
and variance are maintained (typically via exponential moving averages) and
used instead, so that the normalization no longer depends on the current batch.

Batch normalization brings several practical advantages: it allows for larger learning rates and speeds up convergence and it also has a mild regularizing effect, often improving generalization.

In residual networks, BN layers are usually inserted between convolutions and
nonlinearities inside the residual function $\mathcal{F}$. This placement helps
to stabilize the training of very deep architectures and is an essential
component of the original ResNet design.\\
In later chapters, we will adopt a ResNet-like convolutional architecture for the analysis of neutrino images, building on the concepts introduced in this section.

\newpage
\chapter{Quantum Computing Background}
In this chapter we review the quantum-mechanical concepts that are required to understand the Quantum Extreme Learning Machine framework introduced in the following chapters, with particular emphasis on quantum dynamics and measurement as tools for feature generation.
\section{Quantum states and Hilbert space structure}
\label{sec:quantum_states}

The description of quantum systems adopted in this thesis is based on the formalism of state vectors in Hilbert space \cite{Nielsen_Chuang_2010, Sakurai_Napolitano_2020}. In this section we recall the essential concepts required to understand the functioning of Quantum Extreme Learning Machines, with particular emphasis on the role of Hilbert space as a representation space for information.

\begin{figure}[t]
\centering
\begin{subfigure}[t]{0.46\linewidth}
\centering
\begin{blochsphere}[radius=2.1cm, opacity=1]

  \fill[cyan!40, opacity=0.5] (0,0) circle (2.1cm);
  \draw[->] (0,0) -- (2.4,0) node[right] {$x$};
  \draw[->] (0,0) -- (0,2.4) node[above left] {$z$};
  \draw[->] (0,0) -- (-1.6,-1.2) node[left] {$y$};

  \draw[-Latex, very thick] (0,0) -- (1.2,1.6) node[above right] {$|\psi\rangle$};

  \node[above left] at (-0.35,2.25) {$|0\rangle$};
  \node[below left] at (-0.35,-2.25) {$|1\rangle$};
\end{blochsphere}
\caption{Single-qubit state on the Bloch sphere.}
\label{fig:bloch}
\end{subfigure}
\hfill
\begin{subfigure}[t]{0.50\linewidth}
\centering
\begin{tikzpicture}[
  box/.style={draw, rounded corners, align=center, inner sep=6pt},
  arrow/.style={-Latex, thick},
  lab/.style={font=\small}
]
\node[boxc] (q1) {$\mathcal{H}_1 \simeq \mathbb{C}^2$\\\texttt{1 qubit}};
\node[boxc, below=7mm of q1] (q2) {$\mathcal{H}_2=(\mathbb{C}^2)^{\otimes 2}$\\\texttt{2 qubits}};

\node[boxc, below=7mm of q2] (qN) {$\mathcal{H}_N=(\mathbb{C}^2)^{\otimes N}$\\\texttt{$N$ qubits}};
\draw[arrow] (q1) -- node[right,lab] {$\otimes\,\mathbb{C}^2$} (q2);
\draw[arrow] (q2) -- node[right,lab] {$\otimes\,\mathbb{C}^2$} (qN);

\node[lab, right=10mm of q1] (d1) {$\dim(\mathcal{H}_1)=2$};
\node[lab, right=10mm of q2] (d2) {$\dim(\mathcal{H}_2)=4$};
\node[lab, right=10mm of qN] (dN) {$\dim(\mathcal{H}_N)=2^N$};

\end{tikzpicture}
\caption{Exponential growth of Hilbert space with the number of qubits.}
\label{fig:hilbert}
\end{subfigure}

\caption{Geometric representation of a single-qubit state and schematic illustration of the exponential growth of Hilbert space for multi-qubit registers.}
\label{fig:bloch_multiqubit}
\end{figure}
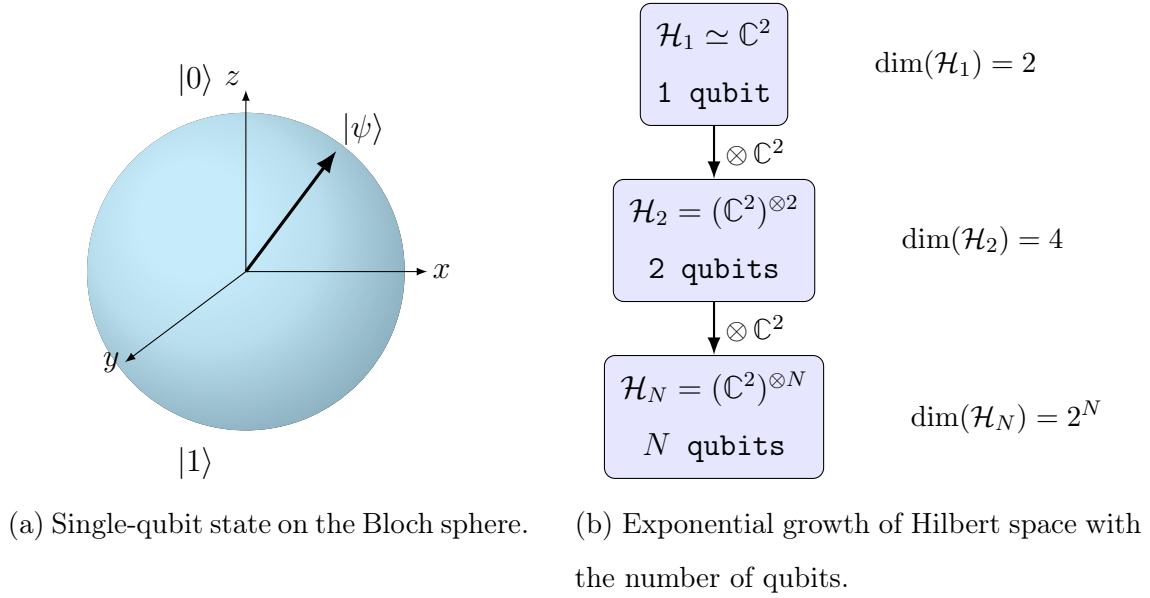

The fundamental unit of quantum information is the qubit, whose state is represented by a normalized vector in a two--dimensional complex Hilbert space, $\mathcal{H} \cong \mathbb{C}^2$. Once the computational basis $\{ |0\rangle, |1\rangle \}$ is fixed, a generic pure state can be written as
\begin{equation}
|\psi\rangle = \alpha |0\rangle + \beta |1\rangle,
\end{equation}
where the complex coefficients $\alpha$ and $\beta$ satisfy the normalization condition $|\alpha|^2 + |\beta|^2 = 1$. A measurement in the computational basis yields the outcomes $|0\rangle$ or $|1\rangle$ with probabilities determined by the Born rule \cite{Sakurai_Napolitano_2020, Nielsen_Chuang_2010}. From a geometrical point of view, pure states of a single qubit can be represented on the Bloch sphere \cite{Nielsen_Chuang_2010}, as shown in Fig.~\ref{fig:bloch}, providing an intuitive picture of how continuous variables may be naturally associated with parameters of a quantum state.

Systems composed of multiple qubits are described through the tensor product of the individual Hilbert spaces. A register of $N$ qubits is therefore associated with the space
\begin{equation}
\mathcal{H}_N = (\mathbb{C}^2)^{\otimes N},
\end{equation}
whose dimension grows exponentially as $2^N$, as illustrated in Fig.~\ref{fig:hilbert}. A generic state can be expressed as a linear combination of the computational basis states,
\begin{equation}
|\Psi\rangle = \sum_{i=0}^{2^N-1} c_i |i\rangle,
\end{equation}
where the complex coefficients $c_i$ fully characterize the state of the system. This exponential growth of the Hilbert space dimension plays a central role in quantum machine learning, as a relatively small number of qubits allows one to explore very high--dimensional representation spaces, enabling highly non--linear transformations of classical data through quantum evolution.

In general, a quantum state can be described by a density matrix $\rho$, which provides a unified framework for representing both pure states and statistical mixtures. In the following, we will mainly consider pure states and ideal unitary evolutions; nevertheless, the density matrix formalism \cite{Nielsen_Chuang_2010, Sakurai_Napolitano_2020} constitutes the more general theoretical framework in which these descriptions are embedded.

A distinctive feature of multi--qubit systems is the possibility of generating entanglement \cite{Nielsen_Chuang_2010}, namely quantum correlations that cannot be reduced to products of local states. A state is said to be separable if it can be written as a tensor product of single--qubit states; otherwise, it is entangled. The presence of entanglement reflects the global structure of the Hilbert space and plays an important role in the models studied in this thesis, where quantum dynamics creates correlations among qubits and contributes to the generation of more expressive data representations.

From a machine learning perspective, Hilbert space can be interpreted as a high--dimensional feature space. The association of classical data with quantum states can be viewed as an embedding into a complex vector space, while unitary evolution acts as a transformation that reorganizes information within this space. In this conceptual framework, the quantum system functions as a generator of non--linear features, whereas the information exploited by classical learning algorithms is extracted through measurement operations, as will be discussed in the following sections.

\section{Quantum circuits and gate-based computation}

One of the most widely used descriptions of quantum computation \cite{Nielsen_Chuang_2010} is the circuit model, in which information processing is implemented through the sequential application of unitary operations, known as quantum gates, followed by measurement. Within this framework, the initial state of a register of qubits is prepared in a reference configuration, typically the computational basis, and subsequently evolved through reversible linear transformations acting on the Hilbert space of the system.

Mathematically, each quantum gate is represented by a unitary operator $U$ that preserves the norm of the state and describes a deterministic and reversible evolution. Single--qubit gates constitute the elementary building blocks of quantum circuits and include, among others, the Pauli operators \cite{Nielsen_Chuang_2010} and rotations around the axes of the Bloch sphere. The Pauli matrices are defined as
\[
\sigma_x =
\begin{pmatrix}
0 & 1 \\
1 & 0
\end{pmatrix}, \qquad
\sigma_y =
\begin{pmatrix}
0 & -i \\
i & 0
\end{pmatrix}, \qquad
\sigma_z =
\begin{pmatrix}
1 & 0 \\
0 & -1
\end{pmatrix},
\]
and generate rotations in the state space through unitary operators of the form
\[
R_{\alpha}(\theta) = e^{-i \theta \sigma_{\alpha}/2}, \qquad \alpha \in \{x,y,z\}.
\]
These operators play a central role both in the circuit model and in the description of spin Hamiltonians, which will be considered in the following chapters.

In order to create correlations between different qubits, it is necessary to introduce two--qubit gates capable of generating entanglement. A paradigmatic example is the controlled-NOT (CNOT) gate, which applies a conditional transformation depending on the state of a control qubit. The set composed of arbitrary single--qubit gates together with at least one universal entangling operation is sufficient to approximate any unitary transformation on a finite number of qubits, making the circuit model computationally universal.

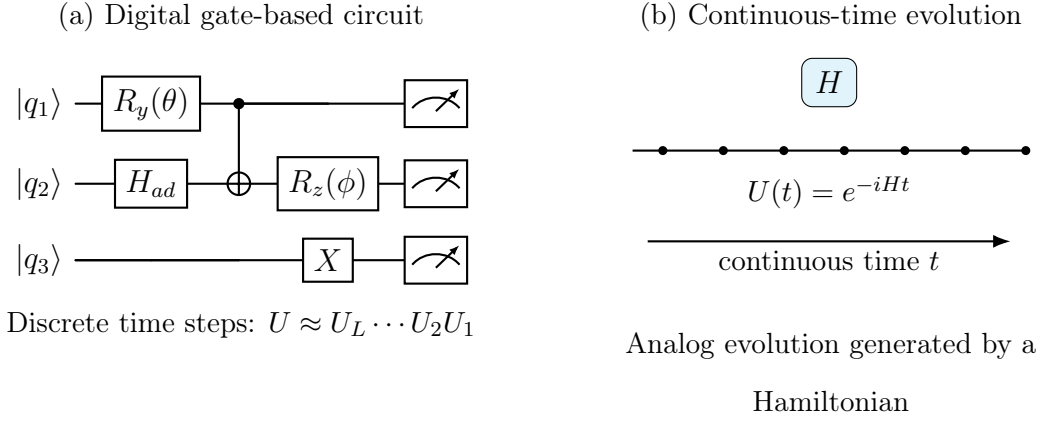
\begin{figure}[t]
\centering

\begin{subfigure}[t]{0.48\linewidth}
\centering
\caption{Digital gate-based circuit}
\vspace{2mm}

\begin{quantikz}[row sep=0.35cm, column sep=0.35cm]
\lstick{$\ket{q_1}$} & \gate{R_y(\theta)} & \ctrl{1} & \qw      & \meter{} \\
\lstick{$\ket{q_2}$} & \gate{H_{ad}}          & \targ{}  & \gate{R_z(\phi)} & \meter{} \\
\lstick{$\ket{q_3}$} & \qw               & \qw      & \gate{X} & \meter{}
\end{quantikz}

\vspace{2mm}
{\small Discrete time steps: $U \approx U_L \cdots U_2 U_1$}
\end{subfigure}
\hfill
\begin{subfigure}[t]{0.48\linewidth}
\centering
\caption{Continuous-time evolution}
\vspace{2mm}

\begin{tikzpicture}[
  dot/.style={circle, fill, inner sep=1.2pt},
  arrow/.style={-Latex, thick},
  lab/.style={font=\small},
  hbox/.style={draw, rounded corners, fill=cyan!10, inner sep=5pt}
]
\draw[thick] (0,0) -- (5.2,0);
\foreach \x in {0.4,1.2,2.0,2.8,3.6,4.4,5.2} {
  \node[dot] at (\x,0) {};
}
\node[hbox] (H) at (2.6,0.9) {$H$};
\node[lab] at (2.6,-0.55) {$U(t)=e^{-iHt}$};

\draw[arrow] (0.2,-1.2) -- (5.0,-1.2);
\node[lab] at (2.6,-1.45) {continuous time $t$};
\end{tikzpicture}

\vspace{2mm}
{\small Analog evolution generated by a Hamiltonian}
\end{subfigure}

\caption{Schematic comparison between gate-based digital quantum computation and continuous-time Hamiltonian evolution. In panel (a), $R_y(\theta)$ and $R_z(\phi)$ denote single-qubit rotations around the $y$- and $z$-axes of the Bloch sphere, $H_{ad}$ denotes the Hadamard gate, and $X$ the Pauli-$X$ gate. In panel (b), $H$ denotes the Hamiltonian generating the continuous-time evolution $U(t)=e^{-iHt}$. See Ref.~\cite{Nielsen_Chuang_2010} for standard definitions of elementary quantum gates.}
\label{fig:digital_vs_continuous}
\end{figure}

Although the discrete gate-based formalism represents the standard framework for digital quantum computation, it is often convenient to consider evolutions generated by continuous-time Hamiltonians. In this case, the evolution of the quantum state is described by a unitary operator of the form
\[
U(t) = e^{-i H t},
\]
where $H$ denotes the Hamiltonian of the system and $t$ the evolution time. Digital quantum circuits can be interpreted as approximations of such continuous dynamics through decompositions into finite sequences of elementary gates. This connection is particularly relevant for the present thesis, where the quantum layer of the Quantum Extreme Learning Machine is mainly described through Hamiltonian dynamics rather than parametrized gate-based circuits. The relation between gate-based digital circuits and continuous-time Hamiltonian evolution is schematically illustrated in Fig.~\ref{fig:digital_vs_continuous}.

The circuit model therefore provides a natural language to describe the quantum processing pipeline: preparation of an initial state, application of unitary transformations, and final measurement. In the following sections this structure will be reinterpreted from a machine learning perspective, where quantum evolution acts as a transformation in feature space and measurement produces the classical quantities exploited by the final classifier.
\section{Quantum dynamics as feature generation}

In the context of quantum machine learning \cite{Biamonte:2016ugo, Schuld:2018gao}, the dynamical evolution of a quantum system can be interpreted as a mechanism for feature generation. Starting from an initial state encoding the input information, unitary dynamics redistributes information within Hilbert space, producing more complex representations that can be exploited by a classical classifier.

From a physical perspective, the evolution of an isolated quantum system is described by a unitary operator generated by a Hamiltonian. Such evolution preserves the global information contained in the quantum state while modifying its distribution across different degrees of freedom. In multi--qubit systems, the dynamics can generate nonlocal correlations and complex structures in state space that are difficult to reproduce using low--dimensional classical transformations.

This viewpoint is closely related to the paradigm of reservoir computing \cite{Fujii2017, jaeger:techreport2001, 10.1162/089976602760407955}, where a dynamical system is used as a high--dimensional feature transformer and training is restricted to a final readout layer. In the quantum setting, the reservoir is represented by the unitary dynamics itself, acting as a fixed transformation in Hilbert space. Unlike variational quantum models \cite{Peruzzo:2013bzg, Farhi:2014ych, Cerezo:2020jpv}, the parameters governing the evolution are not optimized during training but are chosen a priori, allowing for a clearer physical interpretation of the resulting dynamics.

Local Hamiltonians, such as those describing spin chains, provide a natural realization of this framework. In these systems, information initially encoded in local degrees of freedom can propagate along the chain over time, leading to the spreading of correlations and the generation of entanglement between subsystems. The structure of the Hamiltonian and the evolution time therefore determine the complexity of the resulting quantum representation and directly influence the model's ability to separate data classes in feature space.

An important aspect of quantum dynamics is that, although unitary evolution is linear at the level of state vectors, it induces highly non--linear transformations when observed through measurement probabilities. In this sense, the nonlinearity required for machine learning arises from the interplay between unitary evolution and measurement, rather than from explicit activation functions as in classical neural networks.

Within the framework of Quantum Extreme Learning Machines \cite{article, Sakurai:2022ala, Innocenti2023, Xiong:2023qqh}, quantum dynamics can therefore be interpreted as a feature enrichment process, mapping initially simple data into distributed and correlated representations. In the following chapters, it will be shown how dynamical properties of the system, in particular entanglement generation and information propagation, are closely connected to classification performance and to the classical simulability of the model.
To better contextualize the role of quantum dynamics within machine learning
models, it is useful to briefly discuss which physical properties of quantum
systems may contribute to the expressive power of data representations.

\subsection{Expressive power of quantum systems for information processing}

The potential interest of quantum computing for machine learning does not
necessarily rely on asymptotic computational speedups, but rather on the
structure of quantum state spaces and their dynamics. Quantum systems naturally
operate in Hilbert spaces whose dimension grows exponentially with the number
of qubits, providing high--dimensional representations that can be interpreted
as implicit feature spaces \cite{Schuld:2018uel} in which classical information is embedded and
transformed.

In the context of this thesis, this perspective is not interpreted as a claim of
universal quantum advantage, but rather as a guiding principle for understanding
how quantum dynamics can enhance feature representations within the Quantum
Extreme Learning Machine framework.

\section{Quantum measurement and classical readout}

The final stage of the quantum processing pipeline is the measurement of the quantum state and the extraction of classical information. Measurement provides the interface between the quantum system and classical machine learning algorithms, converting the quantum representation generated by the dynamics into a set of classical features that can be processed by standard classifiers \cite{Biamonte:2016ugo, Schuld:2018gao}.

In quantum mechanics, measurement is described by a set of measurement operators acting on the Hilbert space of the system. In the simplest and most commonly adopted scenario, the system is measured in the computational basis, producing a probability distribution over the basis states. For a quantum state $|\Psi\rangle$ expressed in the computational basis, the measurement produces a probability distribution over the basis states, which encodes information about the amplitudes and correlations generated during the unitary evolution.

From the perspective of quantum machine learning, these measurement probabilities can be interpreted as a classical feature vector associated with the input data. This quantum to classical step is what makes the information generated by the unitary evolution accessible to a standard classical classifier.

Different choices of measurement schemes correspond to extracting different types of information from the quantum state. Local measurements probe individual qubits or small subsystems, while global measurements access collective properties of the system. In many--body quantum systems, global observables can capture correlations and structures that are not accessible through purely local measurements. The choice of measurement strategy therefore directly affects the quality and expressiveness of the resulting classical representation.

Within the framework of Quantum Extreme Learning Machines, the readout stage is entirely classical. The quantum system acts as a fixed feature generator, producing a set of features through state preparation, unitary evolution, and measurement, while the classification task is performed by a classical algorithm trained on the measurement outcomes. This separation between a fixed quantum feature generator and a trainable classical readout layer is a defining characteristic of the QELM approach and allows for a clear interpretation of the respective roles of quantum dynamics and classical learning.
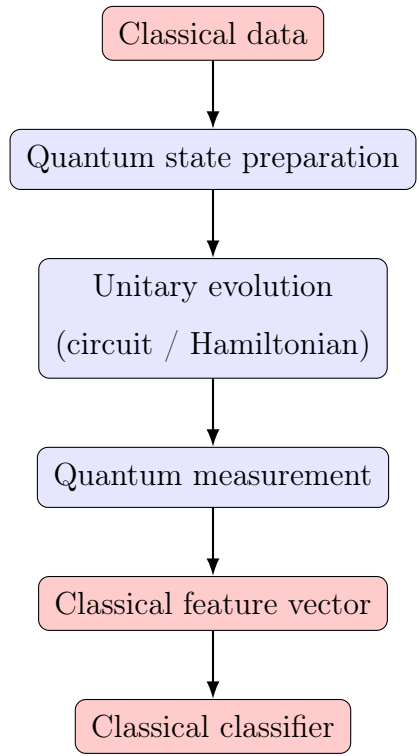
\begin{figure}[t]
\centering
\begin{tikzpicture}[
  node distance=9mm,
  box/.style={draw, rounded corners, align=center, inner sep=6pt, text width=9cm},
  arrow/.style={-Latex, thick}
]
\node[boxcc] (cdata) {Classical data};
\node[boxc, below=of cdata] (prep) {Quantum state preparation};
\node[boxc, below=of prep] (evol) {Unitary evolution\\(circuit / Hamiltonian)};
\node[boxc, below=of evol] (meas) {Quantum measurement};
\node[boxcc, below=of meas] (feat) {Classical feature vector};
\node[boxcc, below=of feat] (clf)  {Classical classifier};

\draw[arrow] (cdata) -- (prep);
\draw[arrow] (prep) -- (evol);
\draw[arrow] (evol) -- (meas);
\draw[arrow] (meas) -- (feat);
\draw[arrow] (feat) -- (clf);
\end{tikzpicture}
\caption{Schematic representation of the quantum--classical processing pipeline underlying Quantum Extreme Learning Machines.}
\label{fig:qelm_pipeline}
\end{figure}

The measurement process thus completes the quantum--to--classical pipeline, schematically summarized in Fig.~\ref{fig:qelm_pipeline}. Together with state preparation and unitary evolution, these elements define a structured mapping from input data to classical feature vectors, which will be explicitly constructed and analyzed in the following chapters.

These elements define the quantum processing pipeline underlying Quantum Extreme Learning Machines, which will be formally introduced and discussed in the next chapter.
\newpage
\chapter{Quantum Extreme Learning Machines}
\label{ch:QELM}
\section{Introduction to Quantum Extreme Learning Machines}
\label{ch:QELM_intro}
As discussed in Chapter~\ref{ch:ml_background}, classical machine
learning provides powerful tools for extracting patterns and predictive
rules from complex data sets. Quantum machine learning (QML) \cite{Schuld:2018gao, Schuld:2018uel, Biamonte:2016ugo, Congarticle, Liu:2022bpb, Cerezo:2022nvi, Jerbi:2021qnz, Huang:2020sqy, McClean:2018jps, Wang:2020yjh, Garcia-Beni:2024hgl, Dawid:2022fga, Perez-Salinas:2019pjx, Franceschetto:2024xqj, book} aims to extend this idea to the setting of quantum devices, seeking to exploit
quantum resources to process and analyse classical or quantum data in a
way that is potentially more efficient or more expressive than purely
classical algorithms. In analogy with traditional machine learning, QML
studies models in which part of the learning process is delegated to a
quantum system, which acts either as a nonlinear feature transformer or
as a predictive model in its own right.

In the era of noisy intermediate-scale quantum (NISQ) devices, most
practically relevant approaches follow a hybrid strategy, in which a
quantum circuit is embedded into a larger classical pipeline \cite{Preskill:2018jim}. Typical
examples include variational quantum circuits and quantum neural
networks, where a parameterized quantum model is trained by a classical
optimizer using gradient-based or gradient-free procedures ~\cite{Schuld:2018gao}.
In these schemes, the parameters of the quantum gates are updated
iteratively, in close analogy with classical neural networks, with the
goal of minimizing a cost function defined on the measurement outcomes.

Quantum Extreme Learning Machines constitute an alternative
family of hybrid models, inspired by classical Extreme Learning
Machines (ELMs) \cite{HUANG2006489,Innocenti2023, Huang_extreme1, Huang_extreme2, Huang_extreme3}. In a classical ELM,
the parameters of the hidden layer are fixed at random and never
trained; only the weights of the output layer are optimized, usually
by means of a simple linear model acting on the feature space. The
underlying idea is to separate the problem of \emph{feature generation} (carried out by a static, possibly overparameterized mapping) from
the problem of \emph{learning}, which is confined to a shallow
classifier operating on top of these features.

In a QELM, the role of the hidden layer is played by a quantum
many-body system that processes classically encoded data through its
unitary dynamics. The classical inputs are first mapped to an initial
state of an $N$-qubit register; this state is then evolved under a fixed
Hamiltonian for a certain time, and finally measured. The resulting
measurement statistics define a new set of classical features, which are
fed into a simple classical output layer, typically a one-layer neural
network (ONN). Crucially, the parameters of the quantum layer (encoding
map, Hamiltonian, and measurement basis) are kept fixed: all training
is restricted to the classical readout \cite{Innocenti2023,DeLorenzis_1, Solanki:2025xji, Assil:2025zhc, QELMSuprano, DeLorenzis_IQIS2025, Vetrano:2024vbh}.

This design places QELMs conceptually close to quantum reservoir
computing (QRC) \cite{Ghosh2019, Ghosh2021, HoanTran2023, Llodra2023, Govia2021, nakajima2019boosting, nokkala2020gaussian, Dudas2023, Gotting2023, Palacios:2024tlc, Llodra:2024qhi, Hou:2025yly, Ivaki:2024dao, Settino:2024qvi, Das:2025gax, Zia:2025fyd}, where a dynamical quantum system acts as a reservoir
that transforms inputs into a high-dimensional feature space, while only
the linear readout is trained \cite{Fujii2017}. In contrast to
fully variational models, no backpropagation through the quantum circuit
is required, and the quantum resources are used solely for feature
generation. This makes QELMs particularly attractive in the NISQ regime,
where circuit depth and control precision are limited and where
classical training remains comparatively cheap.

The present thesis focuses on QELM architectures applied to supervised
classification tasks, with a particular emphasis on image classification
and on the role of entanglement and information propagation within the
quantum layer. The analysis presented here builds on the author’s previous work reported in Refs.~\cite{DeLorenzis_1,DeLorenzis_2}. In particular, we investigate both how the
choice of classical pre-processing and data encoding affects the
performance of QELMs, and how the dynamical properties of the underlying Hamiltonian relate to the expressiveness and (classical) simulability of the resulting model.

\section{General architecture of the QELM model}
\label{sec:architecture_QELM}

Although different implementations may differ in details, all QELM models considered in this thesis share a common architectural pattern.
They are organized as a sequence of classical and quantum stages that
transform an input sample (for instance, an image) into a
classification label. The overall structure can be summarized as a
three-layer hybrid model:
\begin{enumerate}
  \item a \emph{classical pre-processing stage}, which reduces the
        dimensionality of the input and brings it into a format
        compatible with the quantum device;
  \item a \emph{quantum feature map}, implemented by the dynamics of an
        $N$-qubit system evolving under a fixed Hamiltonian;
  \item a \emph{classical readout layer}, which operates on
        measurement-derived features and produces the final prediction.
\end{enumerate}

\subsection*{Workflow of a QELM}
\label{sec:QELMs}
\begin{figure*}[htbp]
    \centering
\includegraphics[width=0.95\textwidth]{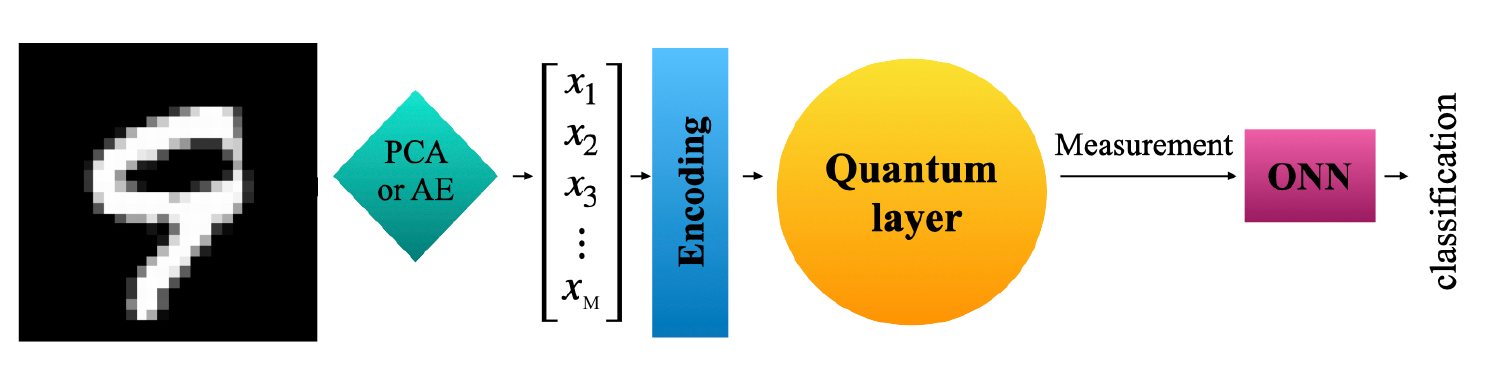}
    \caption{A schematic representation of the QELM. The workflow is as follows: feature reduction through PCA or AE; encoding of classical data into  the quantum initial state; time evolution via the quantum layer; measurement of the evolved quantum state; classification with a classical One-layer Neural Network. Adapted from the author’s previous work, Ref.~\cite{DeLorenzis_1}. Copyright (2025) by the American Physical Society.}
    \label{fig:QELM model}
\end{figure*}

A schematic representation of the QELM architecture is shown in Fig. \ref{fig:QELM model} reproduced from the author’s previous work in Ref.~\cite{DeLorenzis_1}. Given an input sample, the processing pipeline proceeds as follows.
\begin{enumerate}
  \item \textbf{Feature reduction.} High-dimensional inputs, such as images, are first mapped onto a low-dimensional latent vector compatible with the number of available qubits and with the chosen encoding strategy. This compression step is necessary because current NISQ devices still provide limited usable quantum resources: in practice, both the qubit count and, even more importantly, the circuit depth are constrained by noise, finite coherence times, and imperfect gate operations \cite{Preskill:2018jim}. In this thesis, dimensionality reduction is performed using either Principal Component Analysis (PCA) or nonlinear autoencoders (AEs), which are described in more detail in Sec.~\ref{sec:feature_reduction}.
  \item \textbf{Encoding in quantum states.} The latent features are encoded into the initial state of an
  $N$-qubit system through a suitable quantum feature map, producing a pure state $\ket{\psi_0}$.

  \item \textbf{Quantum evolution.} The encoded state is evolved under a fixed Hamiltonian (the quantum layer) for a prescribed evolution time, to a final state in the Hilbert space.

  \item \textbf{Measurement.} The evolved state is measured in a convenient basis, and the resulting measurement statistics are interpreted as a new classical feature vector.

  \item \textbf{Classical output layer.} Finally, the measurement-derived features are fed to a simple one-layer neural network (ONN), whose weights constitute the only trainable parameters of the model and are optimized for the target classification task.
\end{enumerate}

\section{Feature reduction} \label{sec:feature_reduction}
Real-world datasets, and in particular image datasets such as MNIST \cite{mnist}, Fashion-MNIST \cite{fashionmnistnovelimagedataset} and CIFAR-10 \cite{cifar10}, which we used as benchmark datasets in our work, live in very high-dimensional spaces: a single grayscale image of size $28 \times 28$ already corresponds to $784$ real features, while RGB images are even larger. On current and near-term quantum hardware, the number of qubits is severely limited and does not allow one to encode all these features directly into the quantum system. Therefore, the first step consists in compressing the data into a lower-dimensional latent representation.\\ 

\subsection{Principal component analysis}

Principal Component Analysis (PCA) is a widely used technique for dimensionality reduction, feature extraction and data visualization. The basic idea is to transform a set of possibly correlated variables into a new set of linearly uncorrelated variables, which are named as the principal components, ordered by the amount of variance they explain in the data. By retaining only the first few principal components, one obtains a lower-dimensional representation that preserves most of the relevant information.

PCA is particularly useful when dealing with high-dimensional datasets, where it can help suppress noise, reduce computational cost, and reveal underlying structures that may not be immediately evident in the original variables. In addition to these advantages, PCA is often employed as a preprocessing step for subsequent analyses. This is precisely the reason why it has been adopted in the work described in this thesis: since the dataset must be fed to a quantum layer with a limited number of qubits, we cannot afford large quantum dimensionalities and are thus forced to reduce the number of input features, while at the same time preserving as much relevant information as possible.

Before presenting the mathematical formulation of PCA, it is useful to build some intuition in a simple two-dimensional example.

\begin{figure}
	\centering
	\includegraphics[width=0.485\linewidth]{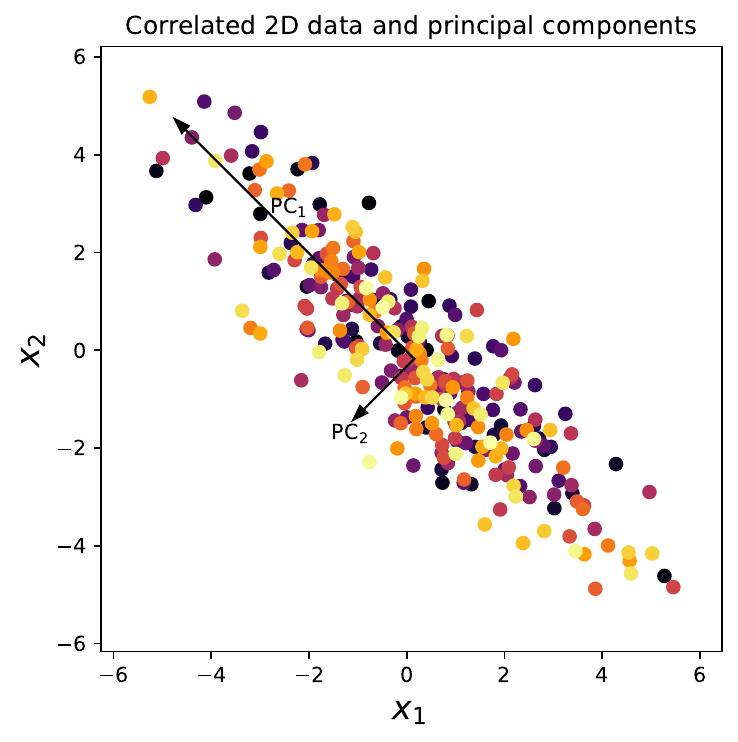}
	\caption{Correlated 2D data and the corresponding principal components. Dropping unexpressive information corresponds to projecting the data onto $\textrm{PC}_1$ and discarding $\textrm{PC}_2$.
	\label{fig:PCAexample}}
\end{figure}

\begin{figure}
	\centering
	\includegraphics[width=1\linewidth]{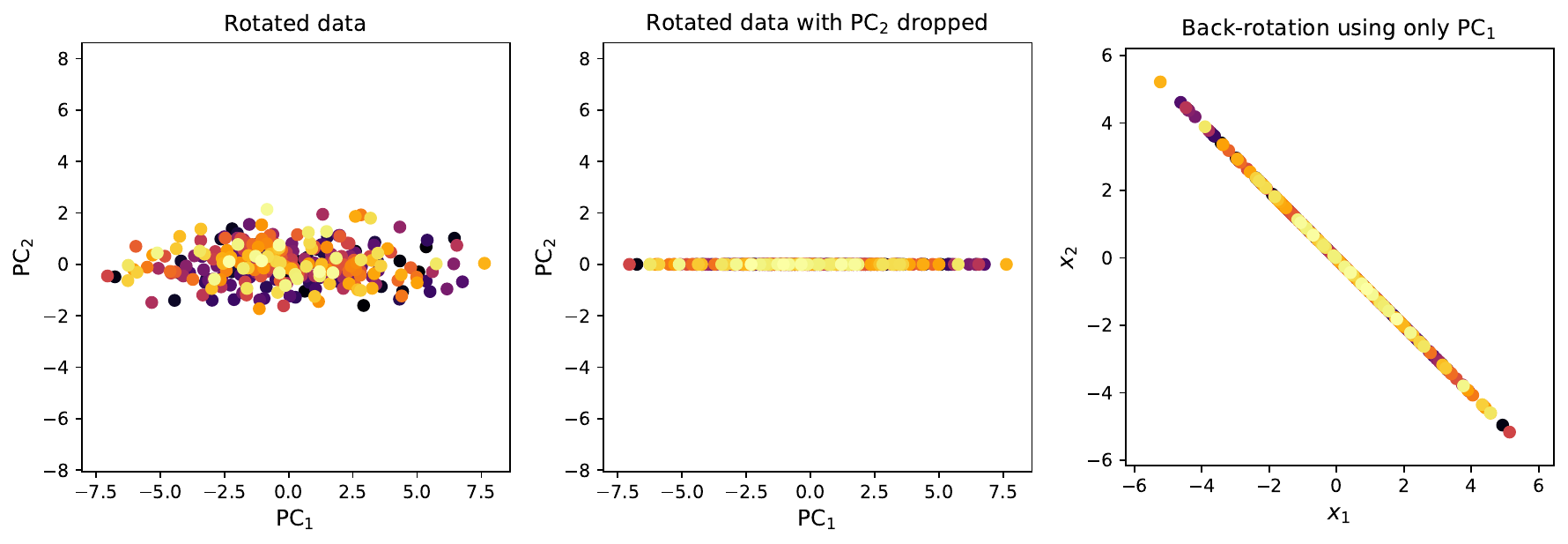}
	\caption{Visualization of PCA as a rotation and projection in two dimensions. Left: data expressed in the principal component basis ($\textrm{PC}_1$, $\textrm{PC}_2$), where the axes are uncorrelated and aligned with the directions of maximum and residual variance. Middle: projection onto the first principal component, obtained by discarding the least expressive direction. Right: projected data rotated back to the original feature space using only $\textrm{PC}_1$.
	\label{fig:PCAsubplots}}
\end{figure}

We may consider, as an example, a dataset consisting of points in a 2D plane, each point representing one event with two features, $x_1$ and $x_2$. The extension to the multi-dimensional case is trivial. If the two features are strongly correlated, the points will tend to cluster along an elongated distribution oriented along some diagonal direction in the $(x_1, x_2)$ plane, rather than being spread uniformly. In a scatter plot, this appears as an oblique ellipse rather than a circular blob, see Fig.~\ref{fig:PCAexample} and \ref{fig:PCAsubplots}.

The idea behind PCA is to apply a linear transformation (a rotation of the coordinate axes) so that this elongated distribution is aligned with the new axes. Concretely, we introduce two new coordinates, $\textrm{PC}_1$ and $\textrm{PC}_2$, corresponding to the first and second principal components. The first principal component $\textrm{PC}_1$ is chosen along the direction of maximum variance of the data (the long axis of the ellipse), while the second component $\textrm{PC}_2$ is orthogonal to it and captures the remaining, smaller variance (the short axis of the ellipse). In the transformed ($\textrm{PC}_1$, $\textrm{PC}_2$) plane, the features are uncorrelated by construction, their covariance is zero and the data distribution appears as an axis-aligned ellipse.

The figure also helps to clarify the meaning of “dropping unexpressive information”. In the 2D example, most of the variability of the data lies along $\textrm{PC}_1$, while $\textrm{PC}_2$ carries comparatively little variance and encodes less information. If we project all points onto the $\textrm{PC}_1$ axis and discard $\textrm{PC}_2$, we effectively collapse the ellipse onto its long axis. The resulting one-dimensional representation loses some information but retains the dominant structure of the dataset. In higher dimensions, PCA performs the same operation: it identifies a set of directions that capture most of the variance and allows us to discard those directions along which the data vary very little. In this sense, PCA replaces a possibly redundant and correlated feature space with a smaller set of uncorrelated features that preserve the most expressive part of the information.

In the following, we provide the details of the implementation of the corresponding algorithm. In practice, PCA is often implemented using the Singular Value Decomposition (SVD) technique. \\
Without loss of generality, we assume that the dataset $X$ is centered, where \(X \in \mathbb{R}^{m \times p}\) with $m$ being the number of datapoints and $p$ the number of features. If that is not the case, we can always subtract the mean $\mathbf{\bar x} = \sum_{i = 1}^m \mathbf{x}^{(i)} /m$. \\
The SVD of $X$ can be written as
\begin{equation}
    X = U S V^{\top},
\end{equation}
where
\begin{itemize}
    \item \(U \in \mathbb{R}^{m \times m}\) is an orthogonal matrix whose columns are called the left singular vectors,
    \item \(V \in \mathbb{R}^{p \times p}\) is an orthogonal matrix whose columns are called the right singular vectors,
\end{itemize}
and $U$ and $V$ are chosen such that \(S \in \mathbb{R}^{m \times p}\) is a (rectangular) diagonal matrix. The corresponding non-zero entries \(s_1 \geq s_2 \geq \dots \geq 0\) are the singular values of \(X\).

The correlations among the features are described by the covariance matrix of the centered data which is given by
\begin{equation}
    \Sigma =  X^{\top} X.
\end{equation}
Substituting the SVD of \(X\), one obtains
\begin{equation}
    \Sigma =  X^{\top} X
      =  V S^{\top} U^{\top} U S V^{\top}
      =  V \left( S^{\top} S \right) V^{\top}.
\end{equation}
Since \(S^{\top} S\) is diagonal with entries \(s_k^2\), this shows that the columns of \(V\) are the eigenvectors
of the covariance matrix \(\Sigma\), and the corresponding eigenvalues are
\begin{equation}
    \lambda_k = s_k^2, \qquad k = 1, \dots, p.
\end{equation}

Therefore, the principal directions used in PCA are given by the right singular vectors of the data matrix,
and the variance along each principal component is proportional to the square of the corresponding singular value. In order to reduce the dimensionality of the original dataset $X$ one can simply multiply it with the matrix $V_d$ as 
\begin{equation}
\tilde X = X V_d, 
\end{equation}
where $V_d$ is defined as the matrix containing the first $d$ principal components, namely the matrix composed of the first $d$ columns of $V$. As a result \(\tilde X \in \mathbb{R}^{m \times d}\) and describes a dataset with $m$ datapoints, each characterized by $d < p$ features.

Notice that PCA is inherently a linear method and may fail to capture nonlinear structures present in the data. In such cases, the autoencoders may be more appropriate. We will explore these possibilities in a subsequent section.

\subsection{Autoencoders for Nonlinear Dimensionality Reduction}

While PCA provides a simple and powerful \emph{linear} method for dimensionality reduction,
many datasets are naturally characterized by nonlinear relations among their features in a high-dimensional space.
In such cases, linear projections may fail to capture the most relevant structures and autoencoders usually provide an efficient alternative that can learn nonlinear low-dimensional representations directly from the data.

An autoencoder is a neural network trained to reproduce its input at the output.
It consists of two main parts: an \emph{encoder} that maps the input to a lower-dimensional
latent space and a \emph{decoder} that reconstructs the input from this latent representation.
By constraining the dimension of the latent space to be smaller than that of the input,
the network is forced to learn a compact representation that preserves the information
necessary for reconstruction.

\begin{figure}
	\centering
    \includegraphics[width=0.7\linewidth]{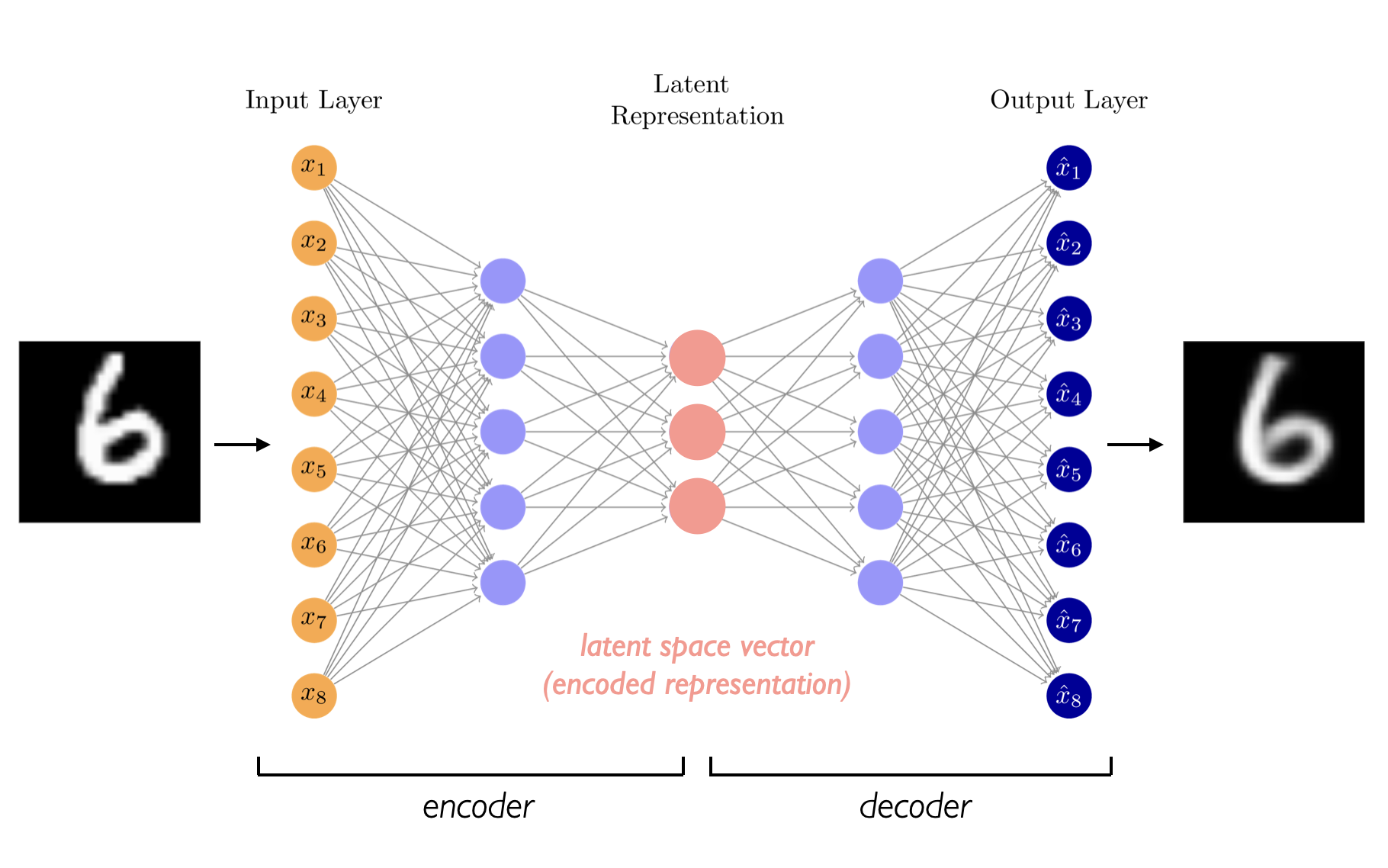}
	\caption{Schematic representation of an autoencoder architecture. The input image is flattened into the input layer and passed through a sequence of hidden layers forming the encoder. The central bottleneck layer encodes the data into a low-dimensional latent representation (latent space vector). The decoder then maps this latent code back to the output layer, which reconstructs an approximation of the original image.
	\label{fig:autoencoder}}
\end{figure}

The overall information flow in an autoencoder is illustrated schematically in Fig.~\ref{fig:autoencoder}. The input data are first mapped by the encoder into
a progressively more compact internal representation, culminating in a
low-dimensional latent vector that is meant to capture the most relevant
features of the data. This latent code is then passed to the decoder, which
aims to reconstruct the original input as accurately as possible. During the
training process, the neural network parameters are adjusted so as to minimize the discrepancy between the input and the reconstruction, thereby forcing the latent representation to retain exactly the information that is most relevant for reproducing the data. In the following, this general scheme will be realized using convolutional autoencoders, where both the encoder and decoder are built from convolutional layers acting directly on images rather than on a flattened feature vector. 

In the following we provide a more detailed description of how an autoencoder works. \\
Let $\mathbf{x} \in \mathbb{R}^p$ denote an input feature vector.
The encoder is a nonlinear model, usually represented by a neural network, 
\begin{equation}
    \mathbf{z} = f_{\phi}(\mathbf{x}) \in \mathbb{R}^{d},
\end{equation}
parameterized by weights and biases $\phi$, which maps $\mathbf{x}$ to a latent vector
$\mathbf{z}$ of dimension $d < p$.
The decoder is another nonlinear model defined by
\begin{equation}
    \hat{\mathbf{x}} = g_{\theta}(\mathbf{z}) \in \mathbb{R}^p,
\end{equation}
and parameterized by $\theta$, which attempts to reconstruct the original input from the
latent representation.
The layer containing $\mathbf{z}$ is often referred
to as the \emph{bottleneck layer}, and the components of $\mathbf{z}$ are interpreted as the learned compressed features.

Given a dataset $X$, the autoencoder parameters $(\phi, \theta)$
are learned by minimizing a reconstruction loss that measures the discrepancy between
the input $\mathbf{x}$ and its reconstruction
$\hat{\mathbf{x}} = g_{\theta}(f_{\phi}(\mathbf{x}))$.
A common choice is the mean squared error
\begin{equation}
    \mathcal{L}(\phi, \theta)
    = \frac{1}{m} \sum_{i=1}^m
    \left\| \mathbf{x}^{(i)} -
    g_{\theta}\big(f_{\phi}(\mathbf{x}^{(i)})\big) \right\|^2.
\end{equation}
The optimization is typically carried out by gradient descent. Once training is complete, the encoder $f_{\phi}$ defines a mapping from the original
feature space to a lower-dimensional latent space, which can be used as a nonlinear
dimensionality reduction technique. The decoder $g_{\theta}$ provides a way to interpret
and visualize latent points by mapping them back to the original space. \\

Autoencoders can be seen as a nonlinear generalization of PCA.
In the special case where both the encoder and decoder consist of a single linear layer
and the loss is mean squared error, the optimal solution spans the same subspace as
the one found by PCA: the encoder learns projections onto the principal components,
and the decoder reconstructs from them. When nonlinear activation functions and
multiple layers are introduced, however, autoencoders gain the ability to approximate
complicated nonlinear data. From the perspective of dimensionality reduction,
PCA finds an \emph{orthogonal linear} subspace that preserves as much variance as possible, while an autoencoder, by contrast, learns a \emph{nonlinear} mapping that preserves information relevant for reconstruction, possibly capturing structures that cannot be represented by any linear subspace of the same dimension.
This greater flexibility comes at the cost of increased model complexity, the need
for training, and a dependence on hyperparameters (architecture, learning rate, etc.). 

As for the PCA, beyond pure visualization or compression, autoencoders are particularly useful as a
preprocessing step. Once trained, the encoder provides a compact
latent representation $\mathbf{z} = f_{\phi}(\mathbf{x})$ that can be used as input to other models. In the context of the present work, this is especially relevant because the data must
be fed to a quantum layer with a limited effective dimensionality: the number of available
qubits constrains the number of input features that can be encoded.
By training an autoencoder to reconstruct the original dataset and then retaining only
the bottleneck representation, we obtain a reduced feature vector of dimension
$d$ that fits the quantum hardware constraints. The goal is to choose
$d$ small enough to be compatible with the number of qubits, yet large
enough to preserve as much relevant information as possible.

\section{Data encoding and quantum representations}
\label{ch:encoding}

The performance of quantum machine-learning models crucially depends on
how classical data are embedded into quantum states. This process,
generally referred to as \emph{data encoding} or \emph{state
preparation}, defines the interface between the classical and quantum
layers of a hybrid algorithm. In the context of the Quantum Extreme
Learning Machine (QELM), the encoding stage determines how the input
features are represented in Hilbert space, thereby influencing the
model's expressive power and its resilience to noise.

Unlike classical neural networks, where data are directly processed as
real-valued vectors, quantum models must translate these values into
amplitudes or phases of quantum states. Different encoding schemes offer
distinct trade-offs between representational capacity, resource
efficiency, and experimental feasibility. Some encodings are designed to
maximize the number of features per qubit, while others focus on
limiting circuit depth or ensuring compatibility with specific hardware
constraints.

This section provides an overview of the encoding strategies employed in
the QELM model, many of which are also widely used in other quantum
neural-network architectures. The discussion begins with the geometric
representation of a qubit on the Bloch sphere, which serves as a
unifying picture for most angle-based encodings. Subsequently, several
common strategies are described, including dense-angle, simple angle,
uniform Bloch-sphere, general, and amplitude encodings. 

\subsection*{Bloch-sphere representation of qubits}

As recalled in Sec.~\ref{sec:quantum_states}, any single-qubit pure state can be represented as a point on the Bloch
sphere \cite{Nielsen_Chuang_2010},
\begin{equation}
\ket{\psi} = \cos\left(\frac{\theta}{2}\right)\ket{0}
           + e^{{\rm i}\phi}\sin\left(\frac{\theta}{2}\right)\ket{1},
\label{eqn:blochsphere_thesis}
\end{equation}
where $\theta \in [0,\pi]$ and $\phi \in [0,2\pi)$ are the polar and
azimuthal angles, respectively. The north and south poles correspond to
$\ket{0}$ and $\ket{1}$, while generic superpositions populate the
surface of the sphere.

Most of the encodings considered in this thesis exploit this geometric
picture by associating (rescaled) classical features to Bloch-sphere
angles. In this way, real-valued vectors are mapped into quantum states
whose amplitudes and phases are directly controlled by the input data.

\subsection*{Dense-angle encoding}

In \emph{dense-angle encoding}~\cite{LaRose:2020mgo}, both Bloch-sphere
angles are used to encode two features per qubit. Given a classical
feature vector $\mathbf{x} = (x_1,\dots,x_M)$, pairs
$(x_{2i-1}, x_{2i})$ are mapped to the polar and azimuthal angles of
the $i$-th qubit:
\begin{equation}
\ket{\mathbf{x}} = \bigotimes_{i=1}^{\lceil M/2 \rceil}
  \left[
  \cos\left(\frac{x_{2i-1}}{2}\right)\ket{0}
  + e^{{\rm i}x_{2i}}\sin\left(\frac{x_{2i-1}}{2}\right)\ket{1}
  \right].
\label{eqn:denseangle_thesis}
\end{equation}
The features are suitably rescaled so that $x_{2i-1}\in[0,\pi]$ and
$x_{2i}\in[0,2\pi)$, ensuring that the encoded states cover a large
portion of the Bloch sphere for each qubit~\cite{Sakurai:2022ala}. This
encoding doubles the feature-to-qubit ratio compared to simple angle
encoding, at the cost of requiring precise control over the qubit phase.

\subsection*{Angle encoding}

A simpler and widely adopted alternative is \emph{angle encoding}, in
which one feature is associated to the polar angle of each qubit, while
the azimuthal angle is fixed (typically $\phi=0$):
\begin{equation}
\ket{\mathbf{x}} = \bigotimes_{i=1}^{M}
\left[
\cos\left(\frac{x_i}{2}\right)\ket{0}
+ \sin\left(\frac{x_i}{2}\right)\ket{1}
\right].
\label{eqn:angleencoding_thesis}
\end{equation}
After a suitable rescaling, $x_i \in [0,\pi]$ for all $i$. This
representation is common in quantum machine-learning models
(e.g.~\cite{LaRose:2020mgo,Schuld:2018gao,Grant:2018oml,Stoudenmire2016SupervisedLW,PhysRevA.101.052309}) because it yields very shallow circuits (typically one single-qubit
rotation per feature) and reduces the need for precise phase control, which can be challenging on noisy hardware.

\subsection*{Uniform Bloch-sphere encoding}

The \emph{uniform Bloch-sphere encoding}~\cite{Mujal_2021} provides a
convenient parametrization of the Bloch-sphere surface by directly
encoding probability amplitudes and phases. Pairs of normalized features
$(x_{2i-1},x_{2i})$ are interpreted as
\begin{equation}
\ket{\mathbf{x}} = \bigotimes_{i=1}^{\lceil M/2 \rceil}
\left[
\sqrt{x_{2i-1}}\ket{0}
+ e^{{\rm i}x_{2i}}\sqrt{1-x_{2i-1}}\ket{1}
\right],
\label{eqn:uniformbloch_thesis}
\end{equation}
where $x_{2i-1}\in[0,1]$ controls the population of $\ket{0}$ and
$x_{2i}\in[0,2\pi)$ sets the relative phase. In practice, the original
data must be normalized or clipped to enforce these ranges. If the
features are drawn from suitable distributions, this parametrization can
lead to an approximately uniform coverage of the Bloch sphere.

\subsection*{General encoding}

A broader class of single-qubit encodings, often referred to as
\emph{general encodings}~\cite{LaRose:2020mgo}, represents feature pairs
as normalized superpositions of the computational-basis states:
\begin{equation}
\ket{\mathbf{x}} = \bigotimes_{i=1}^{\lceil M/2 \rceil}
\frac{1}{\sqrt{x_{2i-1}^2 + x_{2i}^2}}
\left(x_{2i-1}\ket{0} + x_{2i}\ket{1}\right).
\label{eqn:generalencoding_thesis}
\end{equation}
Here, the relative weight of $\ket{0}$ and $\ket{1}$ directly encodes
the two features $(x_{2i-1},x_{2i})$, without explicitly referring to
Bloch-sphere angles. This family encompasses several possible choices of
rescaling and normalization and can be adapted to the statistics of the
dataset.

\subsection*{Amplitude encoding}

Finally, \emph{amplitude encoding} compactly embeds an entire feature
vector into the amplitudes of a multi-qubit state:
\begin{equation}
\ket{\mathbf{x}} = \frac{1}{\|\mathbf{x}\|_2}
\sum_{i=0}^{2^N-1} x_i \ket{i},
\label{eqn:amplitudeencoding_thesis}
\end{equation}
where $\{\ket{i}\}$ denotes the computational basis of $N$ qubits and
the input vector has been padded or reshaped to length $2^N$ if
necessary. In this way, $2^N$ features are encoded using only $N$
qubits, which is exponentially efficient in terms of qubit count.
However, preparing a generic state of the form
Eq.~\eqref{eqn:amplitudeencoding_thesis} typically requires deep and
highly entangling circuits, making amplitude encoding experimentally
demanding on noisy intermediate-scale quantum (NISQ) devices.

\subsection*{Comparison of encoding strategies}

Each encoding strategy strikes a different balance between representational power,
the number of required qubits, and experimental feasibility. Angle-based schemes (dense-angle, simple angle, and uniform
Bloch-sphere encodings) all operate at the level of single-qubit Bloch
coordinates, but differ in how many features they assign to each qubit
and in how demanding they are in terms of phase control and data
preprocessing.

Dense-angle and uniform Bloch-sphere encodings increase the information
density per qubit by exploiting both Bloch angles to encode two features
per qubit. This comes at the price of more stringent requirements on the
precision of phase rotations. Simple angle
encoding, by contrast, associates a single feature to the polar angle of
each qubit and keeps the azimuthal angle fixed. It therefore minimizes
circuit depth (typically one single-qubit rotation per feature) and
reduces the need for high-fidelity phase control, at the cost of using
twice as many qubits for the same number of features.

General and amplitude encodings complete the picture by highlighting the trade-off between flexibility and qubit efficiency: the former can be adapted to the data distribution, whereas the latter minimizes qubit requirements at the price of more demanding state preparation.
A concise summary of the main properties of these encoding schemes is
reported in Table~\ref{tab:encodings_summary}.

\begin{table}[t]
  \centering
  \footnotesize
  \caption{Summary of the main encoding strategies considered in this thesis.}
  \label{tab:encodings_summary}
  \begin{tabular}{p{3.5cm} || p{1.5cm} || p{1.5cm} || p{6.5cm}}
    \hline
    \hline
    Encoding             & Features per qubit & Typical depth & Qualitative remarks \\
    \hline
    \hline
    Angle                & $1$                & Low           & No data-dependent phase control \\
    Dense-angle          & $2$                & Low           & Requires precise phase rotations \\
    Uniform Bloch-sphere & $2$                & Low           & Requires precise phase rotations \\
    General              & $2$                & Low           & Direct parametrization of amplitudes \\
    Amplitude            & $2^N$              & High          & Generic state preparation, deep entangling circuits \\
    \hline
    \hline
  \end{tabular}
\end{table}

In the remainder of this thesis, we will analyse the Quantum Extreme
Learning Machine algorithm and its classification performance as the
encoding strategy is varied, systematically comparing the impact of the
different schemes described in this section on the resulting quantum
representations of the data.

\section{Hamiltonians for the quantum layer}
\label{sec:hamiltonians}

Once the initial state $\ket{\psi_0(\mathbf{z})}$ has been prepared by
encoding the classical features into an $N$-qubit register, the system
is evolved under a fixed many-body Hamiltonian $H$ for a prescribed
evolution time $t$. The unitary evolution,
\begin{equation}
    U(t) = e^{- {\rm i} H t},
\end{equation}
implements a highly nonlinear feature map in Hilbert space, which plays
the role of the ``quantum reservoir'' in the QELM architecture.
Importantly, all Hamiltonians considered in this thesis are
\emph{fixed}: they contain no trainable parameters and are kept constant
throughout training and testing. The only trainable parameters of the
model reside in the weights of the classical output layer.

In the author’s first study reported in Ref.~\cite{DeLorenzis_1}, several Hamiltonians with qualitatively different dynamics are compared in order to disentangle the role of connectivity, integrability, and localization properties on the classification accuracy. Six interacting Hamiltonians, denoted by $H_1,\dots,H_6$, are analyzed, and for one of them also a non-interacting limit is considered. In the author’s second study reported in Ref.~\cite{DeLorenzis_2}, the quantum layer is instead
restricted to a single local and integrable model, an XX chain, but the dependence of the performance on the evolution time is studied in detail.

\paragraph{Long-range driven Ising model $H_1$.}

The first Hamiltonian is a periodically driven long-range Ising-type
model originally proposed in Ref.~\cite{Sakurai:2022ala} and widely
used in quantum reservoir computing. It alternates, in a stroboscopic
fashion, between a global transverse-field rotation and a long-range
Ising interaction:
\begin{equation}
    H_1(t) =
    \begin{cases}
        H_a = B_1 \displaystyle\sum_{i=0}^{N-1} \sigma_x^{(i)}, & 0 \le t < T_1, \\[6pt]
        H_b = \displaystyle\sum_{i,j=0}^{N-1} J^{(1)}_{ij}\, \sigma_z^{(i)} \sigma_z^{(j)}, & T_1 \le t < T,
    \end{cases}
    \label{eq:H1}
\end{equation}
where the Pauli operators $\sigma^{(i)}_{x,z}$ act on site $i$, and
the long-range couplings decay as
\begin{equation}
    J^{(1)}_{ij} = \frac{J_0}{|i-j|^{\alpha}}.
\end{equation}
Following Ref.~\cite{Sakurai:2022ala}, the parameters are set to
$J_0 = 0.06$, $\alpha = 1.51$, and $B_1 = 3.05$, while the period is chosen as $T = 2 T_1 = 1$, which fixes the units of time and inverse energy. In the simulations, the evolution is run for $\Delta t = 50 T$,
so that the reservoir effectively experiences a long driven dynamics.
Physically, $H_1$ generates strongly interacting dynamics on a fully
connected graph, providing a prototypical example of a ``strongly
mixing'' quantum reservoir with long-range connectivity.

\paragraph{Fully connected disordered transverse-field Ising model $H_2$.}

The Hamiltonian $H_2$ describes a static fully connected Ising model in
a transverse field with quenched disorder:
\begin{equation}
    H_2 =
    \sum_{i,j=0}^{N-1} J^{(2)}_{ij}\, \sigma_z^{(i)} \sigma_z^{(j)}
    + \sum_{i=0}^{N-1} B^{(2)}_i\, \sigma_x^{(i)}.
    \label{eq:H2}
\end{equation}
The couplings and local fields are independently sampled from Gaussian
distributions,
\begin{equation}
    J^{(2)}_{ij} \sim \mathcal{N}(0.75, 0.1), \qquad
    B^{(2)}_i \sim \mathcal{N}(1, 0.1),
\end{equation}
and the evolution time is fixed to $\Delta t = 20$ \cite{Domingo:2022fre, DeLorenzis_1}.
This model generates complex disordered dynamics on a fully connected
graph and allows one to assess how randomness in the interaction
strengths affects the reservoir performance. As a useful reference, a
non-interacting limit is also considered by setting all couplings to
zero, $J^{(2)}_{ij}=0$, in which case $H_2$ reduces to independent spins
in random transverse fields and the dynamics becomes trivially
factorized.

\paragraph{Chaotic nearest-neighbour Ising model $H_3$.}

The Hamiltonian $H_3$ is a one-dimensional Ising-like model with
nearest-neighbour interactions and both longitudinal and transverse
fields:\cite{Xiong:2023qqh, DeLorenzis_1}
\begin{equation}
    H_3 =
    J_3 \sum_{i=0}^{N-2} \sigma_z^{(i)} \sigma_z^{(i+1)}
    + B_{3z} \sum_{i=0}^{N-1} \sigma_z^{(i)}
    + B_{3x} \sum_{i=0}^{N-1} \sigma_x^{(i)}.
    \label{eq:H3}
\end{equation}
The parameters are chosen in a quantum-chaotic regime \cite{Xiong:2023qqh}, with
\begin{equation}
    J_3 = -1, \qquad B_{3x} = 0.7, \qquad B_{3z} = 1.5,
\end{equation}
and the evolution time is again fixed to $\Delta t = 20$. In contrast
to $H_1$ and $H_2$, the connectivity here is local (spins are arranged
on a chain), but the competing interactions and fields produce
non-integrable, ergodic dynamics. This Hamiltonian thus provides a
canonical example of a chaotic local quantum reservoir.

\paragraph{Heisenberg XXZ chain $H_4$.}

As a variation of $H_3$, $H_4$ corresponds to a nearest-neighbour
Heisenberg XXZ spin chain in an external $z$-field:
\begin{equation}
    H_4 = -\frac{1}{2} \sum_{i=1}^{N}
    \Big(
        J_{4x}\, \sigma_x^{(i)} \sigma_x^{(i+1)}
      + J_{4y}\, \sigma_y^{(i)} \sigma_y^{(i+1)}
      + J_{4z}\, \sigma_z^{(i)} \sigma_z^{(i+1)}
      + B_{4z}\, \sigma_z^{(i)}
    \Big),
    \label{eq:H4}
\end{equation}
with parameters
\begin{equation}
    J_{4x} = J_{4y} = 2, \qquad J_{4z} = B_{4z} = 0.54,
\end{equation}
and evolution time $\Delta t = 20$.\cite{DeLorenzis_1}
By tuning the anisotropy and field strength one can interpolate between
more integrable-like and more chaotic regimes. In the parameter regime
used in Ref.~\cite{DeLorenzis_1}, $H_4$ realizes a local interacting
reservoir with non-trivial many-body dynamics.

\paragraph{Integrable XX chain $H_5$.}

The fifth Hamiltonian, $H_5$, is the integrable XX model with
nearest-neighbour spin-flip couplings and no external field. In this
thesis we always consider it with periodic boundary conditions, so that
\begin{equation}
    H_5 = \frac{1}{2} \sum_{i=1}^{N}
    \Big(
        \sigma_x^{(i)} \sigma_x^{(i+1)}
      + \sigma_y^{(i)} \sigma_y^{(i+1)}
    \Big),
    \qquad \sigma^{(N+1)} \equiv \sigma^{(1)}.
    \label{eq:H5}
\end{equation}
Via a Jordan--Wigner transformation followed by a Fourier transform,
$H_5$ can be mapped to free fermions hopping on a ring and is therefore
exactly solvable. From the viewpoint of quantum dynamics, it generates
ballistic propagation of quasiparticles and an entanglement growth that
can be characterized analytically. In this thesis, the XX chain serves
as a paradigmatic example of a local and integrable quantum reservoir,
to be contrasted with the chaotic and disordered models introduced
above.

In our first work~\cite{DeLorenzis_1}, the XX chain is used as one
of the candidate Hamiltonians for the quantum layer, and its behaviour
is compared to that of the other models at a fixed evolution time
$\Delta t = 20$. In our second work~\cite{DeLorenzis_2}, the same
Hamiltonian is adopted as the only quantum reservoir, while the
evolution time $t$ is treated as the main control parameter. This
setting provides a convenient framework to systematically investigate
how the properties of the XX-based QELM depend on the duration of the
dynamics. The corresponding analyses and results are presented in the
chapters devoted to the second work.

\paragraph{Disordered long-range $xx$ model with random fields $H_6$.}

The last Hamiltonian considered in Ref.~\cite{DeLorenzis_1} is a
disordered long-range spin model with random $xx$ interactions and
random longitudinal fields, introduced in Ref.~\cite{Martinez2021}. Its explicit form is
\begin{equation}
    H_6 =
    \sum_{i>j=1}^{N} J^{(6)}_{ij}\, \sigma_x^{(i)} \sigma_x^{(j)}
    + \frac{1}{2} \sum_{i=1}^{N} \big( B_6 + D_i \big)\, \sigma_z^{(i)}.
    \label{eq:H6}
\end{equation}
The couplings $J^{(6)}_{ij}$ are drawn from a uniform distribution
\begin{equation}
    J^{(6)}_{ij} \sim \mathcal{U}(-0.5, 0.5),
\end{equation}
while the random longitudinal fields are given by
\begin{equation}
    D_i \sim \mathcal{U}(-W, W).
\end{equation}
Depending on the values of the uniform field $B_6$ and the disorder
strength $W$, the model displays different dynamical regimes, ranging
from ergodic to non-ergodic extended and many-body localized phases,
as characterized in Ref.~\cite{Martinez2021}. In the present context, an ergodic regime is one in which the reservoir dynamics efficiently mixes the encoded information across the system, while a non-ergodic regime retains a stronger memory of the initial state and spreads information less effectively. Within the QELM framework, three parameter
choices are considered:
\begin{itemize}
    \item \emph{Localized regime (region III of Ref.~\cite{Martinez2021})}:
    \(
       B_6 = W = 2 \times 10^{-2}.
    \)
    Here the eigenstates exhibit strong localization and the reservoir
    dynamics is expected to be poor at spreading information across the
    chain.\cite{DeLorenzis_1}
    \item \emph{Intermediate non-ergodic/ergodic regime}:
    \(
       B_6 = 0.03, \quad W = 1.
    \)
    This parameter set lies between non-ergodic and ergodic regions in
    the phase diagram of Ref.~\cite{Martinez2021}, and corresponds to partially
    delocalized dynamics.
    \item \emph{Many-body localized regime (region I of Ref.~\cite{Martinez2021})}:
    \(
       B_6 = 0.03, \quad W = 60.
    \)
    In this case the system is deep in the many-body localized phase,
    where transport and information spreading are strongly suppressed.
\end{itemize}
For all these parameter choices, the evolution time is fixed to
$\Delta t = 20$.\cite{DeLorenzis_1}

\begin{figure*}[htbp]
    \centering
\includegraphics[width=0.5\textwidth]{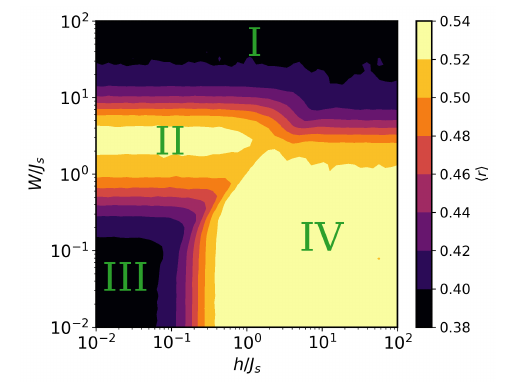}
    \caption{Phase diagram of the long-range disordered $xx$ model of Ref.~\cite{Martinez2021} in the $(B_6=h,W)$ plane, showing four diﬀerent regions: two localization areas (regions I and III, black) and two ergodicity areas (regions II and IV, bright yellow). The localized regime in I corresponds to a MBL paramagnetic phase where eigenvalue statistics is Poissonian, while region III corresponds to a spin-glass phase, where eigenvalue statistics is also Poissonian, but it presents a mobility edge. Reproduced from Ref.~\cite{Martinez2021}.}\label{fig:Zambrini}
\end{figure*}

To better convey the nature of these regimes, one can refer to the
phase diagram presented in Fig. \ref{fig:Zambrini} (Ref.~\cite{Martinez2021}), where the different dynamical
regions (ergodic, non-ergodic extended, many-body localized) are mapped
in the $(B_6,W)$ plane. In the thesis, it is natural to reproduce that
diagram and highlight the three parameter sets used for the QELM
reservoir. 

\medskip

From the viewpoint of this thesis, the set of Hamiltonians
$H_1,\dots,H_6$ (together with their non-interacting and purely-XX
limits) provide a controlled playground to assess how connectivity
(fully connected vs.\ local), degree of integrability (integrable vs.\
chaotic), and localization properties (ergodic vs.\ many-body localized)
affect the performance of Quantum Extreme Learning Machines. This makes
it possible to relate the observed classification behaviour to
underlying physical concepts such as information propagation,
entanglement growth, and classical simulability.

\section{Measurement}
\label{subsec:measurement_classifier}

After the quantum evolution generated by the chosen Hamiltonian,
the final state of the $N$-qubit register is measured in the
computational (Pauli-$Z$) basis. This projective measurement yields
a probability distribution over all bit strings
$\mathbf{s} \in \{0,1\}^{N}$, which can be interpreted as the
populations of the corresponding computational basis states in the
evolved wave function.\footnote{In a real experiment, these
probabilities would be estimated from repeated measurements of
identically prepared states; in our simulations, they are computed
directly from the state vector.}
For each input sample, we thus associate a classical feature vector
\begin{equation}
  \mathbf{p}(t) =
  \big( p_{\mathbf{s}}(t) \big)_{\mathbf{s}\in\{0,1\}^{N}} \in \mathbb{R}^{2^{N}},
  \qquad
  p_{\mathbf{s}}(t)
  = \big|\langle \mathbf{s} \,|\, \psi(t) \rangle \big|^{2},
  \label{eq:measurement_probabilities_comp_basis}
\end{equation}
where $|\psi(t)\rangle = U(t) |\psi_0(\mathbf{z})\rangle$ is the
evolved state corresponding to the encoded feature vector
$\mathbf{z}$ and $U(t) = e^{-iHt}$ is the time-evolution operator.
The probability vector $\mathbf{p}(t)$ constitutes the output of the
quantum layer and the input to the classical classifier in both
Refs.~\cite{DeLorenzis_1,DeLorenzis_2}.%
\subsubsection{Choice of measurement basis}
\label{subsubsec:measurement_basis}

Beyond the evolution time itself, an important design choice in a
QELM is the measurement basis. As discussed in
Appendix~B of Ref.~\cite{DeLorenzis_2},
the basis in which the system is measured should be sufficiently
distinct from the eigenbasis of the Hamiltonian governing the
dynamics, in order to generate expressive and time-dependent
features.

Assume that the QELM dynamics is generated by a time-independent
Hamiltonian $H$ with spectral decomposition
\begin{equation}
  H = \sum_{n} E_{n} \, |E_{n}\rangle \langle E_{n}| ,
  \label{eq:H_spectral}
\end{equation}
where $|E_n\rangle$ and $E_n$ are its eigenstates and eigenvalues.
The encoded initial state can be expanded in this eigenbasis as
\begin{equation}
  |\psi(0)\rangle = \sum_{n} c_{n} \, |E_{n}\rangle .
  \label{eq:psi0_energy_basis}
\end{equation}
After an evolution time $t$, the state becomes
\begin{equation}
  |\psi(t)\rangle
  = \sum_{n} c_{n} \, e^{- i E_{n} t} \, |E_{n}\rangle .
  \label{eq:psit_energy_basis}
\end{equation}

Let the measurement be performed in a basis
$\{|s_k\rangle\}$, which in general does not coincide with the
Hamiltonian eigenbasis. The probability of obtaining outcome $k$ at
time $t$ is given by the Born rule
\begin{equation}
  p_k(t) = |\langle s_k | \psi(t) \rangle|^{2} .
  \label{eq:pk_born}
\end{equation}
By inserting Eq.~\eqref{eq:psit_energy_basis} one obtains
\begin{equation}
  p_k(t)
  = \sum_{n,m}
      \langle s_k | E_{n} \rangle
      \langle E_{m} | s_k \rangle
      e^{- i (E_{n} - E_{m}) t}
      c_{n} c_{m}^{*} ,
  \label{eq:pk_general}
\end{equation}
which explicitly shows how the time dependence of $p_k(t)$ is
controlled by the energy gaps $E_n - E_m$ and by the overlaps
between the measurement basis and the Hamiltonian eigenbasis. This
expression corresponds to the general formula discussed in
Appendix~B of Ref.~\cite{DeLorenzis_2}.

Two limiting cases are particularly illustrative. If the
measurement basis \emph{coincides} with the eigenbasis of the
Hamiltonian, i.e.\ $|s_k\rangle = |E_k\rangle$, all off-diagonal
terms in Eq.~\eqref{eq:pk_general} vanish and the probabilities
become time-independent,
\begin{equation}
  p_k(t) = |c_k|^{2} ,
  \label{eq:pk_eigenbasis}
\end{equation}
so that the feature map is static and the QELM cannot exploit the
quantum dynamics to enrich the representation of the data.

At the opposite extreme, one can consider measurement and
Hamiltonian eigenbases that are \emph{mutually unbiased}, i.e.\
such that
\begin{equation}
  |\langle E_n | s_k \rangle|^{2} = \frac{1}{d}
  \quad \forall\, n,k ,
  \label{eq:MUB_condition}
\end{equation}
where $d$ is the Hilbert-space dimension. In this case the
measurement amplitudes are evenly distributed over all eigenstates,
and the evolution induces maximal interference among the different
energy components. The resulting probabilities $p_k(t)$ display a
rich time dependence and can provide highly nonlinear features for the classical classifier.

In the QELM architectures studied in this thesis, measurements are
performed in the computational basis, while the Hamiltonian
eigenstates of the XX model are delocalized over the chain. Together
with the chosen encoding scheme, this guarantees a substantial
misalignment between the measurement and energy bases, which in turn favours nontrivial interference patterns in the measurement probabilities and enhances the expressiveness of the quantum feature map.
\section{Classical classifier}
\label{sec:classifier}

The measurement step associates to each input sample a classical
probability vector $\mathbf{p}(t)$ of dimension up to $2^{N}$,
cf.~Eq.~\eqref{eq:measurement_probabilities_comp_basis}. This vector
is then processed by a classical one-layer neural network (ONN),
which performs the final classification. In both
Refs.~\cite{DeLorenzis_1,DeLorenzis_2},
the ONN is implemented as a fully connected single layer with
softmax activation.

Let $D$ denote the number of input features (typically $D = 2^{N}$, or
a subset thereof) and let $C$ be the number of classes in the
dataset. The classifier computes, for each sample, a vector of
logits
\begin{equation}
  \mathbf{a}
  = W \, \mathbf{p}(t) + \mathbf{b}
  \in \mathbb{R}^{C},
  \label{eq:logits}
\end{equation}
where $W \in \mathbb{R}^{C \times D}$ and
$\mathbf{b} \in \mathbb{R}^{C}$ are the trainable weights and
biases. The corresponding class probabilities are obtained via the
softmax function,
\begin{equation}
  y_{c}
  = \mathrm{softmax}(\mathbf{a})_{c}
  = \frac{e^{a_{c}}}{\sum_{c'=1}^{C} e^{a_{c'}}} ,
  \qquad c = 1,\dots,C .
  \label{eq:softmax}
\end{equation}
Given a one-hot encoded target label $\mathbf{t}$, the network is
trained by minimizing the categorical cross-entropy loss
\begin{equation}
  \mathcal{L}
  = - \sum_{c=1}^{C} t_{c} \log y_{c} .
  \label{eq:cross_entropy}
\end{equation}
In practice, training is performed using the Adam optimizer as
implemented in the Keras framework, with standard choices of
learning rate and mini-batch size.

Crucially, the parameters $(W,\mathbf{b})$ of the ONN are the
\emph{only} trainable components of the entire QELM architecture.
The quantum reservoir, comprising feature reduction, encoding,
Hamiltonian dynamics, and measurement, remains completely fixed.
This mirrors the philosophy of classical Extreme Learning Machines,
where learning is confined to the output layer while the feature map
is static.
\newpage
\chapter{Quantum Extreme Learning Machines: Methods and Applications to Image Classification}
\label{ch:paper1}
\section{Overview of the study}
\label{sec:paper1_overview}

In the previous chapter we introduced the general framework of Quantum
Extreme Learning Machines (QELMs), emphasizing their hybrid nature and
the role of the quantum layer as a fixed feature generator. In this
chapter we apply that framework to a concrete case study based on results previously published in Ref.~\cite{DeLorenzis_1}, of which the author of this thesis is the first author.

Earlier applications of QELM-like architectures to image classification generally relied on a specific encoding scheme and a fixed choice of quantum dynamics, without a systematic comparison of how these ingredients affect performance. The central goal of the present study is to fill this gap by assessing, within a common pipeline, the impact of the different building blocks of a QELM architecture, particularly feature reduction, encoding strategy, and Hamiltonian choice, on image classification performance.

More specifically, the work focuses on a QELM model in which
grayscale images are first compressed to a low-dimensional latent space,
then encoded into the state of an $N$-qubit register, evolved under a
many-body Hamiltonian, and finally measured to produce classical
features for a one-layer neural-network classifier. Within this
pipeline, several design choices are possible: the dimensionality
reduction can be performed either by Principal Component Analysis (PCA)
or by nonlinear autoencoders; the latent variables can be mapped to
quantum states using different encoding schemes; and the quantum
evolution can be generated by Hamiltonians with distinct connectivity
patterns and dynamical properties. Our study in
Ref.~\cite{DeLorenzis_1} is devoted to exploring this design space in a
controlled and systematic manner.

From a methodological point of view, the analysis proceeds by varying
one ingredient of the QELM at a time while keeping the others fixed.
First, PCA and two different autoencoder architectures are compared as
feature-reduction methods, both in terms of reconstruction quality and
their impact on the final classification accuracy. Second, several
encoding strategies are tested, with particular attention to how the
choice of encoding interacts with the available number of qubits and
with the statistics of the latent features. Third, the dynamics of the
quantum layer is generated by a family of interacting Hamiltonians with
either long-range or local couplings, as well as by a non-interacting
reference model, in order to disentangle the influence of connectivity,
interaction strength, and dynamical complexity on the performance of
the classifier.

Throughout the study, the QELM is benchmarked against purely classical
baselines that operate on the same compressed representations of the
images. This allows one to isolate the contribution of the quantum
reservoir to the overall performance. The main findings, which will be
discussed in detail in the following sections, can be summarized as
follows: the presence of an interacting quantum reservoir systematically
improves the accuracy of the classifier with respect to a purely
classical readout acting on the same latent features; the choice of
encoding has a pronounced effect on the achieved performance; and, perhaps
more surprisingly, Hamiltonians with very different connectivity
structures can lead to comparable discrimination power, provided that
they generate sufficiently interacting dynamics. These results provide a
first systematic benchmark of QELMs for image classification and set the
stage for the more focused analysis of XX dynamics carried out in
Chapter~\ref{chap:Paper_2}.
\section{Datasets and preprocessing}
\label{sec:paper1_datasets}

In this study we consider two standard image-classification benchmarks,
MNIST and Fashion-MNIST, which provide a controlled setting to test the
performance of Quantum Extreme Learning Machines on grayscale images of
handwritten digits and clothing items, respectively. In both cases, the
images are preprocessed and compressed to a low-dimensional latent
representation before being encoded into quantum states, as described in
Sec.~\ref{sec:feature_reduction}. A third dataset, CIFAR-10, will be
introduced in the context of the second study discussed in
Chapter~\ref{chap:Paper_2}.
\subsection*{MNIST}

The MNIST dataset consists of $70\,000$ grayscale images of handwritten
digits from $0$ to $9$, each of size $28\times 28$ pixels. The standard
split used in this work comprises $60\,000$ training examples and
$10\,000$ test examples, with $10$ balanced classes.

\begin{figure*}[htbp]
    \centering
\includegraphics[width=0.9\textwidth]{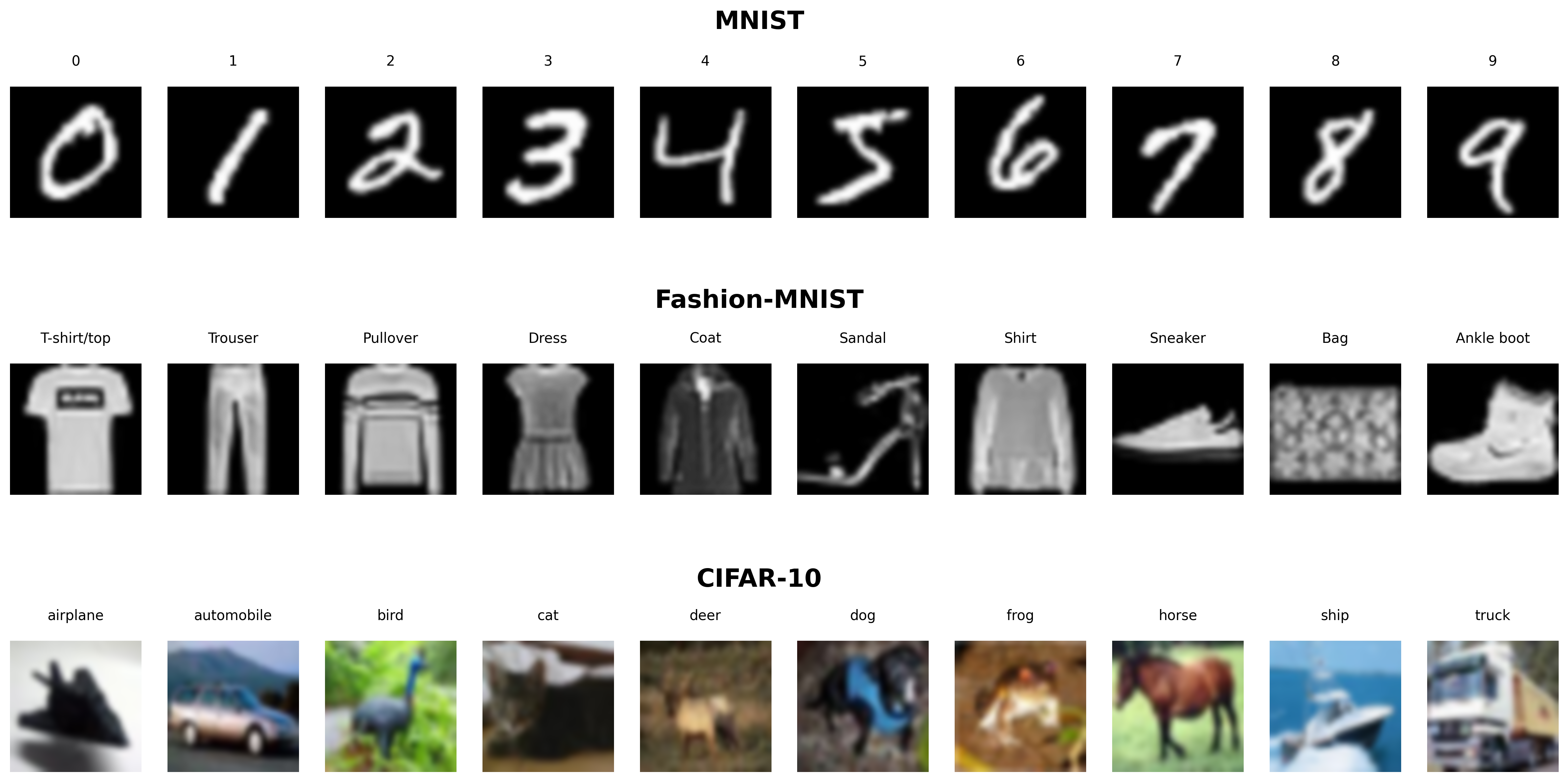}
    \caption{Sample images from three widely used benchmark datasets in this study. The top row displays handwritten digits from MNIST (digits 0–9), the middle row shows clothing items from Fashion-MNIST (10 clothing categories), and the bottom row presents natural object categories from CIFAR-10. Each column corresponds to one class, labeled above each image.}\label{fig:Datasets_sample}
\end{figure*}

Figure~\ref{fig:Datasets_sample} (top) shows representative examples from the
MNIST dataset. Despite their relatively low resolution, these images correspond to
$784$-dimensional data points (one feature per pixel), which motivates
the use of dimensionality-reduction techniques such as PCA and
autoencoders before encoding them into quantum states.
\subsection*{Fashion-MNIST}

Fashion-MNIST is a more challenging drop-in replacement for MNIST, with
the same image format but a different semantic content. It contains
$70\,000$ grayscale images of clothing items (such as T-shirts, trousers,
coats, or shoes), again of size $28\times 28$ pixels and organized into
$10$ balanced classes. As for MNIST, we use $60\,000$ images for training
and $10\,000$ for testing.

Some sample images
from the Fashion-MNIST dataset are shown in
Fig.~\ref{fig:Datasets_sample} (center). Compared to handwritten digits,
these images typically exhibit more complex shapes and textures, which
makes the classification task more demanding and provides a more
stringent test for the QELM architecture.

\subsection*{Preprocessing pipeline}

In both datasets, pixel values are first normalized to the range
$[0,1]$ by dividing the original $8$-bit grayscale intensities by $255$.
No additional denoising or data augmentation is applied, so as to keep
the focus on the effect of feature reduction and of the quantum layer.

When Principal Component Analysis (PCA) is used as a feature-reduction
method, the $28\times 28$ images are flattened into $784$-dimensional
real vectors and then projected onto the first $d$ principal components.
When autoencoders are employed, the $28\times 28$ images are instead
fed to the encoder in their natural tensor form, and only the resulting
$d$-dimensional latent vectors $\mathbf{z}\in\mathbb{R}^d$ are retained.

In both cases, the latent representations are finally rescaled to match
the input range required by the chosen encoding scheme (see
Sec.~\ref{ch:encoding}) and then fed to the QELM quantum layer.
\section{Numerical results}
\label{sec:paper1_results}

In this section we summarise the main numerical findings of our study reported in Refs.~\cite{DeLorenzis_1, DeLorenzis_IQIS2025, DeLorenzis_IQIS2024},  focusing on how the different components of
the QELM pipeline (feature reduction, data encoding, and quantum
dynamics) influence the final classification accuracy on the MNIST and Fashion-MNIST datasets. Throughout this study we adopt a baseline configuration (summarized in Table ~\ref{tab:qelm_baseline})  in which the images are compressed by the autoencoder $AE_1$, the latent features are mapped to quantum states via dense-angle encoding, and the time evolution is generated by the periodically driven
long-range Hamiltonian $H_1$ discussed in
Section~\ref{sec:hamiltonians}. Starting from this baseline, one
ingredient of the architecture is varied at a time while keeping the others fixed, so as to isolate its individual contribution to the overall performance. The number of qubits in the quantum layer is
increased up to $N=12$, which represents the practical limit for classical simulations of the full quantum dynamics in this setup.
\begin{table}[htbp]
\centering
\caption{Baseline configuration used in the QELM numerical analysis. Starting from this setup, one component of the pipeline is varied at a time.}
\label{tab:qelm_baseline}
\begin{tabular}{ll}
\hline
Component & Reference choice \\
\hline
Feature reduction & Autoencoder $AE_1$ \\
Encoding & Dense-angle encoding \\
Hamiltonian & Periodically driven long-range Hamiltonian $H_1$ \\
Classical readout & Feed-forward neural network \\
\hline
\end{tabular}
\end{table}
\subsection{Impact of feature reduction: PCA vs autoencoders}
\label{subsec:results_feature_reduction}

We begin by analysing the effect of different feature-reduction
strategies on the QELM performance. In this comparison, the encoding is
kept fixed to dense-angle encoding and the quantum layer is always
governed by the same time-dependent long-range Hamiltonian. For each
value of $N$, the dimension $d$ of the latent space is chosen to match
the requirements of the encoding (typically $d = 2N$ for dense-angle
encoding), as discussed in Sec.~\ref{sec:feature_reduction}. The three methods under
consideration are linear PCA and the two convolutional autoencoders
$AE_1$ and $AE_2$ introduced below.

\subsection*{Convolutional autoencoders $AE_1$ and $AE_2$}
\label{subsec:paper1_AEs}

In addition to the linear PCA baseline, we employ two convolutional
autoencoders, denoted $AE_1$ and $AE_2$, in order to exploit nonlinear feature
extraction prior to the quantum layer. Both autoencoders are trained in a fully
classical way to reconstruct the input images from a compressed latent
representation.

In both cases, the encoder receives the raw $28\times 28$ grayscale image (with pixel values rescaled to [0,1])
and applies a sequence of convolutional layers with
ReLU activations and max-pooling operations for
progressive spatial downsampling and feature extraction. The final
convolutional feature maps are then flattened and passed through a
fully-connected layer that defines the $d$-dimensional latent vector
$\mathbf{z}$. A sigmoid activation is used at this bottleneck layer, so
that the latent components lie in $[0,1]$ and can be directly employed to match the requirements of the quantum encodings
(cf. Chapter~\ref{ch:encoding}).

The decoder mirrors this structure: the latent vector is first expanded
via a dense layer and reshaped into feature maps, which are then
processed by a series of convolutional and up-sampling layers to
reconstruct an image of size $28\times 28$. Convolutional layers in the
decoder again use ReLU activations, while the final output layer employs
a sigmoid activation to produce pixel intensities in $[0,1]$.

The two autoencoders share the same general layout but differ in their
capacity. $AE_1$ is the more expressive model, with a larger number of
filters and trainable parameters in both encoder and decoder, while $AE_2$
is deliberately more compact. As a result, $AE_1$ is expected to achieve a
lower reconstruction error and to capture more subtle features of the
images, at the cost of a longer training time and a higher risk of
overfitting; $AE_2$, instead, provides a lighter baseline with reduced
capacity.

\begin{figure}[htbp]
    \centering
    \includegraphics[width=1.\textwidth]{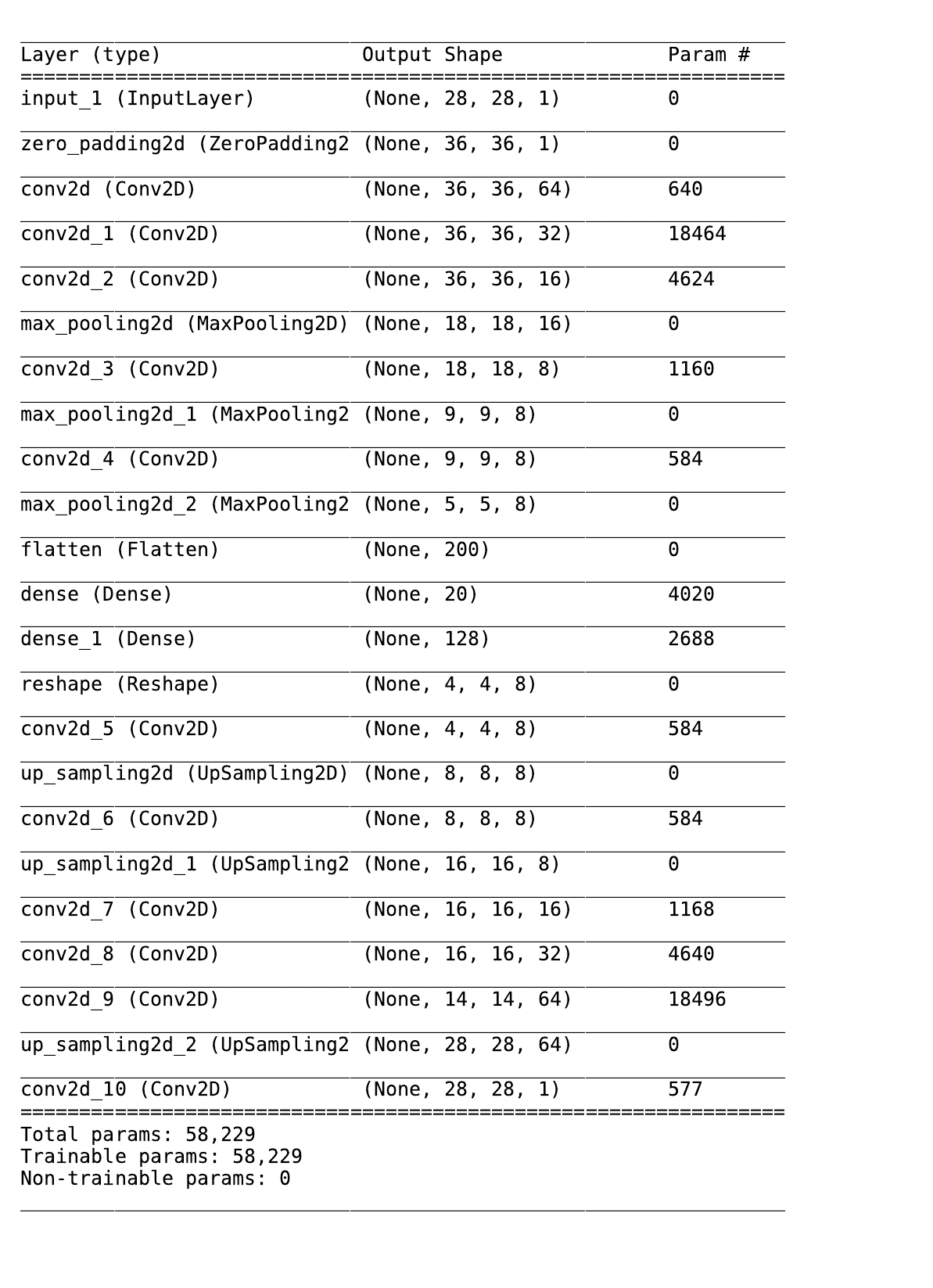}
    \caption{Architecture of the first autoencoder explored in the present analysis. Adapted from the author’s previous work, Ref.~\cite{DeLorenzis_1}. Copyright (2025) by the American Physical Society.}
    \label{fig:AE1}
\end{figure}

\begin{figure}[htbp]
    \centering
    \includegraphics[width=1.\textwidth]{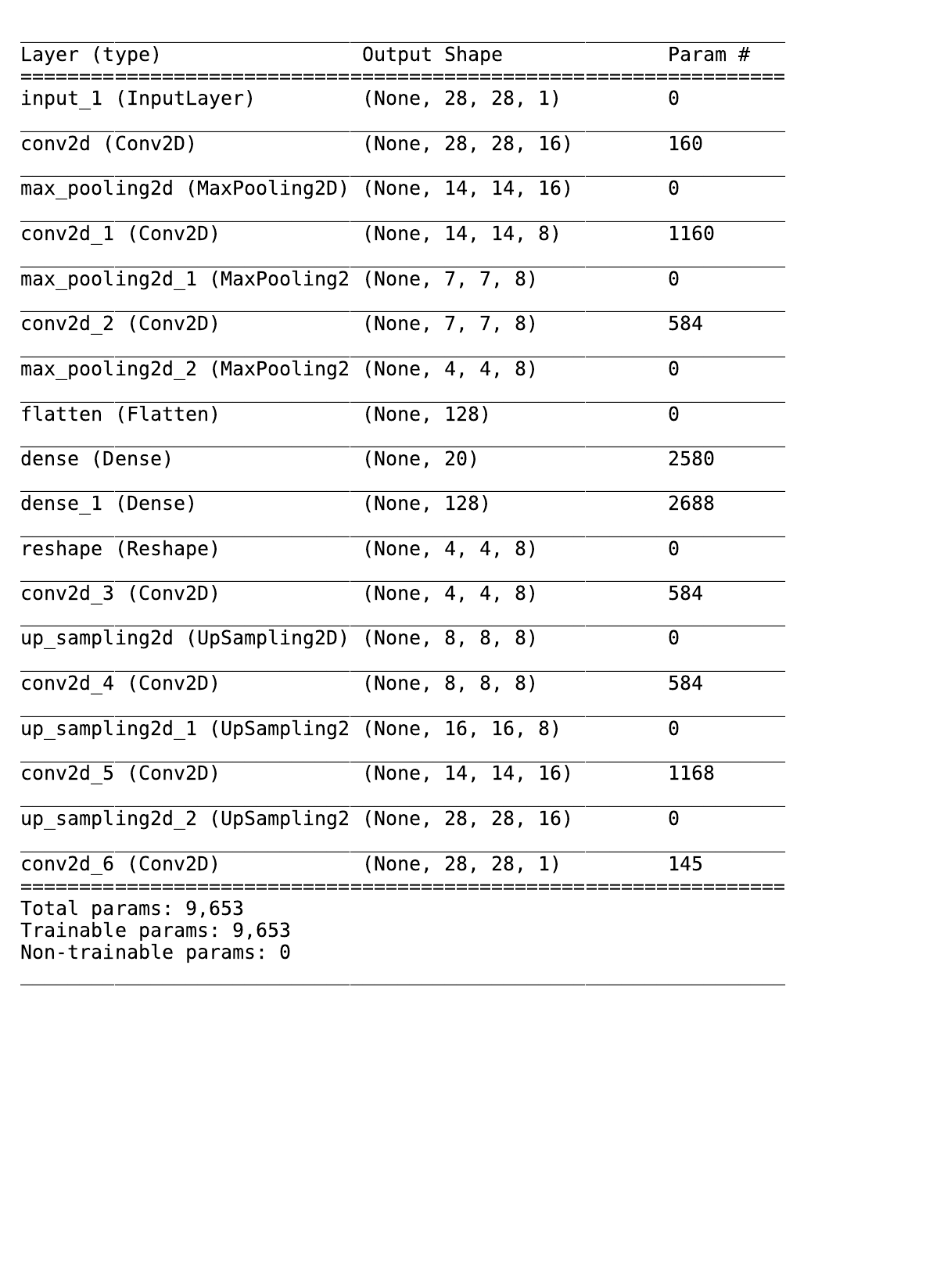}
    \caption{Architecture of the second autoencoder explored in the present analysis. Adapted from the author’s previous work, Ref.~\cite{DeLorenzis_1}. Copyright (2025) by the American Physical Society.}
    \label{fig:AE2}
\end{figure}

Schematic diagrams of the two architectures, together with a summary of
their layers (type, number of filters/neurons),
are shown in the Figs.~\ref{fig:AE1} and ~\ref{fig:AE2}.

\begin{figure*}[htbp]
    \centering
    \begin{subfigure}[t]{0.45\textwidth}
        \centering
        \includegraphics[width=\linewidth]{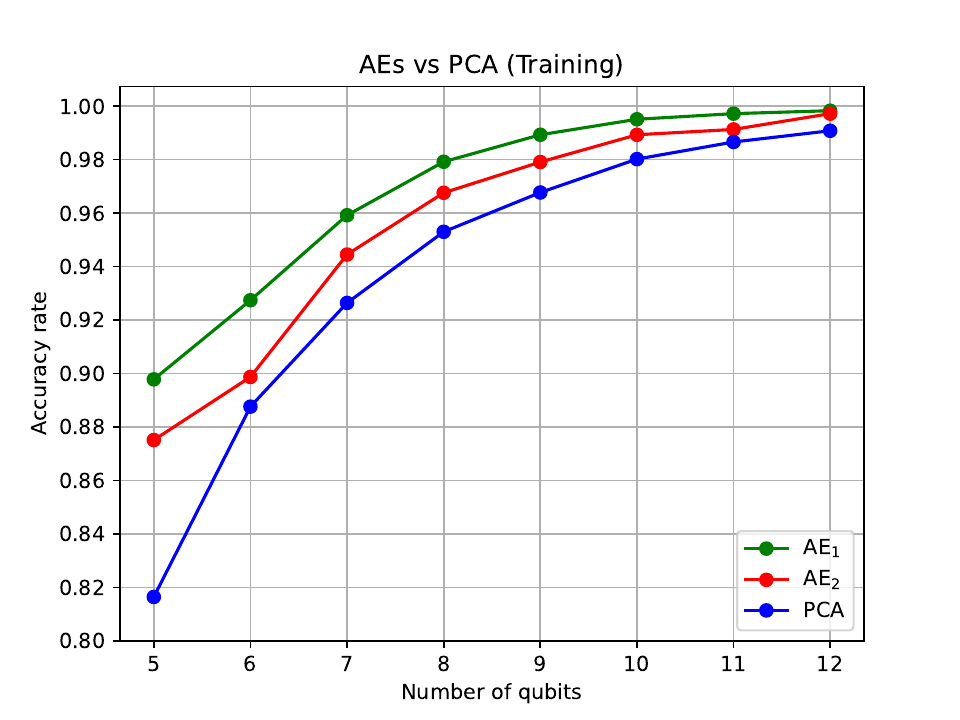}
        \caption{}
    \end{subfigure}
    \hspace{0.02\textwidth}
    \begin{subfigure}[t]{0.45\textwidth}
        \centering
        \includegraphics[width=\linewidth]{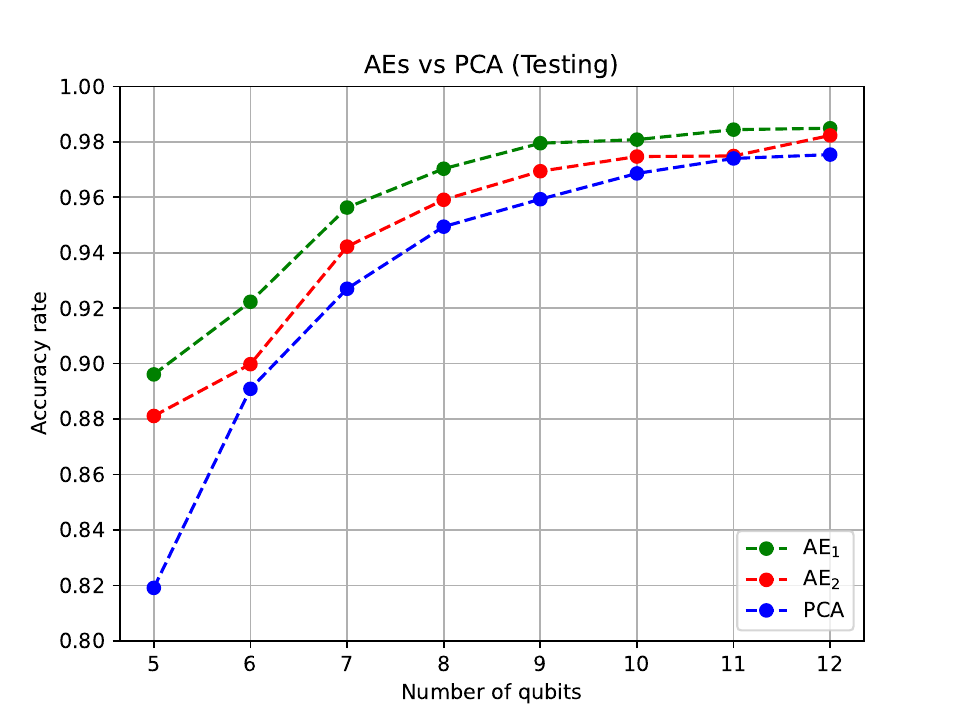}
        \caption{}
    \end{subfigure}
    \caption{Training (left panel) and testing (right panel) accuracy as a function of the number of qubits for different feature reduction schemes: PCA (blue) and two autoencoders ($AE_1$ and $AE_2$, green and red curves respectively). The time evolution and the encoding have been performed with the Hamiltonian $H_1$ and the dense angle, respectively. Adapted from the author’s previous work, Ref.~\cite{DeLorenzis_1}. Copyright (2025) by the American Physical Society.}
    \label{fig:PCA_vs_AEs}
\end{figure*}
\begin{figure*}[htbp]
    \centering
\includegraphics[width=0.55\textwidth]{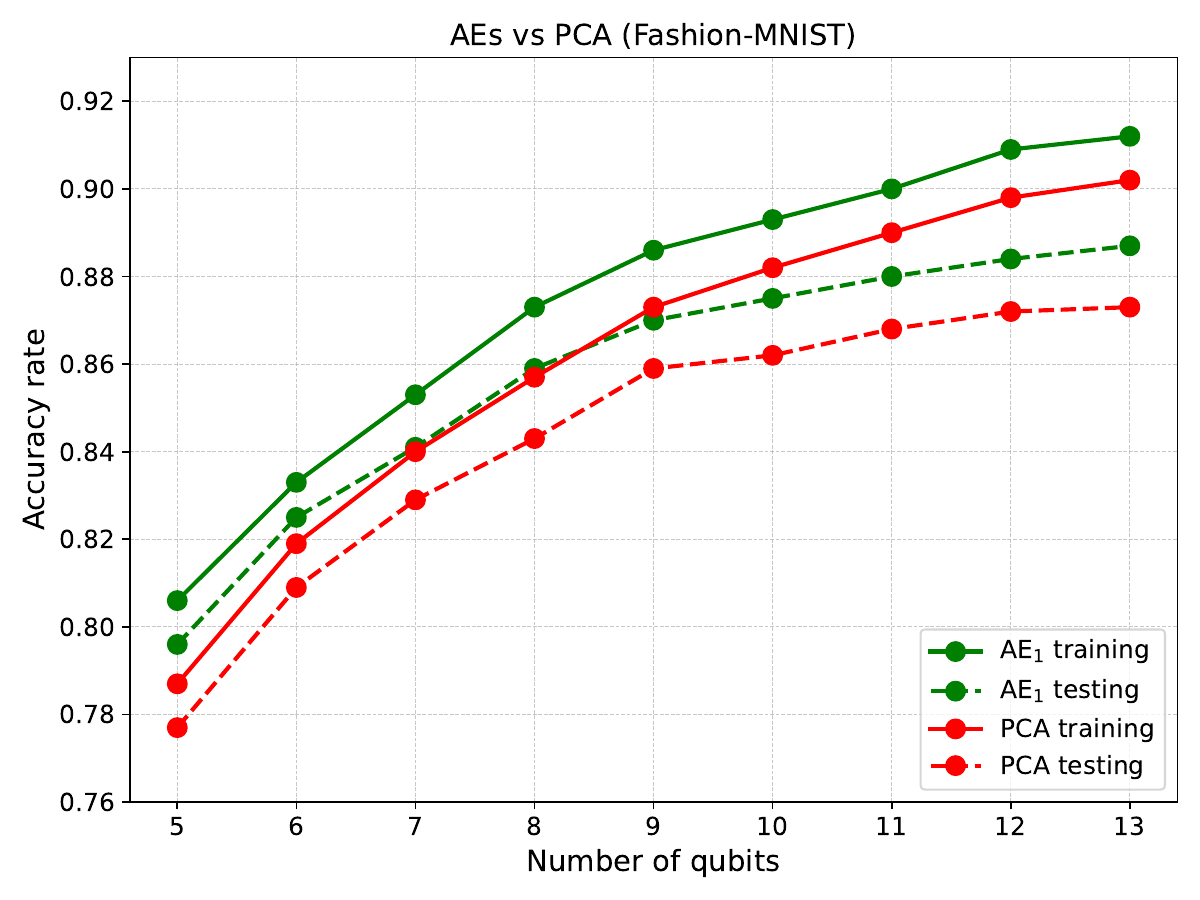}
    \caption{Training (solid line) and testing (dashed line) accuracy as a function of the number of qubits for different feature reduction schemes for the Fashion-MNIST dataset: PCA (red) and autoencoder $AE_1$ (green). The  time evolution and the encoding have been performed with the Hamiltonian $H_1$ and the dense angle encoding, respectively. Adapted from the author’s previous work, Ref.~\cite{DeLorenzis_1}. Copyright (2025) by the American Physical Society.}
    \label{fig:Fashion-MNIST}
\end{figure*}

Figure~\ref{fig:PCA_vs_AEs} (reproduced from
Ref.~\cite{DeLorenzis_1}) shows the training and test
accuracy as a function of the number of qubits for these three
feature-reduction schemes. The linear PCA baseline systematically yields
lower accuracies than the nonlinear autoencoders across the explored
range of system sizes. Among the two AEs, the more expressive
architecture $AE_1$ (with a larger number of trainable parameters)
achieves the best performance, while $AE_2$ still clearly outperforms
PCA but attains slightly lower accuracies than $AE_1$. For all three
methods, the accuracy increases rapidly with $N$ and then saturates
around $N \simeq 9$--$10$, reaching test accuracies close to $98\%$ for
$AE_1$.
As the number of qubits grows and the latent dimension increases, the
gap between the three feature-reduction strategies gradually shrinks,
although PCA remains systematically inferior. This behaviour can be
attributed to the relative simplicity of MNIST: once a sufficiently
large latent space is available, even linear projections retain enough
information for the QELM to achieve high accuracy, whereas nonlinear
autoencoders are particularly advantageous in the regime of strong
compression.
To test the robustness of these conclusions, the same comparison
between PCA and $AE_1$ has been repeated on the more demanding
Fashion-MNIST dataset. The corresponding accuracies, displayed in Fig.~\ref{fig:Fashion-MNIST}, , show the same qualitative
behaviour: across the whole range of qubits explored, $AE_1$ consistently
outperforms PCA, and the gap between the two methods persists.
This indicates that the advantage of nonlinear autoencoders over linear
PCA is not specific to MNIST, but extends to more complex image
distributions as well.

\subsection{Effect of the encoding strategy}
\label{subsec:results_encodings}
We now investigate how the choice of data encoding impacts the
classification accuracy. In this analysis, the feature reduction is
fixed to the autoencoder $AE_1$ and the quantum evolution is again
generated by the Hamiltonian $H_1$. The number of qubits is varied from
$N=5$ to $N=12$, and the latent dimension $d$ is adjusted according to
the encoding scheme (one feature per qubit for simple angle encoding,
two features per qubit for dense-angle, uniform Bloch-sphere, and
general encodings, and $d=2N$ for amplitude encoding), as reported in
Table~\ref{tab:encodings_summary}.

\begin{figure*}[htbp]
    \centering
    \begin{subfigure}[t]{0.45\textwidth}
        \centering
        \includegraphics[width=\linewidth]{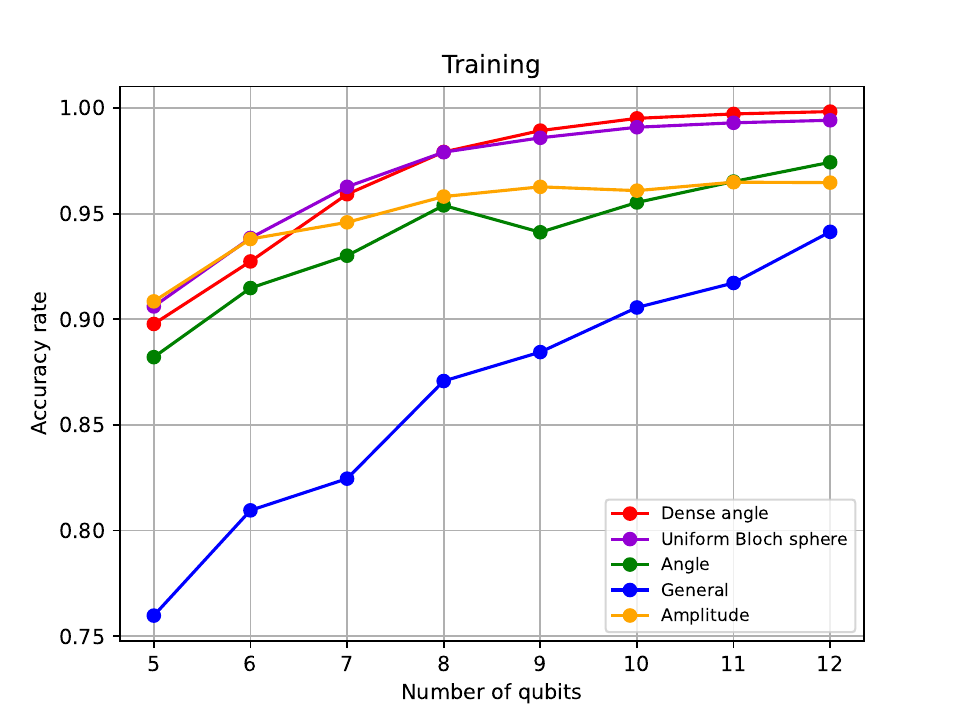}
        \caption{}
    \end{subfigure}
    \hspace{0.02\textwidth}
    \begin{subfigure}[t]{0.45\textwidth}
        \centering
        \includegraphics[width=\linewidth]{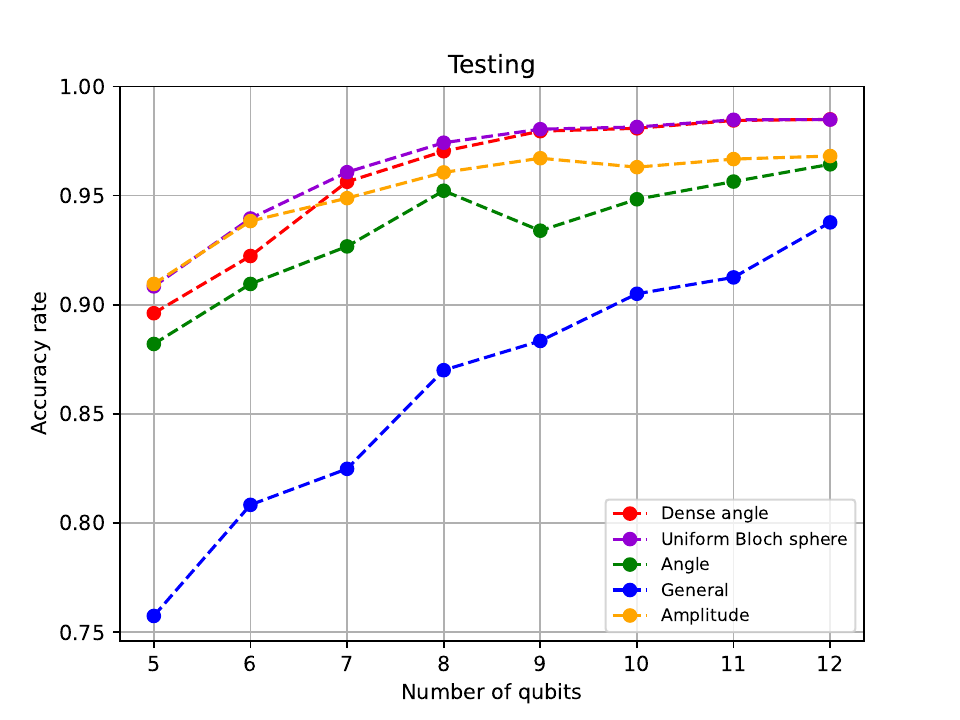}
        \caption{}
    \end{subfigure}
    \caption{Training (left panel) and testing (right panel) accuracy as a function of the number of qubits for different encoding strategies: dense angle (red), uniform Bloch sphere (purple), angle (green), general (blue) and amplitude (orange) encodings. The feature reduction has been performed with the autoencoder $AE_1$ and the time evolution with the Hamiltonian $H_1$. Adapted from the author’s previous work, Ref.~\cite{DeLorenzis_1}. Copyright (2025) by the American Physical Society.}
    \label{fig:Different_encodings}
\end{figure*}

Figure~\ref{fig:Different_encodings} (from Ref.~\cite{DeLorenzis_1})
compares five encodings:
dense-angle, simple angle, uniform Bloch-sphere, general, and amplitude
encoding. Among the “angle-like’’ schemes, dense-angle and uniform
Bloch-sphere encodings clearly stand out as the best performers, both on
the training and on the test sets. Both exploit two Bloch-sphere
parameters per qubit and thus encode two features per qubit, which
translates into higher accuracy than simple angle encoding, where only
the polar angle is varied and a single feature is assigned to each qubit.

The general encoding performs worst, especially for small quantum
layers. This reduced performance can be traced back to the fact that
pairs of different but proportional features are mapped to the same
quantum state, effectively decreasing the distinguishability of samples
in Hilbert space and providing a less informative feature map for the
classifier.

\begin{figure*}[htbp]
    \centering
    \begin{subfigure}[t]{0.45\textwidth}
        \centering
        \includegraphics[width=\linewidth]{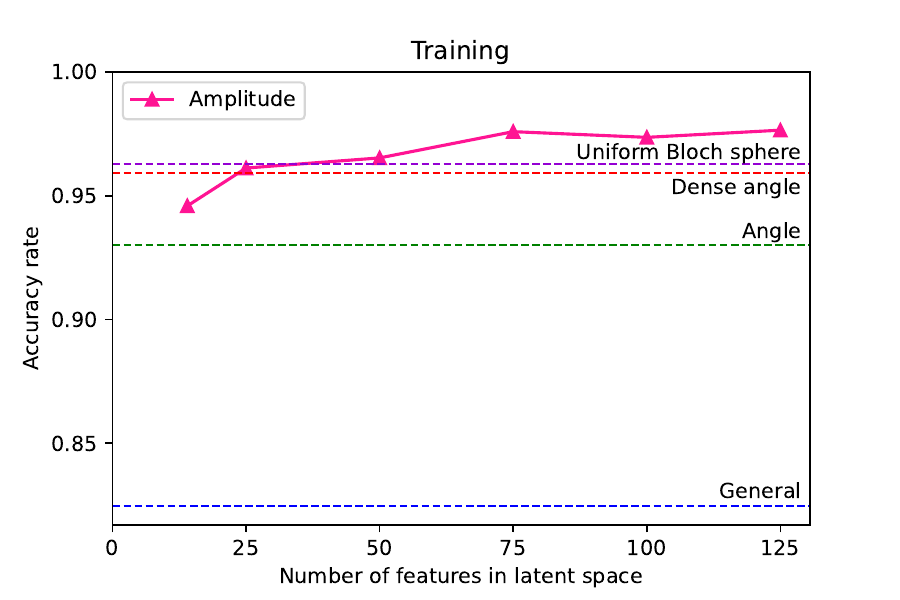}
        \caption{}
    \end{subfigure}
    \hspace{0.02\textwidth}
    \begin{subfigure}[t]{0.45\textwidth}
        \centering
        \includegraphics[width=\linewidth]{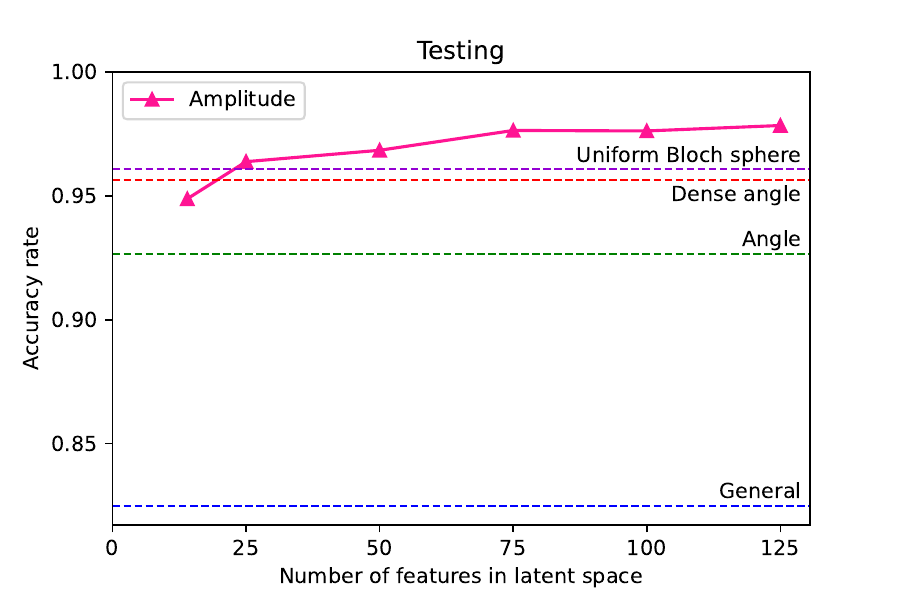}
        \caption{}
    \end{subfigure}
    \caption{Training (left panel) and testing (right panel) accuracy as a function of the number of features in the latent space obtained with the amplitude encoding with $N=7$. The feature reduction has been performed with the autoencoder $AE_1$ and the time evolution with the Hamiltonian $H_1$. Dashed horizontal lines represent the accuracy achieved with the other encoding schemes with the same number of qubits, $N=7$. Adapted from the author’s previous work, Ref.~\cite{DeLorenzis_1}. Copyright (2025) by the American Physical Society.}
    \label{fig:Amplitude encoding}
\end{figure*}

Amplitude encoding requires a separate discussion. In this scheme, the
latent features are mapped to amplitudes of computational-basis states,
so that $N$ qubits can in principle encode up to $2^N$ features. To
enable a meaningful comparison with the other encodings, the study fixes
the number of qubits (for instance $N=7$) and varies instead the number
of encoded features in the latent space. As shown in
Fig.~\ref{fig:Amplitude encoding}, the accuracy of the QELM with
amplitude encoding increases with the latent dimension: for small
feature numbers, it underperforms the best angle-like encodings, but for
sufficiently large latent spaces it can match or surpass their accuracy.
This reflects the fact that amplitude encoding is able to exploit a much
larger portion of the Hilbert space when more features are injected into
the quantum state.

Overall, for the parameter regime and system sizes explored in
Ref.~\cite{DeLorenzis_1}, encodings that exploit both Bloch-sphere
angles per qubit (dense-angle and uniform Bloch-sphere) provide a good
compromise between performance and experimental feasibility: they reach
near-optimal accuracies in the range of latent dimensions realistically
accessible on near-term devices, without requiring the deep, highly
entangling state-preparation circuits typically associated with generic
amplitude encoding.

\subsection{Role of the Hamiltonian and comparison with classical baselines}
\label{subsec:results_hamiltonians}
Finally, we analyse how the choice of the Hamiltonian governing the
quantum layer affects the QELM performance. In this part of the study,
both the feature reduction ($AE_1$) and the encoding (dense-angle) are
kept fixed, while the time evolution is generated by the six
Hamiltonians introduced in Sec.~\ref{sec:hamiltonians}: the
time-dependent long-range model $H_1$, the fully connected disordered
transverse-field Ising model $H_2$, the local chaotic Ising chain $H_3$,
the XXZ chain $H_4$, the integrable XX model $H_5$, and the disordered
long-range $xx$ model $H_6$ in its non-ergodic and transition regimes.
\begin{figure*}[htbp]
    \centering
    \begin{subfigure}[t]{0.45\textwidth}
        \centering
        \includegraphics[width=\linewidth]{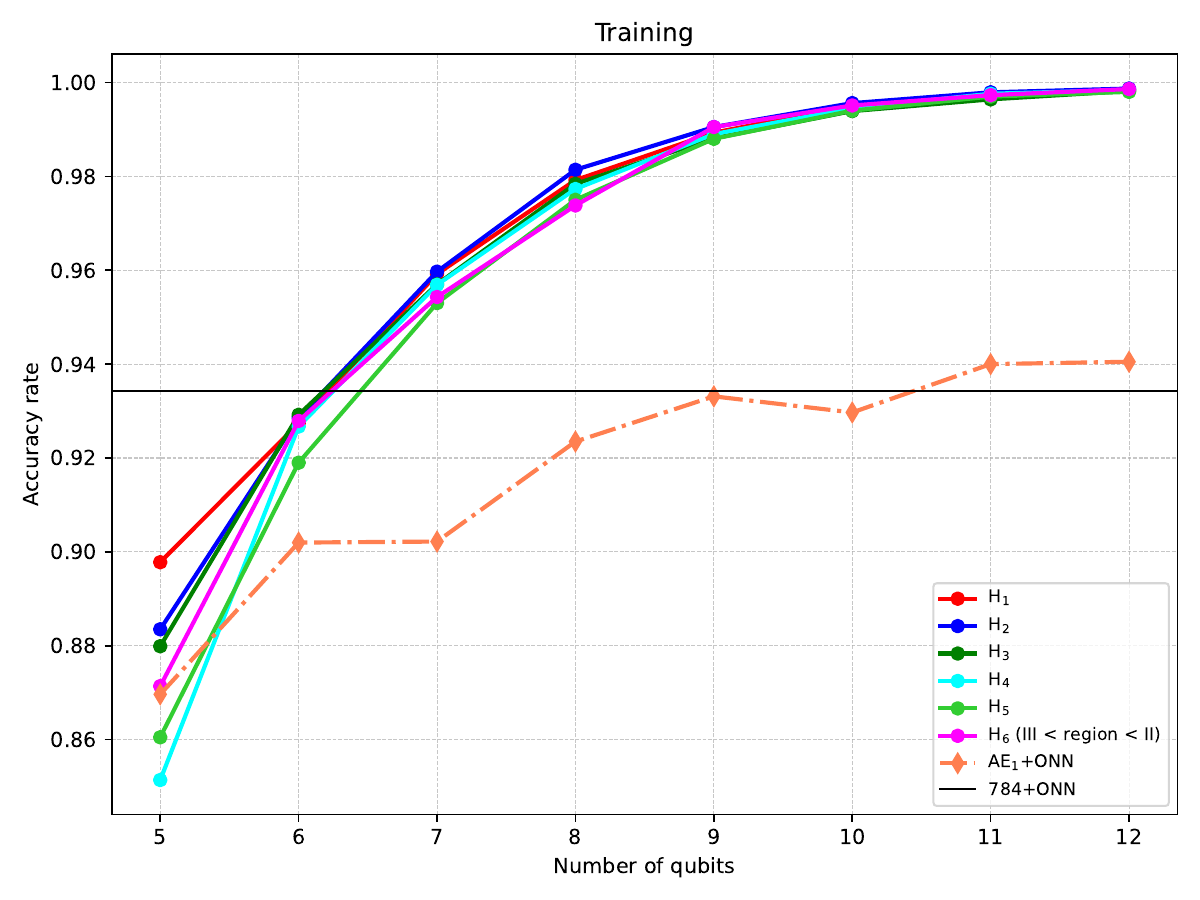}
        \caption{}
    \end{subfigure}
    \hspace{0.02\textwidth}
    \begin{subfigure}[t]{0.45\textwidth}
        \centering
        \includegraphics[width=\linewidth]{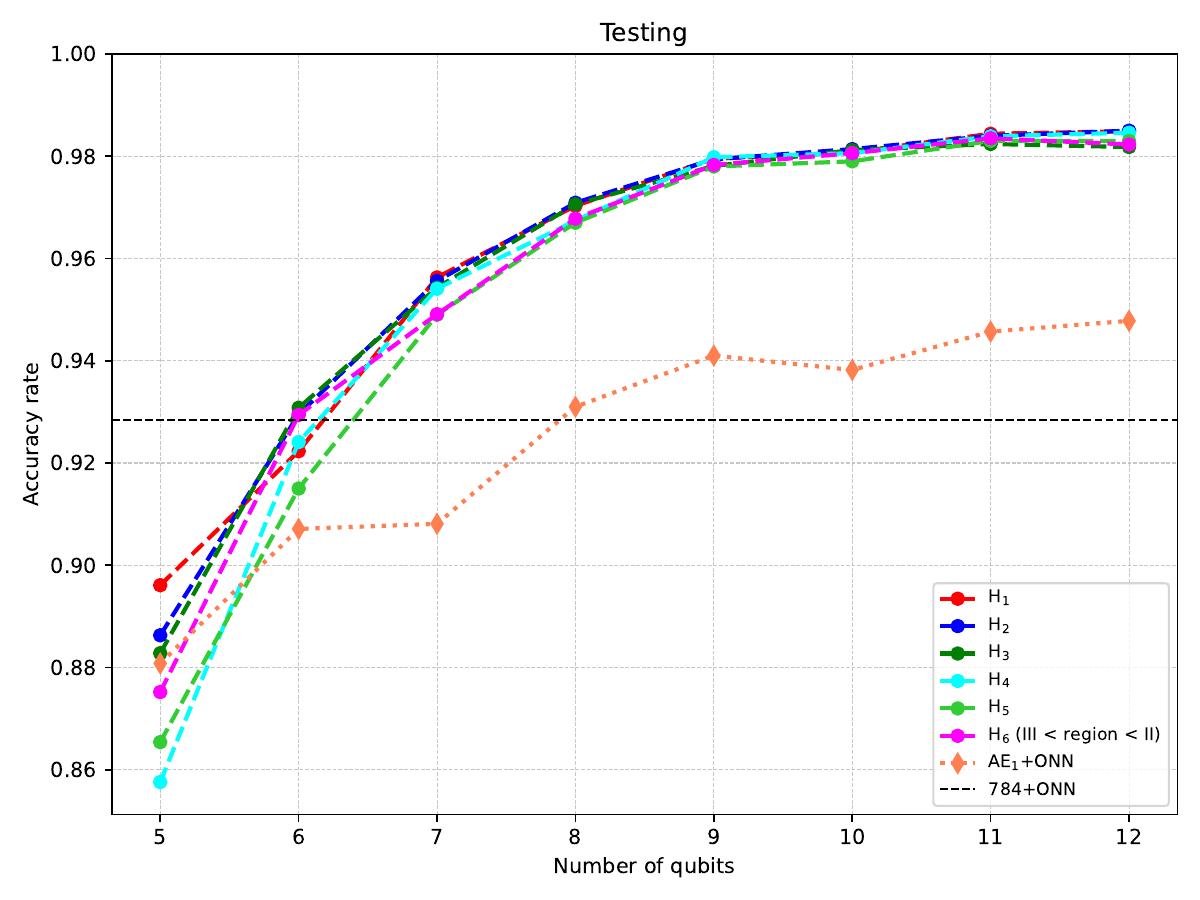}
        \caption{}
    \end{subfigure}
    \caption{Training (left panel) and testing (right panel) accuracy as a function of the number of qubits for six different Hamiltonians. The feature reduction and the encoding have been performed with the autoencoder $AE_1$ and the dense angle encoding, respectively. The horizontal line represents the accuracy obtained with a ONN fed with all the pixel values of the images. Instead, the coral line represents a ONN fed directly with the $2N$ features extracted by the $AE_1$. Adapted from the author’s previous work, Ref.~\cite{DeLorenzis_1}. Copyright (2025) by the American Physical Society.}
    \label{fig:Different_Hamiltonians}
\end{figure*}

Figure~\ref{fig:Different_Hamiltonians} shows the training and test
accuracy as a function of $N$ for these interacting Hamiltonians. In all
cases, the accuracy grows rapidly with the number of qubits and reaches
a plateau around $N \simeq 10$. Remarkably, once this plateau is
reached, the performance is essentially insensitive to the microscopic
details of the Hamiltonian: within numerical uncertainties, all
interacting models yield very similar accuracies for a given $N$.

This lack of sensitivity can be understood by noting that all these
Hamiltonians generate sufficiently complex many-body dynamics, which
spread the initially local information (injected by the encoding) over a
large set of basis states before measurement. As long as the evolution
efficiently redistributes and recombines the encoded features across the
register, the resulting quantum reservoir appears to be expressive
enough for the classification task, irrespective of integrability or
ergodicity in the strict sense.

\begin{figure*}[htbp]
    \centering
    \begin{subfigure}[t]{0.45\textwidth}
        \centering
        \includegraphics[width=\linewidth]{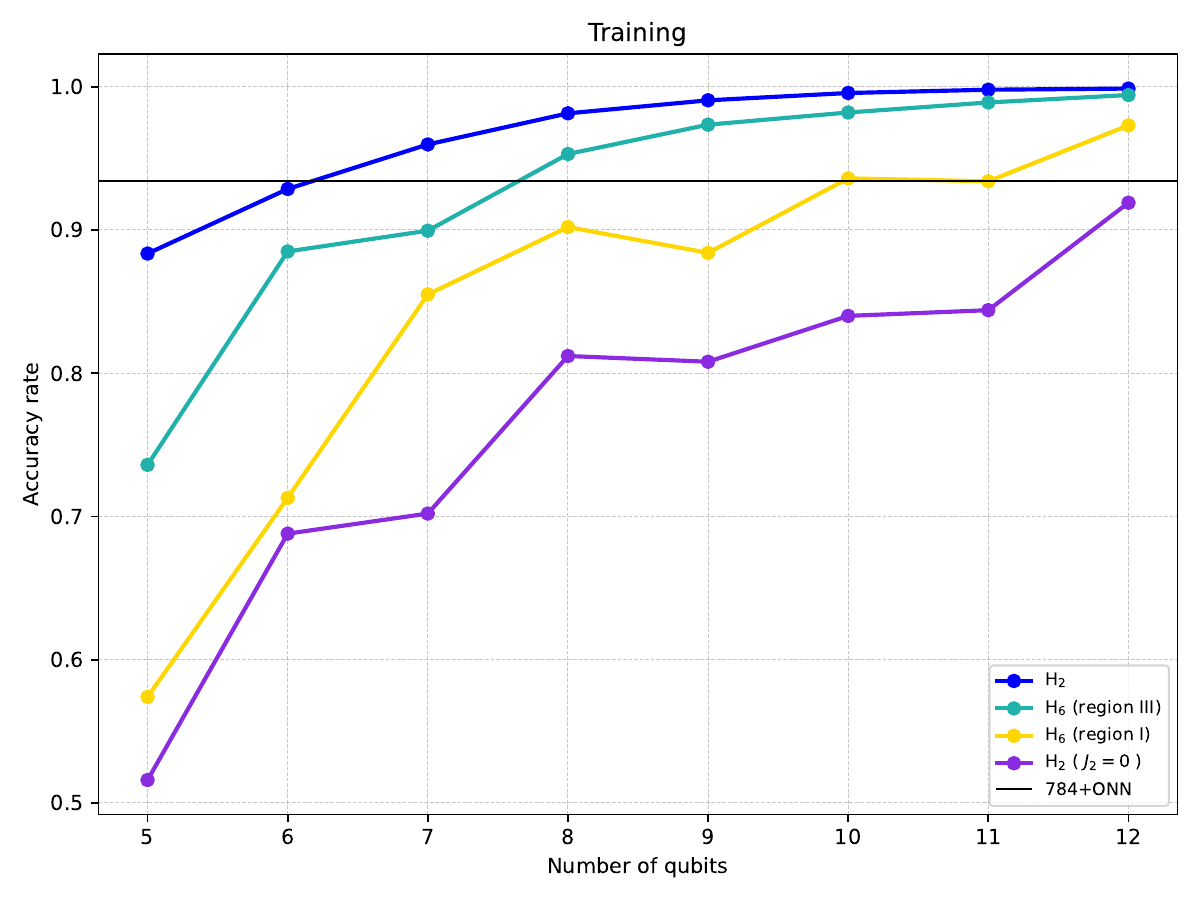}
        \caption{}
    \end{subfigure}
    \hspace{0.02\textwidth}
    \begin{subfigure}[t]{0.45\textwidth}
        \centering
        \includegraphics[width=\linewidth]{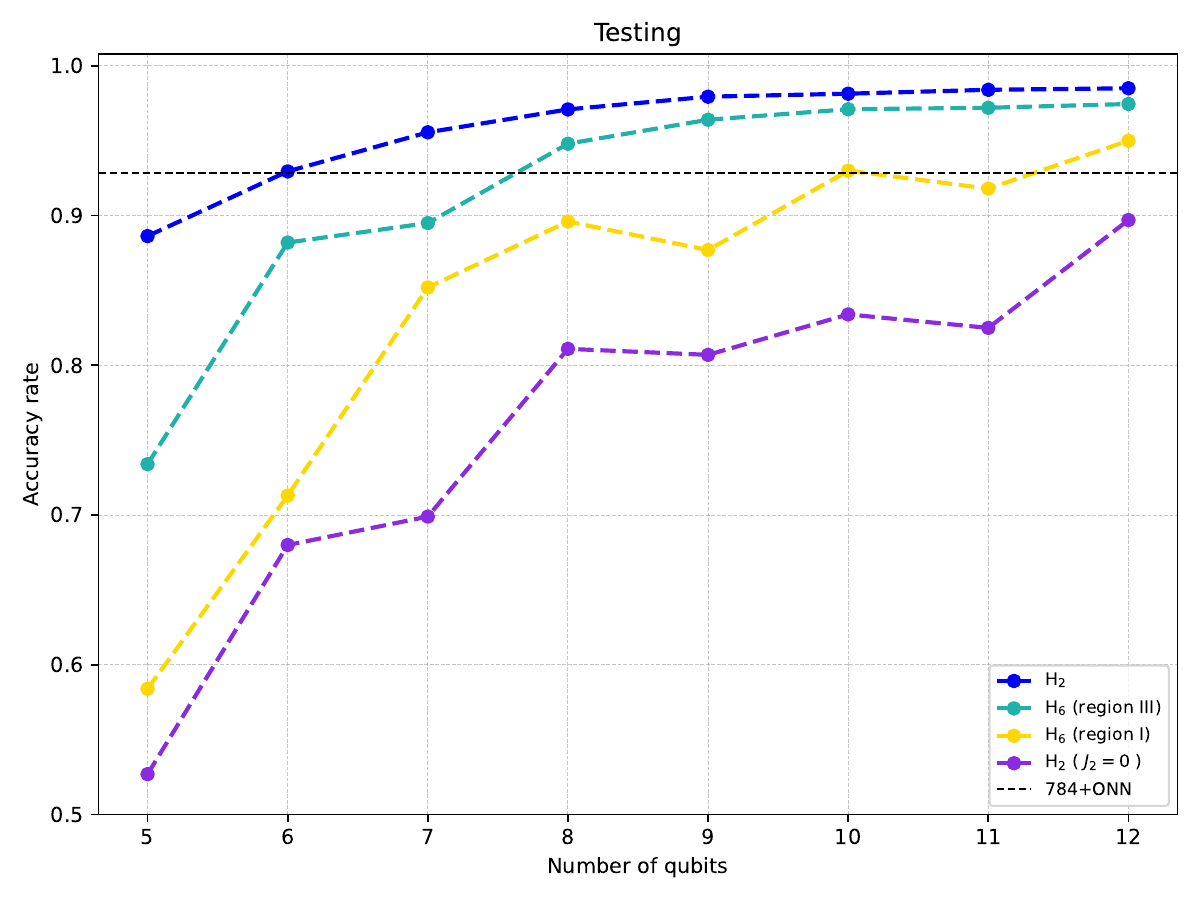}
        \caption{}
    \end{subfigure}
    \caption{Training (left panel) and testing (right panel) accuracy as a function of the number of qubits for other different Hamiltonians. The feature reduction and the encoding have been performed with the autoencoder $AE_1$ and the dense angle encoding, respectively. The horizontal line represents the accuracy obtained with a ONN fed with all the pixel values of the images. Adapted from the author’s previous work, Ref.~\cite{DeLorenzis_1}. Copyright (2025) by the American Physical Society.}
    \label{fig:Different_Hamiltonians_bad}
\end{figure*}

A contrasting behaviour is observed when the dynamics is either
non-interacting or strongly localized. This is illustrated in
Fig.~\ref{fig:Different_Hamiltonians_bad}, which reports the accuracy
for the disordered long-range Hamiltonian $H_6$ in localized parameter
regimes and for the non-interacting limit of $H_2$ (obtained by setting
the interaction strengths to zero). In these cases, the performance
drops significantly and no clear saturation to the same high-accuracy
plateau is observed. When interactions are switched off, the qubits
evolve independently and the information injected at the encoding stage
cannot be efficiently mixed across the register; in localized regimes,
the dynamics fails to delocalize the initial features over the
computational basis. Both effects limit the quality of the quantum
feature map.

To put the quantum results in context, several classical baselines are
also considered. A first baseline consists of a one-layer neural network
(ONN) trained directly on the original $784$-dimensional pixel vectors
(i.e.\ without feature reduction or quantum layer). A second baseline
feeds the same ONN with the $2N$-dimensional latent vectors produced by
$AE_1$, thus mimicking a purely classical architecture operating on the
compressed representations. The corresponding accuracies are shown as
horizontal or coral reference curves in
Figs.~\ref{fig:Different_Hamiltonians}. For small system sizes ($N \simeq
6$), the QELM accuracy is comparable to that of the best classical
baselines; as $N$ increases, however, the quantum model systematically
outperforms both the ONN acting on raw pixels and the ONN acting on
compressed features.

An additional comparison can be drawn by encoding the full $784$
features into $N\geq 10$ qubits using amplitude encoding. In this case,
the QELM achieves test accuracies around $0.97$ for $N=10$--$12$, which
are slightly lower than those obtained with dense-angle encoding and
compressed latent representations for the same number of qubits. This
suggests that, in the present setting, combining dimensionality
reduction with an expressive but relatively shallow quantum feature map
is more advantageous than directly encoding the uncompressed data via
amplitude encoding.

\section{Conclusions of the first QELM study}
\label{sec:paper1_conclusions}
In this chapter, we have presented our study on the application of the general Quantum Extreme Learning Machine framework to image-classification tasks. The results were published in Ref.~\cite{DeLorenzis_1}, with the author of this thesis as first author. The architecture implements a full
quantum machine-learning pipeline, starting from classical images,
compressing them to a low-dimensional latent representation, encoding
the latent features into quantum states, evolving them under a
many-body Hamiltonian, and finally performing classification with a
single-layer classical neural network in the spirit of extreme learning
machines.

The first ingredient that we analysed is the feature-reduction stage,
where we compared a linear Principal Component Analysis with two
nonlinear convolutional autoencoders. As summarised in
Sec.~\ref{subsec:results_feature_reduction}, the autoencoder-based approach
consistently achieves higher classification accuracies than PCA, even on
a relatively simple dataset such as MNIST, and the gap persists when the
model is tested on the more challenging Fashion-MNIST dataset. This
indicates that a nonlinear latent representation is already beneficial
in the strongly compressed regime relevant for near-term QELMs, and
suggests that the advantage of autoencoders should become even more
pronounced for more complex datasets.

We then investigated the impact of different data-encoding strategies
into quantum states. Among the single-qubit schemes considered
(dense-angle, simple angle, uniform Bloch-sphere, and general
encoding), those that exploit both Bloch-sphere angles per qubit
(dense-angle and uniform Bloch-sphere encodings) deliver the best
performance, while remaining comparatively easy to implement on actual
hardware. Amplitude encoding, although experimentally more demanding,
offers a favourable scaling in terms of the number of features per
qubit, and can outperform single-qubit encodings when the number of
available qubits is very limited. Overall, the results point to
angle-like encodings with two features per qubit as a particularly
appealing compromise between expressive power and practical feasibility.

A further aspect of the study concerns the role of the Hamiltonian that
generates the quantum evolution. We benchmarked six different models,
including time-dependent and time-independent dynamics, fully connected
and local interactions, and examples designed to probe the influence of
integrability, ergodicity, and localization. The numerical analysis
shows that, as long as the Hamiltonian induces sufficiently interacting
dynamics that spread information from the initially encoded local
degrees of freedom to the measurement basis, the classification
accuracy is largely insensitive to the microscopic details of the model.
By contrast, in regimes where localization suppresses the delocalisation
of information, the performance deteriorates. Within the precision of
our simulations, all interacting Hamiltonians considered provide very
similar accuracies, suggesting that one can choose the model that is
most convenient from an implementational point of view without
sacrificing performance.

For the final classification step we employed a shallow, fully connected
one-layer neural network, in line with the extreme learning machine
paradigm, so that all trainable parameters are confined to the classical
readout. Interestingly, the accuracies obtained in this setting are
comparable to, or better than, those reported in other quantum
machine-learning approaches to image classification, such as Quantum Capsule Networks \cite{Liu:2022bpb}, Quantum Convolutional Neural Networks \cite{QCNNHur:2021zyz}, the QRC proposed by \cite{Kornjaca2024}, the Quantum Bayes Classifiers proposed in \cite{BayesWang:2024atn} or the approaches proposed in \cite{Slabbert:2024fjv, Sein:2024ael}. This places QELMs
among the competitive hybrid quantum–classical models for near-term
image-classification tasks.

The present study can be extended in several directions. One natural
avenue is to compare QELMs with a broader set of classical baselines,
using multiple performance metrics (e.g.\ accuracy, training time, and
resource consumption) in order to quantify more systematically any
quantum-inspired advantage. Another is to investigate the concrete
implementability of the proposed architectures on real quantum hardware,
including noise effects and hardware-specific constraints. These aspects will be further explored in the subsequent chapter, where we focus on the entanglement structure and classical simulability of QELMs based on the XX Hamiltonian.
The methodological relevance of this study was further recognized by the subsequent selection of the corresponding article for inclusion in\textit{Quantum Frontiers}, a curated collection by \textit{Physical Review Applied}~\cite{QuantumFrontiersAPS}.

\newpage
\chapter{Entanglement and Classical Simulability in QELMs}
\label{chap:Paper_2}

\section{Overview and goals of the study}
\label{sec:paper2_overview}

In the previous chapters we introduced the general framework of Quantum
Extreme Learning Machines (QELMs) and applied it to image-classification
tasks, analysing how different choices of feature reduction, encoding,
and Hamiltonian affect the performance of the model. In this chapter we
build on those results and focus on a more refined question: how do
entanglement and information propagation within the quantum layer relate
to the classification capabilities and classical simulability of the
QELM? The analysis presented here was carried out by the author and forms the basis of Ref.~\cite{DeLorenzis_2}, currently under review, in which the author is the first author.

In contrast to the first study, where several Hamiltonians and encoding
schemes were compared, the QELM architecture considered here is kept
fixed. The classical preprocessing pipeline, the latent
compression stage, and the classical output layer are the same as those
introduced in Chapter~\ref{ch:QELM} and implemented in
Chapter~\ref{ch:paper1}. On the quantum side, we always use dense-angle
encoding to map the latent features to an $N$-qubit register, and the
quantum layer is implemented by a nearest-neighbour XX Hamiltonian with
periodic boundary conditions. Within this fixed architecture, the only
continuous hyperparameter that we vary is the evolution time $t$ of the
quantum dynamics.

The main goals of this chapter are threefold. First, we investigate how
the QELM classification accuracy depends on the evolution time for
different datasets, namely MNIST, Fashion-MNIST, and CIFAR-10. As we shall see, the accuracy exhibits a sharp transition from a low to a high plateau as a function of $t$, with a critical time that is
essentially independent of the system size and whose plateau value coincides with that obtained from Haar-random unitaries acting on the same input states. Second, we analyse the dynamics of entanglement and information spreading in the XX chain, relating the observed accuracy
transition to the onset of local entanglement rather than to fully scrambled, volume-law entangled states. Third, we discuss the classical simulability of the QELM in this regime, by comparing it with shallow random circuits and by exploiting the limited depth and moderate entanglement growth of the dynamics.

Overall, this chapter provides a complementary perspective on QELMs. The first study established that interacting quantum reservoirs can act as
effective feature generators for image classification; here we show that, for the architectures and tasks considered, the relevant quantum
dynamics can be realised by local, integrable models with limited entanglement growth, which suggests that the corresponding QELMs remain
within reach of efficient classical simulation for a broad class of problems.

\section{QELM setup for this study}
\label{sec:paper2_setup}
In order to avoid unnecessary repetition, we do not re-derive the full
QELM architecture here, but rather specify the particular choices made
in our work of Ref.~\cite{DeLorenzis_2} and refer to Chapters~\ref{ch:QELM} and
\ref{ch:paper1} for the general definitions of feature reduction,
encoding, quantum dynamics, and classical readout.

\subsection{Datasets and classical preprocessing}
\label{sec:paper2_datasets}
In this second study we consider three standard image-classification
benchmarks: MNIST, Fashion-MNIST, and CIFAR-10. The first two were already used in the previous chapter, while CIFAR-10 is introduced here as a more challenging dataset with colour images.
\subsection*{MNIST and Fashion-MNIST (brief recap)}
We reuse the MNIST and Fashion-MNIST datasets introduced in
Sec.~\ref{sec:paper1_datasets}, consisting of grayscale $28\times 28$
images of handwritten digits and clothing items, respectively, organized
into $10$ classes. Pixel values are normalized to the range $[0,1]$ in
both cases. When PCA is used as a feature-reduction method, the images
are flattened into $784$-dimensional real vectors and projected onto the
first $d$ principal components. When convolutional autoencoders are
employed, the $28\times 28$ images are instead fed to the encoder in
their natural tensor form, and the resulting $d$-dimensional latent
vectors are used as inputs to the QELM.

\subsection*{CIFAR-10}
\label{subsec:cifar10}

CIFAR-10 provides a more demanding benchmark, with $60\,000$ colour
images of size $32\times 32$ pixels and three colour channels (red, green, and blue), grouped into $10$ classes (airplane, automobile, bird, cat, deer, dog, frog, horse, ship, and truck). We use the standard split of $50\,000$ training and $10\,000$ test images. Compared to MNIST and Fashion-MNIST, CIFAR-10
contains richer textures, backgrounds, and illumination conditions, which makes the classification task significantly harder and offers a
stringent test for the QELM architecture. Representative examples from the CIFAR-10 dataset are shown in Fig.~\ref{fig:Datasets_sample} (bottom).

For CIFAR-10, pixel values are normalized to the range $[0,1]$ by
dividing the original $8$-bit RGB intensities by $255$. When PCA is
used as a feature-reduction method, the $32\times 32\times 3$ images
are flattened into $3072$-dimensional real vectors before projection
onto the first $d$ principal components. When autoencoders are employed,
the images are instead fed to the encoder in their natural tensor form
($32\times 32\times 3$), and the resulting $d$-dimensional latent
vectors are used as inputs to the QELM.

In all three datasets, the $d$-dimensional latent representations are
finally rescaled to match the input range required by the chosen
encoding scheme (see Sec.~\ref{ch:encoding}) and then encoded into
quantum states as described in Sec.~\ref{sec:QELMs}.

\subsection{Quantum layer: XX Hamiltonian and encoding}
\label{subsec:paper2_quantum_layer}

The quantum part of the QELM considered in this chapter is deliberately
kept as simple and structured as possible. The encoding of classical
features into quantum states is always implemented via the dense-angle
encoding introduced in Chapter~\ref{ch:encoding}. Given a latent vector
$\mathbf{z} = (z_1,\dots,z_d)$ of dimension $d=2N$, pairs of components
$(z_{2i-1},z_{2i})$ are mapped to the polar and azimuthal angles of the
$i$-th qubit, yielding an $N$-qubit product state
$\ket{\psi_0(\mathbf{z})}$.

The subsequent time evolution is generated by a one-dimensional XX
Hamiltonian with nearest-neighbour interactions and periodic boundary
conditions,
\begin{equation}
  H = \frac{1}{2}\sum_{i=1}^{N}
      \left(
        \sigma_x^{(i)}\sigma_x^{(i+1)} +
        \sigma_y^{(i)}\sigma_y^{(i+1)}
      \right),
\label{eq:XX_paper2}
\end{equation}
where $\sigma_\alpha^{(i)}$ denotes the Pauli matrix $\sigma_\alpha$ on
site $i$, and the index $i+1$ is understood modulo $N$. This model is
integrable and can be mapped to free fermions via the
Jordan–Wigner transformation, yet its dynamics is capable of generating
non-trivial entanglement and information propagation along the chain.

For a given evolution time $t$, the encoded state is transformed
according to
\begin{equation}
  \ket{\psi_t(\mathbf{z})} = U(t)\ket{\psi_0(\mathbf{z})},
  \qquad
  U(t) = e^{-{\rm i}Ht}.
\end{equation}
Throughout this chapter, all structural aspects of the QELM (feature
reduction, encoding, and measurement) are kept fixed, and the only
continuous hyperparameter that we vary is the evolution time $t$. This
allows us to interpret the QELM as a one-parameter family of feature
maps, and to study how the classification performance, entanglement
properties, and classical simulability depend on the duration of the
quantum dynamics.

\subsection{Measurement and classical classifier}
\label{subsec:paper2_measurement_readout}

After the evolution under the XX Hamiltonian, the $N$-qubit state
$\ket{\psi_t(\mathbf{z})}$ is measured in the computational basis. For
each input sample, we estimate the probabilities
\begin{equation}
  p_t(s\mid\mathbf{z}) = \left|\langle s|{\psi_t(\mathbf{z})}\rangle\right|^2,
\end{equation}
where $s \in \{0,1\}^N$ labels the $2^N$ possible bit strings. These
probabilities are arranged into a classical feature vector
$\mathbf{p}_t(\mathbf{z}) \in \mathbb{R}^{2^N}$, which represents the
point associated with the sample in the probability simplex (or
``probability polytope'').

This feature vector is then passed to a simple classical output layer,
implemented as a one-layer neural network (ONN). The ONN consists of a
fully connected layer mapping $\mathbf{p}_t(\mathbf{z})$ to $C$ output
neurons, where $C$ is the number of classes, followed by a softmax
activation that yields class probabilities. The weights of this layer
are the only trainable parameters of the QELM; they are optimised by
minimising the categorical cross-entropy loss using the Adam optimizer,
with training, validation and test procedures identical to those
described in Chapter~\ref{ch:ml_background}. No backpropagation is
performed through the quantum circuit: the quantum layer acts solely as
a fixed feature generator parametrised by the evolution time~$t$.

In the following sections we will use this setup to study how the
classification accuracy, entanglement dynamics, and geometric structure
of the data in probability space evolve as functions of~$t$ and of the
system size~$N$.
\section{Performance as a function of evolution time}
\label{sec:paper2_performance}

In this section we investigate how the classification performance of the
QELM depends on the evolution time $t$ of the XX Hamiltonian introduced
in Sec.~\ref{subsec:paper2_quantum_layer}. For each dataset and system
size $N$, we compute the training and test accuracy as a function of
$t$, keeping all other elements of the architecture fixed. This allows
us to interpret the QELM as a one-parameter family of feature maps and
to identify the time scales that are most relevant for learning.

\subsection{Accuracy curves for MNIST, Fashion-MNIST and CIFAR-10}
\label{subsec:paper2_accuracy_vs_t}

\begin{figure*}[htbp]
  \centering
  \begin{subfigure}[b]{0.4\textwidth}
    \centering
    \includegraphics[width=\textwidth]{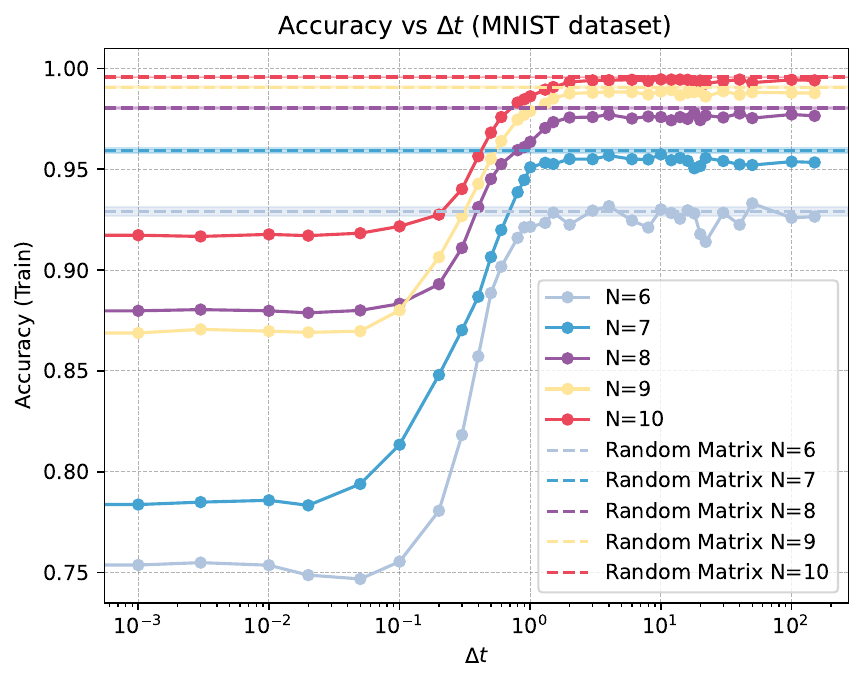}
    \caption{}
  \end{subfigure}
  \hfill
  \begin{subfigure}[b]{0.4\textwidth}
    \centering
    \includegraphics[width=\textwidth]{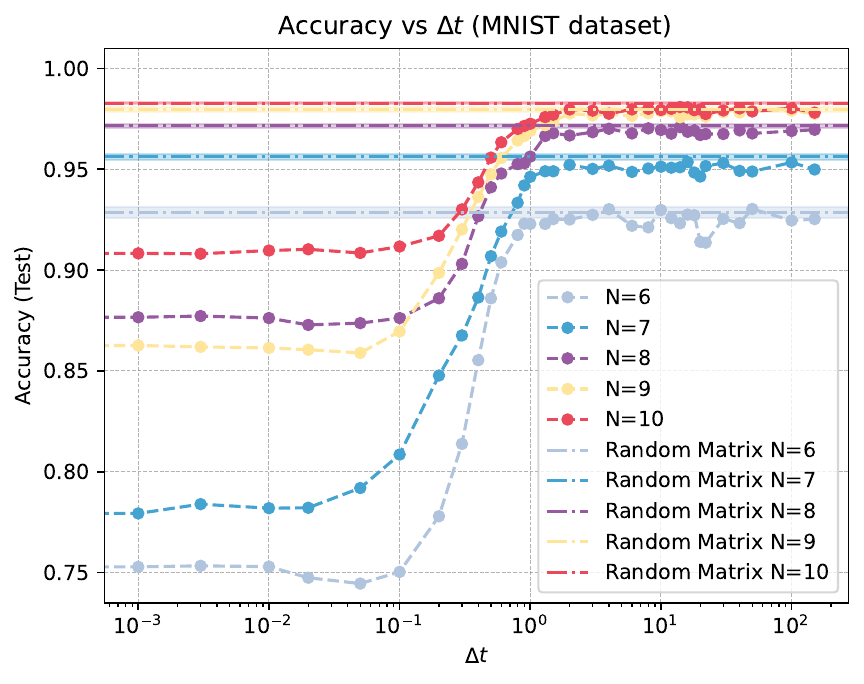}
    \caption{}
  \end{subfigure}

  \caption{Training (left panel) and testing (right panel) accuracy as a function of evolution time using the MNIST dataset. The time evolution and the encoding have been performed with the XX Hamiltonian and dense-angle encoding, respectively. The horizontal dashed lines indicate the performance obtained using a random unitary matrix, averaged over 10 samples with the shaded band denoting the standard deviation. Adapted from the author’s previous work, Ref.~\cite{DeLorenzis_2}. Copyright (2026) by the American Physical Society.}
  \label{fig:MNIST}
\end{figure*}

\begin{figure*}[htbp]
  \centering
  \begin{subfigure}[b]{0.4\textwidth}
    \centering
    \includegraphics[width=\textwidth]{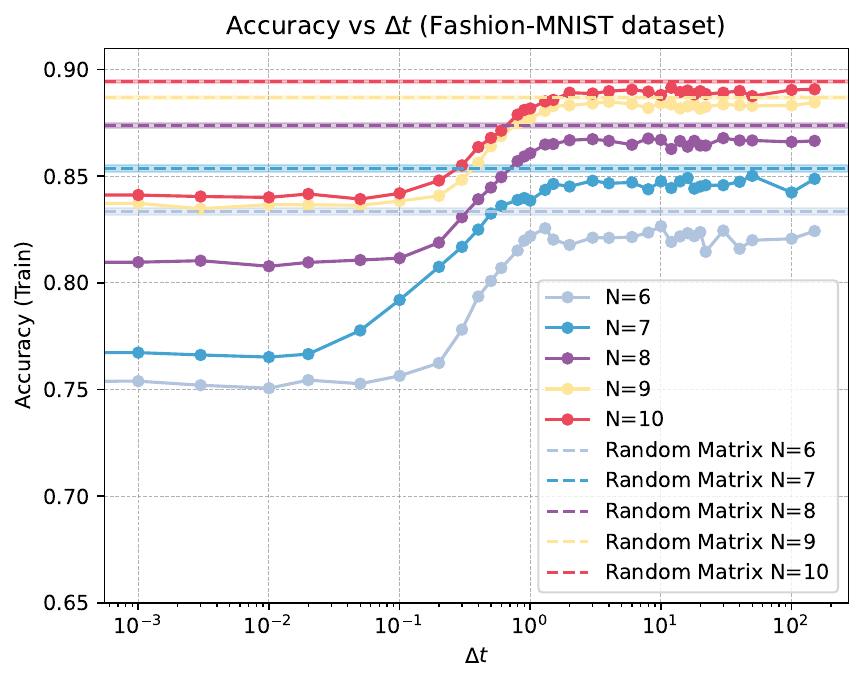}
    \caption{}
  \end{subfigure}
  \hfill
  \begin{subfigure}[b]{0.4\textwidth}
    \centering
    \includegraphics[width=\textwidth]{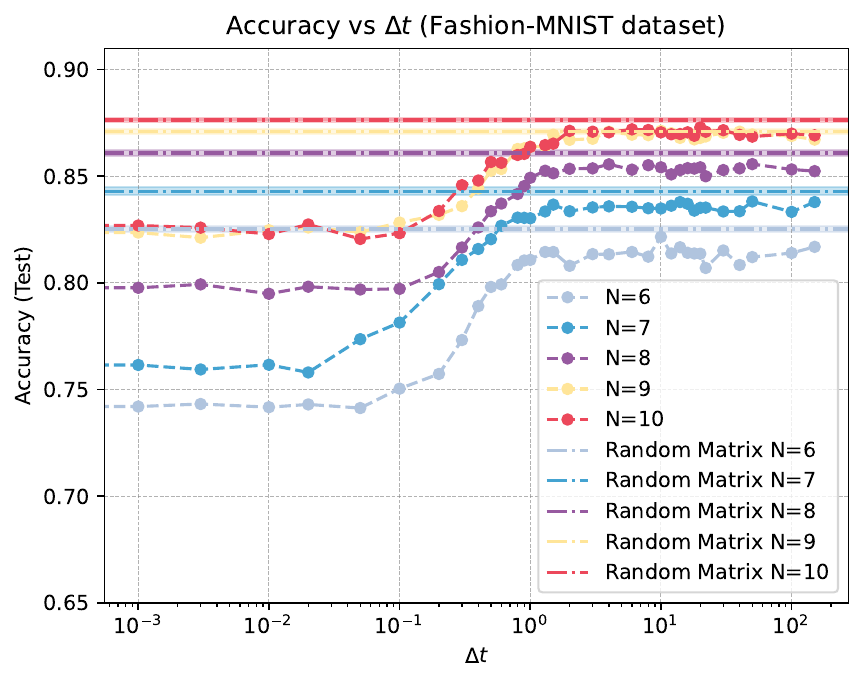}
    \caption{}
  \end{subfigure}

  \caption{Training (left panel) and testing (right panel) accuracy as a function of evolution time using the Fashion-MNIST dataset. The time evolution and the encoding have been performed with the Hamiltonian XX model and the dense angle, respectively. The horizontal dashed lines indicate the performance obtained using a random unitary matrix, specifically by performing 10 measurements and computing the mean and standard deviation. Adapted from the author’s previous work, Ref.~\cite{DeLorenzis_2}. Copyright (2026) by the American Physical Society.}
  \label{fig:FashionMNIST}
\end{figure*}

\begin{figure*}[htbp]
  \centering
  \begin{subfigure}[b]{0.4\textwidth}
    \centering
    \includegraphics[width=\textwidth]{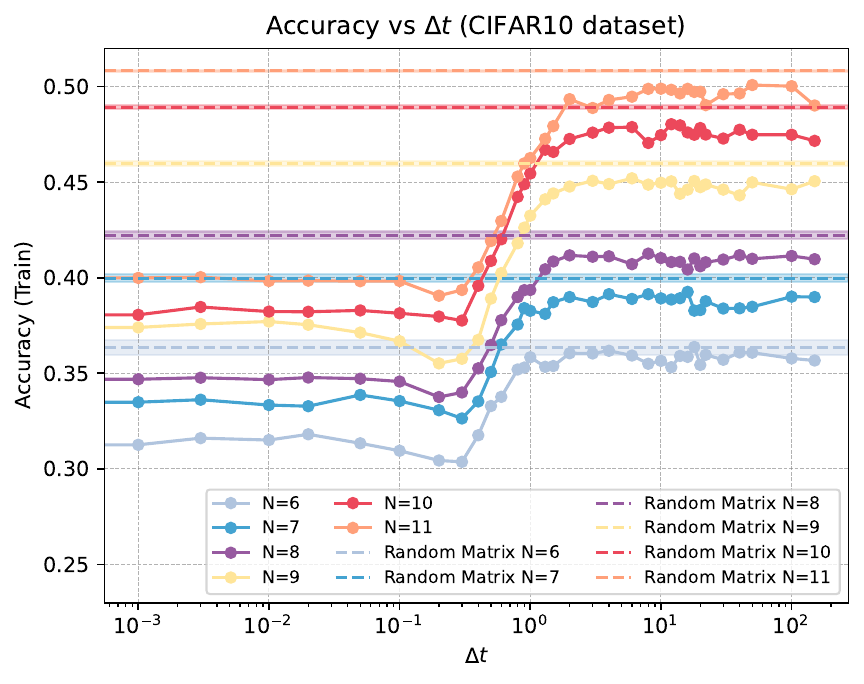}
    \caption{}
  \end{subfigure}
  \hfill
  \begin{subfigure}[b]{0.4\textwidth}
    \centering
    \includegraphics[width=\textwidth]{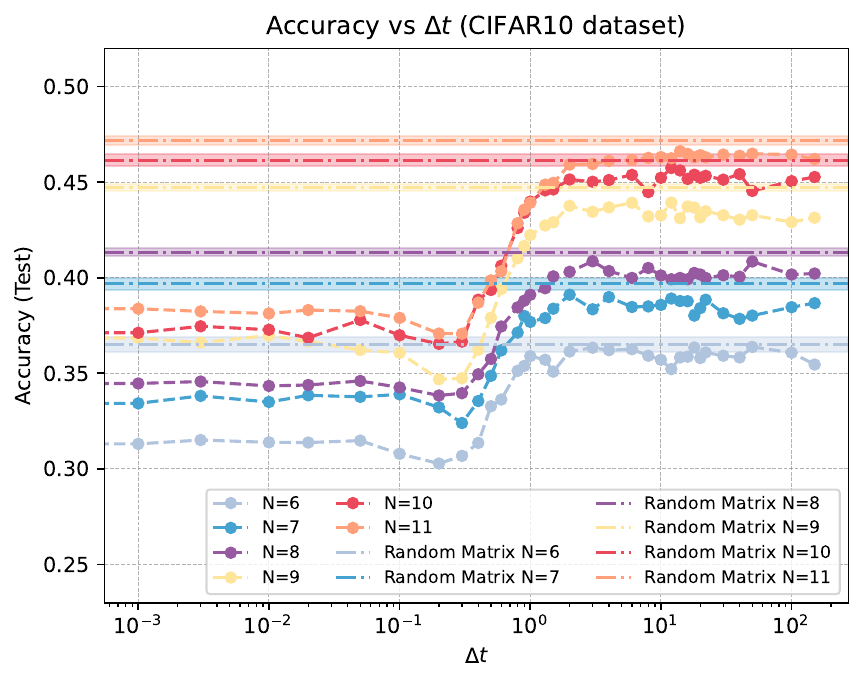}
    \caption{}
  \end{subfigure}

  \caption{Training (left panel) and testing (right panel) accuracy as a function of evolution time using the CIFAR-10 dataset. The time evolution and the encoding have been performed with the Hamiltonian XX model and the dense angle, respectively. The horizontal dashed lines indicate the performance obtained using a random unitary matrix, specifically by performing 10 measurements and computing the mean and standard deviation. Adapted from the author’s previous work, Ref.~\cite{DeLorenzis_2}. Copyright (2026) by the American Physical Society.}
  \label{fig:CIFAR10}
\end{figure*}

Figures~\ref{fig:MNIST}, \ref{fig:FashionMNIST} and
\ref{fig:CIFAR10} show the training (left panels) and test (right
panels) accuracy as a function of the evolution time $t$ for the MNIST,
Fashion-MNIST and CIFAR-10 datasets, respectively. In each plot, the different coloured curves correspond to different numbers of qubits $N$ in the quantum layer: for the simpler datasets (MNIST and Fashion-MNIST) we consider $N$ between $6$ and $10$, while for the more demanding CIFAR-10 dataset we use up to $N=11$ qubits. For reference, the horizontal dashed lines indicate the accuracy obtained when the XX evolution is replaced by a Haar-random unitary acting on the same encoded states. This benchmark helps assess whether the plateau reached by the XX dynamics depends on the specific structure of the Hamiltonian or can already be reproduced by an essentially random feature generator. It is worth stressing that all three datasets correspond to 10-class classification tasks, so that random guessing would yield an accuracy of about $10\%$.


For all datasets and system sizes, the accuracy curves exhibit the same
qualitative behaviour. For very short evolution times, $t \ll 1$, the
accuracy is essentially flat and close to the value obtained when the
quantum evolution is negligible, i.e.\ when the quantum layer acts
almost as the identity. In this regime, the QELM does not fully exploit
the expressive power of the quantum reservoir, and the performance is
mainly determined by the quality of the classical feature reduction and
the expressiveness of the one-layer neural network.

As $t$ increases, a sharp transition occurs around a characteristic time
$t^\ast \sim 1$. In a relatively narrow time window centred at $t^\ast$,
both the training and test accuracies rise rapidly from their initial values to a higher plateau. The location of this transition is remarkably insensitive to the number of qubits: for all $N$ and all datasets considered, the rapid increase in accuracy occurs around the same evolution time $t^\ast$. This suggests that the relevant time scale is set by local properties of the dynamics rather than by the total system size.

For evolution times $t \gtrsim t^\ast$, the accuracy saturates to a
plateau value $A^\ast(N)$ that depends on the number of qubits. In the
case of MNIST, $A^\ast$ quickly approaches high values (close to unity)
as $N$ increases, and the test accuracy reaches the same plateau as the
training accuracy, indicating that overfitting is not significant in
this regime. For Fashion-MNIST, the overall structure of the curves is
the same, but the plateau value is slightly lower, reflecting the
increased difficulty of the classification task. For CIFAR-10, which is
substantially more complex and involves colour images mapped to a
grayscale or compressed representation, the plateau accuracies are
significantly lower, and the dependence on $N$ is more pronounced. In
all cases, however, the three regimes (initial flat region, sharp
transition around $t^\ast$, and saturation plateau) are clearly visible.

It is also worth noting that, beyond the transition time, the accuracy
curves may display small oscillations as a function of $t$, especially
for larger system sizes. These oscillations reflect coherent revivals in
the XX dynamics, but their amplitude is modest and they do not spoil the
overall picture of a well-defined saturation plateau for $t \gtrsim
t^\ast$.

\subsection{Comparison with random unitary dynamics}
\label{subsec:paper2_random_unitaries}

To put the performance of the XX-based QELM into perspective, our study compares it with a hypothetical model in which the quantum layer is replaced by a random unitary transformation drawn from the Haar measure on $\mathrm{U}(2^N)$, i.e. from a distribution that samples unitary operators uniformly at random (see, e.g., Ref.~\cite{Mele:2023ojv, Bengtsson:2006rfv, Mezzadri:2006nac, Page:1993df}). Concretely, for each system size $N$ and each dataset, the authors generate several random
unitary matrices $U_{\mathrm{rand}}$ and compute the classification
accuracy obtained when the encoded states are transformed by
$U_{\mathrm{rand}}$ instead of by the XX time evolution. Since different
random unitaries can lead to slightly different accuracies, the results
are averaged over $10$ independent Haar-random samples; the mean
accuracy is shown as a horizontal dashed line in
Figs.~\ref{fig:MNIST}–\ref{fig:CIFAR10}, with the associated
standard deviation indicated by a shaded band.

A striking observation emerges from this comparison: for each system
size $N$ and each dataset, the saturation value $A^\ast(N)$ reached by
the XX QELM for $t \gtrsim t^\ast$ coincides, within the numerical
uncertainties, with the accuracy achieved by the Haar-random unitaries.
In other words, once the evolution time is sufficiently long for the XX chain to reach its plateau, the classification performance becomes comparable to that obtained with Haar-random unitary processing of the encoded states, indicating that, in the plateau regime, what matters most is the ability of the evolution to spread and reshape the encoded information in feature space, rather than the specific microscopic form of the XX Hamiltonian.

This result is particularly remarkable in view of the properties of the
XX Hamiltonian. The XX model employed in the QELM is \emph{highly specific}: its dynamics is far from generic in the space of all unitaries, it is strictly local (nearest-neighbour interactions on a
one-dimensional ring), translationally invariant, and integrable, as it
can be mapped to free fermions by a Jordan–Wigner transformation followed by a Fourier transform. Haar-random unitaries, by contrast, correspond to maximally complex, fully nonlocal evolutions that have no
special structure.

The fact that such a simple, local, and integrable Hamiltonian can
match the performance of Haar-random dynamics on these classification
tasks suggests that, from the point of view of the classical readout, what matters is not the microscopic complexity of the unitary, but rather the ability of the evolution to spread and mix the encoded
information over a sufficiently large portion of Hilbert space. In the
following sections we will see that this spreading can be understood in terms of the growth of entanglement and the reshaping of the data distribution in probability space, and that it occurs on a time scale
set by local properties of the XX dynamics rather than by global scrambling.
\section{Entanglement dynamics and transition time}
\label{sec:paper2_entanglement}
The sharp change in classification accuracy observed around the
characteristic time $t^\ast$ suggests a dynamical transition in the way
the quantum layer processes information. In order to characterise this transition, we analyse the growth of entanglement
in the XX chain as a function of time and system size. In this section we summarise these results and relate them to the behaviour of the QELM performance.

\subsection{Growth of single-qubit and half-chain entropy}
\label{subsec:paper2_entropy}
A natural way to quantify entanglement in the evolved state
$\ket{\psi_t(\mathbf{z})}$ is through the von Neumann entropy \cite{Nielsen_Chuang_2010, Amico:2007ag, Streltsov:2010kyq, Horodecki:2009zz, Vedral:2002zz, Calabrese:2009qy} of reduced density matrices. The von Neumann entropy quantifies the mixedness of a quantum state and, for pure bipartite states, provides a standard measure of entanglement. Given a bipartition of the $N$-qubit chain
into subsystems $A$ and $B$, with reduced density matrix
$\rho_A(t) = \mathrm{Tr}_B \ket{\psi_t(\mathbf{z})}\bra{\psi_t(\mathbf{z})}$,
the entanglement entropy is defined as
\begin{equation}
  S_A(t) = - \mathrm{Tr}\left[\rho_A(t)\,\log \rho_A(t)\right].
\end{equation}
We considered two cases: (i) $A$ consisting
of a single qubit, leading to the single-site entropy $S_1(t)$; and
(ii) $A$ corresponding to half of the chain, leading to the half-chain
entropy $S_{N/2}(t)$.

\begin{figure*}[htbp]
  \centering
  \begin{subfigure}[b]{0.45\textwidth}
    \centering
    \includegraphics[width=\textwidth]{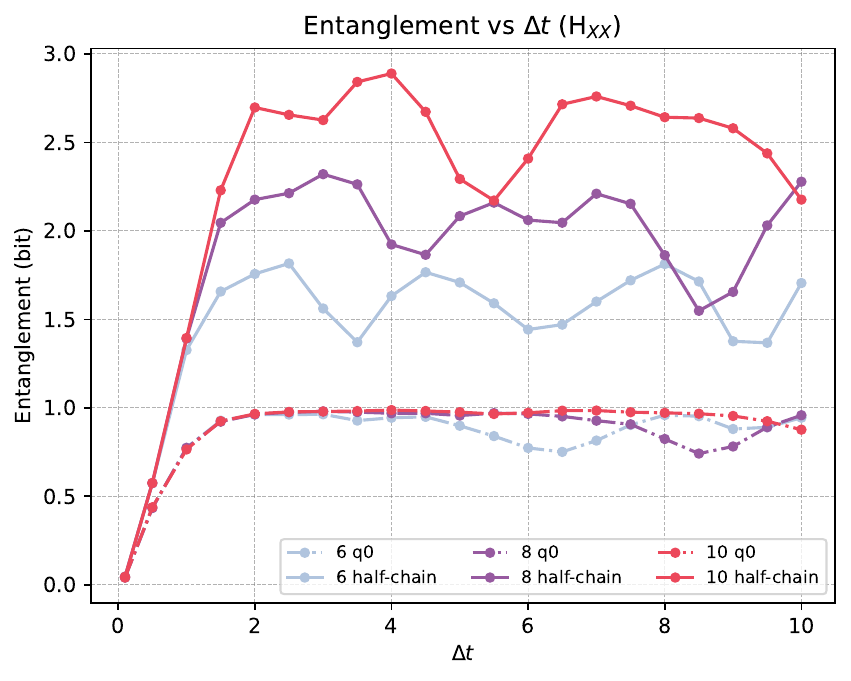}
    \caption{}
  \end{subfigure}
  \hfill
  \begin{subfigure}[b]{0.45\textwidth}
    \centering
    \includegraphics[width=\textwidth]{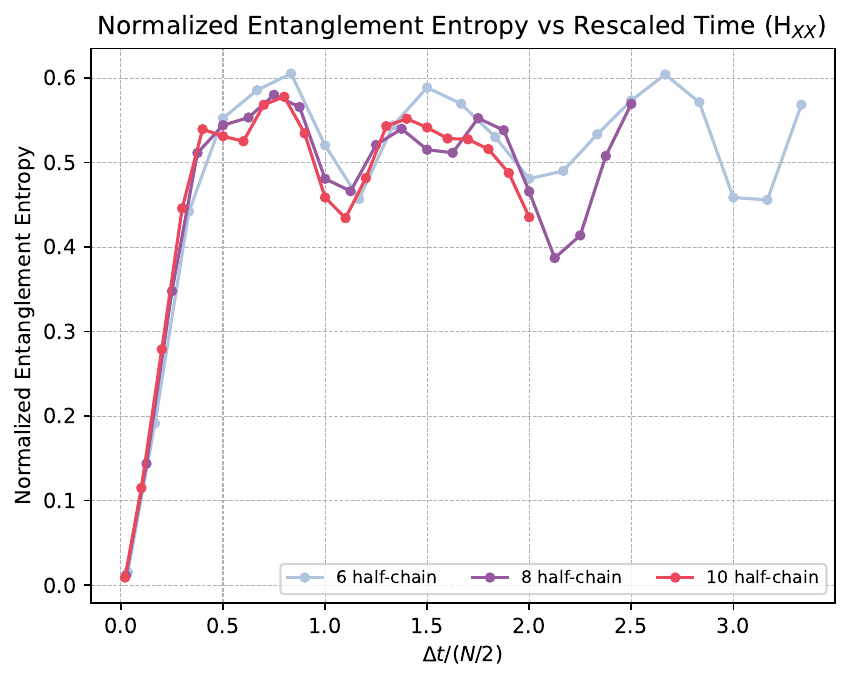}
    \caption{}
  \end{subfigure}

  \caption{a) Von Neumann entropy as function of the evolution time of half-chain (solid lines) and of the single qubit (dash-dotted lines). The colors denote the different sizes $N$ ($N=6, 8, 10$) of the entire system. b) Normalized (with $N/2$) von Neumann entropy  as function of the evolution time (divided by $N/2$) of half-chain. Adapted from the author’s previous work, Ref.~\cite{DeLorenzis_2}. Copyright (2026) by the American Physical Society.}
  \label{fig:Entanglement}
\end{figure*}

Figure~\ref{fig:Entanglement}(a) shows the time evolution of the
single-qubit entropy $S_1(t)$ and of the half-chain entropy $S_{N/2}(t)$
for different system sizes $N$. The single-qubit entropy grows rapidly
from zero and saturates on a time scale $t \sim 1$, independently of the
total number of qubits. The saturation value is close to the maximum
possible entropy for a qubit, indicating that, after a time of order
unity, each spin becomes highly entangled with the rest of the chain.
This behaviour is consistent with the idea that, around $t^\ast$, the
local degrees of freedom have already established strong quantum
correlations with their neighbours.

The half-chain entropy exhibits a rather different behaviour. For each
$N$, $S_{N/2}(t)$ grows approximately linearly in time for intermediate
times, before saturating at a value that scales proportionally to the
subsystem size, i.e.\ to $N/2$. The slope of the linear growth is
essentially independent of $N$, while the saturation value increases
with $N$, as expected for an entanglement entropy that follows a
volume-law within the half-chain. This linear growth followed by
saturation is a hallmark of entanglement spreading under local
Hamiltonian dynamics in one dimension.

The scaling behaviour of the half-chain entropy can be made more
transparent by considering the rescaled quantity $S_N(t)/N$ as a
function of $t/N$. As shown in Fig.~\ref{fig:Entanglement}(b), the
curves for different system sizes collapse onto a single scaling
function of the form
\begin{equation}
  S_N(t) \simeq N\, f\!\left(\frac{t}{N}\right),
\end{equation}
with $f(0)=0$ and $f(x)$ growing linearly for small $x$ before
saturating at a constant value. This scaling reflects the ballistic
propagation of entanglement in the XX chain: the entanglement front
moves with a finite velocity, so that the time needed to entangle
distant regions of the system grows linearly with their separation \cite{Latorre_2009, Calabrese_2005}.

Taken together, these observations show that the entanglement dynamics
in the XX-based QELM is characterised by two distinct time scales. On a
short time scale $t = \mathcal{O}(1)$, independent of $N$, local
entanglement builds up rapidly and single-qubit entropies saturate. On a
longer time scale $t = \mathcal{O}(N)$, the half-chain entropy
approaches its volume-law saturation value as entanglement spreads across the
entire chain. As we will argue below, the sharp transition in
classification accuracy observed at $t^\ast \approx 1$ is associated
with the former, local-entanglement time scale rather than with the
global scrambling of information.

\subsection{Lieb--Robinson velocity and locality of the transition}
\label{subsec:paper2_LR}

The existence of two separate time scales can be understood more
rigorously in terms of Lieb--Robinson bounds. For local quantum
Hamiltonians, these bounds limit the speed at which information and correlations can spread through the system. As a result, the influence of a local perturbation remains exponentially small outside an effective ``light cone'', whose slope defines the Lieb--Robinson velocity $v_{\mathrm{LR}}$~\cite{Lieb:1972wy,Hastings:2005pr, Eisert:2014jea, Nachtergaele:2010lwp}.
As shown in Appendix~A, for the one-dimensional XX model with nearest-neighbour couplings and in the units used in this work, the relevant Lieb--Robinson velocity is $v_{\mathrm{LR}} = 1$.

This result has a direct implication for the QELM dynamics. Starting from a product state prepared by dense-angle encoding, correlations can spread at most over a distance of order $v_{\mathrm{LR}} \cdot t$. Therefore, at times $t \lesssim 1$, each qubit can only become significantly entangled with a small neighbourhood of order one lattice spacing, while regions that are far apart along the chain remain only
weakly correlated. This picture is fully consistent with the entropy
data discussed above: single-site entropies saturate quickly, indicating
strong entanglement with nearby spins, whereas half-chain entropies
require times of order $N$ to reach their maximum values.

The classification results presented in
Sec.~\ref{sec:paper2_performance} show that the QELM accuracy reaches
its plateau for evolution times $t \gtrsim t^\ast \approx 1$, and that
the position of this transition is essentially independent of the system
size $N$. Combining this observation with the Lieb--Robinson bound, we
arrive at an important conclusion: the onset of high classification
accuracy does not require the state to be globally scrambled or
maximally entangled across the chain. Instead, it is already achieved in
a regime where entanglement has developed only locally, over distances
of order one site, and where the quantum dynamics has acted effectively
as a shallow circuit of constant depth in terms of two-qubit gates.

From the perspective of classical simulability, this locality is highly
significant. A dynamics that generates only short-range entanglement and
can be approximated by a circuit of bounded depth is, in many cases,
amenable to efficient classical simulation, for instance via
matrix-product-state or other tensor-network methods. The fact that the
QELM reaches its optimal classification performance within such a
locally entangled, shallow-circuit regime suggests that, at least for
the tasks and system sizes considered here, the observed quantum
advantage is better interpreted as a quantum-inspired enhancement within
a classically simulable regime, rather than as evidence of a strong
computational separation between quantum and classical models.
\section{Data geometry in probability space}
\label{sec:paper2_geometry}

The analysis of the previous sections has shown that the classification
accuracy of the QELM exhibits a sharp transition as a function of the
evolution time $t$, which can be related to the build-up of local
entanglement in the XX chain. In order to gain further intuition on how
the quantum dynamics reshapes the data, we
investigate the geometry of the samples in the feature space actually
seen by the classical readout: the space of measurement probabilities.
In this section we summarise that analysis and compare it with an
alternative description based on local observables.

\subsection{Probability polytope and K-means clustering}
\label{subsec:paper2_probability_polytope}

For a fixed evolution time $t$ and a given input sample with latent
representation $\mathbf{z}$, the quantum layer outputs an $N$-qubit
state $\ket{\psi_t(\mathbf{z})}$, which is then measured in the
computational basis. As discussed in
Sec.~\ref{subsec:paper2_measurement_readout}, this measurement is
equivalently described by a probability vector
\begin{equation}
  \mathbf{p}_t(\mathbf{z}) =
  \bigl( p_t(s_1\mid\mathbf{z}),\dots,p_t(s_{2^N}\mid\mathbf{z}) \bigr)
  \in \mathbb{R}^{2^N},
\end{equation}
where
\begin{equation}
  p_t(s\mid\mathbf{z}) = \left|\langle s |\psi_t(\mathbf{z}) \rangle\right|^2, 
  \qquad s\in\{0,1\}^N.
\end{equation}
By construction, all components are non-negative and sum to one, so that
$\mathbf{p}_t(\mathbf{z})$ lies in the probability simplex
\begin{equation}
  \Delta_{2^N-1} = 
  \left\{
  \mathbf{p}\in\mathbb{R}^{2^N} \,\middle|\,
  p_i \ge 0,\;\sum_{i=1}^{2^N} p_i = 1
  \right\}.
\end{equation}
In the following, we refer to the cloud of points
$\{\mathbf{p}_t(\mathbf{z}_j)\}$ associated with the samples of a given
dataset as the \emph{probability polytope} at time $t$. This is exactly
the feature space that is fed to the classical output layer of the QELM.

To probe how the quantum dynamics modifies the geometry of this
probability polytope, we apply an unsupervised
clustering algorithm, $K$-means, to the set of feature vectors
$\{\mathbf{p}_t(\mathbf{z}_j)\}$ for different values of $t$. The $K$-means algorithm partitions a set of data points $\{\mathbf{x}_j\}_{j=1}^M$ into $K$ clusters $C_1,\dots,C_K$ by minimising the \emph{inertia}, defined as
\begin{equation}
  \mathcal{I} = 
  \sum_{k=1}^K \sum_{\mathbf{x}_j \in C_k}
  \left\| \mathbf{x}_j - \boldsymbol{\mu}_k \right\|_2^2,
\end{equation}
where $\boldsymbol{\mu}_k$ is the centroid of cluster $C_k$ and
$\|\cdot\|_2$ denotes the Euclidean norm. Intuitively, $\mathcal{I}$
measures how tightly the points are grouped around their respective
centroids: smaller inertia corresponds to more compact clusters. Since
$K$-means does not use the class labels, it provides an unsupervised
measure of how ``clustered'' the data are in the chosen feature space.

\begin{figure*}[htbp]
    \centering
\includegraphics[width=0.5\textwidth]{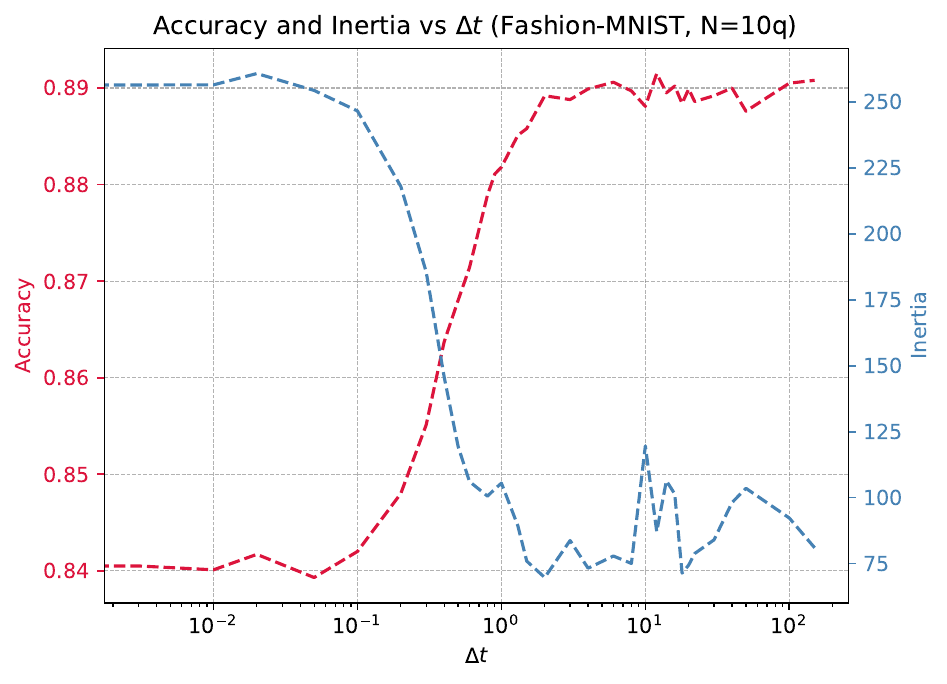}
    \caption{Accuracy (in magenta color) and Inertia (in light blue color) as a function of evolution time using the Fashion-MNIST dataset and $N=10$. Adapted from the author’s previous work, Ref.~\cite{DeLorenzis_2}. Copyright (2026) by the American Physical Society.}
    \label{fig:Inertia}
\end{figure*}

In the context of the QELM, the authors fix $K$ equal to the number of
classes of the dataset (e.g.\ $K=10$ for MNIST) and run $K$-means on the
probability vectors $\mathbf{p}_t(\mathbf{z}_j)$ for each value of $t$.
Figure~\ref{fig:Inertia} shows the behaviour of the $K$-means
inertia $\mathcal{I}(t)$ (light blue curve) and of the test accuracy
(magenta curve) as functions of the evolution time $t$ for the
Fashion-MNIST dataset with $N=10$ qubits.

A clear correlation emerges: as $t$ increases from zero to values of order one, the inertia decreases monotonically, while the test accuracy rises from its initial value to the saturation plateau. This indicates that, under the XX dynamics, the probability vectors associated with images belonging to the same class become progressively more compact and better separated in the probability polytope. In other words, the quantum evolution ``reshapes'' the data distribution in such a way that the classes become more easily separable by the linear classifier realised by the one-layer neural network. The fact that this geometric improvement in cluster compactness tracks the increase in supervised accuracy suggests that the QELM is indeed exploiting global features of the probability distribution to perform classification.

To quantify more directly the agreement between the unsupervised clusters found by $K$-means and the true class labels, the study also computes the \emph{Adjusted Rand Index} (ARI), a standard measure of similarity between two partitions of the same dataset. Given a ground-truth partition into classes and a clustering obtained from $K$-means, the ARI takes values between $-1$ and $1$, with $1$ indicating perfect agreement (up to relabelling of the clusters), $0$ corresponding to random clustering, and negative values indicating anti-correlation. Our results show that, in the probability space, the ARI increases with $t$ and saturates near the transition time $t^\ast$, mirroring the behaviour of the accuracy and inertia. This confirms that the clusters formed in the probability polytope align increasingly well with the true classes as the quantum dynamics builds up local entanglement.

\subsection{Failure of local observables}
\label{subsec:paper2_local_observables}

The previous analysis suggests that the geometry of the data in
probability space becomes more favourable for classification as
entanglement grows. A natural question is whether a similar improvement
occurs in a more restricted, ``local'' feature space defined by
single-site observables. To address this, we
construct, for each evolved state $\ket{\psi_t(\mathbf{z})}$, a vector
of local expectation values
\begin{equation}
  \mathbf{m}_t(\mathbf{z}) =
  \bigl(
    \langle \sigma_x^{(1)} \rangle_t,\,
    \langle \sigma_y^{(1)} \rangle_t,\,
    \langle \sigma_z^{(1)} \rangle_t,\,
    \dots,\,
    \langle \sigma_x^{(N)} \rangle_t,\,
    \langle \sigma_y^{(N)} \rangle_t,\,
    \langle \sigma_z^{(N)} \rangle_t
  \bigr)
  \in \mathbb{R}^{3N},
\end{equation}
where $\langle \sigma_\alpha^{(i)} \rangle_t$ denotes the expectation
value of the Pauli operator $\sigma_\alpha$ on site $i$ at time $t$.
These local magnetisations form an alternative feature space of
dimension $3N$, which captures only single-qubit observables and ignores
higher-order correlations between spins.

\begin{figure*}[htbp]
  \centering
  \begin{subfigure}[b]{0.45\textwidth}
    \centering
    \includegraphics[width=\textwidth]{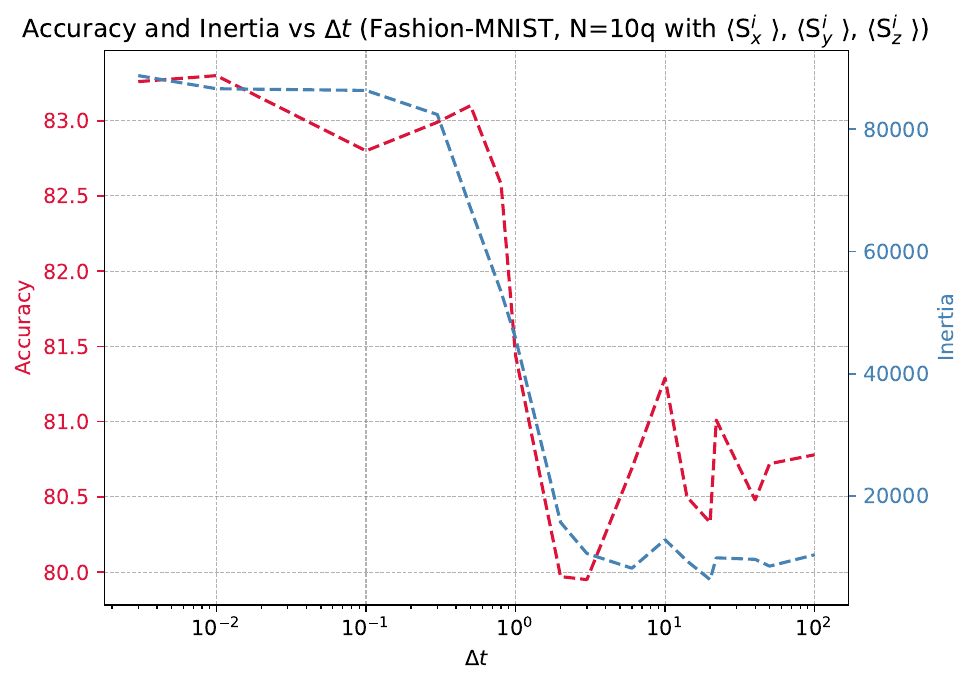}
    \caption{}
  \end{subfigure}
  \hfill
  \begin{subfigure}[b]{0.45\textwidth}
    \centering
    \includegraphics[width=\textwidth]{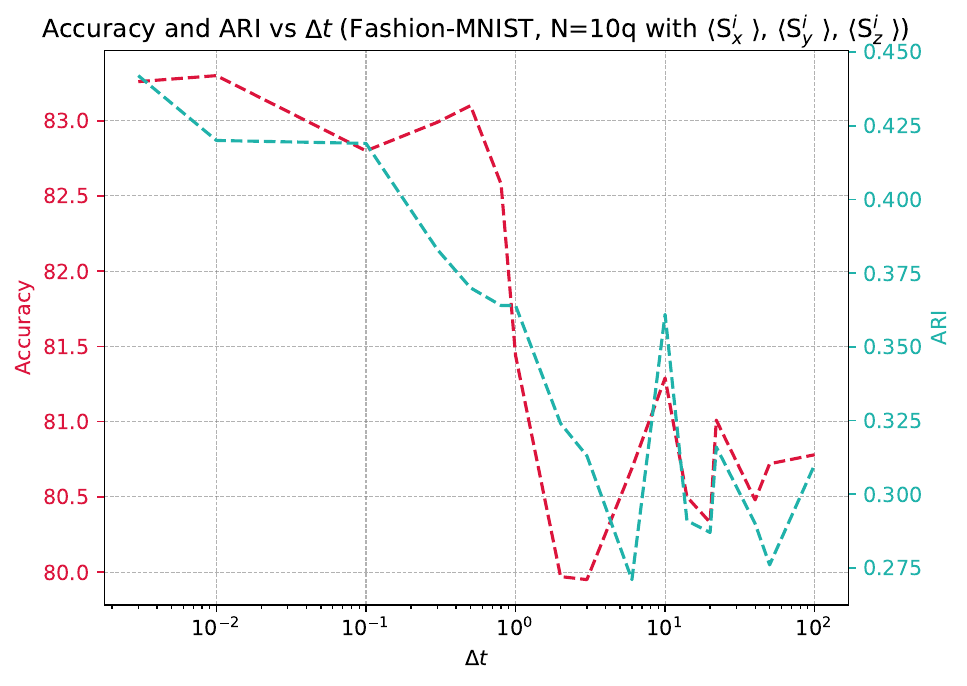}
    \caption{}
  \end{subfigure}

  \caption{a) Accuracy (in magenta color) and Inertia (in light blue color) as a function of evolution time using the Fashion-MNIST dataset and $N=10$.  b) Accuracy (in magenta color) and ARI (in light green color) as a function of evolution time using the Fashion-MNIST dataset, for $N=10$, obtained from the local expectation values only. Adapted from the author’s previous work, Ref.~\cite{DeLorenzis_2}. Copyright (2026) by the American Physical Society.}
  \label{fig:Expectation_values}
\end{figure*}

The same $K$-means analysis is then repeated in this local-observable
space, with $K$ fixed to the number of classes as before. The resulting
inertia $\mathcal{I}_{\mathrm{loc}}(t)$ and ARI as functions of time are
shown in Fig.~\ref{fig:Expectation_values}. A strikingly different
behaviour is observed compared to the probability space. As $t$
increases, both the inertia and the ARI tend to \emph{decrease}. The
reduction in inertia indicates that the local-observable vectors become
more compactly clustered, but the simultaneous decrease in ARI shows
that these clusters no longer align with the true class labels.
Furthermore, the test accuracy of a classifier trained directly on the
local features also deteriorates with time.

In other words, in the space of local observables the data still undergo
a form of ``clustering'' under the XX dynamics, but this clustering is
not correlated with the semantic structure of the labels: samples from
different classes are pulled closer together in the local-feature space,
and the class information is effectively scrambled. This is in stark
contrast with what happens in the probability space, where the
entanglement-driven evolution makes the clusters more compact and better
aligned with the true classes.

This comparison highlights a key aspect of the QELM architecture: the
use of full measurement probabilities as features allows the model to
exploit global correlations and higher-order interference patterns that
are invisible to local observables. The quantum dynamics, by distributing
the encoded information over many basis states in a correlated way,
creates a representation in the probability polytope where samples from
the same class occupy well-defined regions separated by low-density
boundaries. A shallow classical readout can then successfully separate
these regions. At the same time, the information contained in local
one-point functions becomes less and less informative about the class
labels, reflecting the scrambling of local degrees of freedom induced by
the entangling dynamics.
\section{Classical simulability and shallow random circuits}
\label{sec:paper2_simulability}

The results discussed so far indicate that the QELM based on the XX
Hamiltonian reaches its optimal classification performance in a regime
characterised by local entanglement, limited propagation of information,
and an effective circuit depth that does not grow with the system size
$N$. In this section we discuss the implications of this behaviour for
the classical simulability of the model and compare the analog XX
dynamics with a digital random-circuit construction that reproduces
similar accuracies with a small number of layers.

\subsection{Limited entanglement and tensor-network simulability}
\label{subsec:paper2_tn_simulability}

As shown in Sec.~\ref{sec:paper2_entanglement}, the entanglement
generated by the XX evolution exhibits two distinct time scales. On a
short time scale $t = \mathcal{O}(1)$, independent of $N$, the
single-qubit entropies rapidly grow and saturate, signalling the onset
of strong local entanglement between each spin and its immediate
neighbours. On a longer time scale $t = \mathcal{O}(N)$, the half-chain
entropy approaches a volume-law saturation value, reflecting the
spreading of entanglement across the entire chain. The classification
accuracy of the QELM, however, reaches its plateau already at a
characteristic time $t^\ast \approx 1$, which does not depend on $N$
within the range of sizes considered.

From the perspective of quantum circuits, this means that the relevant
dynamics for classification can be implemented by a circuit of constant
depth in terms of local two-qubit gates. Indeed, an evolution under a
nearest-neighbour Hamiltonian such as the XX model can be approximated to a given time $t$ by a Trotter decomposition, i.e. by splitting the continuous-time evolution operator into a product of elementary short-time unitary steps. For local Hamiltonians, the number of such layers grows linearly with $t$, so that stopping the evolution at $t^\ast=\mathcal{O}(1)$ corresponds to a sequence of $\mathcal{O}(1)$ layers, independent of $N$ (see, e.g., Refs.~\cite{Lloyd:1996aai,Childs:2019hts, Bosse:2024maw, Ostmeyer:2022lxs}). The fact that the QELM attains its best performance
within this shallow-circuit regime suggests that the amount of
entanglement and the complexity of the generated quantum states remain
moderate, even as $N$ increases.

This observation has important consequences for classical simulability.
One-dimensional local circuits of constant or slowly growing depth can
often be simulated efficiently using tensor-network methods, such as
matrix-product states (MPS), provided that the bipartite entanglement
entropy across any cut remains bounded or grows at most logarithmically
with system size. In the present case, the entanglement entropy across a
half-chain cut at times $t \lesssim t^\ast$ remains well below its
volume-law saturation value, and the effective circuit depth required to
reach the accuracy plateau does not scale with $N$. These are precisely
the conditions under which MPS-based algorithms are expected to be
effective.

Therefore, although the QELM does exploit genuinely quantum resources,
in the sense that entanglement plays a crucial role in reshaping the
data in probability space, the regime in which it operates for the
classification tasks studied here is compatible with efficient classical
simulation. This suggests that the observed performance enhancements
should be interpreted as quantum-inspired improvements within a
classically simulable regime, rather than as evidence of a strong
computational advantage over classical models.

\subsection{Digital random-gate model}
\label{subsec:paper2_random_circuits}

\begin{figure*}[htbp]
    \centering
\includegraphics[width=0.5\textwidth]{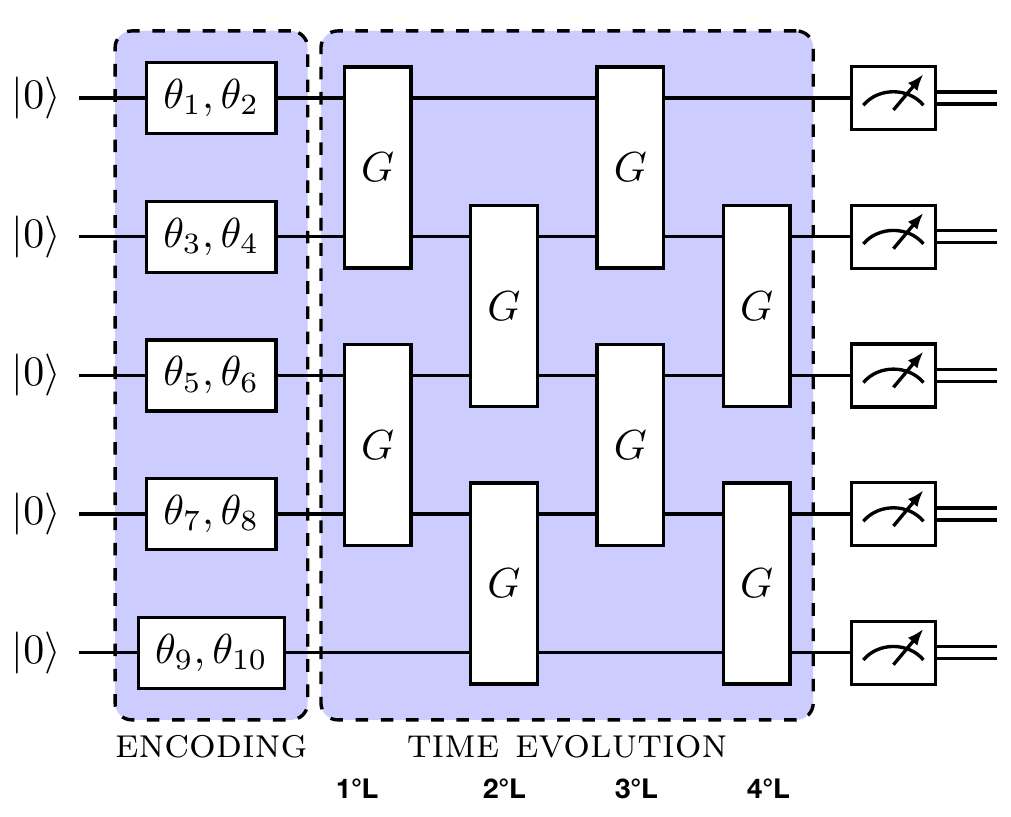}
\caption{A schematic representation of an algorithm with 5 qubits, in which a unitary random matrix is applied for pairs of qubits contiguous. Adapted from the author’s previous work, Ref.~\cite{DeLorenzis_2}. Copyright (2026) by the American Physical Society.}\label{fig:Schematic}
\end{figure*}

To further support this interpretation, we
consider a digital random-circuit model designed to mimic the effect of
the analog XX evolution while making the circuit depth explicit. The
setup is schematically illustrated in Fig.~\ref{fig:Schematic}. Starting from the same encoded product state
$\ket{\psi_0(\mathbf{z})}$ as in the XX-based QELM, the quantum layer is
replaced by a sequence of $L$ layers of random two-qubit gates acting on
nearest neighbours along the chain.

Each layer consists of two sub-layers arranged in a brickwork pattern.
In the first sub-layer, two-qubit gates $G_{1,2}, G_{3,4}, \dots$
are applied in parallel to disjoint pairs of neighbouring qubits
$(1,2), (3,4), \dots$. In the second sub-layer, the pattern is shifted
by one site, and gates $G_{2,3}, G_{4,5}, \dots$ act on pairs
$(2,3), (4,5), \dots$. This ensures that, over two consecutive
sub-layers, every bond in the chain participates in at most one gate.
The individual gates $G_{i,i+1}$ are drawn independently at random from
a suitable ensemble of two-qubit unitaries, for instance the Haar
measure on $\mathrm{U}(4)$. The total depth of the circuit is thus
$L$ layers, each of which has the same local connectivity pattern as a
single Trotter step of the XX evolution.

The output of the random circuit for a given depth $L$ is an $N$-qubit
state $\ket{\psi^{(L)}(\mathbf{z})}$, which is measured in the
computational basis as in the analog QELM. The resulting probability
vectors $\mathbf{p}^{(L)}(\mathbf{z})$ are fed to the same classical
output layer (ONN) used in the XX-based model, and the classification
accuracy is evaluated as a function of $L$.

\begin{figure*}[htbp]
  \centering
  \begin{subfigure}[b]{0.45\textwidth}
    \centering
    \includegraphics[width=\textwidth]{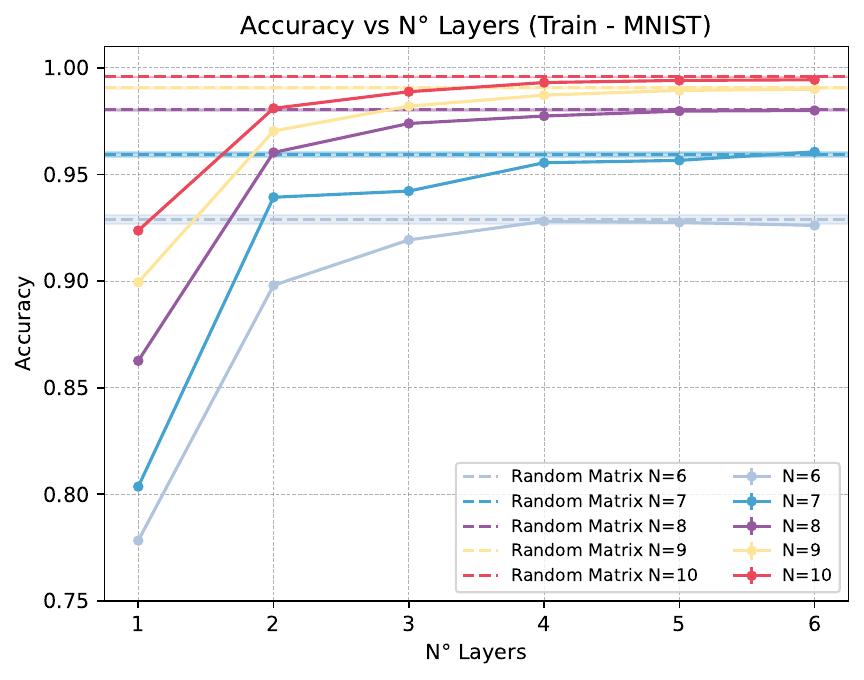}
    \caption{}
  \end{subfigure}
  \hfill
  \begin{subfigure}[b]{0.45\textwidth}
    \centering
    \includegraphics[width=\textwidth]{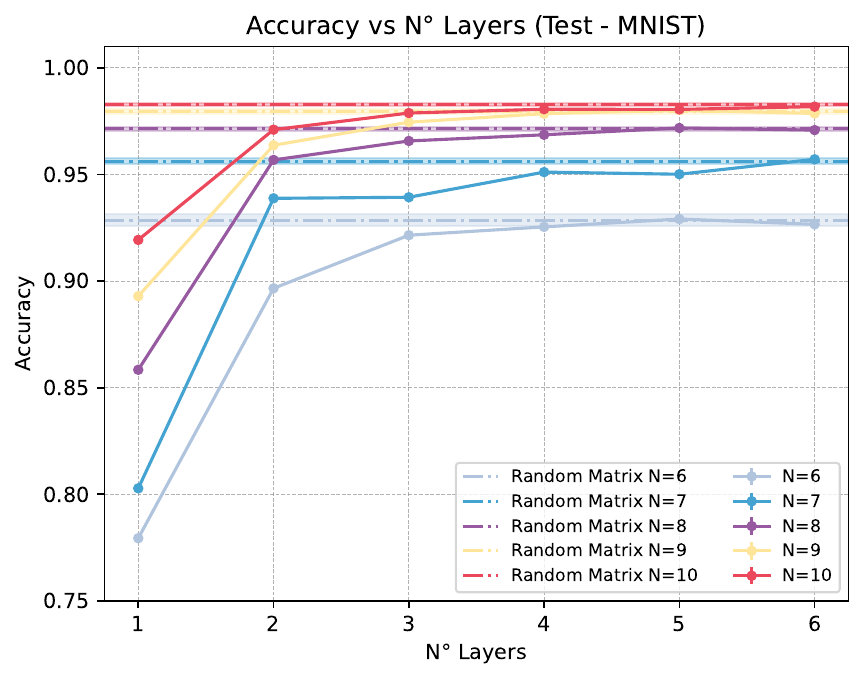}
    \caption{}
  \end{subfigure}

  \caption{Training (left panel) and testing (right panel) accuracy as a function of the number of layers using the MNIST dataset. Adapted from the author’s previous work, Ref.~\cite{DeLorenzis_2}. Copyright (2026) by the American Physical Society.}
  \label{fig:MNIST_layers}
\end{figure*}

\begin{figure*}[htbp]
  \centering
  \begin{subfigure}[b]{0.45\textwidth}
    \centering
    \includegraphics[width=\textwidth]{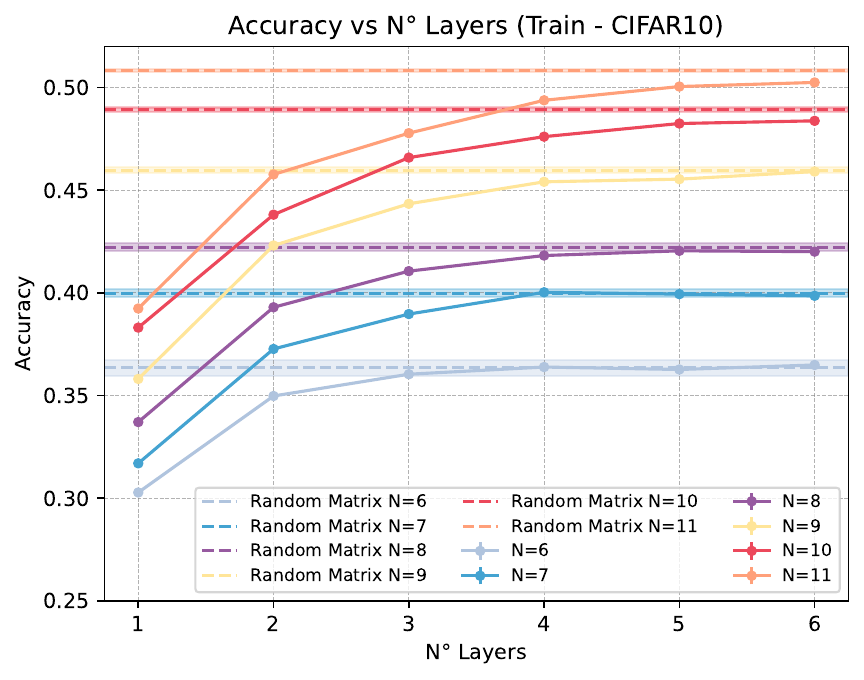}
    \caption{}
  \end{subfigure}
  \hfill
  \begin{subfigure}[b]{0.45\textwidth}
    \centering
    \includegraphics[width=\textwidth]{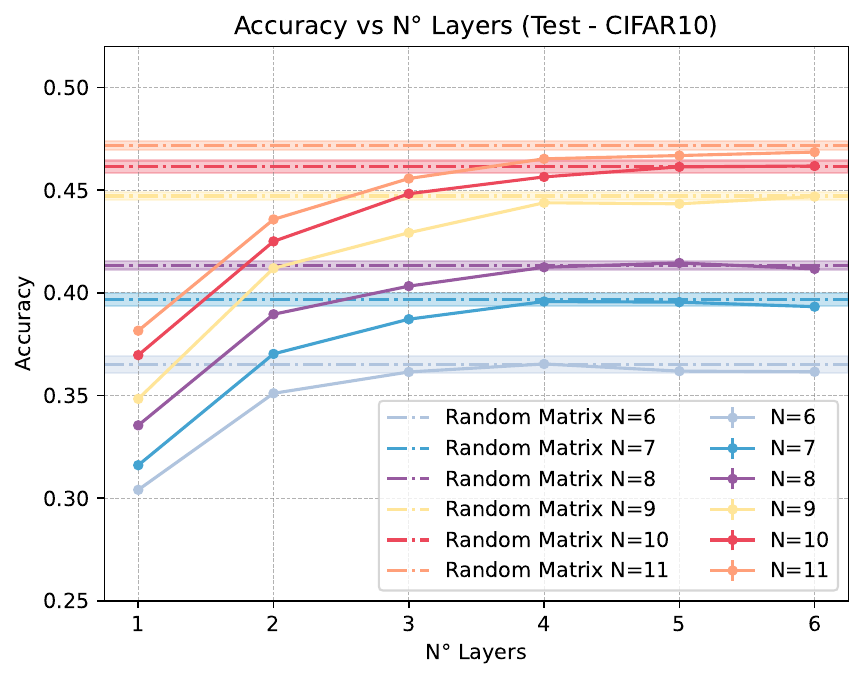}
    \caption{}
  \end{subfigure}

  \caption{Training (left panel) and testing (right panel) accuracy as a function of the number of layers using the CIFAR-10 dataset. Adapted from the author’s previous work, Ref.~\cite{DeLorenzis_2}. Copyright (2026) by the American Physical Society.}
  \label{fig:CIFAR-10_layers}
\end{figure*}

Figures~\ref{fig:MNIST_layers} and
\ref{fig:CIFAR-10_layers} show the test accuracy of this random-gate
QELM as a function of the number of layers $L$ for the MNIST and
CIFAR-10 datasets, respectively, and for different values of $N$. In
both cases, the accuracy rises from its initial value at $L=0$ and
saturates after a small number of layers, typically between $L=4$ and
$L=6$, with only a weak dependence on $N$. Increasing the depth beyond
this range does not lead to significant further improvements and may
even slightly degrade the performance due to overfitting or numerical
noise.

This behaviour closely parallels what was observed for the analog XX
dynamics as a function of the evolution time $t$: the classification
accuracy exhibits a rapid increase over a few effective ``entangling
steps'' and then reaches a plateau that does not require deep circuits.
Moreover, for fixed $N$ and dataset, the plateau accuracy obtained with
the random-gate model is comparable to that of the XX-based QELM,
indicating that the detailed structure of the Hamiltonian is not
essential. What matters is that the circuit implements a sufficiently
rich, yet shallow, entangling transformation of the encoded features.

From a classical-simulation viewpoint, these digital random circuits are
amenable to the same tensor-network techniques as the analog XX
evolution: they are one-dimensional, have strictly local interactions,
and operate at depths that are constant or very mildly increasing with
$N$. The fact that they reproduce the performance of the analog QELM
provides further evidence that the classification tasks considered in
this work can be solved by quantum architectures that remain within the
regime of classical simulability, at least for the system sizes and
datasets explored here.

In summary, the analysis of both the entanglement dynamics and the
random-circuit model suggests that the QELM operates as a shallow,
locally entangling feature map that can be efficiently approximated by
tensor-network methods. While this does not diminish the conceptual
interest of using quantum dynamics to generate useful representations of
classical data, it highlights that the search for genuine quantum
advantages in machine learning should likely focus on architectures or
tasks that go beyond this shallow, locally entangled regime.
\section{Summary and outlook}
\label{sec:paper2_summary}

In this chapter we have analysed in detail the behaviour of Quantum
Extreme Learning Machines when the quantum layer is fixed to a
nearest-neighbour XX chain with periodic boundary conditions and dense-angle
encoding. The results presented here form the basis of Ref.~\cite{DeLorenzis_2}, accepted for publication in Physical Review Applied. In particular,  we have shown that the classification accuracy on MNIST, Fashion-MNIST and CIFAR-10 exhibits a sharp change as a function of the evolution time $t$. For each dataset and for all system sizes considered, the accuracy remains almost flat at short times, undergoes a rapid rise around a characteristic time $t^\ast \sim 1$, and then saturates to a plateau. Remarkably, this transition time does not
depend on the number of qubits, while the plateau value increases with
$N$ and matches, within statistical uncertainties, the accuracy obtained
from fully random (Haar-distributed) unitaries acting on the same
encoded inputs.

The dynamical origin of this behaviour was clarified by studying the
growth of entanglement and the geometry of the data in feature space.
The onset of high accuracy coincides with the regime in which local
entanglement between each spin and its neighbours is already large,
whereas the entanglement across half of the chain is still far from its
asymptotic volume-law value. Consistently with Lieb--Robinson bounds for
the XX model, the time $t^\ast$ is just sufficient for information to
reach nearest neighbours, but not distant sites, implying that the
effective depth of the underlying circuit is of order one and does not
grow with $N$. Within this locally entangled regime, the measurement
probabilities in the computational basis reorganise into clusters that
are more compact and better aligned with the class labels, as revealed
by $K$-means inertia and, in the probability space, by the agreement
between unsupervised clusters and true labels. In contrast, the same
analysis performed on local observables shows that one-point functions
become less informative about the classes as time increases, indicating
that class information is progressively scrambled at the local level
while being encoded in global correlations.

A further piece of evidence comes from the comparison with a digital
random-circuit model, in which the XX evolution is replaced by a brickwork
of random two-qubit gates acting on neighbouring qubits. In this setting
the only control parameter is the number of layers, i.e.\ the circuit
depth. The numerical results demonstrate that, for all datasets, the
test accuracy reaches essentially the same plateau as in the analog XX
case after only a few layers (typically four to six), with a very weak
dependence on system size. This confirms that the QELM performance in
the present setting can be reproduced by shallow random circuits with
local interactions, which are natural candidates for efficient classical
simulation using tensor-network methods.

Altogether, these findings portray QELMs based on the XX chain as
powerful \emph{entangling feature maps} that enhance classical
representations without requiring deep circuits or strongly entangled
many-body states. The improvement in accuracy is tightly connected to
the emergence of moderate, predominantly local quantum correlations that
reshape the data into more class-separable manifolds in probability
space. At the same time, the limited depth and the structure of the
entanglement place these models close to the boundary of classical
simulability. From this perspective, QELMs constitute both a useful
hybrid quantum--classical architecture for near-term devices and a
convenient testbed to explore where, and under which conditions, genuine
quantum advantages in machine learning may arise beyond the shallow,
locally entangled regime investigated in this work.
\chapter{Neutrinos, Cherenkov Light and Water Cherenkov Detectors}

\section{Neutrinos and their experimental relevance}

Neutrinos are among the most elusive particles in the Standard Model of particle physics \cite{pdg, Griffiths:1987tj}. They are electrically neutral fermions belonging to the lepton family and appear in three flavor states, namely the electron neutrino, muon neutrino, and tau neutrino \cite{Griffiths:1987tj}. For many years they were assumed to be massless, but the observation of neutrino oscillations established that neutrinos have non-zero masses and that flavor states are quantum superpositions of mass eigenstates \cite{pdg, Griffiths:1987tj}. This discovery represents one of the clearest indications that the Standard Model, in its minimal form, is incomplete.

Because of their weak interactions and tiny masses \cite{pdg}, neutrinos can travel vast distances through matter with only a very small probability of interaction. This property makes them extremely valuable messengers in both particle physics and astrophysics: they carry information from otherwise inaccessible environments such as the solar core, supernovae, the Earth’s atmosphere, nuclear reactors, and accelerator beams. At the same time, exactly these properties make them very challenging to detect experimentally.

In modern neutrino physics, experimental programs are driven by several major open questions. Precision measurements of neutrino mixing parameters \cite{pdg}, the determination of the neutrino mass ordering, and the search for leptonic CP violation are among the central goals of current and future long-baseline experiments. These measurements are essential not only for completing the picture of neutrino mixing, but also for testing whether neutrino properties may point toward physics beyond the Standard Model. The Hyper-Kamiokande program is one of the leading efforts in this direction.
\subsection{Main properties of neutrinos}

Neutrinos are spin-$1/2$ leptons that participate in weak interactions and gravity, but not in electromagnetic or strong interactions. Their lack of electric charge prevents direct ionization signatures of the kind produced by charged particles, and their weak coupling strongly suppresses interaction rates in matter. In the Standard Model framework, neutrinos are produced and detected in flavor eigenstates associated with the charged leptons $e$, $\mu$, and $\tau$. However, propagation in space is governed by mass eigenstates, and the mismatch between these two bases gives rise to flavor oscillations.

The experimental discovery of neutrino oscillations showed that neutrino flavor is not conserved during propagation and therefore that at least two neutrino mass eigenstates must be non-zero. This result has profound implications: it demonstrates that the Standard Model requires extension, and it makes neutrinos unique probes of fundamental physics at very small mass scales. 

Another important feature of neutrinos is their penetrative power. Since they interact only weakly, they can escape dense media that are opaque to photons or charged particles. This is why neutrino observations provide complementary information to electromagnetic measurements. Solar-neutrino and atmospheric-neutrino experiments, for example, have been decisive in establishing the oscillation framework, while accelerator-based experiments allow controlled studies of flavor transitions over known baselines and energies.

\subsection{Why neutrinos are difficult to detect}

The main experimental challenge in neutrino physics arises from the very small interaction cross section of neutrinos with matter. A neutrino can traverse enormous amounts of material before interacting, which means that detectors must either observe extremely intense neutrino fluxes, instrument very large target masses, or ideally combine both strategies. This is one of the reasons why neutrino detectors are often built on very large scales and placed underground, where backgrounds from cosmic rays can be reduced.

A second difficulty is that neutrinos are not observed directly. Instead, experiments detect the charged particles produced when a neutrino interacts with nuclei or electrons in the detector medium. The measured quantities are therefore indirect signatures, such as scintillation light, Cherenkov light, ionization, or delayed coincidence signals, depending on the detector technology. Reconstructing the original neutrino properties from these final-state observables requires careful event reconstruction, background rejection, and often detailed detector simulations.

The importance of backgrounds depends strongly on the neutrino energy and on the experimental context. In low-energy measurements, such as solar-neutrino experiments, radioactivity, detector noise, and other environmental backgrounds can significantly affect the signal and require low-background environments, high-purity detector media, and precise calibration. In higher-energy experiments, however, neutrino interactions may produce clear visible signatures, for example well-defined Cherenkov rings in water detectors. In such cases, the experimental challenge shifts from background suppression to the reconstruction and interpretation of complex event topologies, for which pattern-recognition and particle-identification algorithms are essential tools for extracting the physical content of recorded events.

\section{From neutrino interactions to detectable signals}



\subsection{Indirect detection of neutrinos}

The indirect detection of neutrinos begins with a weak interaction between the neutrino and a target particle in the detector, typically a nucleus or an electron. Such an interaction transfers part of the neutrino energy and momentum to the final state, producing one or more outgoing particles that may be experimentally observable. The detector response is therefore determined not by the neutrino alone, but by the full interaction process and by the propagation of the secondary particles through the detector medium.

This has important consequences for data analysis. First, the visible signal depends on the interaction channel, the detector material, and the detector geometry. Second, the reconstruction of neutrino properties is inevitably affected by nuclear effects, final-state interactions, detector resolution, and backgrounds. For this reason, neutrino measurements rely heavily on event simulation, calibration, and pattern-recognition techniques, all of which are needed to connect the recorded detector response to the underlying physical process. Rather than measuring a neutrino track directly, the experiment measures a set of observables from which the neutrino event must be reconstructed statistically and topologically.

A further consequence of indirect detection is that classification and reconstruction tasks become central components of the experimental workflow. The detector must distinguish neutrino-induced events from non-neutrino backgrounds, separate different interaction topologies, and identify the nature of the visible final-state particles. In large-volume detectors, where enormous numbers of optical or electronic channels may contribute to a single event, this task naturally leads to the use of increasingly sophisticated reconstruction algorithms and machine-learning methods.

\subsection{Main interaction modes and event complexity}

At the energies relevant for this study, neutrino interactions can proceed through different channels, which lead to qualitatively different final states and hence to different visible topologies in the detector.

In charged-current quasi-elastic interactions, the final state is relatively simple and is often dominated by a single outgoing charged lepton. In water Cherenkov detectors, such events are therefore commonly associated with simple single-ring topologies.

In resonant interactions, the neutrino produces a charged lepton together with additional hadronic activity, typically including a pion. These events are correspondingly more complex and may already give rise to multi-ring signatures even though they originate from a single interaction vertex.

At higher energies, deep inelastic scattering becomes important. In this regime, the final state may contain a charged lepton together with several hadrons, including multiple pions. Depending on threshold and kinematics, only part of this activity may be visible, so that the observed topology need not reflect the full complexity of the underlying interaction.

This distinction is important for the pile-up problem studied later in this thesis. A single neutrino interaction can already produce a non-trivial visible pattern, especially in resonant and deep inelastic events, whereas pile-up corresponds to the superposition of two or more independent interactions within the same readout window. The experimental challenge is therefore not simply to identify events with multiple rings, but to distinguish complex single-vertex topologies from genuinely overlapping multi-vertex events.

\subsection{Charged particles as observable signatures}

Among the particles produced in neutrino interactions, charged particles are of particular importance because they can generate measurable detector signals while traversing matter. Unlike the incoming neutrino, a charged lepton or hadron can deposit energy, emit light, or trigger sensor responses along its path. In this sense, charged particles act as the experimentally accessible proxies of the original neutrino interaction. Their trajectories, energies, and light-emission patterns contain the information from which the event is reconstructed.

In water Cherenkov detectors, which are the focus of this work, the most relevant signatures arise when relativistic charged particles travel through water faster than the phase velocity of light in the medium. Under these conditions they emit Cherenkov radiation, which can be detected by photomultiplier tubes surrounding the active volume. The spatial and temporal distribution of this light forms the basis for event reconstruction and particle identification. Therefore, although the physical process of interest is initiated by a neutrino, the experimentally recorded signal is ultimately determined by the charged particles produced in the interaction and by the light they generate in the detector.

The central role of charged secondaries also explains why the topology of the visible event carries so much information. Different final states may produce different light distributions, ring structures, or time patterns, and these differences can be exploited to distinguish among interaction channels or event classes. This idea is particularly relevant for the classification problem studied later in this thesis, where the observable structure of the detector response is used to discriminate between different neutrino-event topologies. The next section introduces the physical origin of Cherenkov radiation, which is the key detection mechanism underlying this type of measurement. 

\section{Cherenkov radiation}

Cherenkov radiation is the physical mechanism that makes large water Cherenkov detectors sensitive to neutrino interactions. Although the neutrino itself is invisible to the detector, the charged particles produced in its interaction with matter may generate optical signals while traversing the medium. When these particles are sufficiently energetic, they emit a characteristic cone of coherent light, which can then be recorded by photosensors surrounding the detector volume. The reconstruction of this light pattern provides access to the topology and kinematics of the underlying event. 

\subsection{Physical origin of Cherenkov light}

The physical origin of Cherenkov radiation can be understood by considering the propagation of a charged particle through a dielectric medium. As the particle moves, it polarizes the surrounding atoms or molecules. These induced excitations subsequently relax and re-emit electromagnetic radiation. If the particle velocity is lower than the phase velocity of light in the medium, the emitted wavefronts do not add coherently and no macroscopic directional radiation is produced. By contrast, when the particle velocity exceeds the phase velocity of light in the medium, the emitted wavefronts interfere constructively along a well-defined direction, giving rise to Cherenkov radiation. The condition for emission is therefore
\[
v > \frac{c}{n},
\]
where \(c\) is the speed of light in vacuum and \(n\) is the refractive index of the medium. 

This effect is analogous, in geometric terms, to the formation of a shock wave when an object moves through a medium faster than the propagation speed of sound. In the electromagnetic case, however, the radiation arises from the coherent superposition of the fields emitted by the polarized medium along the particle trajectory. In transparent materials such as water, this radiation falls partly in the visible range and can therefore be detected with optical sensors. For relativistic charged particles in water, Cherenkov emission becomes one of the dominant observable signatures used for event reconstruction. 

\subsection{Cherenkov angle and emission cone}

The constructive-interference condition not only determines whether Cherenkov radiation is emitted, but also fixes its geometry. The emitted light forms a cone around the particle direction, with opening angle \(\theta_C\) given by
\[
\cos \theta_C = \frac{1}{\beta n},
\]
where \(\beta = v/c\). This relation shows that the Cherenkov angle depends both on the particle velocity and on the optical properties of the medium. As the particle becomes highly relativistic, \(\beta \to 1\), and the angle approaches a medium-dependent maximum value. In water, where \(n \approx 1.33\) in the optical range, this maximum angle is about \(42^\circ\). 

The conical structure of the emission is of fundamental importance for detector design and reconstruction. Since the light is emitted at a predictable angle with respect to the particle trajectory, the observed hit pattern on the detector walls carries directional information. In addition, the amount and timing of the recorded light depend on the particle type, energy, and path length in the detector. This makes Cherenkov radiation not only a threshold effect for detecting relativistic charged particles, but also a tool for inferring the topology of the underlying interaction. 

\subsection{Ring formation in detector geometries}

In a water Cherenkov detector, the conical emission of light intersects the photosensitive detector surfaces and gives rise to characteristic ring-like patterns. In an idealized geometry, a single relativistic charged particle traveling through the detector produces a cone of Cherenkov light whose intersection with the detector wall appears as a ring. The radius, sharpness, and light distribution of this ring depend on the particle direction, vertex position, and propagation length, as well as on optical effects in the medium and the detector geometry itself. 

This ring-forming mechanism is the basis of event imaging in water Cherenkov experiments. Simple event topologies may correspond to a single dominant ring, whereas more complex topologies can generate overlapping or distorted patterns. The distinction between different visible structures is therefore directly related to the physical content of the event and plays a central role in reconstruction and classification. 

\begin{figure}[t]
    \centering
    \includegraphics[width=0.72\textwidth]{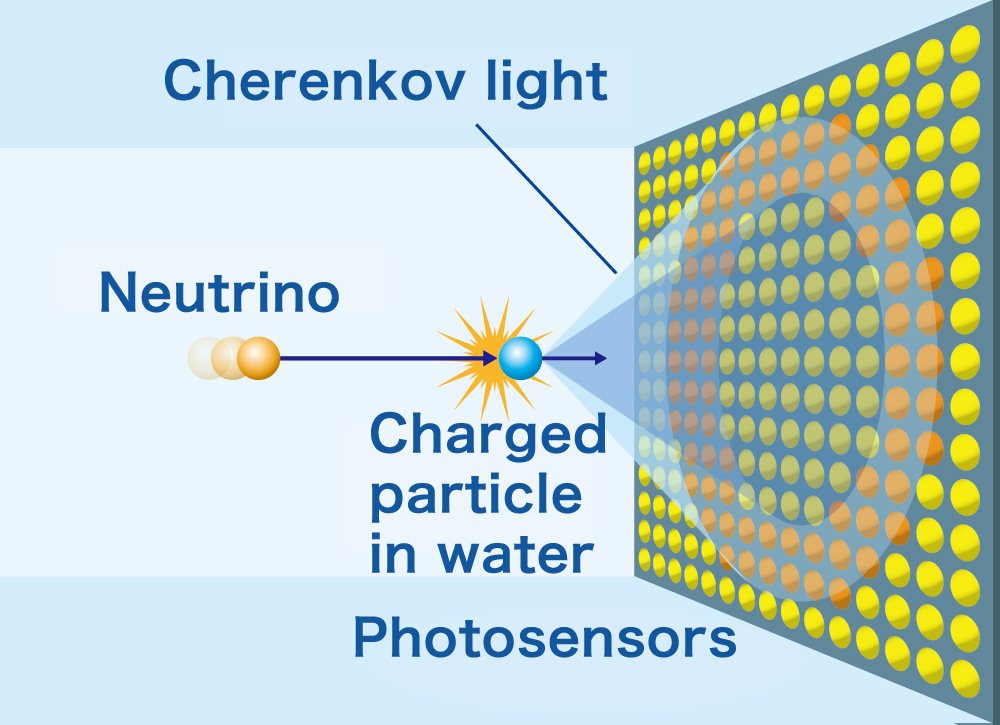}
    \caption{Schematic illustration of a Cherenkov light cone produced by a charged
    particle traversing the detector medium. The light cone intersects the cylindrical
    detector surface, producing a ring-like pattern of detected photons on the
    photosensors. This geometric projection underlies the ring structures used for
    event reconstruction.}
    \label{fig:iwcd_ring_projection}
\end{figure}

A schematic illustration of this mechanism is shown in Fig.~\ref{fig:iwcd_ring_projection}.

In the context of this thesis, the mapping between Cherenkov-light distributions and image-like detector representations provides the foundation for the machine-learning and quantum-inspired analyses developed in the following chapter.
\section{Water Cherenkov detectors}

\subsection{Photomultiplier tubes and optical readout}

The optical readout of a water Cherenkov detector is typically based on photomultiplier tubes (PMTs) installed on the detector walls and facing the active volume. These sensors convert incoming Cherenkov photons into electrical pulses that can be digitized by the readout electronics. In practice, each PMT acts as an individual measurement channel, recording whether light was observed and, if so, with what intensity and at what time.

The role of the PMT system is crucial, since the event reconstruction relies on the combined information from many channels distributed over the detector surface. The recorded hit pattern reflects both the underlying event topology and the optical propagation of light in the detector medium. Modern developments in optical detection emphasize not only light collection, but also increasingly precise timing information, since fast photon sensors can improve the separation of different light components and enhance reconstruction capabilities. 

\subsection{Charge and timing information}

Two of the most important observables recorded in water Cherenkov detectors are the charge and the arrival time associated with each PMT hit. The measured charge is related to the amount of light detected by a given sensor and therefore encodes information about the local light intensity. The recorded time, on the other hand, reflects when the photons reached the PMT and is sensitive to the event geometry, the particle direction, and the optical path followed by the light. Event reconstruction in water Cherenkov detectors is therefore commonly based on the combined use of PMT hit charges and times.

This combined charge-time information is particularly valuable because the two observables carry complementary physical content. Charge patterns help identify the overall morphology of the event and the relative brightness of different detector regions, while timing patterns provide additional constraints on the interaction vertex, particle propagation, and the distinction between prompt and delayed light contributions. In practical terms, a water Cherenkov event can thus be viewed as a structured set of charge and time measurements over a large array of sensors. This is precisely the type of information that, after suitable geometrical mapping, can be represented in image form and used as input for the analysis methods discussed in the following chapter. 

\section{Hyper-Kamiokande and the IWCD}

The experimental context relevant to the second part of this thesis is provided by the
Hyper-Kamiokande programme \cite{Hyper-Kamiokande:2018ofw, Hyper-Kamiokande:2025asb, Hyper_2015, HyperK}, whose long-baseline component is a next-generation neutrino experiment based on a large water Cherenkov detector in Japan. Within this programme, the
Intermediate Water Cherenkov Detector (IWCD) \cite{Scott:2016kdg} plays a particularly important role,
since it provides a water-target detector based on the same detection principle as the
far detector while operating sufficiently close to the beam source to characterize
neutrino interactions before oscillation effects dominate. In this sense, the IWCD
forms a natural bridge between the beamline environment and the oscillation
measurements performed at large distance.

\subsection{The Hyper-Kamiokande experiment}

\begin{figure}
    \centering
    \includegraphics[width=0.9\linewidth]{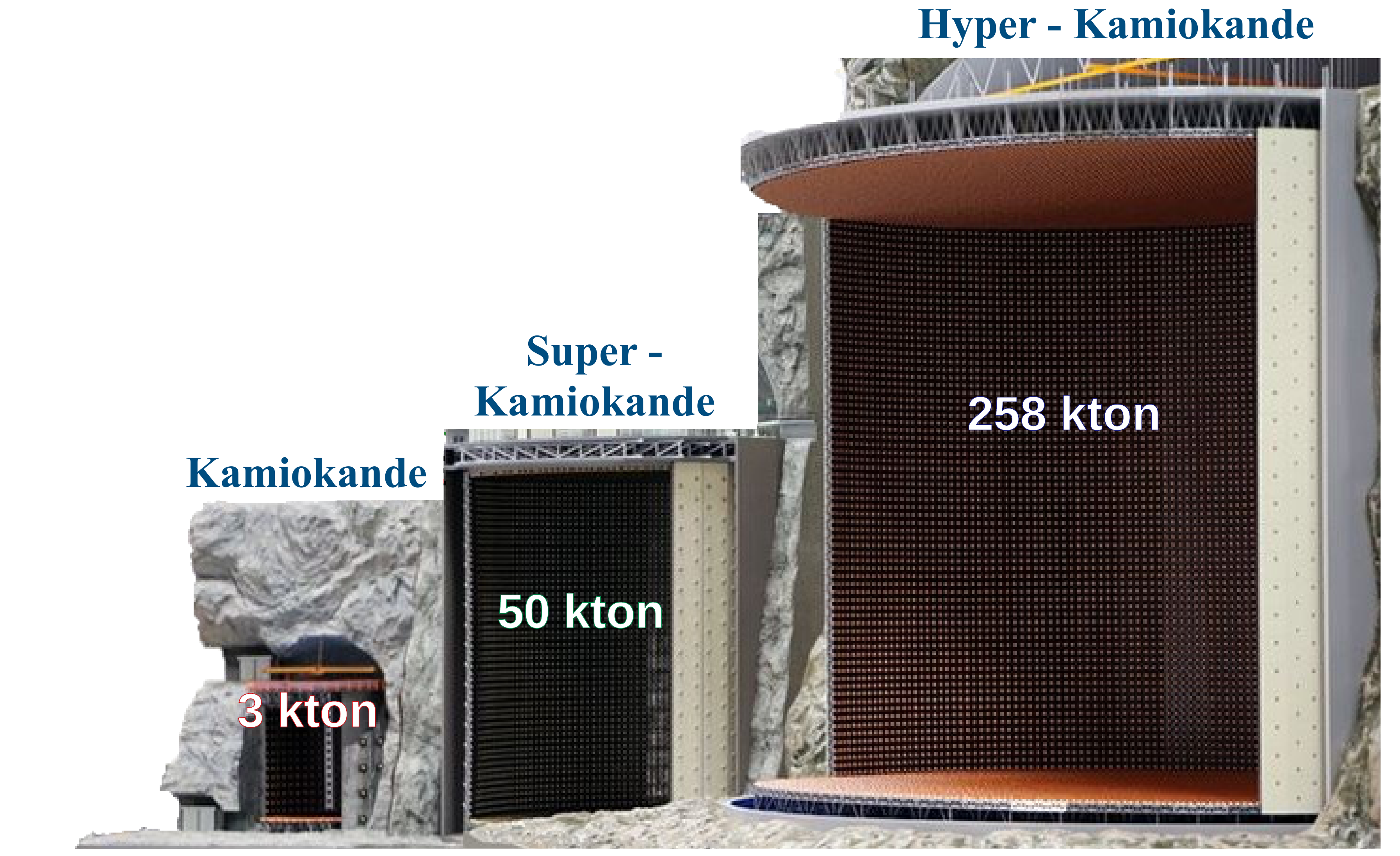}
    \caption{Schematic comparison of the Kamiokande, Super-Kamiokande, and
    Hyper-Kamiokande detectors, illustrating the growth in detector scale across the
    three generations. Source: adapted from material of the Hyper-Kamiokande
    Collaboration.} \label{fig:kamiokande_comparison}
\end{figure}

Hyper-Kamiokande is the natural evolution of the Kamiokande \cite{Kamiokande-II:1987idp} and
Super-Kamiokande \cite{SuperKamiokande:2002weg, Super-Kamiokande:2004orf, Super-Kamiokande:1998kpq} experiments, extending the water Cherenkov technique to a much
larger detector scale and a broader physics programme. Compared with its
predecessors, Hyper-Kamiokande features a substantial increase in target mass,
thereby enabling significantly larger event samples and improved sensitivity to
accelerator-neutrino oscillations, atmospheric and solar neutrinos, supernova
neutrinos, and rare processes such as proton decay. The increase in scale across the
Kamiokande, Super-Kamiokande, and Hyper-Kamiokande detector generations is
illustrated in Fig.~\ref{fig:kamiokande_comparison}.
The detector is located underground in Japan, approximately 295~km from the
J-PARC neutrino beam source in Tokai, and operates in a long-baseline configuration
optimized for oscillation studies. The neutrino beam is directed towards the detector
with an off-axis angle chosen to concentrate the flux near the energy region of
maximum oscillation sensitivity. In this respect, Hyper-Kamiokande continues and
extends the long-baseline programme previously developed through K2K \cite{K2K:2006yov} and T2K \cite{T2K:2011qtm, T2K:2023smv},
while significantly enlarging both detector scale and scientific reach.

\begin{figure}[t]
    \centering
    \includegraphics[width=0.78\textwidth]{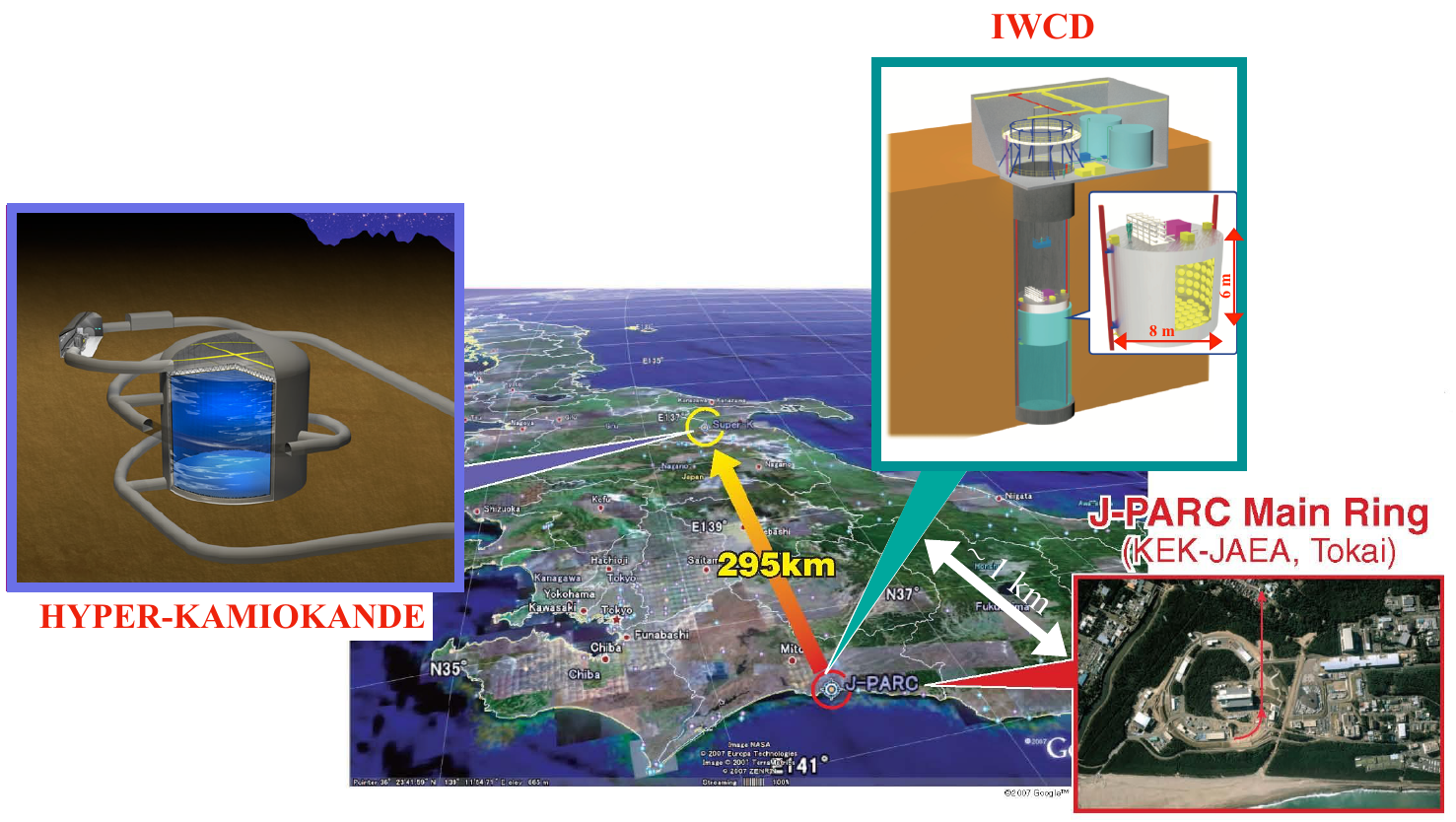}
    \caption{Schematic layout of the Hyper-Kamiokande long-baseline configuration,
    showing the relative positions of J-PARC, the Intermediate Water Cherenkov
    Detector (IWCD), and the Hyper-Kamiokande far detector. Source: adapted from
    material of the Hyper-Kamiokande Collaboration.}
    \label{fig:hk_layout}
\end{figure}

A central feature of the Hyper-K strategy is that it should not be regarded as a
single detector, but rather as an integrated experimental programme combining a
powerful far detector with a system of near and intermediate detectors. The far
detector provides the large fiducial mass required for precision oscillation
measurements, while the detectors located closer to the beam constrain the neutrino
flux, interaction models, and detector-related systematic effects. This experimental
layout is shown schematically in Fig.~\ref{fig:hk_layout}.
\subsection{Role of the Intermediate Water Cherenkov Detector}

Within this programme, the Intermediate Water Cherenkov Detector is designed to
reduce systematic uncertainties in the oscillation analysis \cite{Scott:2016kdg}. Since it uses water as
target medium and the same Cherenkov detection principle as the far detector, it provides measurements that are more directly comparable to those performed at Hyper-Kamiokande than those from detectors based on different target materials.
This feature is especially important for constraining interaction-model uncertainties and for improving the consistency between near-detector information and far-detector
oscillation measurements.
The IWCD is a cylindrical water Cherenkov detector with an inner volume approximately $8\,\mathrm{m}$ in diameter and $6\,\mathrm{m}$ in height, instrumented with multi-PMT optical modules, each containing 19 small photomultiplier tubes. Its design allows data taking at multiple off-axis positions.

A distinctive aspect of the IWCD concept is its capability to sample different off-axis beam angles. In the J-PARC beam, the neutrino energy spectrum depends on the off-axis angle, and data taken at several positions can therefore probe different
narrow-band flux configurations. Linear combinations of these measurements can then
be used to better constrain the energy dependence of neutrino interactions, reducing the reliance on interaction models. This strategy originates in the nuPRISM concept \cite{Scott:2016kdg}
and is one of the main reasons why the IWCD is expected to provide an important reduction of oscillation-systematic uncertainties.
Beyond this role in flux and cross-section control, the IWCD also provides a very
useful environment for reconstruction and event-classification studies in a realistic water Cherenkov setting. Because it is smaller than the far detector but based on the
same physical detection mechanism, it offers a controlled framework in which detector
response, event topology, and analysis strategies can be investigated in detail. For the purposes of this thesis, this makes the IWCD a particularly suitable detector for studying image-based and quantum-inspired classification methods applied to neutrino-induced Cherenkov patterns.
\begin{figure}[t]
    \centering
    \includegraphics[width=0.42\textwidth]{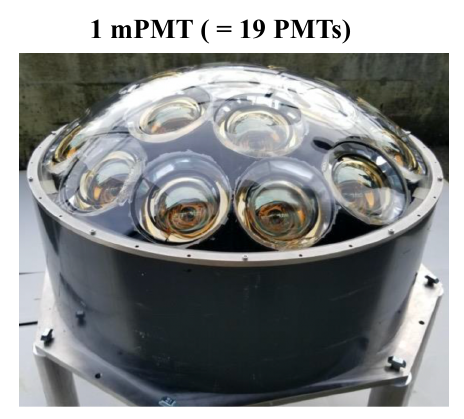}
    \caption{Photograph of a multi-PMT optical module used in the IWCD. The
    module contains 19 small photomultiplier tubes within a single enclosure.
    Source: adapted from material of the Hyper-Kamiokande Collaboration.}
    \label{fig:iwcd_mpmt}
\end{figure}

\begin{figure}[t]
    \centering
    \includegraphics[width=0.8\textwidth]{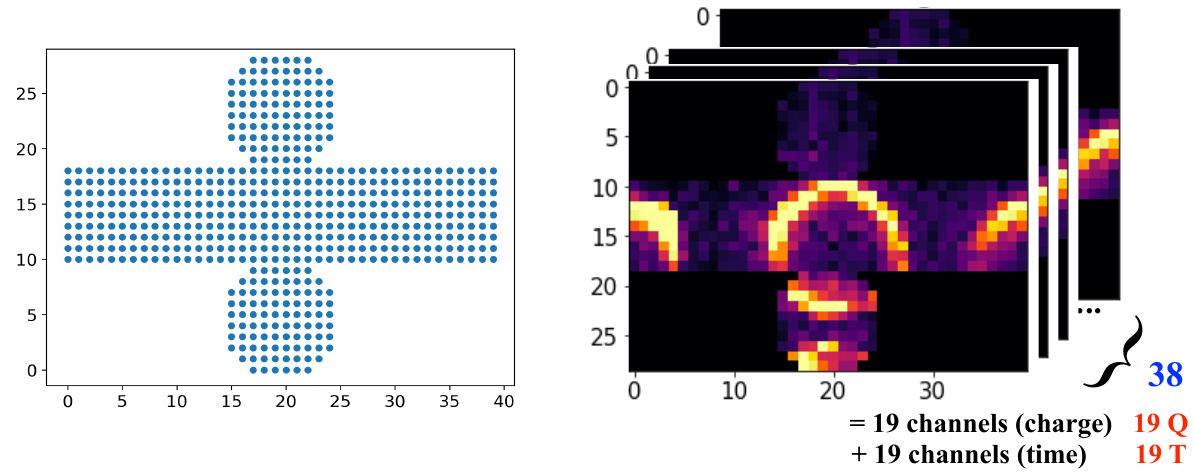}
    \caption{Schematic illustration of the image-like representation adopted for the
    IWCD detector response. On the right, the cylindrical detector surface is
    unwrapped into a two-dimensional map in $(\phi,z)$ coordinates, with points
    indicating the positions of the multi-PMT modules. On the left, example image
    channels are shown, corresponding to the charge and timing information recorded
    by the photomultipliers within each module.}
    \label{fig:iwcd_unwrapped}
\end{figure}

\subsection{Detector geometry, instrumentation, and image-like detector response}

From the detector point of view, the IWCD is a cylindrical water Cherenkov detector
instrumented with a photosensor system based on multi-PMT technology and designed
to operate at multiple off-axis positions. Although it is much more compact than the
far detector, it retains the essential ingredients of water Cherenkov reconstruction: a
water target, optical readout through photosensors, and event characterization through
the spatial and temporal distribution of Cherenkov light.

A representative multi-PMT module is shown in Fig.~\ref{fig:iwcd_mpmt}. Each
module contains multiple small photomultiplier tubes housed within a single optical
unit. Compared with a single large PMT channel, such modules provide a finer
granularity of optical information and can enrich the detector response with additional
timing and directional structure.

For the analyses developed in the next chapter, this detector response is recast into an
image-like representation. The cylindrical detector surface is mapped onto a
two-dimensional grid in $(\phi,z)$ coordinates, so that each photosensor is assigned a
position on an unwrapped view of the detector. Charge information, and when used
timing information, are then binned on this grid to form one or more image channels,
in close analogy with standard grayscale or multi-channel images. A schematic
illustration of this construction is shown in Fig.~\ref{fig:iwcd_unwrapped}.

The importance of the IWCD for the present work therefore lies not in the full
engineering detail of the detector, but in the fact that it provides a realistic water
Cherenkov environment in which neutrino interactions are observed through
structured optical patterns distributed over many channels. These patterns carry
topological and temporal information and naturally lend themselves to a
multi-channel image representation. The following chapter builds on this detector
context and focuses on how such image-like inputs are used for event classification.
\newpage
\chapter{Machine Learning and Quantum Machine Learning for Neutrino Physics}
\section{Motivation for Machine Learning in Cherenkov Event Classification}
In water Cherenkov detectors, such as the Intermediate Water Cherenkov Detector designed for the Hyper-Kamiokande experiment, the correct identification of event topologies (for example electrons vs muons, electrons vs photons or single-vertex vs pile-up events) is crucial for achieving precision in the determination of the physical parameters of interest. In this context, traditional reconstruction techniques based on physics-driven models, such as the fiTQun algorithm \cite{Missert:2017qdz, Patterson:2009ki}, have long represented the standard approach for estimating kinematic quantities and identifying particle types. However, fiTQun relies on likelihood functions and detector response models that may become limited when dealing with complex events or high particle fluxes.

In recent years, machine learning algorithms have increasingly been employed to address these challenges, yielding very promising results within the context of water Cherenkov detectors \cite{Prouse:2021fkx, SuperK_ml, Jamieson:2022scv}. In particular, significant progress has been achieved by the WatChMaL.org (Water Cherenkov Machine Learning) collaboration \cite{WatChMaL}, an international effort dedicated to the development and application of machine learning techniques for large-scale water Cherenkov experiments such as IWCD and Hyper-Kamiokande \cite{Atta:2026pam}.
\begin{figure}
    \centering
    \includegraphics[width=0.6\linewidth]{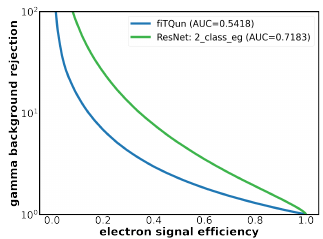}
    \caption{Comparison of the classification performance of a ResNet-18 architecture and the classical fiTQun algorithm for the separation of electron events (signal) and gamma events (background) in a water Cherenkov detector. The gamma background rejection is shown as a function of the electron signal efficiency. The ResNet-18 curve exhibits improved background rejection at fixed signal efficiency compared to fiTQun. Figure adapted from \cite{Mondal:2024agd}.}
    \label{fig:motivation_ml_eg}
\end{figure}

Within this framework, convolutional neural networks such as ResNet-18 have been shown to be capable of learning directly from the spatial and temporal patterns of photomultiplier signals, extracting non-trivial information from either raw data or image-based representations of the events. This approach allows for a more efficient exploitation of the large amount of information encoded in Cherenkov light signals, leading to superior performance both in terms of classification accuracy and computational speed \cite{Atta:2026pam}.

A clear evidence of these advantages emerges from results presented at several international workshops and conferences by members of the Hyper-Kamiokande and WatChMaL collaborations \cite{Mondal:2024agd, Prouse:2023vli, Atta:2026pam}, where deep-learning architectures based on convolutional neural networks have been compared with classical reconstruction algorithms for water Cherenkov detectors. In these studies, ResNet models executed on GPUs achieve inference times of about $1$--$2$ ms per event, corresponding to speed-ups of $3.2\times 10^4$--$5.2\times 10^4$ relative to traditional likelihood-based reconstruction methods such as fiTQun.

In addition to the significant gain in processing speed, these approaches also exhibit competitive or superior performance in classification quality. A relevant example \cite{Mondal:2024agd} is provided by the separation between electron events (signal) and gamma events (background), shown in Fig. \ref{fig:motivation_ml_eg}, where the gamma background rejection is reported as a function of the electron signal efficiency. In this comparison, ResNet-based architectures show improved rejection performance relative to the available fiTQun baseline, although the comparison should be interpreted with some caution, since the likelihood-based reference is not presented as a fully optimized dedicated e--$\gamma$ separator \cite{Atta:2026pam}.

Overall, these results strongly motivate the integration of machine learning techniques, and in particular CNN architectures such as ResNet-18, into the data analysis workflows of Cherenkov detectors, where fast inference and the ability to learn complex patterns represent crucial advantages over classical approaches.
\section{Image representation and classification task}

In this chapter, the term \textit{neutrino images} refers to image-like representations of IWCD events. Building on the physical and experimental context introduced in Chapter~7, we focus here on the specific image representation adopted for the detector response and on the corresponding classification task, namely the discrimination between single-vertex and \textit{pile-up} events.

\subsection{IWCD image representation}

As discussed in the previous chapter, neutrino interactions in the IWCD are observed
through structured patterns of Cherenkov light distributed over the detector surface.
For the purposes of the present analysis, this detector response is represented in a
form suitable for image-based learning methods. The cylindrical detector surface is
mapped onto a two-dimensional grid in $(\phi,z)$ coordinates, and the optical
information recorded by the photosensors is organized into image channels.

In this way, the spatial distribution of the detector response is preserved, while the
charge and timing observables can be treated as separate channels of the input. This
representation provides a natural interface between the detector output and the
convolutional architectures studied in the following sections.

\subsection{Event topologies: single-vertex vs pile-up}
\label{subsec:single_vs_pileup}
In this work, we focus on a binary classification problem aimed at distinguishing between two broad classes of event topologies. The IWCD operates in the high-rate environment of the J-PARC neutrino beam, where neutrinos arrive in narrow bunches and the large detector mass leads to a non-negligible interaction rate. Although water Cherenkov detectors provide very good timing resolution, the finite time required for Cherenkov light to propagate across the detector implies that the optical signals produced by different neutrino interactions may overlap within the same acquisition window. The recorded event may therefore correspond either to a single neutrino interaction or to the superposition of two or more independent interactions. This motivates the distinction between the following two classes:
\begin{itemize}
    \item \textbf{Single-vertex (SV) events}, in which exactly one neutrino interaction occurs within the fiducial volume and within the relevant time window. Although these events originate from a single interaction vertex, the observed topology can nevertheless be non-trivial. A single neutrino interaction may produce multiple charged particles in the final state, such as a primary charged lepton accompanied by hadronic activity, leading to the formation of one or more Cherenkov rings and hence to multi-ring topologies.
    \item \textbf{Pile-up events}, in which two or more neutrino interactions occur in the detector within the same readout window. These interactions are physically independent and are characterised by distinct and uncorrelated interaction vertices. The resulting Cherenkov light patterns are a superposition of contributions from multiple vertices and particle tracks, often producing partially or fully overlapping rings on the cylindrical surface of the detector.

\end{itemize}

\begin{figure*}[htbp]
    \centering
\includegraphics[width=0.6\textwidth]{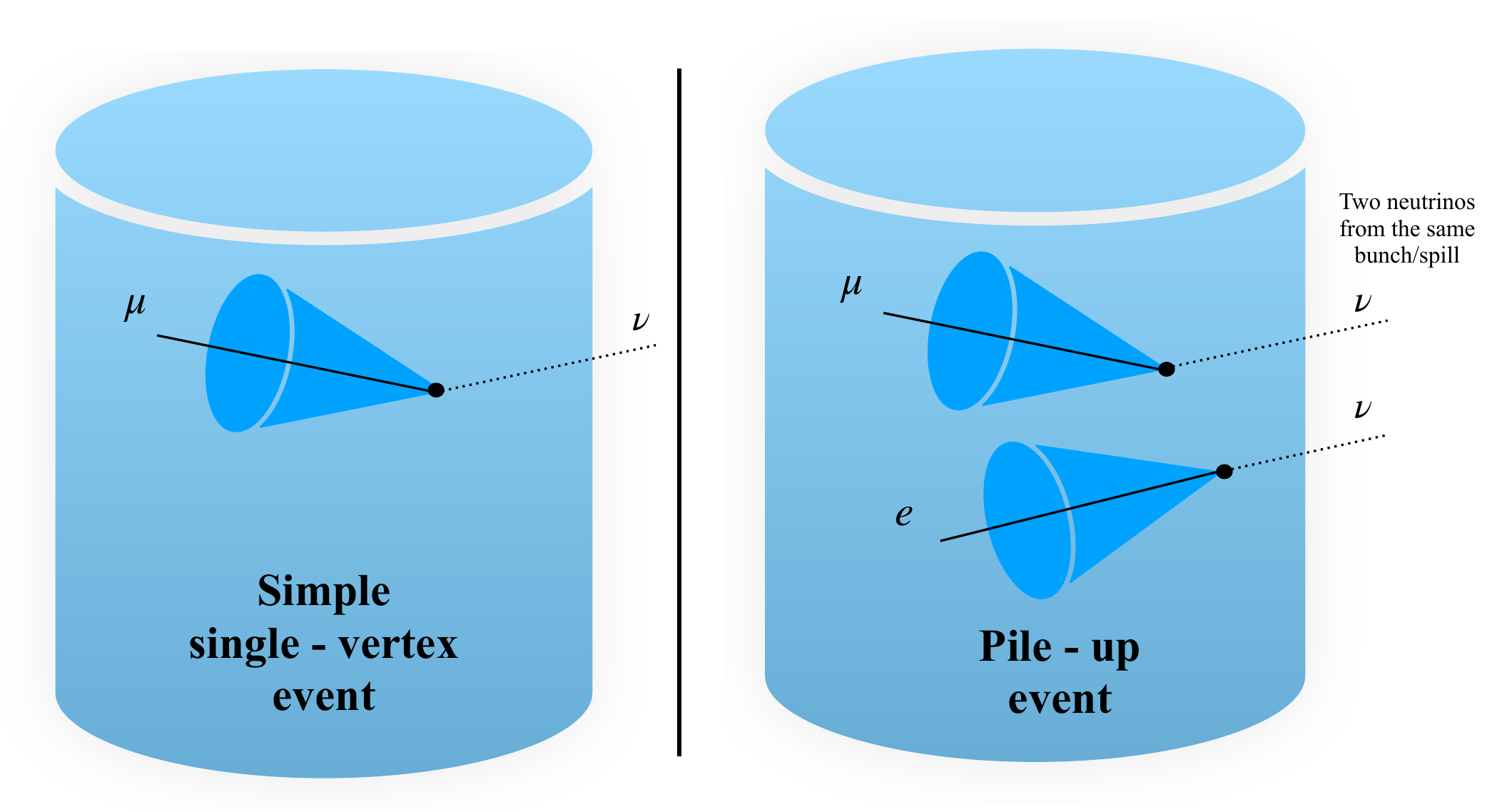}
    \caption{Schematic illustration of idealised event topologies in a water Cherenkov detector. A simple single-vertex event is shown on the left, where a single neutrino interaction produces one charged particle and a single Cherenkov ring. On the right, a pile-up event is depicted, in which two independent neutrino interactions occur within the same readout window, giving rise to two spatially separated interaction vertices and two distinct Cherenkov rings.}
    \label{fig:Pile_up1}
\end{figure*}
\begin{figure*}[htbp]
    \centering
\includegraphics[width=0.9\textwidth]{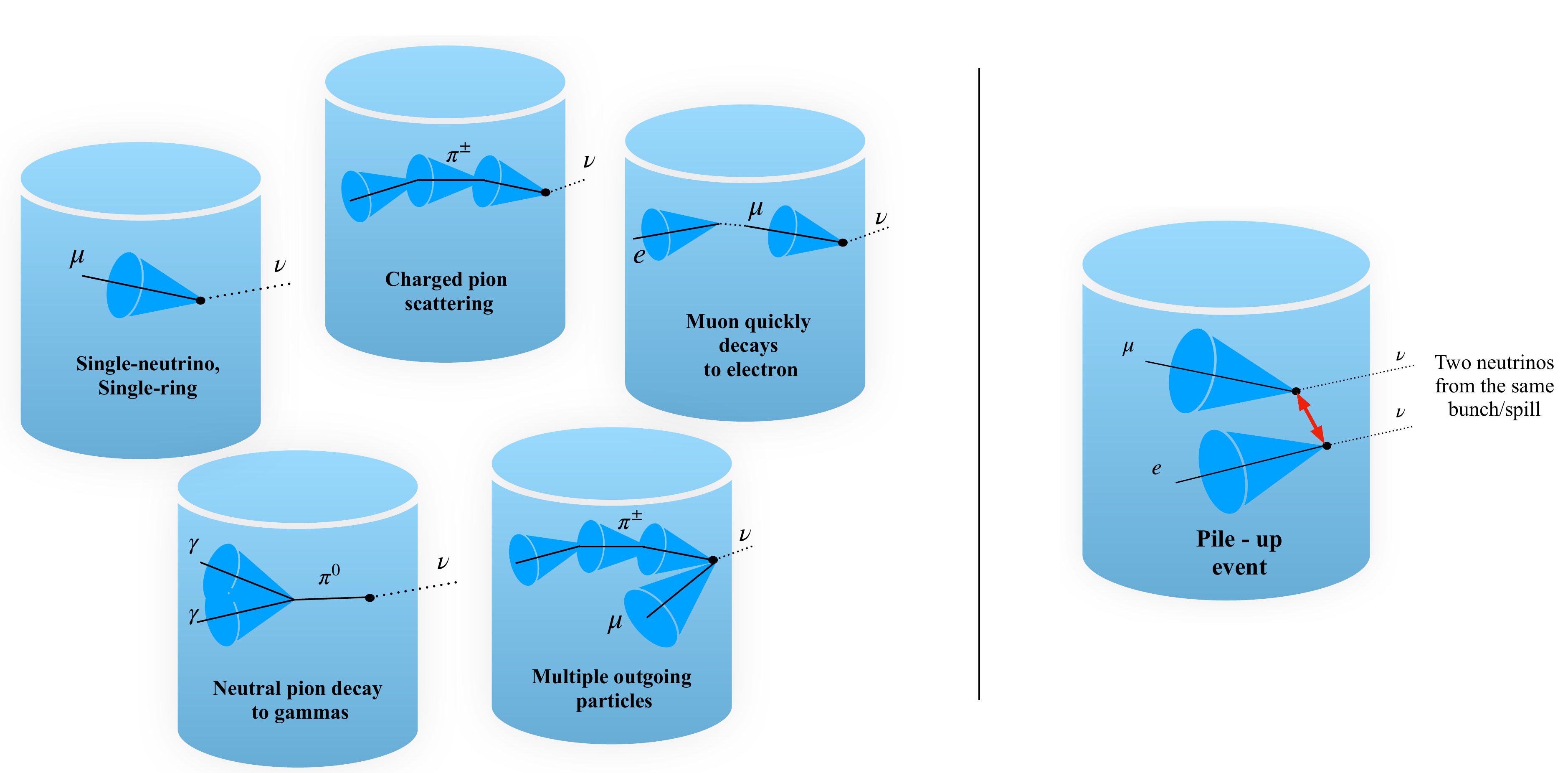}
    \caption{Examples of more realistic event topologies illustrating the challenge of separating single-vertex and pile-up events. Single-vertex interactions (left) can produce complex multi-ring patterns originating from a single interaction vertex, while pile-up events (right) arise from multiple independent interaction vertices. In both cases, several Cherenkov rings may be present, leading to similar topological features despite the different physical origin of the events.}
    \label{fig:Pile_up2}
\end{figure*}
The separation between single-vertex and pile-up events is therefore a non-trivial problem, as the mere presence of multiple Cherenkov rings is not sufficient to uniquely identify pile-up. This is illustrated schematically in Fig. \ref{fig:Pile_up1}, where a simple single-vertex event, characterised by a single interaction vertex producing a single Cherenkov ring, is contrasted with a pile-up event in which two distinct interaction vertices each give rise to a separate ring. In such an idealised scenario, the classification task would be straightforward.

However, the situation becomes significantly more complex in realistic detector conditions. As shown in Fig. \ref{fig:Pile_up2}, single-neutrino interactions can produce a variety of multi-ring topologies. These single-vertex multi-ring events can exhibit light patterns that closely resemble those of genuine pile-up events, making a classification based solely on ring multiplicity ineffective.

From a physical perspective, the key discriminating feature lies in the spatio-temporal structure of the event. In single-vertex interactions, all particle tracks originate from a common interaction point and share a coherent timing structure, whereas in pile-up events the interaction vertices are physically independent, separated in space and time, and show no intrinsic correlation. Effective classification strategies must therefore exploit spatial and timing information, rather than relying exclusively on topological complexity.

In this thesis, machine learning techniques are explored as a means to address the classification of these event topologies.

\section{Simulated data and image dataset}
\label{sec:iwcd_simulated_data}
\subsection{Event generation and detector simulation}
For this analysis, dedicated simulated datasets containing both single-vertex and pile-up event topologies were produced. The simulation chain is based on two main components: event generation and detector response simulation. The latter is performed using the WCSim package \cite{WCSimGitHub}, which models particle propagation in water and the response of the IWCD detector, including the geometry, photosensor configuration, and data acquisition settings.

In all datasets considered, the interaction vertex position, corresponding to the initial position of the generated tracks, is sampled uniformly within the fiducial volume of the detector: the \(x\) and \(z\) coordinates are drawn uniformly within a circle of radius \(R = 400\ \mathrm{cm}\), while the \(y\) coordinate is sampled uniformly in the interval \([-300,\,300]\ \mathrm{cm}\). The vertex time is drawn from a Gaussian distribution with a mean of \(200\ \mathrm{ns}\) and a standard deviation of \(25\ \mathrm{ns}\).

The detector response is simulated using a fixed Data Acquisition (DAQ) configuration, which defines the trigger conditions and the readout window used for all datasets considered in this analysis. The main DAQ parameters are summarised in Table~\ref{tab:daq_parameters}. A relatively high NDigits threshold is adopted, so that only events with a sufficiently large number of digitized PMT hits within a fixed time window are recorded. The detector activity is then saved in predefined pre-trigger and post-trigger windows around the trigger time.\\
\begin{table}[t]
\centering
\caption{Main DAQ parameters used in the WCSim simulation for all datasets considered in this analysis \cite{WCSimGitHub}.}
\label{tab:daq_parameters}
\begin{tabular}{l c}
\hline
\textbf{Parameter} & \textbf{Value} \\
\hline
Digitizer timing precision & 0.1 (time units in WCSim) \\
TriggerSaveFailures mode & 1 \\
Dummy trigger time (TriggerSaveFailures) & 100 ns \\
Pre-trigger window (TriggerSaveFailures) & $-400$ ns \\
Post-trigger window (TriggerSaveFailures) & $+2950$ ns \\
NDigits trigger threshold (count) & $2.5 \times 10^{6}$ \\
NDigits trigger window & 200 ns \\
\hline
\end{tabular}
\end{table}

\noindent
\textbf{Particle-gun dataset (No-Ph)} \\
The \textit{particle-gun} dataset, hereafter referred to as \textbf{no-physics (No-Ph)}, is constructed by generating final-state particle tracks directly within the detector volume, starting from one or more assigned interaction vertices, without explicitly modelling the underlying neutrino--nucleus interaction. This choice is motivated by the aim of reducing bias from a specific physics generator, so that the classifier learns the detector signature of single-vertex versus pile-up events themselves, rather than generator-dependent features. This approach provides full control over the event topology and kinematics and is particularly well suited for studies focused on topological and spatio-temporal features.

For each event, a variable number of tracks is generated according to predefined constraints. The particle species considered are electrons, muons, neutral pions (\(\pi^0\)), and charged pions (\(\pi^-\)). Track directions are sampled isotropically. The total visible energy per event is limited to a maximum value of \(E_{\mathrm{vis}}^{\mathrm{max}} = 2000~\mathrm{MeV}\). For each track, the energy is sampled from a uniform distribution in the interval \([0,\,E_{\mathrm{remaining}}]\), where \(E_{\mathrm{remaining}}\) denotes the visible energy still available after the generation of the previous tracks in the same event.

The dataset includes both single-vertex (SV) and multi-vertex pile-up (MV) event configurations. For single-vertex events, the maximum number of interaction vertices per event is set to one, while for pile-up events up to two vertices are allowed. The maximum number of tracks per event is five. A minimum of two tracks is required for single-vertex events, while pile-up events may contain vertices associated with a single track.

The dataset is constructed with an equal fraction of single-vertex and pile-up events. A total of five million events is used for training, while an independent sample of one million events is reserved for testing. In this dataset, energy distributions are intentionally chosen to be flat, in order to prevent energy from acting as a discriminating variable during the training phase.\\

\noindent
\textbf{Neutrino-Interaction Datasets (GENIE and NEUT)} \\
In addition to the particle-gun dataset, two independent datasets were produced using the \textsc{GENIE} and \textsc{NEUT} neutrino event generators \cite{Genie, Neut}. These datasets explicitly model the physics of neutrino--nucleus interactions and provide a realistic description of the particle content and kinematics of the tracks emerging from neutrino interactions.

The GENIE and NEUT datasets are generated independently using their respective simulation frameworks, and are used exclusively for testing and validation purposes. This choice allows the performance of the classification models, trained on the No-Ph dataset, to be evaluated under realistic physical conditions.

Both datasets include single-vertex and pile-up event configurations. For single-vertex events, the maximum number of interaction vertices per event is set to one, while for pile-up events up to two interaction vertices are allowed. A minimum of two tracks per vertex is required for single-vertex events, whereas pile-up events may include vertices associated with a single track. In all cases, the events are generated with two tracks originating from one or two vertices.

The particle species appearing in these datasets include electrons, positrons, muons, neutral pions (\(\pi^0\)), charged pions (\(\pi^-\) and \(\pi^+\)), protons, and photons. The kinematic properties of the particles are determined by the underlying interaction models implemented in the generators and are not imposed a priori.

For both the GENIE and NEUT datasets, the composition in terms of event topology is fixed to approximately 32\% single-vertex events and 68\% pile-up events. Each dataset consists of one million events, which are used as independent test samples to assess the robustness and generalisation capability of the classification algorithms.

\subsection{Construction of the image dataset}
Building on the image representation introduced in the previous section, each simulated IWCD event is represented as a fixed-size two-dimensional image, where each multi-PMT module corresponds to a single pixel. Given the geometry and instrumentation of the IWCD, the resulting images have a spatial resolution of \(29 \times 40\) pixels.

Each multi-PMT module contains 19 individual PMTs, each of which records both the collected charge and the photon arrival time associated with the detected Cherenkov light. This information is encoded in the image representation through multiple channels. In the most detailed configuration, each pixel is associated with 19 charge channels and 19 timing channels, corresponding to the individual PMTs within a multi-PMT module, for a total of 38 channels per pixel.

Depending on the specific input configuration used in the analysis, different subsets or combinations of these channels are employed. When only charge information is used, the input representation is denoted as \textbf{19Q}, while \textbf{19T} refers to the use of timing information only. The notation \textbf{19Q+19T} indicates that both charge and timing channels are included, resulting in a total of 38 channels.

In addition to the full per-PMT information, more compact representations are also considered. In these cases, the charge or timing information from the 19 PMTs within a multi-PMT module is summarised by their mean value and standard deviation. Such configurations are denoted as \textbf{2Q} or \textbf{2T}, respectively, while \textbf{2Q+2T} indicates the use of both charge- and timing-based summary statistics.

In summary, the information collected for a single IWCD event, involving approximately 500 multi-PMT modules, is projected onto a \(29 \times 40\) pixel image with a variable number of channels depending on the chosen input configuration. This flexible representation allows for a systematic study of the impact of spatial, charge, and timing information on the performance of the machine learning models discussed in the following sections.

The image-construction pipeline was implemented within the open-source WatChMaL framework \cite{WatChMaL}, including the use of timing information in the event representation.

\subsection{Image preprocessing and data augmentation}

The image representation described in the previous section is further processed to account for the cylindrical geometry of the detector and to improve the robustness of the machine learning models during the training phase.\\

\noindent
\textbf{Handling of the cylindrical geometry: double cover} \\
The IWCD detector has an intrinsically cylindrical geometry, and the angular coordinate \(\phi\) is periodic. When the cylindrical surface is unwrapped into a planar image representation, this periodicity introduces artificial boundaries at the left and right edges of the image. Standard convolutional neural networks do not natively account for such periodic boundary conditions, which can lead to edge effects and discontinuities in the learned features.

To mitigate this issue, a \textit{double cover} representation is adopted. In this approach, the image is extended along the angular direction by duplicating the original image, effectively providing a continuous representation across the \(\phi = 0\) and \(\phi = 2\pi\) boundary. Operationally, this is achieved by duplicating and reconfiguring the unwrapped image along the angular direction, effectively implementing circular boundary conditions. This allows convolutional filters to operate consistently along the periodic direction and reduces artefacts associated with the unwrapping of the cylindrical surface.

\begin{figure}
    \centering
    \includegraphics[width=0.8\linewidth]{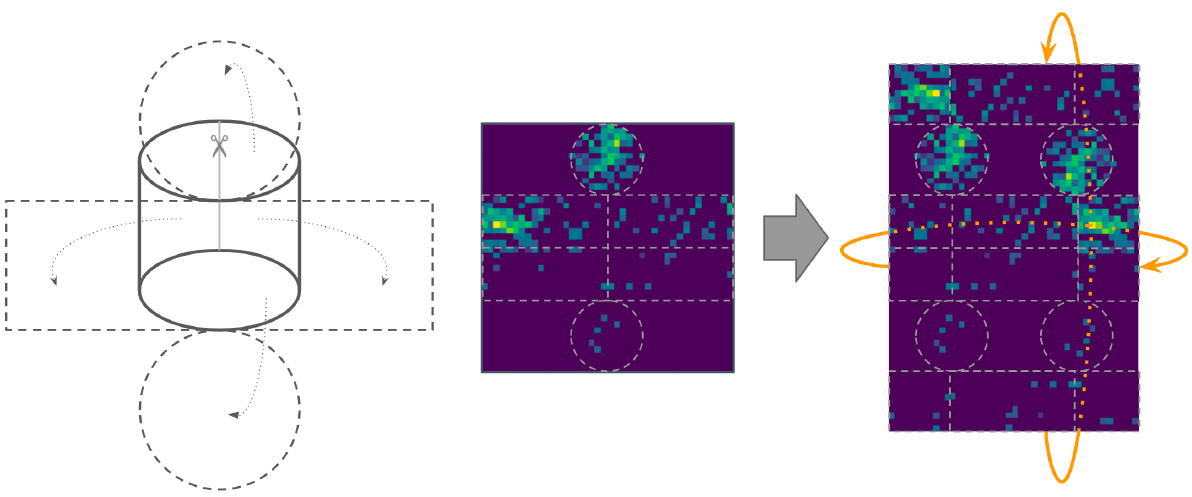}
    \caption{Schematic representation of the double cover construction used to handle the cylindrical geometry of the detector. The unwrapped detector surface is divided into sections, duplicated, and reconfigured to form a double cover, effectively implementing circular boundary conditions along the angular direction. Figure adapted from \cite{Prouse:2023vli}.}
    \label{fig:double_cover}
\end{figure}

A schematic illustration of this procedure is shown in Fig.\ref{fig:double_cover}, where the unwrapped detector surface is duplicated and reconfigured to form a double cover, ensuring continuity across the angular boundaries.\\
\\
\noindent
\textbf{Data augmentation} \\
In addition to the geometric treatment described above, data augmentation techniques are applied during the training phase in order to improve the generalisation capability of the models and to reduce their sensitivity to specific orientations of the event images. The transformations employed preserve the physical content of the events, modifying only their spatial representation.

In particular, the following transformations are considered and applied randomly during the training phase:
\begin{itemize}
  \item \textbf{Horizontal reflection}, corresponding to a reflection along the angular direction;
  \item \textbf{Vertical reflection}, corresponding to a reflection along the detector axis;
  \item \textbf{Front--back reflection}, which exchanges opposite regions of the detector surface.
\end{itemize}

These transformations exploit the approximate symmetries of the detector and of the underlying physics, and do not alter the event topology nor the spatio-temporal correlations relevant for the classification task. The random application of such transformations increases the effective variability of the training dataset without introducing physically unmotivated distortions. Unless otherwise specified, data augmentation transformations are applied exclusively during the training phase, while validation and test samples are left unchanged.

\section{Architecture of ResNet-18 for neutrino images}

This work employs a deep convolutional neural network derived from the Residual Network family, specifically a ResNet-18--based architecture \cite{he2016deep, he2016identity}. The architecture used in our analysis and presented in this chapter preserves the canonical ResNet-18 depth and block configuration, but incorporates modifications to the input and output layers adapted to the classification problem under investigation.

\begin{figure}[t]
\centering
\resizebox{\linewidth}{!}{%
\begin{tikzpicture}[node distance=4mm and 6mm]
\scriptsize

\node[op] (input) {Input};
\node[op, right=of input] (conv1) {Conv $1{\times}1$\\$C_{in}\!\to\!64$\\s=1};
\node[op, right=of conv1] (bn1) {BN};
\node[op, right=of bn1] (relu1) {ReLU};

\node[stage, right=8mm of relu1] (l1) {L1\\$2{\times}$RB\\64\\s=1};
\node[stage, right=of l1] (l2) {L2\\$2{\times}$RB\\128\\s=2};
\node[stage, right=of l2] (l3) {L3\\$2{\times}$RB\\256\\s=2};
\node[stage, right=of l3] (l4) {L4\\$2{\times}$RB\\512\\s=2};

\node[op, right=8mm of l4] (gap) {AdaAvgPool\\$1{\times}1$};
\node[op, right=of gap] (flat) {Flatten};
\node[op, right=of flat] (fc) {FC\\$512\!\to\!C_{out}$};
\node[op, right=of fc] (out) {Output};

\draw[arrow] (input) -- (conv1);
\draw[arrow] (conv1) -- (bn1);
\draw[arrow] (bn1) -- (relu1);
\draw[arrow] (relu1) -- (l1);
\draw[arrow] (l1) -- (l2);
\draw[arrow] (l2) -- (l3);
\draw[arrow] (l3) -- (l4);
\draw[arrow] (l4) -- (gap);
\draw[arrow] (gap) -- (flat);
\draw[arrow] (flat) -- (fc);
\draw[arrow] (fc) -- (out);

\end{tikzpicture}%
}
\caption{Overall ResNet-18--based architecture used in this thesis. The stem is Conv $1\times1$ + BN + ReLU; Residual stages follow a $[2,2,2,2]$ ResidualBlock layout with channel progression $64\!\to\!128\!\to\!256\!\to\!512$.}
\label{fig:resnet18_overview}
\end{figure}
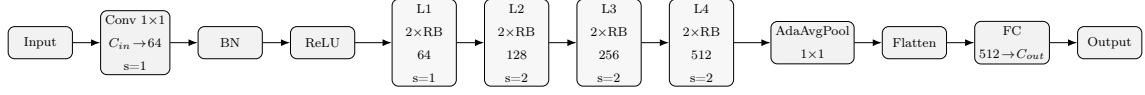

An overview of the complete network architecture is illustrated in Figure~\ref{fig:resnet18_overview}. The model is implemented in PyTorch and consists of four residual stages, each containing two residual blocks, yielding a total depth of 18 learnable layers when convolutional and fully connected layers are counted.

\subsection{Network Stem}

The network stem performs the initial transformation of the input tensor into a feature representation. Unlike the canonical ImageNet ResNet-18 architecture, which uses a $7 \times 7$ convolution with stride $s = 2$ followed by max pooling, the present implementation employs a simplified stem:
\begin{enumerate}
    \item A $1 \times 1$ convolution mapping the input channels to 64 feature maps, with stride $s = 1$, no padding and no bias,
    \item Batch normalization,
    \item ReLU activation.
\end{enumerate}

\begin{figure}[t]
\centering
\resizebox{0.75\linewidth}{!}{%
\begin{tikzpicture}[node distance=4mm and 8mm]
\scriptsize

\node[op] (in) {Input};
\node[op, right=of in] (c) {Conv $1{\times}1$\\$C_{in}\!\to\!64$};
\node[op, right=of c] (bn) {BN};
\node[op, right=of bn] (re) {ReLU};
\node[op, right=of re] (to) {to L1};

\draw[arrow] (in) -- (c);
\draw[arrow] (c) -- (bn);
\draw[arrow] (bn) -- (re);
\draw[arrow] (re) -- (to);

\end{tikzpicture}%
}
\caption{Input stem: Conv $1\times1$ followed by BN  and a ReLU. The output is passed to the first residual stage L1.}
\label{fig:stem}
\end{figure}
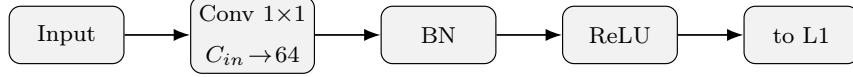

Although a max-pooling layer was used in the original model, it is not applied during the forward pass. Consequently, early-stage spatial resolution is preserved, which is advantageous for tasks requiring fine-grained spatial detail. The stem design is illustrated in Figure~\ref{fig:stem}.

\subsection{Residual Stages}

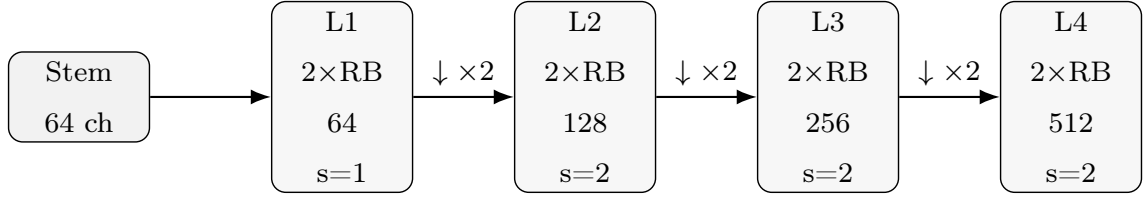
\begin{figure}[t]
\centering
\resizebox{\linewidth}{!}{%
\begin{tikzpicture}[node distance=4mm and 10mm]
\scriptsize

\node[op] (stem) {Stem\\64 ch};
\node[stage, right=12mm of stem] (l1) {L1\\$2{\times}$RB\\64\\s=1};
\node[stage, right=of l1] (l2) {L2\\$2{\times}$RB\\128\\s=2};
\node[stage, right=of l2] (l3) {L3\\$2{\times}$RB\\256\\s=2};
\node[stage, right=of l3] (l4) {L4\\$2{\times}$RB\\512\\s=2};

\draw[arrow] (stem) -- (l1);
\draw[arrow] (l1) -- node[above, font=\scriptsize] {$\downarrow \times2$} (l2);
\draw[arrow] (l2) -- node[above, font=\scriptsize] {$\downarrow \times2$} (l3);
\draw[arrow] (l3) -- node[above, font=\scriptsize] {$\downarrow \times2$} (l4);

\end{tikzpicture}%
}
\caption{Residual stages and downsampling schedule. Downsampling by a factor of two occurs at the first block of layers 2--4 via stride-2 convolutions.}
\label{fig:stages}
\end{figure}

Following the stem, the network is organized into four sequential residual stages, denoted as \texttt{layer1} through \texttt{layer4}. Each stage contains two Residual Blocks (RBs), yielding the characteristic $[2,2,2,2]$ configuration of ResNet-18. A stage-wise depiction of feature map resolution and channel expansion is provided in Figure~\ref{fig:stages}.

\paragraph{Stage Configuration}

\begin{itemize}
    \item \textbf{Layer 1}: 64 output channels, two blocks, stride $s=1$, no spatial downsampling.
    \item \textbf{Layer 2}: 128 output channels, two blocks, stride $s=2$ in the first block.
    \item \textbf{Layer 3}: 256 output channels, two blocks, stride $s=2$ in the first block.
    \item \textbf{Layer 4}: 512 output channels, two blocks, stride $s=2$ in the first block.
\end{itemize}

Spatial downsampling is achieved via strided convolutions in both the main and shortcut paths.

\subsection{Residual Basic Block}

\begin{figure}[t]
\centering
\resizebox{0.95\linewidth}{!}{%
\begin{tikzpicture}[node distance=4mm and 7mm]
\scriptsize

\node[op] (x) {$x$};

\node[op, right=10mm of x] (c1) {Conv $3{\times}3$\\s=$s$};
\node[op, right=of c1] (b1) {BN};
\node[op, right=of b1] (r1) {ReLU};
\node[op, right=of r1] (c2) {Conv $3{\times}3$\\s=1};
\node[op, right=of c2] (b2) {BN};

\node[plus, right=10mm of b2] (add) {$+$};
\node[op, right=of add] (r2) {ReLU};
\node[op, right=of r2] (y) {$y$};

\draw[arrow] (x) -- (c1);
\draw[arrow] (c1) -- (b1);
\draw[arrow] (b1) -- (r1);
\draw[arrow] (r1) -- (c2);
\draw[arrow] (c2) -- (b2);
\draw[arrow] (b2) -- (add);
\draw[arrow] (add) -- (r2);
\draw[arrow] (r2) -- (y);


\node[op, below=10mm of b2] (proj) {Downsampling proj (if needed)\\Conv $1{\times}1$, s=$s$ + BN};
\draw[dashedarrow] (x) |- (proj.west);
\draw[dashedarrow] (proj.east) -| (add.south);

\end{tikzpicture}%
}
\caption{Residual block used in the network: two $3\times3$ convolutions with BN and ReLU, summed with an identity shortcut. When stride or channel dimensionality changes, the shortcut uses a $1\times1$ projection (dashed).}
\label{fig:basicblock}
\end{figure}
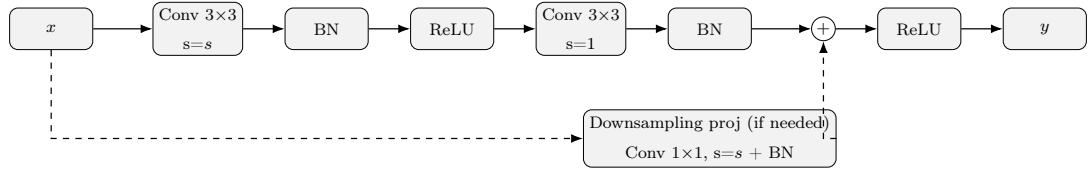

The fundamental computational unit of the network is the Residual Block (RB), shown schematically in Figure~\ref{fig:basicblock}. This block corresponds to the original residual block design used in ResNet-18 and ResNet-34 architectures. \\
Each block consists of:
\begin{enumerate}
    \item A $3 \times 3$ convolution with configurable stride $s$, followed by batch normalization and a ReLU activation ($s=1$ is typically used, except for the layers 2-4 in which case $s=2$).
    \item A second $3 \times 3$ convolution with stride one, followed by batch normalization.
    \item An identity shortcut connection, optionally containing a downsampling projection when $s=2$ in the first 3$\times$3 convolution in order to match the dimensionality.
    \item A final ReLU activation.
\end{enumerate}

The two 3$\times$3 convolutions in the residual block are characterized by a padding of one pixel and no bias term. We remark that this operator preserves spatial resolution for unit stride and performs spatial downsampling when $s=2$.

As stated above, when necessary, the shortcut path applies a downsampling projection to ensure dimensional compatibility. The downsampling is implemented with a strided 1$\times$1 convolution with no padding and no bias, followed by a BatchNormalization. 

\subsection{Output head}

\begin{figure}[t]
\centering
\resizebox{0.85\linewidth}{!}{%
\begin{tikzpicture}[node distance=4mm and 10mm]
\scriptsize

\node[op] (in) {From L4\\512 ch};
\node[op, right=of in] (pool) {AdaAvgPool\\$1{\times}1$};
\node[op, right=of pool] (flat) {Flatten};
\node[op, right=of flat] (fc) {FC\\$512\!\to\!C_{out}$};
\node[op, right=of fc] (out) {Output};

\draw[arrow] (in) -- (pool);
\draw[arrow] (pool) -- (flat);
\draw[arrow] (flat) -- (fc);
\draw[arrow] (fc) -- (out);

\end{tikzpicture}%
}
\caption{Output head: adaptive global average pooling followed by a fully connected layer producing $C_{out}$ outputs.}
\label{fig:head}
\end{figure}
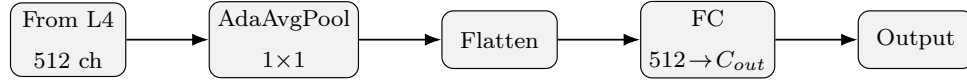

After the last residual stage, the network transitions to the final output head. 
\begin{enumerate}
\item Adaptive average pooling reduces each feature map to a single spatial element, producing 512 representations of dimension $1 \times 1$, independently of the input resolution. 
\item The pooled features are the flattened.
\item A fully connected layer (FC) maps the 512-dimensional vector to the desired number of output channels. 
\end{enumerate}
This process is illustrated in Figure~\ref{fig:head}.

\subsection{Training setup}

The classification models are trained using the Adam optimizer with a learning rate of 0.01. Training is performed for a total of 20 epochs using a batch size of 512 for both training and validation samples. The task is formulated as a two-class classification problem, and the training objective is defined by the cross-entropy loss (\texttt{torch.nn.CrossEntropyLoss}), which in the binary case is equivalent to the binary cross-entropy discussed in the machine learning background chapter.

During training, the input images are processed using the double cover representation to account for the cylindrical geometry of the detector. Random data augmentation transformations, as described in the previous section, are applied to the training samples only. Validation and test datasets are evaluated without data augmentation.

\subsection{Classification performance of ResNet-18}

\noindent
\textbf{Global performance metrics} \\
In this section, we present the performance of the classification models for the discrimination between single-vertex and pile-up events; these results were also presented at the conference reported in Ref.~\cite{DeLorenzis_Stanford}. The evaluation is performed using a set of metrics introduced in Chapter~2. Particular emphasis is placed on the ROC curves, which provide a direct representation of the trade-off between single-vertex efficiency and pile-up rejection.

The classification results obtained on the No-Physics dataset are summarised in Table~\ref{tab:noph_results}. The best overall performance is achieved when both charge and timing information are used in the non-compact configuration (19Q$+$19T), yielding an AUC of 0.987. A modest performance degradation is observed when adopting the compact representation (2Q$+$2T), indicating that a large fraction of the discriminating information is preserved even when the input representation is compressed.

Comparing charge-only and timing-only configurations, timing information clearly provides stronger discrimination power than charge alone, highlighting the importance of spatio-temporal correlations in separating single-vertex from pile-up events.

\begin{table}[htbp]
\centering
\caption{Classification performance on the No-Physics (No-Ph) dataset for different input channel configurations.}
\label{tab:noph_results}
\begin{tabular}{lcccc}
\hline
\textbf{Configuration} & \textbf{Loss} & \textbf{Accuracy} & \textbf{F1 score} & \textbf{AUC} \\
\hline
19Q + 19T (non-compact) & 0.128 & 0.954 & 0.953 & 0.987 \\
2Q + 2T (compact)      & 0.136 & 0.951 & 0.950 & 0.986 \\
19T                    & 0.142 & 0.949 & 0.948 & 0.985 \\
19Q                    & 0.212 & 0.916 & 0.913 & 0.970 \\
\hline
\end{tabular}
\end{table}

The results obtained on the GENIE dataset, reported in Table~\ref{tab:genie_results}, are comparable to those observed for the No-Physics samples, despite the increased physical realism of the events. The non-compact 19Q$+$19T configuration reaches an AUC of 0.990, demonstrating that a classifier trained on simplified topological patterns generalises well to realistic neutrino interaction final states.

As observed for the No-Physics dataset, timing information plays a dominant role, while compact representations remain competitive with the non-compact configuration.

\begin{table}[htbp]
\centering
\caption{Classification performance on the GENIE dataset for different input channel configurations.}
\label{tab:genie_results}
\begin{tabular}{lcccc}
\hline
\textbf{Configuration} & \textbf{Loss} & \textbf{Accuracy} & \textbf{F1 score} & \textbf{AUC} \\
\hline
19Q + 19T (non-compact) & 0.114 & 0.954 & 0.966 & 0.990 \\
2Q + 2T (compact)      & 0.126 & 0.950 & 0.963 & 0.987 \\
19T                    & 0.126 & 0.950 & 0.963 & 0.988 \\
19Q                    & 0.180 & 0.924 & 0.943 & 0.976 \\
\hline
\end{tabular}
\end{table}

\begin{figure}
    \centering
    \includegraphics[width=0.7\linewidth]
    {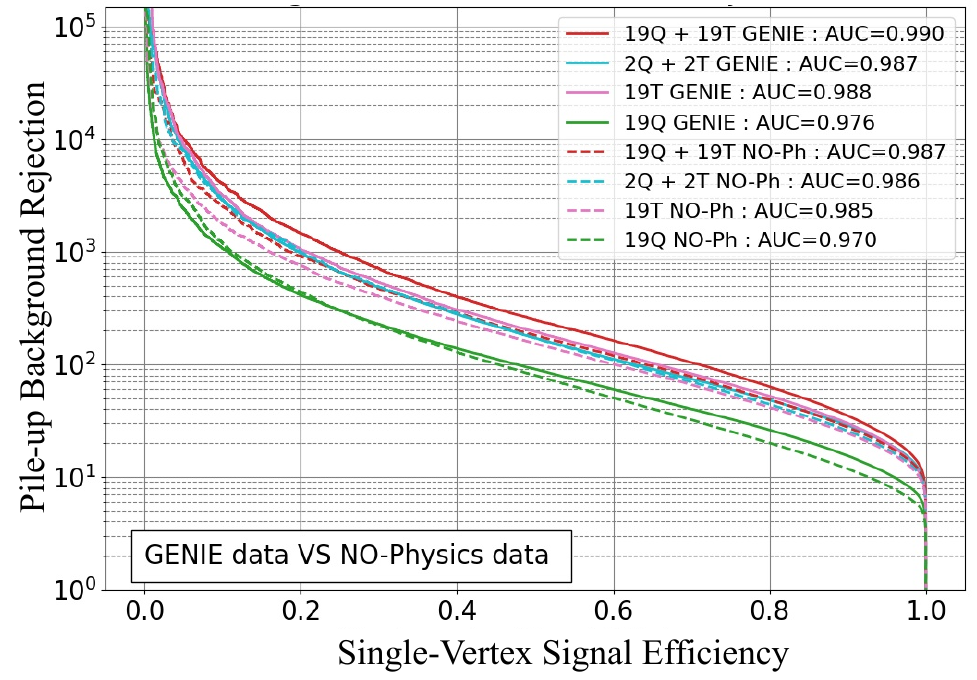}
    \caption{Receiver operating characteristic (ROC) curves for the single-vertex versus pile-up classification using different input channel configurations. The performance is evaluated on the No-Physics and GENIE datasets, illustrating the impact of charge and timing information, as well as the consistency of the results across the two datasets.}
    \label{fig:roc_moreconf}
\end{figure}

Fig. \ref{fig:roc_moreconf} shows the ROC curves obtained for different input channel configurations, evaluated on both No-Physics and GENIE datasets. The figure illustrates the relative impact of charge and timing information on the classification performance, as well as the consistency between the two datasets across a wide range of operating points.

\begin{figure}
    \centering
    \includegraphics[width=0.7\linewidth]{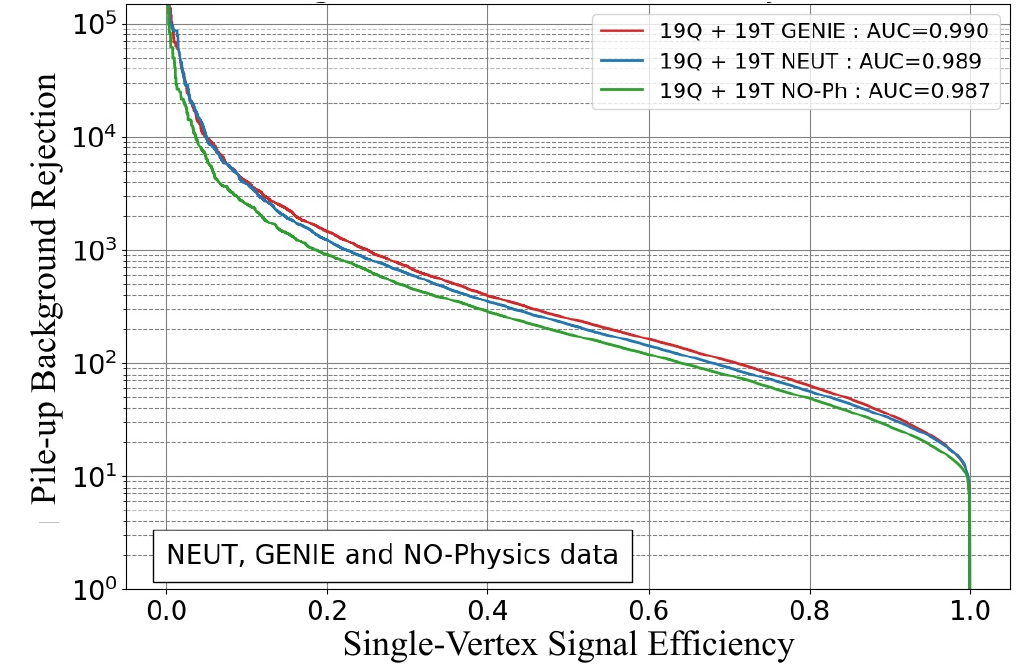}
    \caption{Comparison of ROC curves for the single-vertex versus pile-up classification using the non-compact 19Q$+$19T configuration, evaluated on the No-Physics, GENIE, and NEUT datasets. The close agreement between the curves demonstrates the robustness of the classification performance with respect to the underlying neutrino interaction model.}
    \label{fig:roc}
\end{figure}

A direct comparison between No-Physics, GENIE, and NEUT datasets is shown in Fig. \ref{fig:roc} for the 19Q$+$19T configuration. The close agreement between the ROC curves demonstrates the robustness of the classification performance with respect to the underlying neutrino interaction model.

\begin{figure}[htbp]
    \centering
    \begin{subfigure}[t]{0.48\textwidth}
        \centering
        \includegraphics[width=\textwidth]{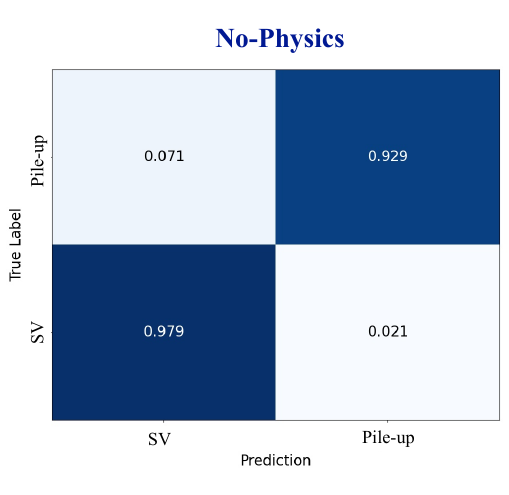}
        \caption{No-Physics dataset}
        \label{fig:confmat_nophysics}
    \end{subfigure}
    \hfill
    \begin{subfigure}[t]{0.48\textwidth}
        \centering
        \raisebox{1mm}{\includegraphics[width=\textwidth]{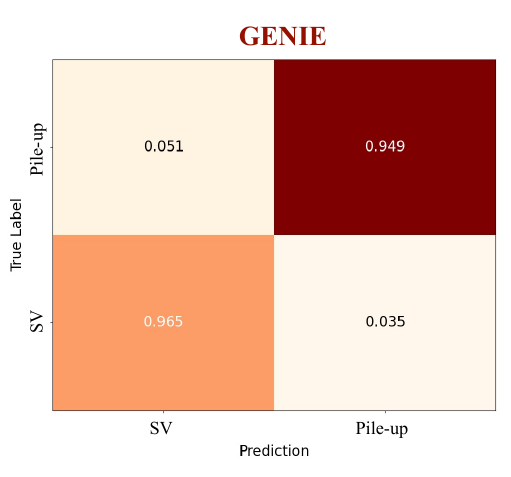}}
        \caption{GENIE dataset}
        \label{fig:confmat_genie}
    \end{subfigure}
    \caption{Confusion matrices for the single-vertex versus pile-up classification using the 19Q$+$19T input configuration, evaluated on the No-Physics (left) and GENIE (right) datasets. The matrices are normalised to the total number of events in each true class.}
    \label{fig:confusion_matrices}
\end{figure}

For completeness, the confusion matrices corresponding to the 19Q$+$19T input configuration are shown in Fig. \ref{fig:confusion_matrices} for the No-Physics and GENIE datasets. These matrices provide a concrete illustration of the classification performance for a fixed decision threshold, complementing the information conveyed by the ROC curves.

In both cases, the majority of single-vertex and pile-up events are correctly classified, with only a small fraction of misclassified events. The observed patterns are consistent with the high classification performance quantified by the ROC curves and the AUC values reported above. The confusion matrices are normalised per true class, such that each row sums to unity.\\


\noindent
\textbf{Dependence of the classification performance on event topology} \\
To further characterise the behaviour of the classifier, the classification accuracy is studied as a function of several event-level quantities, considering a selected subset of events with a maximum of two reconstructed tracks.

Within this selection, single-vertex events consist of a single interaction vertex that can produce up to two tracks, while pile-up (multi-vertex) events are composed of two distinct interaction vertices, each producing a single track, resulting in a total of two tracks per event.
\begin{figure*}[htbp]
    \centering
    \begin{subfigure}[t]{0.48\textwidth}
        \centering
        \includegraphics[width=\textwidth]{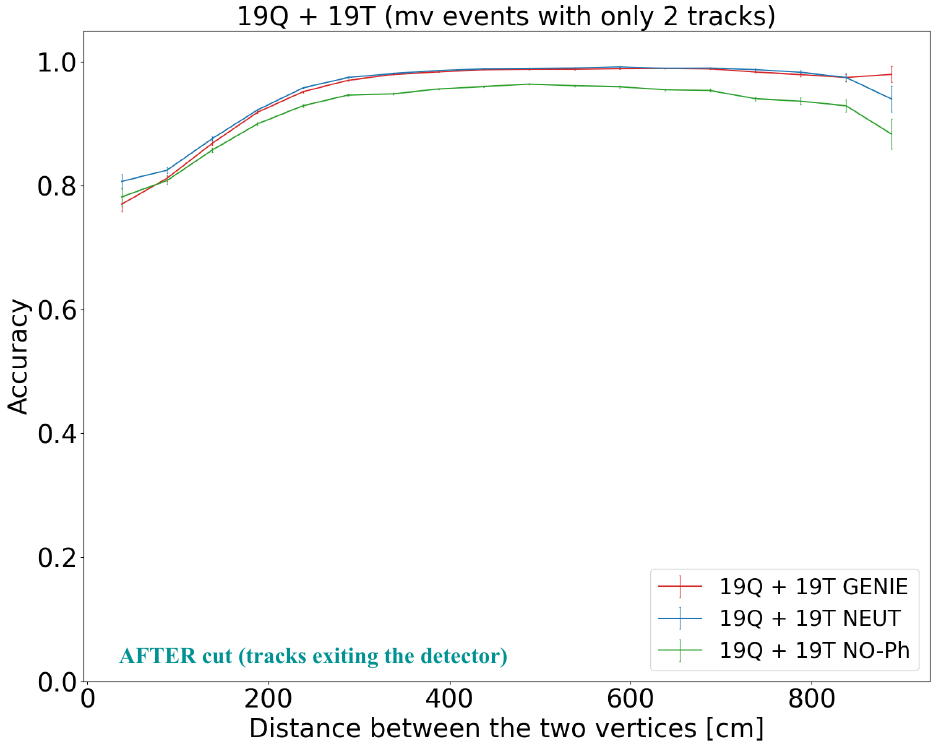}
        \caption{Accuracy as a function of the distance between the two vertices.}
        \label{fig:acc_vtxdist}
    \end{subfigure}
    \hfill
    \begin{subfigure}[t]{0.48\textwidth}
        \centering
        \includegraphics[width=\textwidth]{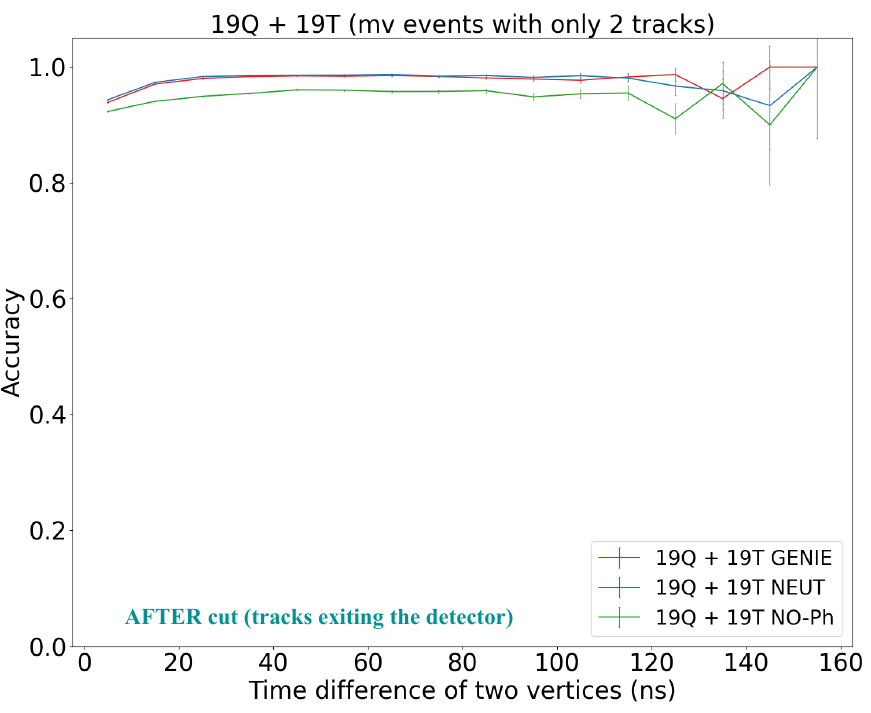}
        \caption{Accuracy as a function of the time difference between the two vertices.}
        \label{fig:acc_deltat}
    \end{subfigure}

    \vspace{2mm}

    \begin{subfigure}[t]{0.48\textwidth}
        \centering
        \includegraphics[width=\textwidth]{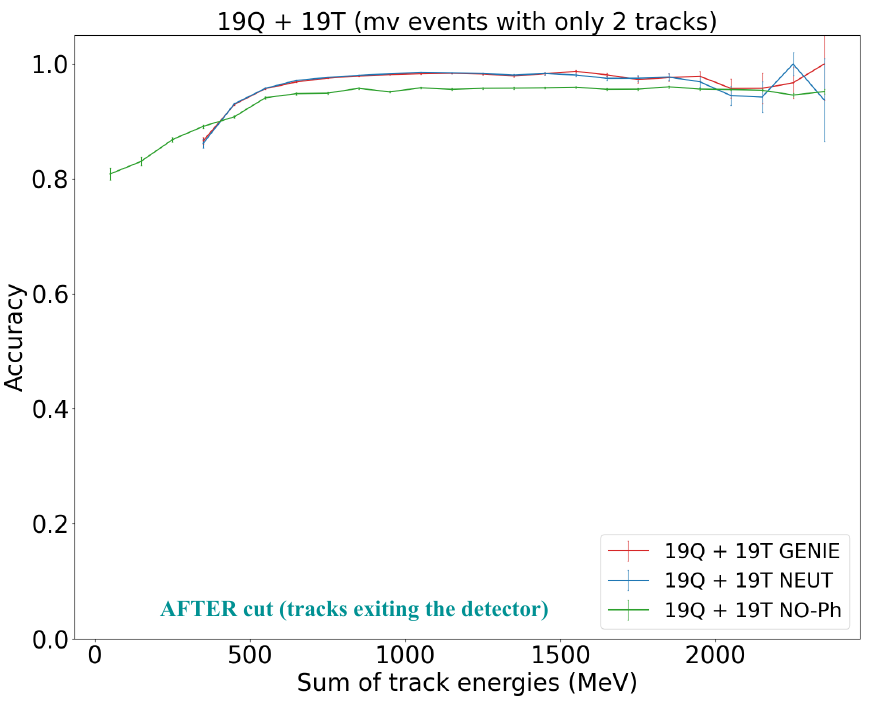}
        \caption{Accuracy as a function of the sum of the track energies.}
        \label{fig:acc_sumE}
    \end{subfigure}
    \hfill
    \begin{subfigure}[t]{0.48\textwidth}
        \centering
        \includegraphics[width=\textwidth]{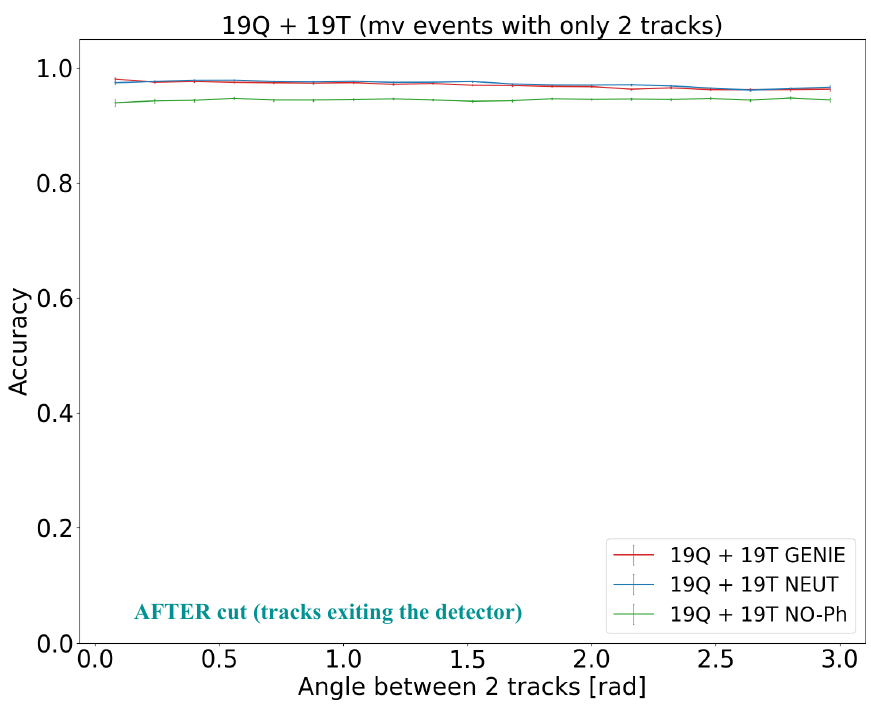}
        \caption{Accuracy as a function of the opening angle between the two tracks.}
        \label{fig:acc_angle}
    \end{subfigure}

    \caption{Classification accuracy for the 19Q$+$19T configuration as a function of event-level quantities, shown for multi-vertex events with exactly two tracks. Results are compared across the No-Physics, GENIE, and NEUT datasets.}
    \label{fig:acc_vs_physicsvars}
\end{figure*}

This choice enables a controlled comparison between single-vertex and multi-vertex events with similar track multiplicity, allowing the impact of the spatial and temporal separation between the interactions, as well as geometric effects related to the detector boundaries, on the classification performance to be isolated.

Fig \ref{fig:acc_vs_physicsvars} shows the classification accuracy as a function of several event-level variables for the selected two-track sample, with particular emphasis on quantities related to the spatial and temporal separation of the interactions and on the kinematics of the tracks. A clear dependence on the distance and time difference between the two vertices is observed: the classification performance improves as the separation between the interactions increases, indicating that well-separated pile-up events are more easily distinguishable from single-vertex topologies. For small separations, both in space and in time, the accuracy decreases, reflecting the increased difficulty in discriminating strongly overlapping interactions.
\begin{figure*}[htbp]
    \centering
    \begin{subfigure}[t]{0.45\textwidth}
        \centering
        \includegraphics[width=\textwidth]{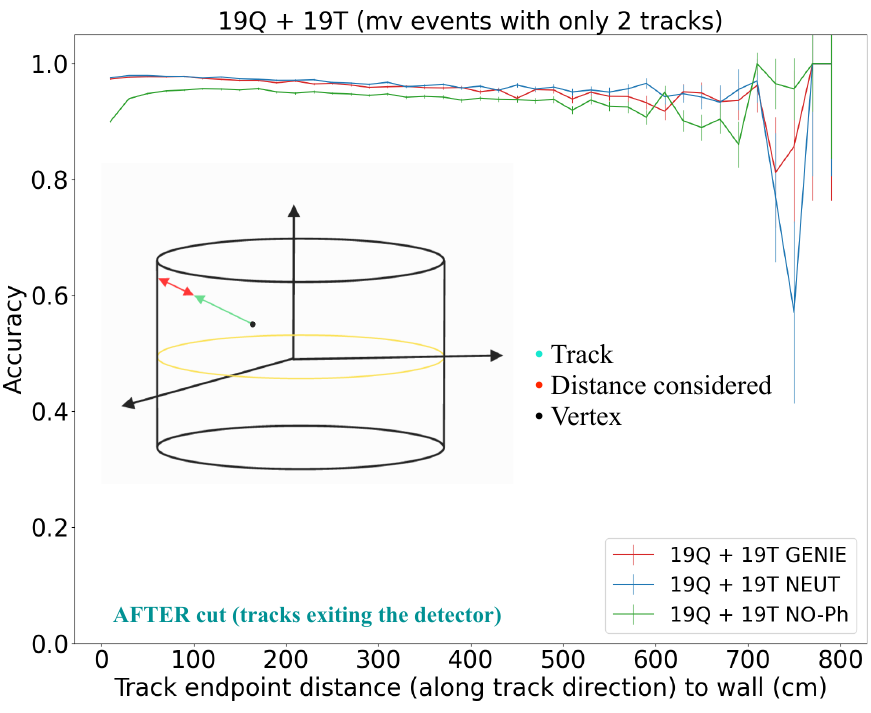}
        \caption{Track endpoint distance to wall (along track direction).}
        \label{fig:acc_endpointwall}
    \end{subfigure}
    \hfill
    \begin{subfigure}[t]{0.45\textwidth}
        \centering
        \includegraphics[width=\textwidth]{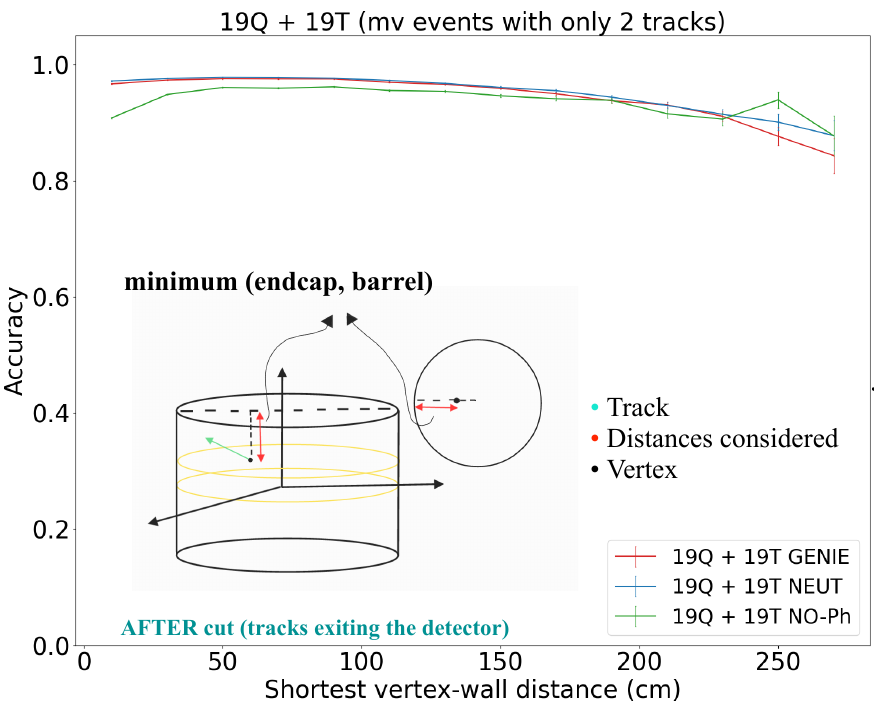}
        \caption{Shortest vertex--wall distance.}
        \label{fig:acc_vtxwall}
    \end{subfigure}
    \hfill
    \begin{subfigure}[t]{0.45\textwidth}
        \centering
        \includegraphics[width=\textwidth]{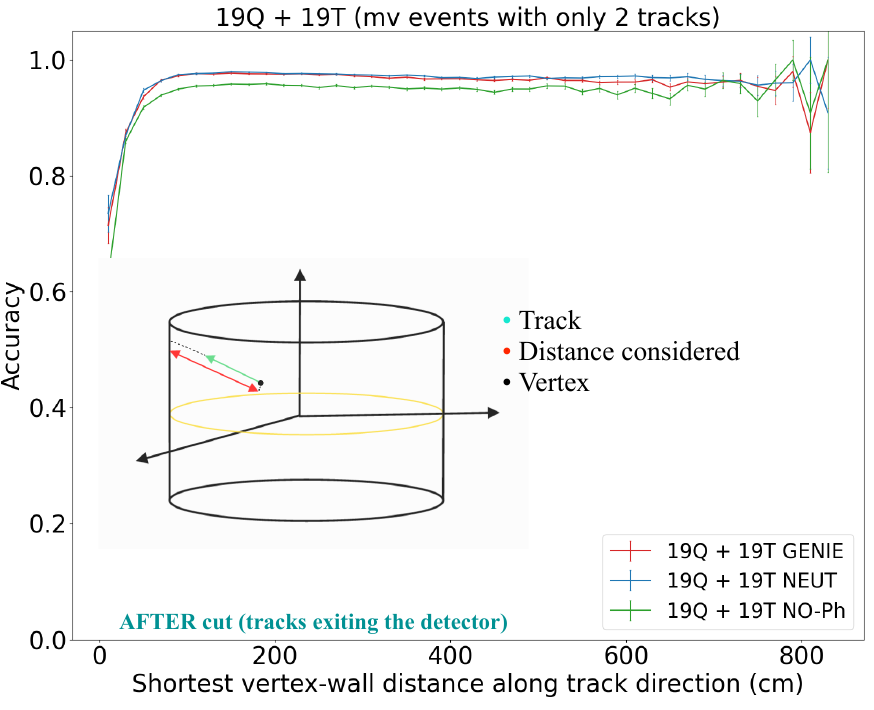}
        \caption{Shortest vertex--wall distance along track direction.}
        \label{fig:acc_vtxwall_dir}
    \end{subfigure}

    \caption{Classification accuracy for the 19Q$+$19T configuration as a function of geometry-related quantities. The dependence on distances to the detector wall highlights boundary effects that can impact the reconstruction of Cherenkov light patterns.}
    \label{fig:acc_vs_geomvars}
\end{figure*}

Fig. \ref{fig:acc_vs_geomvars} instead reports the classification accuracy as a function of geometry-related variables associated with the position of the vertices and tracks with respect to the detector boundaries. In this case, a moderate degradation of the performance is observed for events occurring close to the detector walls, consistent with acceptance effects and with a partial loss of information due to incomplete Cherenkov light patterns.

A qualitatively consistent behaviour is observed across the No-Physics, GENIE, and NEUT datasets in both figures, confirming that the classifier exploits robust topological and spatio-temporal features rather than relying on generator-specific details.

In addition to the one-dimensional studies, a two-dimensional representation of the classification accuracy is considered to investigate the combined dependence on pairs of event-level variables, such as the spatial and temporal separation of the interaction vertices, or the correlation between timing and total track energy.
These distributions provide further insight into regions of reduced performance and help disentangle correlations that are not visible in the one-dimensional projections.
The corresponding two-dimensional accuracy maps for the No-Physics, GENIE and NEUT datasets are shown in Appendix \ref{app:2d_accuracy}.\\

\noindent
\textbf{Summary of classification performance}
The results presented in this section demonstrate that the ResNet-18 architecture provides a robust and effective solution for the discrimination between single-vertex and pile-up events in the IWCD. High classification performance is achieved across all datasets considered, with AUC values close to unity when both charge and timing information are used.

Timing information is found to play a dominant role in the discrimination, consistently outperforming charge-only configurations. At the same time, compact input representations based on summary statistics retain most of the discriminating power of the full per-PMT encoding, highlighting the redundancy present in the detailed channel information.

The classifier trained on simplified No-Physics samples generalises well to more realistic neutrino interaction datasets generated with GENIE and NEUT, indicating that the learned features are driven by fundamental topological and spatio-temporal characteristics rather than by generator-specific details.

Studies of the classification performance as a function of event-level variables further confirm the physical interpretability of the results. The accuracy improves with increasing spatial and temporal separation between interaction vertices, while a moderate degradation is observed for events occurring close to the detector boundaries or for strongly overlapping interactions. Two-dimensional analyses reinforce these trends and reveal consistent behaviour across all datasets.

Taken together, these results demonstrate that convolutional neural networks can effectively exploit the spatial and timing structure of Cherenkov light patterns to address the pile-up identification problem in high-rate water Cherenkov detectors, providing a solid foundation for future extensions to more complex event topologies and experimental conditions.

\subsection{Quantum Extreme Learning Machines for neutrino images}

As a preliminary proof-of-concept study, the Quantum Extreme Learning Machine framework described in Chapters 4--6 was applied to the binary classification problem between single-vertex and pile-up events in the IWCD dataset, namely the same physics task addressed in the previous section by means of ResNet architectures. Unlike ResNet-18, which operates directly on the full detector images and can exploit their complete spatial and multichannel structure, the QELM approach requires a preliminary reduction of the classical input dimensionality before quantum encoding. In the present study, the model was therefore tested not on the full available sample, but on a balanced subset of 60\,000 NEUT-simulated events, whereas in the previous part the NEUT and GENIE datasets used as independent test samples consist of one million events each. For this reason, the comparison with ResNet must be interpreted with caution: the results reported here are purely preliminary and are mainly intended to assess the transferability of the QELM approach to realistic neutrino-image data.

Three different input configurations were considered, all with total dimension equal to 20 features: (i) a representation built using charge information only, (ii) a representation built using timing information only, and (iii) a mixed time+charge representation, obtained by concatenating 10 timing features and 10 charge features. In this first analysis, dimensionality reduction was performed by means of Principal Component Analysis (PCA). The resulting vectors were then encoded into 10-qubit quantum states through Bloch-sphere encoding, evolved under an XX-type Hamiltonian, and finally projected onto the computational basis. The probability vector obtained from measurement was then used as input to a final classifier consisting of a single sigmoid readout. Since the goal of this section is only to provide an initial feasibility check of the method, the evaluation was carried out through a simple train/test split.

The results obtained for the three considered configurations are reported in Table~\ref{tab:qelm_prelim_results}.

\begin{table}[htbp]
\centering
\caption{Preliminary QELM performance for the single-vertex/pile-up classification task on a balanced subset of 60\,000 NEUT events.}
\label{tab:qelm_prelim_results}
\begin{tabular}{lcccccc}
\hline
Configuration & AUC train & ACC train & F1 train & AUC test & ACC test & F1 test \\
\hline
Charge only    & 0.6600 & 0.6159 & 0.6058 & 0.6558 & 0.6110 & 0.6049 \\
Time only      & 0.6560 & 0.6131 & 0.6215 & 0.6482 & 0.6112 & 0.6241 \\
Time + charge  & 0.6659 & 0.6227 & 0.6288 & 0.6575 & 0.6187 & 0.6286 \\
\hline
\end{tabular}
\end{table}

Overall, the results show that the QELM is able to extract a non-trivial amount of discriminating information even from strongly compressed representations of IWCD images. Indeed, all three configurations perform above random guessing, with AUC values around 0.65. In addition, the training and test metrics are very close in all cases, suggesting the absence of strong overfitting in the setup considered here and indicating a good stability of the full preprocessing, encoding, and classification pipeline.

From the comparison among the three input modalities, the mixed time+charge configuration emerges as the one providing the best overall performance, with test values equal to AUC $= 0.6575$, ACC $= 0.6187$, and F1 $= 0.6286$. The improvement with respect to the single-modality cases remains moderate, but systematic, and suggests that charge and timing retain complementary information even after a drastic reduction to only 20 features. At the same time, the timing-only representation yields a slightly higher F1 value than the charge-only case, in qualitative agreement with what was observed in the ResNet study, where timing information appeared to be particularly relevant for the separation between single-vertex and pile-up events.

A more detailed class-wise view is provided by the row-normalized confusion matrices, summarized in Table~\ref{tab:qelm_classwise_recalls}. The charge-only configuration gives the highest recall for class 0, whereas the timing-only configuration achieves the highest recall for class 1. The mixed time+charge representation provides the most balanced behavior across the two classes, improving the class-0 recall with respect to the timing-only case while preserving the same class-1 recall. This is consistent with the slightly better global metrics obtained for the mixed configuration.

\begin{table}[htbp]
\centering
\caption{Class-wise recalls extracted from the row-normalized confusion matrices for the three QELM input configurations.}
\label{tab:qelm_classwise_recalls}
\begin{tabular}{lcc}
\hline
Configuration & Recall class 0 & Recall class 1 \\
\hline
Charge only    & 0.629 & 0.593 \\
Time only      & 0.580 & 0.642 \\
Time + charge  & 0.595 & 0.642 \\
\hline
\end{tabular}
\end{table}

However, the performance achieved by the QELM remains significantly below that of the ResNet-18 reported in the previous section, where the use of full detector images leads to AUC values close to unity and to a much more effective discrimination. This difference is not unexpected. In the present case, the QELM does not receive as input the full $29\times 40$ image with all available channels, but a strongly compressed representation consisting of only 20 classical features, obtained on a subset of 60\,000 events. Such a compression inevitably implies a loss of spatial, topological, and inter-pixel correlation information that ResNet is instead able to exploit naturally through its convolutional layers. For this reason, the results presented here should not be interpreted as a competitive benchmark with respect to the classical deep-learning baseline, but rather as a first feasibility check of the application of QELM to realistic neutrino data.

An important aspect to emphasize is that, in this first analysis, dimensionality reduction was performed exclusively through PCA, namely through a linear compression of the images. It will therefore be interesting in future work to study the behavior of the QELM when feature extraction is performed by means of a convolutional autoencoder, which may be better suited to preserve the local structure of the images and the spatial correlations among pixels, representing an essential part of the physical information carried by Cherenkov events. Similarly, it will be necessary to explore the role of the number of qubits, in order to assess whether a quantum encoding in larger Hilbert spaces may improve the discriminating power of the model.

Finally, beyond the comparison in terms of classification performance alone, it will also be important to carry out a systematic comparison from the computational point of view. Although in the present setup the QELM does not reach the performance of ResNet, the training of the final readout is extremely fast, and this could represent a significant advantage in scenarios where training time or total computational cost play an important role. A more complete assessment should therefore jointly consider classification quality, feature dimensionality, number of qubits employed, and training/inference times.

These preliminary results show that the QELM can be transferred, at least at the proof-of-principle level, from standard benchmark settings to the classification of realistic neutrino images. Although it still remains far from the performance achieved by deep convolutional architectures operating on the full information content, the method yields a non-trivial separation between the two classes and a stable behavior between training and test, thus providing a promising starting point for future developments.
\newpage
\chapter{Conclusions}
In this thesis, we have explored the use of machine-learning techniques
along two complementary directions: the development of quantum-inspired
models for supervised learning, and the application of deep-learning
methods to the analysis of data from neutrino detectors.

In the first part of this work, we investigated Quantum Extreme Learning
Machines as a hybrid quantum--classical framework for classification
tasks. By encoding classical data into quantum states and processing
them through fixed quantum dynamics, QELMs provide a natural way to
generate nonlinear feature representations in high-dimensional spaces.
Our analysis has shown that the performance of these models depends
sensitively on several key design choices, including the data-encoding
strategy, the feature-reduction method, and the structure of the
underlying Hamiltonian. In particular, we highlighted the role of
quantum dynamics in shaping the geometry of the data in feature space,
and explored the connection between entanglement, expressivity, and
classical simulability. These results contribute to a better
understanding of the regimes in which quantum-inspired approaches may
offer advantages over classical methods.

In the second part of the thesis, we focused on the application of deep
learning to neutrino physics, considering the classification of images
produced by water Cherenkov detectors. We developed convolutional
architectures, including residual networks, tailored to the
characteristics of detector data, and evaluated their performance on
realistic simulated datasets. The results show that deep-learning models
are able to effectively capture the relevant features of complex event
topologies, providing accurate and robust classification even in
challenging scenarios. This confirms the potential of modern
machine-learning techniques as powerful tools for data analysis in
next-generation neutrino experiments.

Although developed in different contexts, the two parts of this work
share a common methodological perspective: both quantum-inspired models
and deep neural networks can be interpreted as mechanisms for
transforming input data into representations in which relevant
information becomes more accessible to simple classifiers. From this
viewpoint, the study of representation learning provides a useful
framework for relating and comparing different approaches to the
analysis of complex data.

Despite the promising results obtained, several limitations remain.
Quantum Extreme Learning Machines, while currently constrained by the
size of simulable quantum systems and by the challenges associated with
realistic implementations on NISQ hardware, represent a conceptually
rich framework for studying the role of quantum dynamics in the
construction of data representations. At the same time, the application
of deep learning to neutrino data requires careful treatment of
systematic uncertainties, detector effects, and possible discrepancies
between simulated and real data.

These considerations naturally point to several directions for future
research. On the quantum side, it will be particularly interesting to
explore more scalable architectures, alternative encoding strategies,
and possible implementations on noisy intermediate-scale quantum
devices. On the classical side, further progress may come from improving
the robustness and interpretability of deep-learning models, as well as
their integration with traditional reconstruction and experimental
analysis techniques. More generally, the interaction between classical
and quantum approaches appears to be a particularly fertile direction
for future research.

In conclusion, this thesis shows how machine learning, in both its
classical and quantum forms, can provide powerful tools for addressing
complex problems in fundamental physics. In particular, the work on
QELMs highlights the potential of quantum models as tools for
constructing nontrivial data representations, while the application of
deep learning to neutrino images confirms the effectiveness of these
techniques in realistic experimental settings. Taken together, the
results obtained suggest that the interface between machine learning
and physics is likely to play an increasingly central role in the
development of new tools for the analysis and interpretation of
scientific data.
\newpage


\appendix
\chapter*{Appendix A}
\markboth{Appendix A}{Appendix A}
\addcontentsline{toc}{chapter}{Appendix A}

\setcounter{section}{0}
\renewcommand{\thesection}{A.\arabic{section}}
\setcounter{figure}{0}
\renewcommand{\thefigure}{A.\arabic{figure}}
\section{Lieb--Robinson bounds and application to the XX chain}

In this appendix we briefly review the general form of Lieb--Robinson bounds and explain how they apply to the one-dimensional XX model considered in the main text. The goal is to justify the estimate of the relevant propagation velocity used in the discussion of the QELM transition time.

Consider a quantum spin system defined on a lattice $\Lambda$, with local Hilbert space $\mathcal{H}_i$ associated with each site $i \in \Lambda$, and Hamiltonian
\begin{equation}
    H=\sum_{X \subset \Lambda} \Phi(X),
\end{equation}
where $\Phi(X)$ denotes an interaction term supported on a finite subset $X \subset \Lambda$.

We assume that the interactions are local, in the sense that the operators $\Phi(X)$ are uniformly bounded, $\|\Phi(X)\| \leq J$ for some constant $J>0$, and either have finite range $r_0$ or decay sufficiently fast with distance.

Under these assumptions, Lieb--Robinson bounds imply that, for any two observables $A$ and $B$ supported on disjoint regions $X,Y \subset \Lambda$, one has
\begin{equation}
    \|[A(t),B]\| \leq c\,\|A\|\,\|B\|\, e^{-\mu (d(X,Y)-v|t|)},
\end{equation}
where $c$, $\mu$, and $v$ are positive constants, and $d(X,Y)$ denotes the distance between the supports of the two observables. The quantity $v$ is the Lieb--Robinson velocity, which sets the effective maximal speed for the propagation of correlations and information across the system. In particular, it grows linearly with the interaction scale $J$.

Physically, this result means that the effect of a local perturbation remains exponentially suppressed outside an effective light cone of slope $v$, so that stronger local couplings allow faster spreading of correlations.

\subsection{The XX model}

We now specialize to the one-dimensional XX model with nearest-neighbour couplings,
\begin{equation}
\begin{split}
H &= J \sum_{j=1}^{N} \left(\sigma_j^x \sigma_{j+1}^x+\sigma_j^y \sigma_{j+1}^y\right) \\
  &= J \sum_{j=1}^{N} \left(\sigma_j^+ \sigma_{j+1}^-+\sigma_j^- \sigma_{j+1}^+\right),
\end{split}
\end{equation}
where
\begin{equation}
    \sigma_j^{\pm}=\frac{\sigma_j^x \pm i \sigma_j^y}{2}.
\end{equation}

Through the Jordan--Wigner transformation, this spin model can be mapped onto a system of free fermions,
\begin{equation}
    H=J\sum_{j=1}^{N} \left(c_j^\dagger c_{j+1}+ c_{j+1}^\dagger c_j\right).
\end{equation}

Because the Hamiltonian is quadratic, the Heisenberg evolution of the fermionic operators remains linear:
\begin{equation}
    c_j(t)=\sum_k u_{jk}(t)\, c_k,
\end{equation}
with coefficients
\begin{equation}
    u_{jk}(t)= i^{\,j-k} J_{j-k}(2Jt),
\end{equation}
where $J_n$ denotes the Bessel function of the first kind.

It follows that the anticommutator between fermionic operators at different sites satisfies
\begin{equation}
    \|\{c_j(t), c_k^\dagger\}\|
    = |u_{jk}(t)|
    = |J_{j-k}(2Jt)|
    \leq C\, e^{-\mu (|j-k|-2Jt)},
\end{equation}
which is an explicit realization of the Lieb--Robinson bound. In this case, the corresponding velocity is
\begin{equation}
    v_{\mathrm{LR}}=2J.
\end{equation}

\subsection{Relation to the group velocity}

The same result can be understood from the single-particle dispersion relation. Introducing Fourier modes,
\begin{equation}
    c_j=\frac{1}{\sqrt{N}} \sum_k e^{ikj} a_k,
\end{equation}
with momenta $k=2\pi m/N$, $m \in \mathbb{Z}_N$, the Hamiltonian becomes diagonal:
\begin{equation}
    H=\sum_k \epsilon(k)\, a_k^\dagger a_k,
\end{equation}
with dispersion relation
\begin{equation}
    \epsilon(k)=2J\cos k.
\end{equation}

The associated group velocity is therefore
\begin{equation}
    v_g(k)=\frac{d\epsilon(k)}{dk}=-2J\sin k.
\end{equation}
Its maximal absolute value is
\begin{equation}
    v_{\max}=\max_k |v_g(k)| = 2J,
\end{equation}
which coincides with the Lieb--Robinson velocity:
\begin{equation}
    v_{\max}=v_{\mathrm{LR}}.
\end{equation}

Thus, for the XX chain, the effective light-cone velocity can be identified with the maximal propagation speed of the elementary excitations.

\subsection{Propagation of a localized excitation}

To make this picture more explicit, consider a localized excitation created at site $j$ at time $t=0$,
\begin{equation}
    |\psi(0)\rangle = c_j^\dagger |0\rangle.
\end{equation}
Its time evolution is
\begin{equation}
    |\psi(t)\rangle = c_j^\dagger(t)|0\rangle
    = \sum_k i^{\,k-j} J_{k-j}(2Jt)\, c_k^\dagger |0\rangle.
\end{equation}

\begin{figure}
    \centering
    \includegraphics[width=0.5\linewidth]{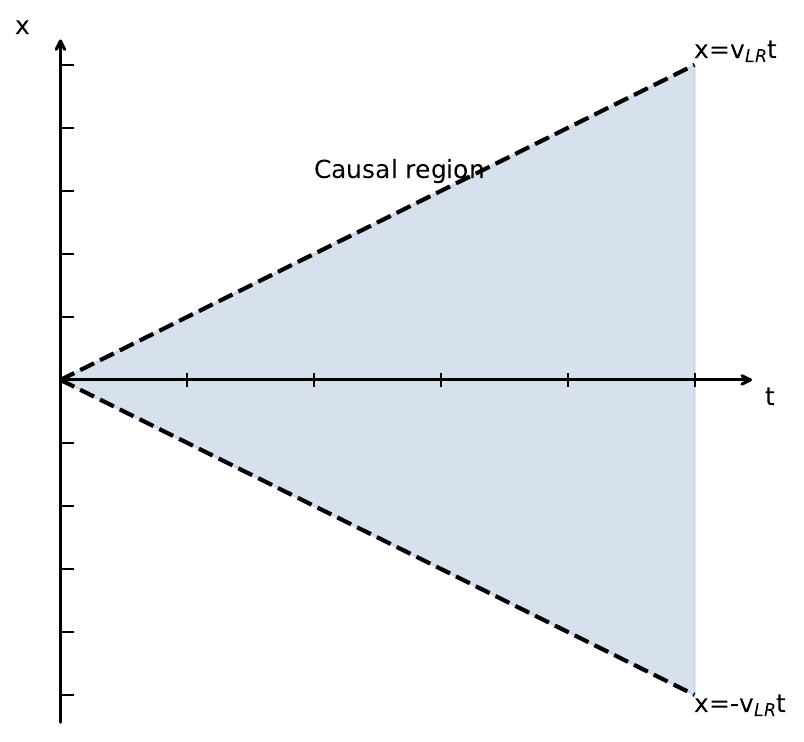}
    \caption{Lieb--Robinson light cone for the XX chain, with velocity $v_{\mathrm{LR}}=2J$. Outside this region, the commutators of local observables are exponentially suppressed.}
    \label{fig:cone}
\end{figure}

The amplitude at site $k$ is therefore controlled by $|J_{k-j}(2Jt)|$. This quantity is peaked around
\begin{equation}
    |k-j| \approx 2Jt,
\end{equation}
and is exponentially suppressed outside that region. The excitation thus propagates ballistically with characteristic velocity $v=2J$, consistently with the Lieb--Robinson light-cone picture (see Fig.~\ref{fig:cone} and Ref.~\cite{Amico}).

\subsection{Connection with the QELM accuracy saturation}

The results presented in Figs.~\ref{fig:MNIST}, \ref{fig:FashionMNIST} and
\ref{fig:CIFAR10} show that the QELM accuracy increases with the total quantum evolution time and then saturates to a plateau. Larger quantum reservoirs generally lead to higher plateau accuracies.

Two observations are relevant here. First, within the range of system sizes explored in this thesis, the saturation time remains on a relatively short timescale, without clear evidence of scaling proportionally with the full system size. Second, this behaviour can be related to the distance over which information is able to propagate in the quantum layer.

In particular, the different system sizes reach their maximal accuracy at approximately
\begin{equation}
    t \simeq 1.
\end{equation}
Using the Lieb--Robinson velocity derived above, this corresponds to an information-spreading distance of order
\begin{equation}
    d \sim v_{\mathrm{LR}}\, t = (2J)t.
\end{equation}
For the choice of units adopted in the main text, $J=1/2$, so that at $t=1$ one finds
\begin{equation}
    d \sim (2 \cdot \tfrac{1}{2}) \cdot 1 = 1.
\end{equation}

This estimate suggests that the onset of the accuracy plateau is associated with information propagating only over a short, local distance. In this sense, the relevant timescale for useful feature generation is controlled by local spreading rather than by full traversal of the entire chain.

\clearpage
\chapter*{Appendix B}
\markboth{Appendix B}{Appendix B}
\addcontentsline{toc}{section}{Appendix B}

\setcounter{section}{0}
\renewcommand{\thesection}{B.\arabic{section}}
\setcounter{figure}{0}
\renewcommand{\thefigure}{B.\arabic{figure}}

\section{Two-dimensional accuracy studies}
\label{app:2d_accuracy}

In this appendix, we present two-dimensional studies of the classification accuracy aimed at further characterising the behaviour of the ResNet-18 classifier. These analyses complement the one-dimensional results discussed in Section~6.5.6 by explicitly highlighting correlations between pairs of event-level variables.

The studies are performed on a restricted subset of events with a maximum of two reconstructed tracks. Within this selection, single-vertex events consist of a single interaction vertex producing up to two tracks, while pile-up events are composed of two distinct interaction vertices, each producing a single track. This choice allows for a controlled comparison between single-vertex and pile-up events with similar track multiplicity, isolating the impact of spatial and temporal separation between interactions.

Unless otherwise stated, the classification decision is obtained using a fixed operating point corresponding to a threshold of 0.5 on the classifier output score. The accuracy is computed in bins of the relevant variables, and identical binning and colour scales are used across the No-Physics, GENIE, and NEUT datasets to facilitate a direct comparison.

\subsection{Vertex separation in space and time}

\begin{figure}[htbp]
    \centering
    \includegraphics[width=1.\textwidth]{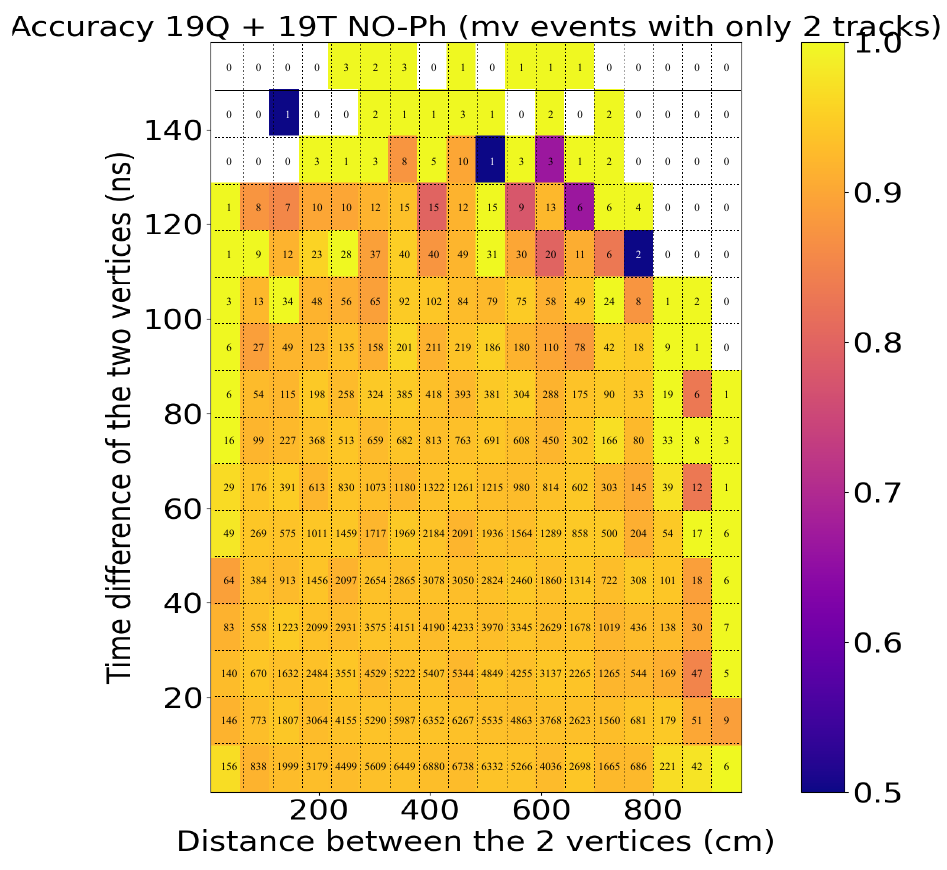}
    \caption{Two-dimensional classification accuracy as a function of the spatial distance and time difference between the two interaction vertices for events with two reconstructed tracks. Results for the No-Physics dataset.}
    \label{fig:2d_time_distance_nophysics}
\end{figure}

\begin{figure}[htbp]
    \centering
    \includegraphics[width=1.\textwidth]{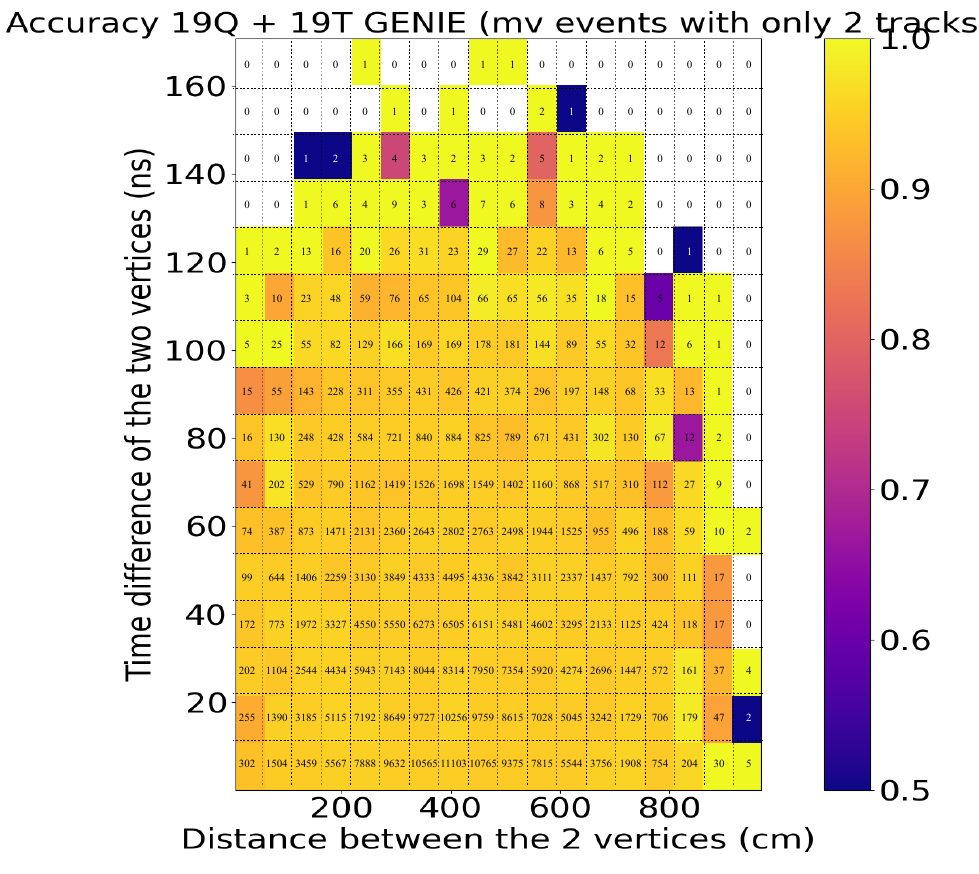}
    \caption{Two-dimensional classification accuracy as a function of the spatial distance and time difference between the two interaction vertices for events with two reconstructed tracks. Results for the GENIE dataset.}
    \label{fig:2d_time_distance_genie}
\end{figure}

\begin{figure}[htbp]
    \centering
    \includegraphics[width=1.\textwidth]{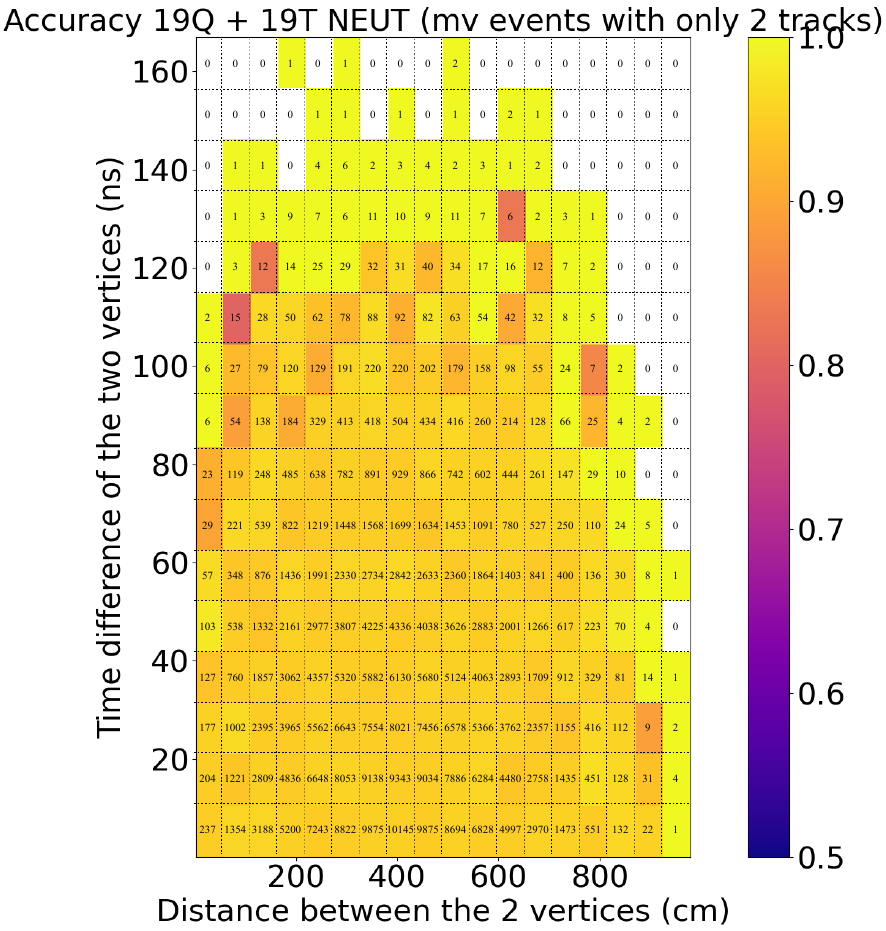}
    \caption{Two-dimensional classification accuracy as a function of the spatial distance and time difference between the two interaction vertices for events with two reconstructed tracks. Results for the NEUT dataset.}
    \label{fig:2d_time_distance_neut}
\end{figure}

Figures \ref{fig:2d_time_distance_nophysics}, \ref{fig:2d_time_distance_genie} and \ref{fig:2d_time_distance_neut} show the classification accuracy as a function of the spatial distance between the two interaction vertices and their time separation. This representation highlights the combined effect of spatial and temporal overlap on the classifier performance.

Across all datasets, the accuracy is observed to improve as the separation between vertices increases, reflecting the increased distinguishability of pile-up events when interactions are well separated in space and time. A mild degradation is visible in regions corresponding to small separations, where the Cherenkov light patterns from the two interactions strongly overlap.


\subsection{Time separation and total visible energy}

\begin{figure}[htbp]
    \centering
    \includegraphics[width=1.\textwidth]{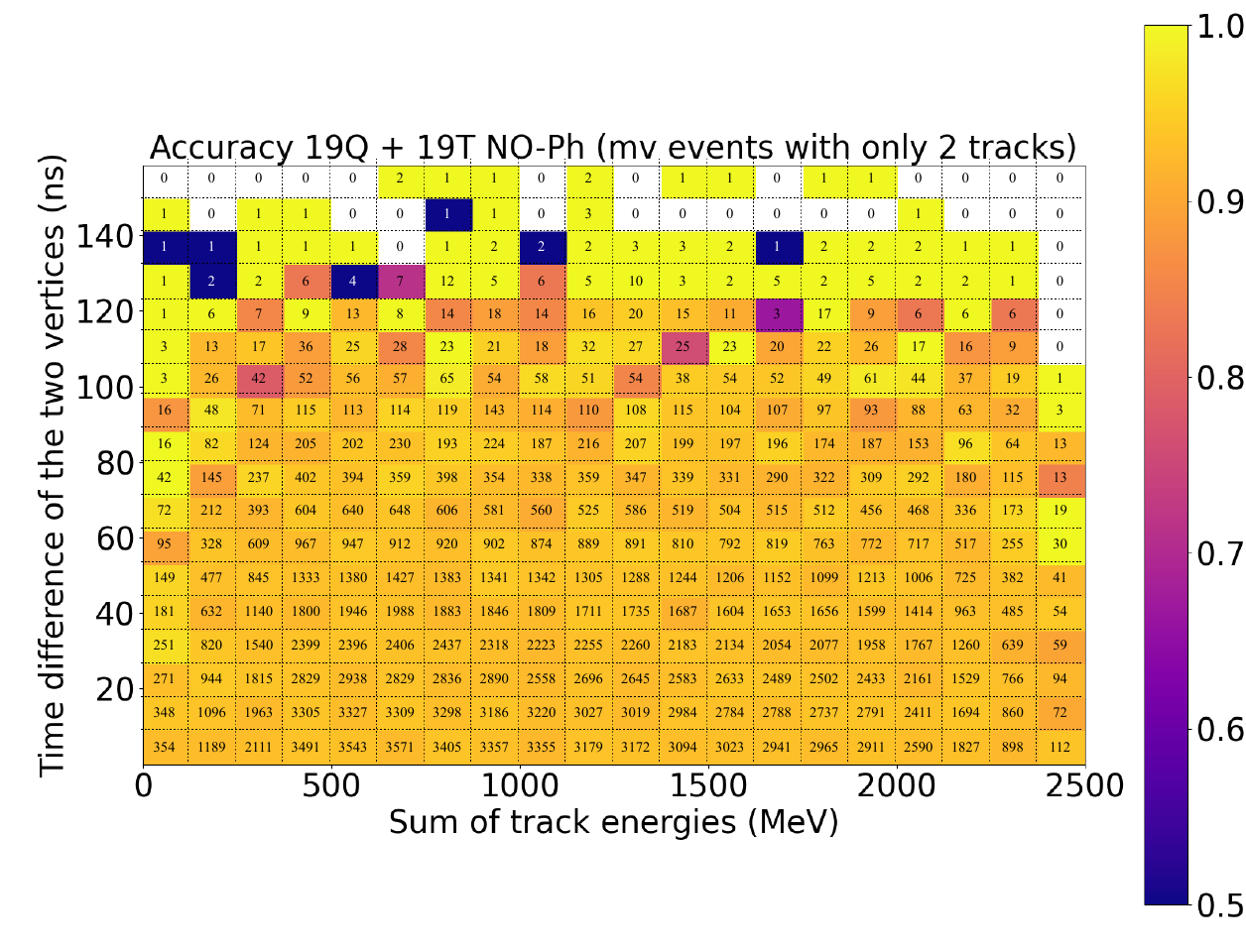}
    \caption{Two-dimensional classification accuracy as a function of the time difference between the two interaction vertices and the sum of reconstructed track energies for events with two reconstructed tracks. Results for the No-Physics dataset.}
    \label{fig:2d_time_energy_nophysics}
\end{figure}

\begin{figure}[htbp]
    \centering
    \includegraphics[width=1.\textwidth]{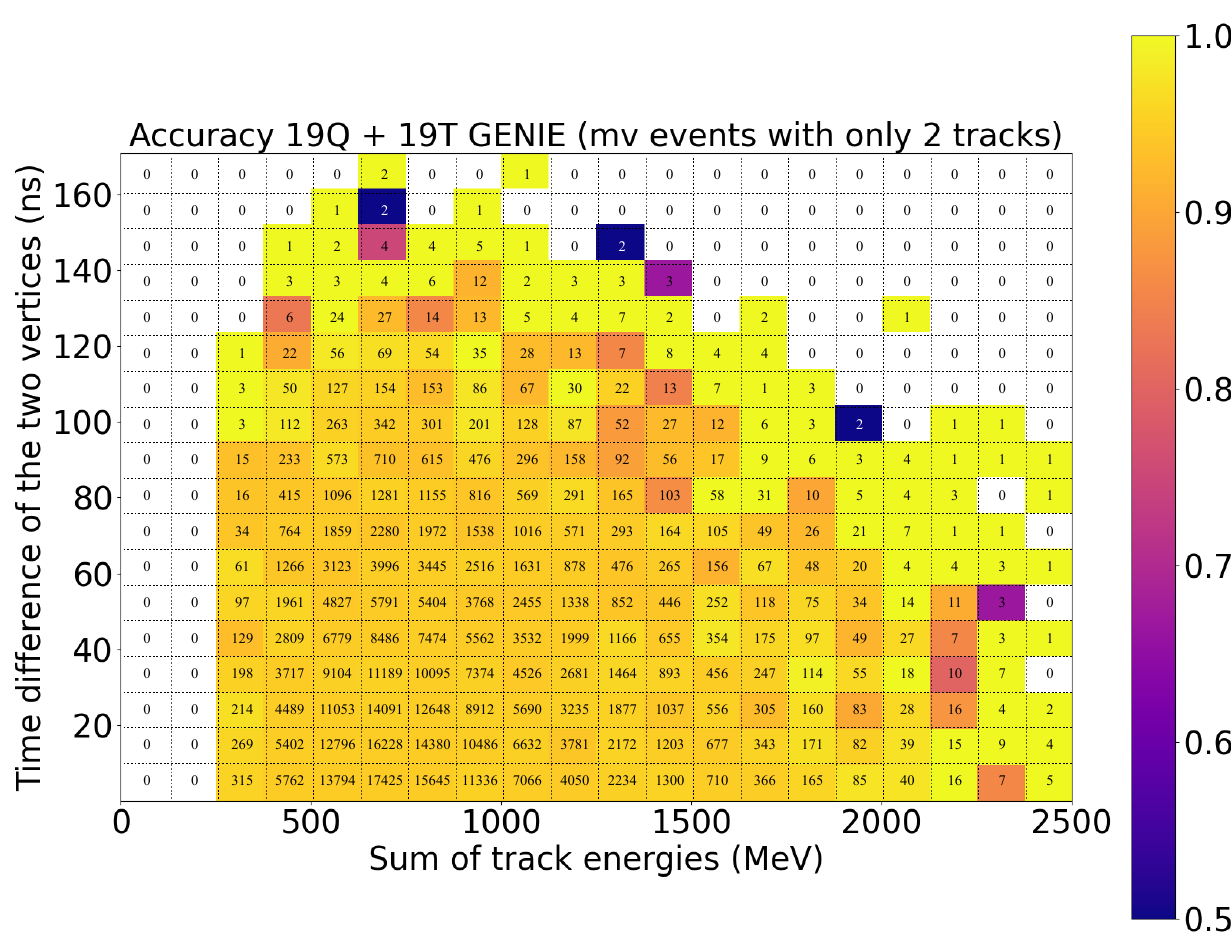}
    \caption{Two-dimensional classification accuracy as a function of the time difference between the two interaction vertices and the sum of reconstructed track energies for events with two reconstructed tracks. Results for the GENIE dataset.}
    \label{fig:2d_time_energy_genie}
\end{figure}

\begin{figure}[htbp]
    \centering
    \includegraphics[width=1.\textwidth]{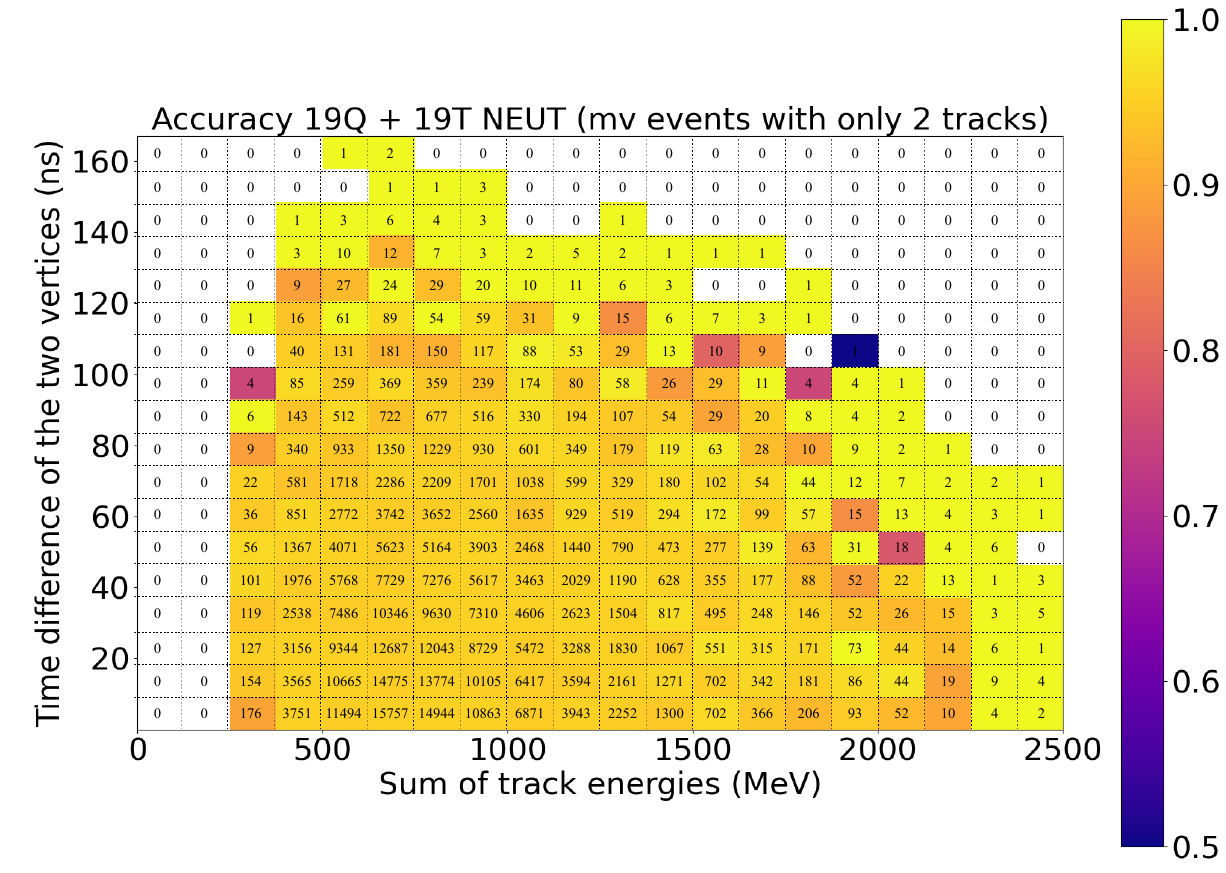}
    \caption{Two-dimensional classification accuracy as a function of the time difference between the two interaction vertices and the sum of reconstructed track energies for events with two reconstructed tracks. Results for the NEUT dataset.}
    \label{fig:2d_time_energy_neut}
\end{figure}

Figures \ref{fig:2d_time_energy_nophysics}, \ref{fig:2d_time_energy_genie} and \ref{fig:2d_time_energy_neut}  present the classification accuracy as a function of the time difference between the two vertices and the sum of the reconstructed track energies. This study probes the interplay between temporal separation and the overall event activity.

The accuracy remains high over a broad range of energies, indicating that the classifier performance is largely driven by spatio-temporal features rather than by the absolute energy scale. A modest decrease in accuracy is observed at small time separations, particularly at high energies, where increased light yield can enhance overlap effects between interactions.

Overall, the two-dimensional studies confirm the trends observed in the one-dimensional projections and demonstrate that the classifier exhibits stable performance over a wide region of the parameter space. The close agreement between the No-Physics, GENIE, and NEUT datasets further indicates that the classification strategy is robust with respect to the underlying modelling of neutrino interactions.
\bibliographystyle{unsrt}
\bibliography{references}

\chapter*{Acknowledgements}
\markboth{Acknowledgements}{Acknowledgements}
I have reached the end of this PhD journey, which has undoubtedly been intense, complex, and at times challenging, but also deeply formative, stimulating, and rewarding. I would therefore like to express my sincere gratitude to all the people who have been by my side throughout these years and who, in different ways, have contributed to this path.

My first and heartfelt thanks go to my supervisors, Pilar Casado, Thorsten Lux, and Arnau Riera, who have accompanied me throughout this journey with constant guidance, availability, and care.

I am particularly grateful to Pilar for her attentive guidance, professionalism, dedication, and rigor, which have always characterized her work and advice, representing a fundamental point of reference for me.

To Thorsten, I am thankful for his patience, support, and numerous suggestions, both from a scientific and a personal perspective, which have helped me face and overcome the challenges encountered along the way.

To Arnau, I am grateful for the valuable discussions, as well as for his calm and balanced approach, which has helped me face this journey with greater serenity and optimism.

I am grateful to Qilimanjaro Quantum Tech and AGAUR (Agència de Gestió d’Ajuts Universitaris i de Recerca) for their support throughout these years. I would like to extend special thanks to Marta Estarellas and Victor Canivell for making this opportunity possible and for their trust in this project, as well as to the entire Qilimanjaro team, in particular those who supported me with the computational aspects and cluster management.

I am also grateful to all the collaborators with whom I worked during my PhD, for their scientific contributions, the continuous exchange, and the many interactions that accompanied the development of this work. In particular, I would like to thank Francesco Plastina, Nicola Lo Gullo, and Jacopo Settino, with whom I shared a valuable collaboration and many stimulating conversations.

I would also like to extend my sincere thanks to all members of the Hyper-Kamiokande, IWCD, WCTE, and WatChMaL collaborations, in particular those I had the opportunity to meet during shifts at CERN. I would like to thank Patrick De Perio, Mark Scott, Mark Hartz, and Lauren Anthony, and especially Nick Prouse for his availability, constant support, and the many valuable suggestions on the WatChMaL code.

A special thought goes to my professors at the University of Salento, Claudio Corianò, Edoardo Gorini, and Margherita Primavera, who supported me from the very beginning of this journey and who, at different stages, contributed to my education through encouragement, trust, and valuable guidance. Their support has been deeply meaningful to me and has represented an important reference throughout my academic path.

To my PhD colleagues, Merlin Varghese and Loris Martinez, I extend my heartfelt appreciation for sharing this experience with me and all the moments that came with it. To my friends, and especially to those who have patiently stood by me with affection and understanding over these years, I offer my sincere gratitude. A very special thought goes to Simonetta Adelfio, who first encouraged me to embark on the path of a PhD. I am also grateful to the people I met at the University of Calabria, whose kind and encouraging words made me feel at home.

My deepest thanks go to my family: to my parents, to my sisters Federica, Francesca, and Roberta, and to my brother-in-law Remo, for always being by my side with love, patience, and support throughout this journey.

Finally, my most special and heartfelt thanks go to my partner, Luigi Delle Rose, for his constant support, patience, love, and for the many evenings spent talking and reflecting together. Without his presence, his optimism, and his unwavering belief in me, I would never have made it to the end of this journey. Thank you for everything.
\addcontentsline{toc}{chapter}{Acknowledgements}
\end{document}